\documentclass[11pt,a4paper]{article}

\usepackage{amsfonts, amsmath, amssymb, amsthm, color, enumitem, float, mathrsfs, mathtools, multirow, nccmath, rotating, tikz, subfig}

\usepackage{lmodern}[lmr]

\usepackage[hidelinks]{hyperref}

\usepackage[margin=0.60in]{geometry}

\usepackage[round]{natbib}
\usepackage[doublespacing]{setspace}

\usepackage{xr}
\def\be{\begin{equation}}
\def\ee{\end{equation}}
\def\bea{\begin{eqnarray}}
\def\eea{\end{eqnarray}}

\author{}
\title{}

\DeclareMathOperator*{\argmin}{\arg\!\min}
\DeclareMathOperator*{\argmax}{\arg\!\max}

\DeclareMathOperator*{\diag}{\normalfont\textrm{diag}}

\newtheorem{corollary}{Corollary}
\newtheorem{definition}{Definition}
\newtheorem{lemma}{Lemma}

\newtheorem{remark}{Remark}
\newtheorem{theorem}{Theorem}
\newtheorem{assumption}{Assumption}

\numberwithin{corollary}{section}
\numberwithin{equation}{section}
\numberwithin{lemma}{section}
\numberwithin{proposition}{section}
\numberwithin{remark}{section}
\numberwithin{theorem}{section}

\allowdisplaybreaks[4]

\begin{document}

\setlength{\abovedisplayskip}{7pt}
\setlength{\belowdisplayskip}{7pt}

\begin{titlepage}

\begin{center}
{\Large \bf Localized Neural Network Modelling of Time Series:\\ A Case Study on US Monetary Policy}

\medskip

{\small {\sc Jiti Gao$^\dag$, Fei Liu$^\sharp$, Bin Peng$^\dag$ and Yanrong Yang$^*$}

\medskip

$^\dag$Monash University, $^\sharp$Nankai University and $^*$The Australian National University}

\bigskip\bigskip

\today

\bigskip

\begin{abstract}
In this paper, we investigate a semiparametric regression model under the context of treatment effects via a localized neural network (LNN) approach. Due to a vast number of parameters involved, we reduce the number of effective parameters by (i) exploring the use of identification restrictions; and (ii) adopting a variable selection method based on the group-LASSO technique. Subsequently, we derive the corresponding estimation theory and propose a dependent wild bootstrap procedure to construct valid inferences accounting for the dependence of data. Finally, we validate our theoretical findings through extensive numerical studies. In an empirical study, we revisit the impacts of a tightening monetary policy action on a variety of economic variables, including short-/long-term interest rate, inflation, unemployment rate, industrial price and equity return via the newly proposed framework using a monthly dataset of the US.

\medskip


\medskip

\end{abstract}
\end{center}

\end{titlepage}

\section{Introduction}\label{S1}

Neural network (NN) architecture has received increasing attention over the last several decades. On relevant topics, a large number of papers have been published in different journals such as \textit{Econometrica}, \textit{The Annals of Statistics}, and \textit{Journal of Machine Learning Research}, etc. by experts from different disciplines. Apparently, we cannot exhaust the literature, but refer interested readers to \cite{BMR2021} and \cite{FMZ2021} for extensive reviews from a methodological point of view.

NN usually includes three ingredients: input layer,  hidden layer(s), and output layer. We now briefly comment on them one by one. The input layer is possibly the easiest one to understand, as it includes regressors only. The hidden layer(s) involve lots of activation functions and parameters mapping linear combinations of the regressors to a certain range of the real line.  Usually, sparsity has to be imposed to ensure a reasonable number of effective parameters and a small set of active activation functions (e.g., \citealp{SH2020, WL2021}). Once the parameters are estimated, one can load the test dataset to evaluate the performance of NN. Finally, the output layer receives the outcome, of which there are two notable types (i.e., quantitative and qualitative). Against this background, a semiparametric regression model under the context of treatment effects (e.g., \citealp{BCH2014}) naturally includes these two types of output in one framework, so it offers a nice structure to start the following semiparametric regression model:
\begin{eqnarray}\label{EQ1.1}
y = z\alpha_0 + g(\mathbf{x})+\varepsilon,
\end{eqnarray} 
where $\mathbf{x} =(x_1,\ldots, x_d)^\top$ is a $d\times 1$ vector of control variables, $z$ is a treatment/policy variable subject to the influence of $\mathbf{x}$ via the structure $z= I(G(\mathbf{x} )-\eta\ge 0 )$ with $I(\cdot)$ being the indicator function, both $\varepsilon$ and $\eta$ are idiosyncratic error components, and $g$ and $G$, defined on $[-a,a]^d\to \mathbb{R}$ with $a$ being fixed\footnote{We consider the fixed $a$ case in the main text, and explain how to allow them to be defined on $\mathbb{R}^d$ in Appendix \ref{App.2}.}, are unknown functions.

Model (\ref{EQ1.1}) belongs to a class of partially linear models studied extensively in the relevant literature, see, \cite{robinson1988}, \cite{hlg2000}, \cite{Gao}, \cite{LR2007}, \cite{ttg2010}, and \cite{BCH2014}, for example. Existing estimation and inferential methods are mainly based on nonparametric kernel and series methods for the case where the dimensionality of $\mathbf{x}$ is small. In the current big--data environment where the dimensionality of $\mathbf{x}$ is large, there are newly proposed methods, including machine learning based methods. 

The main features of model (\ref{EQ1.1}), which has been proposed and discussed in \cite{BCH2014}, are that model (\ref{EQ1.1}) involves a binary structure for $z$, and \cite{BCH2014} develop a series based approach for the estimation of $\alpha$. By contrast, this paper proposes using a localized NN (LNN) method for the estimation of $\alpha$ and $g(\cdot)$ simultaneously. In addition, this paper also develops an easily implementable dependent wild bootstrap method for the inference of both $\alpha$ and $g(\cdot)$.

Until very recently, the investigation on NN architecture mainly focuses on some fixed design regression models (e.g., \citealp{Cybenko1989} and many follow-up studies since then), or uses independent and identically distributed (i.i.d.) data (e.g., \citealp{BK2019,SH2020}; and many references therein). There are only limited studies available for us to understand NN with dependent data from theoretical perspective (see, \citealp{cs1998,chen2007}; for example), although NN based methods have been widely used to study time series data in practice (e.g., \citealp{HOR1996,crs2001,GU2021429,Gu2020}; just to name a few). We would like to contribute along this line of research, and thus assume the following time series data are observable:
\begin{eqnarray}\label{EQ1.2}
\{ (y_t, z_t,\mathbf{x}_t )\, |\, t\in [T]\},
\end{eqnarray}
where, for a positive integer $T$, $[T]$ stands for $\{ 1,\ldots,T\}$. 

Meanwhile, it seems that so far the majority of the literature focuses on prediction errors, and barely talks about how to build feasible inferential procedures, such as constructing confidence intervals. A few exceptions known to us are \cite{DFLSP2021}, \cite{FLM2021}, \cite{CLMZ2022}, and \cite{hhll24} for example on estimation and testing for the average treatment effect rather than on $g(\cdot)$ that we are also interested in this paper. 

Therefore, one important objective and contribution of this paper is that we develop NN based approach to addressing both estimation and inferential issues for both $\alpha_0$ and  $g(\cdot)$ using \eqref{EQ1.2}. We also show how $G(\cdot)$ can be recovered via NN practically in Appendix \ref{App.2}. All things considered, we draw Figure \ref{Fig1} for the purpose of illustration, in which  only the dark area of the hidden layer is activated. A few questions arise naturally:
\begin{enumerate}[leftmargin=*, noitemsep] 
\item Why are there only a small of number of functions getting activated ?

\item How does sparsity come to play? If the least absolute shrinkage and selection operator (i.e., LASSO) is employed, how do we define the set of true parameters ?

\item Provided a set of dependent time series data, can any inference (such as a confidence interval) be established ? and so forth.
\end{enumerate}

{\small

\begin{figure}[htp!]
\centering
\begin{tikzpicture}

\tikzstyle{annot} = [text width=4em, text centered]

\node[circle, minimum size = 6mm, fill=orange!30] (Input-2) at (0,-2) {$x_1$};
\node[circle, minimum size = 6mm, fill=orange!30] (Input-4) at (0,-4) {$\vdots$};
\node[circle, minimum size = 6mm, fill=orange!30] (Input-6) at (0,-6) {$x_d$};
 
\node[rectangle, minimum size = 6mm, fill=teal!10] (Hidden-1) at (5,-1.5) {$\begin{array}{c}\sigma: \mathbf{x}\to [0,1] \\ \vdots \\ \sigma: \mathbf{x}\to [0,1] \end{array}$};

\node[rectangle, minimum size = 6mm, fill=teal!50] (Hidden-4) at (5,-4) {$\begin{array}{c}\sigma: \mathbf{x}\to [0,1]\\ \vdots \\ \sigma: \mathbf{x}\to [0,1] \end{array}$};

\node[rectangle, minimum size = 6mm, fill=teal!10] (Hidden-8) at (5,-6.5) {$\begin{array}{c} \sigma: \mathbf{x}\to [0,1] \\ \vdots \\ \sigma: \mathbf{x}\to [0,1] \end{array}$};
 

\node[rectangle,  minimum size = 5mm, fill=purple!50] (Output-4) at (10,-4) {Outcome};

\draw[->, shorten >=1pt] (Input-2) -- (Hidden-1);  
\draw[->, shorten >=1pt] (Input-4) -- (Hidden-1);  
\draw[->, shorten >=1pt] (Input-6) -- (Hidden-1);  

\draw[->, shorten >=1pt] (Input-2) -- (Hidden-4);  
\draw[->, shorten >=1pt] (Input-4) -- (Hidden-4);  
\draw[->, shorten >=1pt] (Input-6) -- (Hidden-4);   

\draw[->, shorten >=1pt] (Input-2) -- (Hidden-8);  
\draw[->, shorten >=1pt] (Input-4) -- (Hidden-8);  
\draw[->, shorten >=1pt] (Input-6) -- (Hidden-8);  

\draw[->, shorten >=1pt] (Hidden-1) -- (Output-4);  
\draw[->, shorten >=1pt] (Hidden-4) -- (Output-4);  
\draw[->, shorten >=1pt] (Hidden-8) -- (Output-4);

\node[annot,above of=Hidden-1, node distance=2.30cm] (hl) {Hidden layer};
\node[annot,above of=Input-2, node distance=2.30cm] {Input layer};
\node[annot,above of=Output-4, node distance=4.50cm] {Output layer};
\end{tikzpicture}
\caption{One Layer Neural Network}\label{Fig1}
\end{figure}

}

Another challenge which arises with the complexity of NN architecture is the transparency of algorithms (see Appendix \ref{App.1} for a brief survey of the existing software packages). One main reason is the lack of practical guidelines for establishing a feasible version. Our literature review highlights that social science studies using NN approach rarely provide detailed descriptions of their numerical implementation. While we concur with \cite{Athey2019} that machine learning will have a transformative impact on social science, transparent algorithms are crucial to ensure the practical relevance and utility of the findings derived from these approaches.

Having those said, our contributions are as follows. 

(i) We establish an approximation procedure that approximates polynomials via NN in a local sense rather than a global sense. 

(ii) We then explore the use of identification restrictions and establish the LNN based approach under a set of mild conditions. 

(iii) We show that some closed--form expressions can be obtained for the estimators of the parameters of interest. 

(iv) Accordingly, asymptotic distributions are derived, and a dependent wide bootstrap procedure is proposed for inferential purposes.   

(v) As shown in Theorems 2.1 and 2.2 and their discussions in Section 2 below, the LNN based estimation and inferential methods outperform such results associated with existing methods.

(vi) We validate our theoretical findings through extensive numerical studies. 

(vii) In an empirical study, we revisit the impacts of a tightening monetary policy action on a variety of economic variables, including short-/long-term interest rate, inflation, unemployment rate, industrial price and equity return via the newly proposed framework using a monthly dataset of the US.

The rest of the paper is organized as follows. In the main text, 

(a) Section \ref{S2} introduces LNN architecture, proposes a group--LASSO based estimation procedure, and then establishes the corresponding asymptotic properties to infer $\alpha_0$ and $g(\cdot)$, respectively;
 
(b) We provide extensive simulation studies in Section \ref{S3} to examine the finite-sample performance; 

(c) Section \ref{S4} presents an empirical study that investigates the average effects of the US monetary policy change on macroeconomic and financial variables; 

(d) Section \ref{Sec6} concludes.
\medskip

In the online supplementary appendices,

(e) Appendix \ref{App.1} includes some discussions on issues associated with practical implementation and also presents a detailed algorithm;

(f) Appendix \ref{App.2} discusses the estimation of a fully nonparametric model which is a special case of \eqref{EQ1.1}, explains how to relax the restriction about $a$, and infers $G(\cdot)$ of $z_t$ via LNN;

(g) Appendix \ref{App.3} includes additional simulations; 

(h) We finally give the proofs in Appendix \ref{App.4}.  
\medskip
 
To close this section, we introduce some notation and mathematical symbols. Vectors and matrices are always expressed in bold font. Further, $\| \cdot \|$ denotes the Euclidean norm of a vector or the Frobenius norm of a matrix; $ \mathbf{0}_{a}$ and $ \mathbf{1}_{a}$ are respectively $a\times 1$ vectors of zeros and ones for $a\in \mathbb{N}$ and $\mathbf{I}_a$ denotes an $a\times a$ identity matrix; for a vector of nonnegative integers $\pmb{\mu} = (\mu_1,\ldots, \mu_d)^\top \in \mathbb{N}_0^{d}$ in which $\mathbb{N}_0=0\cup  \mathbb{N}$, let $\pmb{\mu}! =\mu_1!\cdots \mu_d!$; $\mathtt{c}$, $\mathtt{C}$ and $O(1)$ always stand for fixed constants, and may be different at each appearance; $\to_P$ and $\to_D$ stand for convergence in probability and convergence in distribution, respectively. 

For a function $m\,:\, [-a,a]^d\mapsto \mathbb{R}$, let $\|m \|_{\infty} = \sup_{\mathbf{x}\in [-a,a]^d}|m(\mathbf{x})|$. If the partial derivative of $m(\mathbf{x})$ exists, we write $\frac{\partial^{|\pmb{\mu}|} m(\mathbf{x})}{\partial \mathbf{x}^{\pmb{\mu}}} =\frac{\partial^{|\pmb{\mu}|} m(\mathbf{x})}{\partial x_1^{\mu_1}\cdots \partial x_d^{\mu_d}}$ for short, where $|\pmb{\mu}| =\sum_{j=1}^d \mu_j$. Additionally, let  
\begin{eqnarray*}
[(1, \mathbf{x}^\top)\mathbf{a} ]^q =\sum_{|\mathbf{r}|=q}\binom{q}{\mathbf{r}} \cdot a_0^{r_0}\prod_{k=1}^d a_k^{r_k}   x_k^{r_k}
\end{eqnarray*}
for $\mathbf{a}= (a_0, a_1,\ldots, a_d)^\top \in \mathbb{R}^{d+1}$, $\mathbf{r} = (r_0,r_1,\ldots, r_d)^\top\in \mathbb{N}_0^{d+1}$, and $ \mathbf{x}\in \mathbb{R}^{d}$. Denote that for given $q\geq 1$

\begin{eqnarray}\label{EQ1.3}
\mathscr{P}_q= \left\{ \text{Linear span of the monomials } \prod_{k=1}^d x_k^{n_k} \text{ with } 0\le |\mathbf{n}|\le q\right\},
\end{eqnarray}
where $\mathbf{n}=(n_1,\ldots, n_d)^\top\in \mathbb{N}_0^d$, and the dimension of $\mathscr{P}_q$ is apparently $\text{dim}\mathscr{P}_q  =\binom{d+q}{d}\coloneqq d_q$. Accordingly, let

\begin{eqnarray}\label{EQ1.4}
\mathbf{m}(\mathbf{x}\, |\, \mathbf{x}_0) = (m_1(\mathbf{x}\, |\, \mathbf{x}_0),\ldots, m_{d_q}(\mathbf{x}\, |\, \mathbf{x}_0))^\top,
\end{eqnarray} 
where $m_j(\mathbf{x} \, |\, \mathbf{x}_0)$'s are the basis monomials (centered at $\mathbf{x}_0$) of $ \mathscr{P}_q$.  Denote a set 

\begin{eqnarray} \label{EQ1.5}
C_{\mathbf{x}_0,h} =\{\mathbf{x} \, | \, |x_j-x_{0,j}|\le h \text{ for } j\in [d] \},
\end{eqnarray}
where $x_j$ and $x_{0,j}$ are the $j^{th}$ elements of $\mathbf{x}$ and $\mathbf{x}_0$ respectively, and $h$ is a bandwidth.  Finally, let $ \mathbf{H}=\diag\{H_1,\ldots, H_{d_q}\}$ with $H_j =\prod_{k=1}^dh^{-n_{j,k}}$, and let $\mathbf{n}_j =(n_{j,1},\ldots, n_{j,d})^\top$ include the corresponding power terms of  $m_j(\mathbf{x} \, |\, \mathbf{x}_0)$  defined in \eqref{EQ1.4}.

\section{Methodology and Asymptotic Theory}\label{S2}

In this section, we first introduce LNN  in Section \ref{S2.1} and then infer $\alpha_0$ and $g(\cdot)$ in Section \ref{S2.3}. Notably, several new and useful results are established in Lemmas \ref{LEM2.1}-\ref{LEM2.3} to show how to approximate $g(\cdot)$ by LNN. These results contribute to the current literature in at least the following two points:

1. In contrast to the current literature that usually allows for all parameters to be estimated from data, we start by presenting some identification conditions, which can help reduce the number of effective parameters significantly. 

2. One key idea behind NN architecture is that it can approximate polynomial terms, of which as well understood the linear combination can further approximate unknown functions by standard nonparametric analysis (e.g., \citealp{BK2019, SH2020}). 

In this paper, we do the same, but the difference is that we introduce a bandwidth parameter $h$ below. The reason is that approximating polynomials via NN can only be achieved in a local sense rather than a global sense (cf., Lemmas \ref{LEM2.1}-\ref{LEM2.3} below). This is why we use the terminology LNN. More importantly, $h$ has a direct control on active and non-active activation functions of the hidden layer.  
\medskip

To proceed, we recall the notation defined in Section \ref{S1} and explain how NN architecture works conceptually. In the literature of machine learning, one prefers to target the entire area that the unknown function is defined on, and normally uses a training set to pre-specify a lot of parameters, which do not change with respect to the observations of the test set. By doing so, one just pays some price when calculating the parameters in the first time, and no longer needs to update them when loading the test set.  We are now ready to proceed. 

\subsection{LNN Architecture}\label{S2.1}

First, we define a family of sufficiently smooth functions, and formally state the first assumption of the paper.

\begin{definition}[Continuity]\label{Def1}
Let $p=q+s$ for some $q\in \mathbb{N}$ and $0<s\le 1$. A function $m\,:\, [-a,a]^d\mapsto \mathbb{R}$ is called $(p, \mathscr{C})$-smooth, if for every $\pmb{\mu} \in \mathbb{N}_0^d$ with $|\pmb{\mu}|=q$ the partial derivative $\frac{\partial^{|\pmb{\mu}|} m(\mathbf{x})}{\partial \mathbf{x}^{\pmb{\mu}}}$ exists and satisfies that for all $\mathbf{x}, \mathbf{z}\in [-a,a]^d$, $ \big|\frac{\partial^{|\pmb{\mu}|} m(\mathbf{x})}{\partial \mathbf{x}^{\pmb{\mu}}}  -\frac{\partial^{|\pmb{\mu}|} m(\mathbf{z})}{\partial \mathbf{z}^{\pmb{\mu}}}  \big| \le \mathscr{C}  \| \mathbf{x} - \mathbf{z}\|^s.$
\end{definition}

\begin{assumption}\label{Ass.1}
\item 
\begin{enumerate}[leftmargin=*, noitemsep] 
\item Let $g(\mathbf{x})$ be $(p,\mathscr{C})$-smooth, and $\max_{|\pmb{\alpha} |\le q}  \| \frac{ \partial^{|\pmb{\alpha}|} g(\mathbf{x}) }{\partial\mathbf{x}^{\pmb{\alpha}}   }  \|_{\infty}\le \mathtt{c}$ for $0<c<\infty$.

\item Let Sigmoidal function $\sigma(\cdot)\, : \, \mathbb{R}\to [0,1]$ satisfy that 

\begin{enumerate}[leftmargin=*, noitemsep]
\item $\sigma(\cdot)$ is at least $q+1$ times continuously differentiable with bounded derivatives.

\item A point $u_\sigma\in \mathbb{R}$ exists, where all derivatives up to the order $q$ of $\sigma(\cdot)$ are different from zero.

\end{enumerate}
\end{enumerate}
\end{assumption}
Assumption \ref{Ass.1}.1 is widely adopted in the literature of nonparametric regression (e.g., \citealp{LR2007}). The main point is that each component in Taylor expansion of $g(\cdot)$ is bounded and also sufficiently smooth. Assumption \ref{Ass.1}.2 nests a wide class of activation functions commonly used in the literature as special cases, e.g., Sigmoidal squasher (i.e., $\sigma(x) = \frac{1}{1+\exp(-x)}$), Error function (i.e., $\sigma(x) =\frac{2}{\sqrt{\pi}}\int_0^x \exp(-w^2)dw$), etc. We refer the interested reader to \cite{DUBEY202292} for a comprehensive review on different activation functions, and to Appendix \ref{App.1} for a detailed example.  We acknowledge the growing literature on the rectified linear unit (ReLU) function (e.g., \citealp{SH2020, FLM2021}). As ReLU and Sigmoidal functions require different approximation theories, we therefore focus on the Sigmoidal function in this paper.

\begin{lemma} \label{LEM2.1}
Let Assumption \ref{Ass.1}.2 hold. For $\forall x_0\in \mathbb{R}$, there exist $\pmb{\gamma}=(\gamma_1,\ldots, \gamma_{q+1})^\top$ and $\pmb{\beta}=(\beta_1,\ldots, \beta_{q+1})^\top$ with $\gamma_k= \frac{(-1)^{q+k-1} {\mathtt{C}}^q}{\sigma^{(q)} (u_\sigma)} \binom{q}{k-1} $ and $\beta_k= \frac{k-1}{\mathtt{C}}$, for $k\in [q+1]$, such that

\begin{eqnarray*}
\sup_{|x-x_0|\le h}\left| \sum_{k=1}^{q+1} \gamma_k \sigma (\beta_k \cdot (x-x_0)+ u_\sigma) - (x-x_0)^q \right| =O\left( h^{q+1} \right),
\end{eqnarray*}
where $\mathtt{C}$ is a constant, and $h$ is a bandwidth.
\end{lemma}

\begin{remark}\label{rmk1}
\normalfont
\item 
\begin{enumerate}[leftmargin=*, noitemsep] 
\item Lemma \ref{LEM2.1} yields a recursive relationship, as one can repeatedly invoke Lemma \ref{LEM2.1} to replace $(x-x_0)$ inside the activation. It will then yield a NN with multiple hidden layers. However, we do not see any benefit of doing so unless $g(\cdot)$ of \eqref{EQ1.1} has certain specific structure.  

\item  Lemma \ref{LEM2.1} is independent of data. Only the order of the polynomial term to be approximated depends on the smoothness of $g(\cdot)$.  The quantities $\pmb{\gamma}$, $\pmb{\beta}$ and $u_\sigma$ are fully decided by the activation function and the polynomial term, so they are known prior to regression. The constant $\mathtt{C}$ raises an issue of identifiability, so we simply let $\mathtt{C}=1$ throughout the rest of this paper. $h$ will be decided by the sample size later, and is introduced to control which activation functions will be activated. 
\end{enumerate}
\end{remark}

We then show the feasibility of LNN architecture.

\begin{lemma}[Feasibility]\label{LEM2.2}
Suppose that Assumption \ref{Ass.1}.2 holds.  For $\forall \mathbf{x}_0 \in \mathbb{R}^d$, we define $p(\mathbf{x}\, |\, \mathbf{x}_0, \pmb{\lambda})=\pmb{\lambda}^\top\mathbf{m}(\mathbf{x}\, |\, \mathbf{x}_0),$ where $\pmb{\lambda} =(\lambda_1,\ldots, \lambda_{d_q})^\top$ with $d_q=\binom{d+q}{d}$ and $\|\pmb{\lambda} \|\le \mathtt{c}$.  Then there exists a localized neural network of the form:  
\begin{eqnarray*}
s(\mathbf{x}\, |\, \mathbf{x}_0, \widetilde{\pmb{\lambda}}) =(\widetilde{\pmb{\lambda}}\otimes \pmb{\gamma})^\top\pmb{\sigma}(\mathbf{x}\, |\, \mathbf{x}_0 ) \quad\text{with}\quad \pmb{\sigma}(\mathbf{x}\, |\, \mathbf{x}_0 ) =\Big\{\sigma \big(\pi_{j0}+\sum_{k=1}^d\pi_{jk}(x_k-x_{k0}) \big) \Big\}_{d_q(q+1)\times 1} 
\end{eqnarray*}
such that $\sup_{\mathbf{x}\in C_{\mathbf{x}_0,h} }|s(\mathbf{x}\, |\, \mathbf{x}_0, \widetilde{\pmb{\lambda}})-p(\mathbf{x}\, |\, \mathbf{x}_0, \pmb{\lambda})|= O(h^{q+1})$,
where $\widetilde{\pmb{\lambda}}= \mathbf{D}^\top \pmb{\lambda}$ with $\mathbf{D}$ being a rotation matrix,  and $\pmb{\pi}_j=(\pi_{j0},\cdots,\pi_{jd})^\top$ satisfies

\begin{eqnarray*}
(\pmb{\pi}_1,\ldots, \pmb{\pi}_{d_q (q+1)})&=&\frac{\diag\{ h, \mathbf{I}_d\}\mathbf{W} ( \mathbf{I}_{d_q}\otimes \pmb{\beta}^\top)}{d+1} + \bigg[\begin{matrix}
u_\sigma\\
 \mathbf{0}_{d}
\end{matrix}\bigg] \otimes \mathbf{1}_{d_q(q+1)}^\top,
\end{eqnarray*}
in which $\mathbf{W}$ is a user chosen matrix satisfying that $\max_{j}\|\mathbf{w}_{j}\|\le \sqrt{d+1}$ and $\mathbf{w}_{j_1}\ne \mathbf{w}_{j_2}$ for any given $j_1, j_2\in [d_q]$, and $\mathbf{w}_{j}$ stands for the $j^{th}$ column of $\mathbf{W}$.
\end{lemma}

\begin{remark}\label{rmk2}
\normalfont 
\item 
\begin{enumerate}[leftmargin=*, noitemsep] 
\item We use vector form to rewrite each element of $\pmb{\sigma}(\mathbf{x}\, |\, \mathbf{x}_0 ) $, i.e., $\pmb{\sigma}(\mathbf{x}\, |\, \mathbf{x}_0 ) =\Big\{\sigma \big(\pi_{j0}+\sum_{k=1}^d\pi_{jk}(x_k-x_{k0}) \big) \Big\}_{d_q(q+1)\times 1} $ which is exactly what Sigmoidal activation function does (i.e., mapping a liner combination of regressors plus a location parameter to $[0,1]$). By design, $\pmb{\sigma}(\mathbf{x}\, |\, \mathbf{x}_0 )$ naturally and automatically explores different interaction terms of the regressors.

\item Lemma \ref{LEM2.2} infers that LNN architecture rotates the parameters of interest (i.e., $\pmb{\lambda}$) with a pre-determined $d_q\times d_q$ full rank matrix $\mathbf{D}$, which can be easily constructed practically. We present the details in Appendix \ref{App.1} for the sake of space.

\item Without loss of generality, we can let $\mathbf{w}_j= \frac{\sqrt{d+1}}{q} \mathbf{w}_j^*$, where $\{\mathbf{w}_j^*\}$ are the vectors corresponding to the powers of the distinctive terms in the expansion of $(1+x_1+\cdots+x_d)^q$. Thus,  $\|\mathbf{w}_j\|\le  \frac{\sqrt{d+1}}{q}  |\mathbf{w}_j^*|=\sqrt{d+1},$ which ensures Lemma \ref{LEM2.1} can be invoked.
\end{enumerate}
\end{remark}

Using Lemma \ref{LEM2.2}, the LNN method requires us to focus on a compact set $[-a, a]^d$, as we need to partition it into lots of small cubes.  These small cubes may have different names in each appearance. For example, they are referred to as (hyper-) cubes in  \cite{BK2019} and \cite{SH2020}, and are referred to as localization in \cite{FLM2021}. Each cube  is corresponding to an effective sample set, which is jointly determined by the point of interest and the bandwidth. Finally, we note that in Appendix \ref{App.2}, we show that the main results remain valid when $a\rightarrow \infty$ along with the sample size. 

That said, for a given integer $M\ge 1$, we subdivide $[-a, a]^d$ into $M^d$ cubes of side length $2h\coloneqq\frac{2a}{M}$, and number these cubes by $C_{\mathbf{x}_{0\mathbf{i}},h} $ with $\mathbf{i}\in [M]^d$:
\begin{eqnarray}\label{EQ2.1}
C_{\mathbf{x}_{0\mathbf{i}},h} = \{\mathbf{x} \, | \, |x_j- x_{0\mathbf{i},j}|\le h \text{ for } j\in [d] \},
\end{eqnarray}
where $\mathbf{x}_{0\mathbf{i}}$ represents the center of $C_{\mathbf{x}_{0\mathbf{i}},h} $, and $x_j$ and $x_{0\mathbf{i},j}$ are the $j^{th}$ elements of $\mathbf{x}$ and $\mathbf{x}_{0\mathbf{i}}$ respectively.  

Let $\widetilde{s}(\mathbf{x} \, |\, \widetilde{\pmb{\Lambda}}) = \sum_{\mathbf{i}\in [M]^d} I_{\mathbf{i},h}(\mathbf{x}) s(\mathbf{x} \,|\, \mathbf{x}_{0\mathbf{i}},  \widetilde{\pmb{\lambda}}_{\mathbf{i}})$, $\widetilde{\pmb{\Lambda}} =\{\widetilde{\pmb{\lambda}}_{\mathbf{i}}\, |\, \mathbf{i}\in [M]^d\}$ and $I_{\mathbf{i},h}(\mathbf{x}) =I(\mathbf{x}\in C_{\mathbf{x}_{0\mathbf{i}},h} )$. We then establish the following important lemma.

\begin{lemma}[LNN Architecture]\label{LEM2.3}
Suppose that Assumption \ref{Ass.1} holds. Then we can approximate $g(\mathbf{x})$ by $ \widetilde{s}(\mathbf{x} \, |\, \widetilde{\pmb{\Lambda}})$ such that

\begin{eqnarray*}
\| g(\mathbf{x}) -  \widetilde{s}(\mathbf{x} \, |\, \widetilde{\pmb{\Lambda}})  \|_{\infty}=O(h^p).
\end{eqnarray*}

\end{lemma}

\begin{remark}\label{rmk3}
\normalfont The use of the indicator function is consistent with the sparsity setting of \cite{SH2020}, where the author argues that ``\textit{the network sparsity assumes that there are only few non-zero/active network parameters}". In fact, our study clarifies the definition of ``\textit{the network sparsity}" by (i) defining non-active activation functions (such as those in Figure \ref{Fig1}) which is realized through the use of indicator function $I_{\mathbf{i},h}(\mathbf{x})$; and (ii) pointing out the number of effective parameters. 

In Section \ref{S2.3} below, we further explore the second point, as it allows us to  define a set of true parameters when using thresholding techniques. 
\end{remark}

\subsection{Estimation and Inference for $\alpha_0$ and $g(\cdot)$}\label{S2.2}

We are now ready to work on our estimation method. To accommodate a potentially large number of explanatory variables, we adopt a sparse structure, and impose the following assumption.

\begin{assumption}\label{Ass.2}
Suppose that there exists an integer $d_0\leq d$ and a function $g_c\,:\, [-a,a]^{d_0}\to \mathbb{R}$ such that $g(\mathbf{x}_0) \equiv g_c(\mathbf{x}_{c,0})$, where $\mathbf{x}_{c,0}$ contains the first $d_0$ elements of $\mathbf{x}_0$.  
\end{assumption}

Here, to be clear, we still require $d_0$ and $d$ to be fixed in theory as in \cite{BK2019}, \cite{SH2020} and \cite{FLM2021}. As pointed out in \cite{fan1996local}, when it comes to estimate unknown functions, such settings where $d\geq 4$ and the same size is not big enough can cause the so--called ``curse of dimensionality". In our numerical studies, with a relatively large $d_0$, our approach still achieves good finite sample properties.

Under Assumption \ref{Ass.2}, it is clear to see that the coefficient vector $\pmb{\lambda}$ of Lemma \ref{LEM2.2} inherits the sparsity of $g(\mathbf{x}_t)$. Specifically, we denote the true space as
\begin{eqnarray*}
\mathscr{P}^0_q= \left\{ \text{Linear span of the monomials } \prod_{k=1}^{d_0} x_k^{n_k} \text{ with } 0\le |\mathbf{n_0}|\le q\right\},
\end{eqnarray*}
where $\mathbf{n}_0=(n_1,\ldots, n_{d_0})^\top\in \mathbb{N}_0^{d_0}$. We then define $\overline{\mathscr{P}}_q$ as the complement space, such that $\mathscr{P}^0_q \cup \overline{\mathscr{P}}_q=\mathscr{P}_q$ and $\mathscr{P}^0_q \cap \overline{\mathscr{P}}_q=\emptyset$. Simple algebra shows that the dimensions of $\mathscr{P}^0_q$ and $\overline{\mathscr{P}}_q$ are  $\text{dim}\mathscr{P}^0_q  =\binom{d_0+q}{d_0}\coloneqq d^0_q$ and $\text{dim}\overline{\mathscr{P}}_q  = d_q-d^0_q$ respectively. In connection with \eqref{EQ1.4}, we suppose further that $\{m_1(\mathbf{x} \, |\, \mathbf{x}_0),\ldots, m_{d^0_q}(\mathbf{x} \, |\, \mathbf{x}_0)\}$ constitute a basis for $\mathscr{P}^0_q$ without loss of generality\footnote{This implies that we can express $m_j(\mathbf{x} \, |\, \mathbf{x}_0)=m_{c,j}(\mathbf{x}_c \, |\, \mathbf{x}_{c,0})$ for $j=1,\ldots, d_q^0$.  However, this requirement is purely for notational simplicity and  is not  necessary for the validity of our estimation and theoretical development.}.  Consequently,  the sparsity of  $\pmb{\lambda}$ is expressed as $\{\lambda_{j}=0\, |\,j=d_q^0+1,\ldots,d_q\}.$ Moreover, similar arguments to those presented in Lemma \ref{LEM2.3} can be applied to show that the true function $g_c(\mathbf{x}_c)$ can be approximated by the oracle NN architecture $\widetilde{s}_c(\mathbf{x}_c \, |\, \widetilde{\pmb{\Lambda}}_c)$: $\| g_c(\mathbf{x}_c) -  \widetilde{s}_c(\mathbf{x}_c \, |\, \widetilde{\pmb{\Lambda}}_c)  \|_{\infty}=O(h^p)$, where $\widetilde{\pmb{\Lambda}}_c$ is the oracle counterpart of $\widetilde{\pmb{\Lambda}}$.

Drawing upon the sparsity structure of $\pmb{\lambda}$, we propose using the group-LASSO strategy (\cite{YL2006}) to formulate the following penalized estimation:
\begin{eqnarray}\label{EQ2.2}
(\widetilde{\alpha},\widetilde{\pmb{\Theta}}) = \argmin_{\alpha\in \mathbb{R}, \ \widetilde{s} \in \mathcal{S}}\left( \widetilde{Q}  (\alpha,\pmb{\Theta})+\sum_{j=1}^{d_q}\psi_j\|\pmb{\Theta}_{D,j}\|\right),
\end{eqnarray}
where $\{\psi_j\}$ denote the tuning  parameters, $\widetilde{Q}  (\alpha,\pmb{\Theta})= \sum_{t=1}^T[y_t-z_t\alpha -\widetilde{s}(\mathbf{x}_t \, |\, \pmb{\Theta}) ]^2$, $\mathcal{S} = \{ \widetilde{s}(\mathbf{x} \, |\, \pmb{\Theta})\}$ with $\widetilde{s}(\mathbf{x} \, |\, \cdot)$ being defined in Lemma \ref{LEM2.3}, $\pmb{\Theta} =\{\pmb{\theta}_{\mathbf{i}}\, |\, \mathbf{i}\in [M]^d\}$ with $\pmb{\theta}_{\mathbf{i}}$'s being $d_q\times 1$ vectors, and  $\pmb{\Theta}_{D,j}$ is an $M^d\times 1$ vector with the $j^{th}$ component being $\mathbf{D}^{\top,-1}\pmb{\theta}_{\mathbf{i}}$. 

Let $\widetilde{\pmb{\theta}}_{\mathbf{i}}$ and $\widetilde{\pmb{\Theta}}_{D,j}$ be the corresponding estimators of $\pmb{\theta}_{\mathbf{i}}$ and $\pmb{\Theta}_{D,j}$. For the purpose of comparison, we also define the following oracle estimators: $ (\widetilde{\alpha}_c,\widetilde{\pmb{\Theta}}_c)$ of the form:

\begin{equation*}
(\widetilde{\alpha}_c,\widetilde{\pmb{\Theta}}_c)=\argmin_{\alpha\in \mathbb{R}, \widetilde{s}_c\in \mathcal{S}_{c}}  \widetilde{Q}_c  (\alpha,\pmb{\Theta}_c),
\end{equation*}
where $\widetilde{Q}_c  (\alpha,\pmb{\Theta}_c)= \sum_{t=1}^T[y_t-z_t\alpha -\widetilde{s}_c(\mathbf{x}_{c,t} \, |\, \pmb{\Theta}_c) ]^2$, and $\pmb{\Theta}_c$, $\widetilde{s}_c(\mathbf{x}_{c,t} \, |\, \pmb{\Theta}_c)$ and $\mathcal{S}_{c}$ respectively represent the oracle counterparts of $\pmb{\Theta}$, $\widetilde{s}(\mathbf{x}_{t} \, |\, \pmb{\Theta})$ and $\mathcal{S}$ (i.e., assuming the sparsity is known). Although the oracle estimators $\widetilde{\alpha}_c$ and $\widetilde{\pmb{\Theta}}_c$ are infeasible practically, they can be approximated by the group-LASSO estimators with asymptotically negligible biases. To facilitate the rest of our development, we impose the following time series structure on \eqref{EQ1.2} in Assumption 3 below. 

\begin{assumption}\label{Ass.3}
\item 
\begin{enumerate}[leftmargin=*, noitemsep] 
\item $\{(\mathbf{x}_t,\eta_t,\varepsilon_t)\, |\, t\in [T]\}$ are strictly stationary and $\alpha$-mixing with mixing coefficient  
\begin{eqnarray*}
\alpha(t) = \sup_{A\in \mathcal{F}_{-\infty}^0, B\in  \mathcal{F}_t^\infty} |P(A)P(B) -P(AB)|
\end{eqnarray*}
satisfying $\sum_{t= 1}^{\infty} \alpha^{\nu/(2+\nu)}(t) <\infty$ for some $\nu>0$, where $\mathcal{F}_{-\infty}^0$ and $\mathcal{F}_{t}^\infty$ are the $\sigma$-algebras generated by $\{(\mathbf{x}_s,\eta_s,\varepsilon_s): s \leq 0\}$ and $\{(\mathbf{x}_s,\eta_s,\varepsilon_s): s \geq t\}$, respectively. 

\item The probability density function of $\mathbf{x}_1$, say $ f_{\mathbf{x}}(\cdot)$,   and the function  $G(\mathbf{x})$ are Lipschitz continuous and bounded on $[-a,a]^d$.  Additionally, $f_{\mathbf{x}}(\cdot)$ is bounded away from 0 on $[-a,a]^d$.

\item $E[\varepsilon_1 \,| \, \mathbf{x}_1,\eta_1]=0$, $E[\varepsilon_1^2 \,| \, \mathbf{x}_1, \eta_1]=\sigma_\varepsilon^2$  almost surely (a.s.), and $E[|\varepsilon_1|^{2+\nu} \,| \, \mathbf{x}_1, \eta_1]\le\mathtt{c}$ a.s., where $\nu$ is the same as involved in Assumption \ref{Ass.3}.1.
\end{enumerate}
\end{assumption}

Assumption \ref{Ass.3} is standard in the literature of time series analysis (see, \cite{Gao}, for example). Heterogeneity may also be introduced by, for example, $E[\varepsilon_1^2 \,| \, \mathbf{x}_1,\eta_1]=\psi(\mathbf{x}_1,\eta_1)$, which however makes notation even more complicated than what we need to involve.  $\{\mathbf{x}_t\}$ may also have more complex structures such as linear processes, locally stationarity, deterministic trends, etc. Surely, the corresponding development and asymptotic results will need to be modified accordingly, but it will not add too many credits to the original idea of this paper. We therefore focus on the current setting.

To proceed, we introduce some additional notation.  Let $\mathbf{H}_c$, $\mathbf{x}_{c,0\mathbf{i}}$, and  $\mathbf{m}_c(\mathbf{x}_{c,t}\, |\, \mathbf{x}_{c,0\mathbf{i}})$ be the oracle counterparts of $\mathbf{H}$, $\mathbf{x}_{0\mathbf{i}}$, and  $\mathbf{m}(\mathbf{x}_{t}\, |\, \mathbf{x}_{0\mathbf{i}})$. Also, let $\Phi_\eta(x)$  and $f_{\mathbf{x}_c}(\mathbf{x}_{c,\mathbf{i}})$ denote the CDF of $\eta_t$ and the density function of $\mathbf{x}_{c,t}$, respectively.  Also define  
\begin{eqnarray}\label{notnew}
\widetilde{z}_t&=& z_t-\sum_{\mathbf{i}\in [M]^{d_0}} I_{\mathbf{i},h} (\mathbf{x}_{c,t}) \mathbf{M}_{c,\mathbf{i}}^\top \pmb{\Sigma}_{c,\mathbf{i}}^{-1}\mathbf{H}_c\mathbf{m}_c(\mathbf{x}_{c,t}\, |\, \mathbf{x}_{c,0\mathbf{i}}),\nonumber \\
\mathbf{M}_{c,\mathbf{i}}&=&\int_{[-a,a]^{d-d_0}}\Phi_\eta(G(\mathbf{x}_{c,0\mathbf{i}},\mathbf{z}))    f_{\mathbf{x}}(\mathbf{x}_{c,0\mathbf{i}},\mathbf{z})\mathrm{d}\mathbf{z} \int_{[-1,1]^{d_0}}    \mathbf{m}_c(\mathbf{x}_c\, |\, \mathbf{0}) \mathrm{d} \mathbf{x}_c,
\nonumber\\
\pmb{\Sigma}_{c,\mathbf{i}}&=&f_{\mathbf{x}_c}(\mathbf{x}_{c,\mathbf{i}}) \int_{[-1,1]^{d_0}}  \mathbf{m}_c(\mathbf{x}_c\, |\, \mathbf{0})  \mathbf{m}_c(\mathbf{x}_c\, |\, \mathbf{0})^\top  \mathrm{d} \mathbf{x}_c.
\end{eqnarray}

\begin{assumption}\label{Ass.4}
\item 
\begin{enumerate}[leftmargin=*, noitemsep] 

\item Assume that   $\inf_{\mathbf{x}\in [-a,a]^d} \lambda_{\min}(\pmb{\Omega}^\ast_0(\mathbf{x}))>0$, where{\small
\begin{eqnarray*}
\pmb{\Omega}^\ast_0(\mathbf{x})=\left(
\begin{array}{c c}
2^d\Phi_\eta(G(\mathbf{x}))  f_{\mathbf{x}}(\mathbf{x})& \Phi_\eta(G(\mathbf{x}))    f_{\mathbf{x}}(\mathbf{x}) \int_{[-1,1]^d}    \mathbf{m}(\mathbf{x}\, |\, \mathbf{0})^\top  \mathrm{d} \mathbf{x}\\
\Phi_\eta(G(\mathbf{x}))    f_{\mathbf{x}}(\mathbf{x}) \int_{[-1,1]^d}    \mathbf{m}(\mathbf{x}\, |\, \mathbf{0}) \mathrm{d} \mathbf{x} &
f_{\mathbf{x}}(\mathbf{x}) \int_{[-1,1]^d}  \mathbf{m}(\mathbf{x}\, |\, \mathbf{0})  \mathbf{m}(\mathbf{x}\, |\, \mathbf{0})^\top  \mathrm{d} \mathbf{x}
\end{array}
\right).
\end{eqnarray*}}
\item Suppose that $\sigma_{c,z}^2=\lim_{T\rightarrow\infty}\frac{1}{T}\sum_{t=1}^T\sum_{s=1}^TE[\widetilde{z}_t\widetilde{z}_s]E[\varepsilon_t\varepsilon_s]$ is a positive constant. 

\item  
Let {\small $0<\lim\sup_{T\rightarrow \infty} Th^{d_0 +2p}<\infty$ when $d_0<d$, and $\lim_{T\rightarrow \infty} Th^{d_0 + 2p} =0$ when $d_0=d$, $\lim_{T\rightarrow \infty} \min_{d_q^0+1\leq j\leq d_q}\{\psi_jH_j\}T^{-\frac{1}{2}}=\infty$ and $\lim_{T\rightarrow \infty} \max_{j\in[d^0_q]} \{\psi_jH_j\}T^{-\frac{1}{2}} =0$}. 
\end{enumerate}
\end{assumption}

Assumption 4 imposes a set of regularity conditions. Assumptions 4.1 and 4.2 ensure the positive definiteness of the asymptotic covariances, and Assumption 4.3 regulates the bandwidth and the group-LASSO tuning parameters, which can be easily fulfilled. The first part of Assumption 4.3 requires that when the dimensionality of the true covariates satisfies $d_0<d$, there is no need to impose an under--smoothing condition, and it is required to impose the under--smoothing condition when $d_0=d$. For the second part of Assumption 4.3, detailed justifications are available from Appendix A.1 of the online supplementary document for more details.

Given these extra conditions, we establish the following lemma.

\begin{lemma}\label{LEM2.4}
Under Assumptions \ref{Ass.1}-\ref{Ass.4}, we have 
\begin{enumerate}[leftmargin=*, noitemsep] 
\item $P(\|\widetilde{\pmb{\Theta}}_{D,j}\|=0)\rightarrow1$,  for $j=d_q^0+1,\ldots,d_q$;
\item $|\widetilde{\alpha}- \widetilde{\alpha}_c|=O_P(h^p)+o_P (T^{-\frac{1}{2}})$.
\end{enumerate}
\end{lemma}

Lemma \ref{LEM2.4}.1 indicates that the group-LASSO method can correctly identify the sparsity of $\pmb{\lambda}$. Lemma \ref{LEM2.4}.2 shows $\widetilde{\alpha}$ and $ \widetilde{\alpha}_c$ are asymptotically equivalent up to a rate of an order of $O_P(h^p)+o_P (T^{-\frac{1}{2}} )$.

Let $\mathbf{Z}=(z_1,\cdots,z_T)^\top$ and $\widetilde{\mathbf{X}}_{c,\mathbf{i}} = (\widetilde{\mathbf{x}}_{c,\mathbf{i}, 1}, \cdots, \widetilde{\mathbf{x}}_{c,\mathbf{i}, T})^\top$, where $$\widetilde{\mathbf{x}}_{\mathbf{i},c,t}  =I_{\mathbf{i},h} (\mathbf{x}_{c,t})(\mathbf{I}_{d^0_q}\otimes \boldsymbol{\gamma}^\top)  \pmb{\sigma}_c( \mathbf{x}_{c,t}\, |\,\mathbf{x}_{0,c,\mathbf{i}})$$ and $\pmb{\sigma}_c( \mathbf{x}_{c,t}\, |\,\mathbf{x}_{0,c,\mathbf{i}})$ denotes the oracle counterpart of $\pmb{\sigma}(\mathbf{x}\, |\, \mathbf{x}_0 ) $.
Building upon Lemma \ref{Lem_Ora} of the online supplement, we can now establish the asymptotic distributions of the proposed estimators.

\begin{theorem}\label{THM2.1}
 Suppose that Assumptions \ref{Ass.1}-\ref{Ass.4} hold. 
\begin{enumerate}[leftmargin=*, noitemsep] 
\item[1.] As $T\rightarrow\infty$,
\begin{equation*}
\sqrt{T}\widetilde{M}_{c,z}(\widetilde{\alpha}-\alpha_0+O_P(h^p))\to_D N (0,1),
\end{equation*}
where $\widetilde{M}_{c,z}=T^{-1}\sigma_{c,z}^{-1}\mathbf{Z}^\top\big(\mathbf{I}_{T}-\sum_{\mathbf{i}\in [M]^{d_0}} \widetilde{\mathbf{X}}_{c,\mathbf{i}}\bigl[\widetilde{\mathbf{X}}_{c,\mathbf{i}}^\top \widetilde{\mathbf{X}}_{c,\mathbf{i}}\bigr]^{-1}\widetilde{\mathbf{X}}_{c,\mathbf{i}}^\top\big) \mathbf{Z}$.

\item[2.] 
For $\forall \mathbf{x}_{0}\in [-a,a]^{d}$, let  $\widetilde{g}(\mathbf{x}_{0})= \sum_{\mathbf{i}\in [M]^{d}} \widetilde{\mathbf{x}}_{ 0,\mathbf{i}}^\top \widetilde{\pmb{\theta}}_{\mathbf{i}}$, where $\widetilde{\mathbf{x}}_{0, \mathbf{i}}=I_{\mathbf{i},h} (\mathbf{x})(\mathbf{I}_{d_q}\otimes \pmb{\gamma}^\top)  \pmb{\sigma}( \mathbf{x}\, |\,\mathbf{x}_{0\mathbf{i}})$. Then, as $T\rightarrow\infty$,
\begin{equation*}
\sqrt{Th^{d_0}} \widetilde{\sigma}_{\mathbf{x}_{c,0}}^{ -1}(\widetilde{g}(\mathbf{x}_{0}) - g_c(\mathbf{x}_{c,0}) +O_P(h^p)) \to_D N (0, 1 ),
\end{equation*}
where $\widetilde{\sigma}_{\mathbf{x}_{c,0}}^2 = \sigma_\varepsilon^2\sum_{\mathbf{i}\in [M]^{d_0}} I_{\mathbf{i},h}(\mathbf{x}_{c,0}) \mathbf{m}_c(\mathbf{x}_{c,0}\, |\, \mathbf{x}_{c,0\mathbf{i}})^\top\mathbf{H}_c \pmb{\Sigma}_{c,\mathbf{i}}^{-1}\mathbf{H}_c\mathbf{m}_c(\mathbf{x}_{c,0}\, |\, \mathbf{x}_{c,0\mathbf{i}}).$
\end{enumerate}
\end{theorem}

Note that the rate of convergence, $\sqrt{T h^{d_0}}$, is faster than that of the conventional kernel estimator of an order of $\sqrt{T h^d}$ due to the fact that $Th^d = Th^{d_0} \, h^{d- d_0} = o\left(T h^{d_0}\right)$ when $d_0<d$. Meanwhile, the order of the bias term is also smaller than that of the conventional kernel estimator due to the construction of the LNN architecture and the definition of $p>2$. This is the main reason that the proposed LNN method outperforms over the conventional kernel method in finite--sample studies particularly when $d_0\geq 4$. Table 1 of Section 3 further supports the finite--sample superiority of the proposed LNN method.

When establishing the second result of Theorem \ref{THM2.1}, the terms $E[\varepsilon_1\varepsilon_{1+t}]$ for $t\ge 1$ all vanish in the asymptotic covariance matrix due to the partition of \eqref{EQ2.1} and the use of the indicator function. As a result, the confidence interval associated with the second result of Theorem \ref{THM2.1} can be easily achieved. 

Because $\sigma_{c,z}$ defined in Assumption 4 involves a long--run variance component, however, the serial dependence is not asymptotically negligible when inferring $\widetilde{\alpha}$. Therefore, we propose using the following dependent wild bootstrap method.

\begin{enumerate}[leftmargin=*, noitemsep] 
\item Based on the estimation residuals $\widehat{\varepsilon}_t=y_t - z_t\widetilde{\alpha} - \widetilde{g}(\mathbf{x}_{t}) $, generate the bootstrap random errors: $\varepsilon^\ast_t =\widehat{\varepsilon}_t\varsigma_t$, where $\varsigma_t$ is an $\ell$-dependent time series  that satisfies $E[\varsigma_t]=0$, $E[\varsigma_t^2]=1$, $E[\varsigma_t^4]<\infty$, and $E[\varsigma_t\varsigma_s]=K\left(\frac{t-s}{\ell}\right)$ with $\ell\rightarrow\infty$ as $T\rightarrow\infty$. Here $K(\cdot)$ is a symmetric and  Lipschitz continuous kernel function defined on $[-1,1]$, and satisfies $K(0)=1$ and $\int_{-\infty}^\infty K(u)e^{-iux}du\geq 0$ for all $x\in \mathbb{R}$. 

\item Construct the dependent bootstrap variables as $y_t^\ast = z_t\widetilde{\alpha} + \widetilde{g}(\mathbf{x}_{t})+\varepsilon^\ast_t$. With $\{y_t^\ast, z_t, \mathbf{x}_t\}$, we can re-estimate $\alpha$ and $g(\mathbf{x}_{0})$ and obtain the bootstrap group-LASSO estimators $\widetilde{\alpha}^\ast$ and $\widetilde{g}^\ast(\mathbf{x}_0)$.

\item We repeat above two steps for a sufficiently large number of times and obtain the bootstrap draws.
\end{enumerate}

\begin{theorem}\label{Thm_Bootstrap}
Let Assumptions \ref{Ass.1}-\ref{Ass.4} hold. Assume further there exists a positive number $\nu^\ast>\nu$ such that $E[|\varepsilon_1|^{2+\nu^\ast} \,| \, \mathbf{x}_1, \eta_1]\le\mathtt{c}$ a.s., and $\ell$ satisfies that $\ell h^{2p}\rightarrow0$, $\frac{\ell}{\sqrt{T}}\rightarrow0$, and $\ell \, T^{\frac{2(\nu-\nu^\ast)}{(2+\nu)(2+\nu^\ast)}}\rightarrow 0$, where $\nu$ is defined in Assumption \ref{Ass.3}. 

Then, as $T\rightarrow\infty$,
{\small
\begin{enumerate}[leftmargin=*, noitemsep] 
\item $\sup_{w}\Bigl|{\rm Pr}^\ast\bigl[\sqrt{T}(\widetilde{\alpha}^\ast-\widetilde{\alpha})\leq w \bigr]-{\rm Pr}\bigl[\sqrt{T}(\widetilde{\alpha}-\alpha_0)\leq w \bigr]\Bigr|=o_P(1)$,
\item $\sup_{w}\Bigl|\text{\normalfont Pr}^* (\sqrt{Th^{d_0}} \widetilde{\sigma}_{\mathbf{x}_{c,0}}^{-1}[\widetilde{g}^*(\mathbf{x}_0) - \widetilde{g}(\mathbf{x}_0)]\le w ) - \Pr (\sqrt{Th^{d_0}} \widetilde{\sigma}_{\mathbf{x}_{c,0}}^{-1}[\widetilde{g}(\mathbf{x}_0) - g_c(\mathbf{x}_{0,c})]\le w )\Bigr|=o_P(1)$,
\end{enumerate}
where ${\rm Pr}^\ast$} denotes the probability measure conditional on the observed sample. 
\end{theorem}

\begin{remark}
\item 
\normalfont

\begin{enumerate}[leftmargin=*, noitemsep] 
\item It is a well-known problem with the LASSO-type estimators that simple residual-based bootstrap methods fail to consistently estimate their distributions unless some thresholding techniques are applied to handle zero components \citep[see][for example]{Chatterjee2011}. However, in the case of group-LASSO estimation, which shares a similar idea to the adaptive-LASSO, it automatically incorporates soft-thresholding penalties. Consequently, there is no need for additional truncation. Similar discussions can be found in Section 4 of \cite{Chatterjee2011}.
\item The serial dependence in $\varepsilon_t$ only complicates the inference for $\alpha_0$. For a purely nonparametric model without treatment components, a straightforward wild bootstrap method can be employed to mimic the distribution of $\widetilde{g}(\mathbf{x}_0)$. More detailed discussions are provided in Appendix \ref{App.21} of the online supplement.

\end{enumerate}
\end{remark}

To close this section, we emphasize that as have stated in Section \ref{S1}, we provide the detailed study about a fully nonparametric regression in Appendix \ref{App.2} (i.e., letting $\alpha_0 =0$). Although it is a special case of \eqref{EQ1.1}, the investigation yields some useful insights and allows us to further clarify some features of LNN compared to the existing literature. Also, we infer $G(\cdot)$ of the binary structure of $z_t$ in Appendix \ref{App.2}, which is also of great interest in both theory and practice.

\subsection{Further Discussion}\label{S2.3}
 
Up to this point, we would like to point out that those questions raised in Section \ref{S1} have all been answered, so Figure \ref{Fig1} can be understood better. We summarize some key points which may have been discussed previously here and there, and further discuss some remaining issues.

\medskip

\noindent \textbf{Sparsity} --- It is now clear that without sparsity, the total number of activation functions is

\begin{eqnarray*}
M^d\cdot d_q(q+1) =\left( \frac{a}{h}\right)^d \cdot d_q(q+1),
\end{eqnarray*}
of which only $d_q(q+1)$ neurons are activated when loading test data.  Among the activated activation functions, the number of effective parameters is $d_q$, while the rest of the parameters are predetermined. Provided sparsity, utilizing identification conditions allows us to define the true set of parameters, so we can further reduce the effective parameters via the group-LASSO approach.

\medskip

\noindent  \textbf{Multiple Hidden Layers} --- Lemma \ref{LEM2.1} yields a recursive relationship, as one can repeatedly invoke Lemma \ref{LEM2.1} to replace $(x-x_0)$ inside the activation. It will then yield a LNN architecture with multiple hidden layers. Although having multiple hidden layers is achievable, at this stage it is not clear to us why we should do so.  As discussed in Remark \ref{rmk1}, this step is completely independent of data, so we do not see any benefit of doing so unless $g(\cdot)$ (or $G(\cdot)$) has certain specific structure. Under some extra structure on $g(\cdot)$ (or $G(\cdot)$), however, the necessity of developing LNN architecture with multiple hidden layers deserves extra attention in future research.

\medskip

\noindent \textbf{Dependence \& Trending} --- LNN automatically eliminates some correlation of observations from different time periods when establishing the asymptotic distribution for the unknown functions. In this paper, we assume that the regressors $\{\mathbf{x}_t\, |\, t\in [T]\}$ are strictly stationary and mixing. In fact, they can have other more complex structures, such as linear processes, locally stationarity, heterogeneity, deterministic trends, etc.  As a result, many climate models (such as those in \citealp{MUDELSEE2019310}) may be better captured. For such cases, one may need to revise the assumptions and proofs accordingly depending on detailed research questions.

\medskip

\noindent \textbf{Data-Splitting} --- One may further connect the above results with the data-splitting technique of \cite{Chernozhukov2018}, and simultaneously explore the sparsity of the binary structure in $z_t$.

\section{Simulation}\label{S3}

In this section, we conduct simulations to validate the theoretical findings, focusing specifically on the semiparametric model with sparsity. Additional simulation results are provided in Appendix \ref{App.3} of the online supplement to exam additional theoretical results of Appendix \ref{App.2}, and to demonstrate the newly proposed method works reasonably well even without involving sparsity. 

As discussed in Section \ref{S2}, many parameters are involved in the LNN architecture. It would be extremely difficult to systematically check every single one in this paper due to the page constraints, so we have to be selective.  Also, we are constrained by computing power.  That said, the following quantities are pre-fixed without loss of generality.

\begin{itemize}[leftmargin=*, noitemsep] 
\item Throughout, we use the Sigmoidal squasher, $\sigma(w)= 1/(1+\exp(-w))$, as the activation function. 

\item $\pmb{\pi}_j$'s are generated in exactly the same way as mentioned in Remark \ref{rmk2}.

\item Let $R=200$ for the bootstrap procedure. 
\end{itemize}

We consider the  semiparametric treatment effects model \eqref{EQ1.1}, where $\varepsilon_t = 0.5\varepsilon_{t-1} + N(0, 0.75)$, the $j^{th}$ element of $\mathbf{x}_t$ is independently generated as $x_{t,j}\sim U(-1, 1)$, and $z_t$ is independently generated from Bernoulli ($\cos(|\mathbf{x}_{c,t}^\top \mathbf{1}_{d_0}/d_0|)$) to allow for dependence between $z_t$ and $\mathbf{x}_{0,t}$.  We simply let  $\alpha_0=1$ and $g(\mathbf{x}_t)=g_c(\mathbf{x}_{c,t}) = 1+\sin (\mathbf{x}_{c,t}^\top \mathbf{1}_{d_0})$, where $\mathbf{x}_{c,t}$ contains the first $d_0$ elements in $\mathbf{x}_t$. The bandwidth $h$ is set as $h=a/M$, where $M$ is the integer closest to $a/h_1$ with $h_1=1.5\cdot T^{-1/(d+2p-0.5)}$. Here, $-0.5$ is to ensure $\sqrt{T} h^{p+d/2}\to 0$ holds. In fact, $h$ is very close to $h_1$, and the current setup is simply to guarantee $M$ is a large positive integer. When designing the simulations, our impression is that the results are not sensitive to the choices of $g(\mathbf{x})$ and the bandwidth. Therefore, in what follows, we only vary the values of $q$, $d$($d_0$), $u_\sigma$. Specifically, we use $d\in\{2, 8, 14\}$, $d_0=d/2$, $q\in\{3,4\}$ and $u_\sigma\in\{-0.5,0.5\}$.

To measure the estimate of $g(\cdot)$, we select a few test points from $[-a,a]^d$. Ideally, we would like to select $L$ points from each dimension, so it gives $L^d$ points to evaluate in total. However,  it will create  a lot computational overhead, so we select the points $\mathbf{x}_{L,j} \coloneqq  (-a+\frac{2aj}{L+1} )\mathbf{1}_d$ for $j\in[L]$. At each test point, we construct the estimate  along with the $95\%$ confidence interval using the method of Section \ref{S2.2}. To evaluate the finite sample performance, we compute the root mean squared errors (RMSEs) and coverage rates (CRs) after $n$ replications:
\begin{eqnarray*}
&&\text{RMSE}_\alpha =\left\{\frac{1}{n}\sum_{i=1}^n[\widetilde{\alpha}_i-\alpha_0]^2\right\}^{1/2}, \quad  \text{RMSE}_g = \left\{ \frac{1}{nL  }\sum_{i=1}^n \sum_{j=1}^L  [\widetilde{g}_i(\mathbf{x}_{L,j}) -g(\mathbf{x}_{L,j})]^2 \right\}^{1/2},\nonumber \\
&&\text{CR}_\alpha =\frac{1}{n }\sum_{i=1}^n  I(\widetilde{\alpha}_i -\alpha_0 \in \text{CI}_{\alpha,ij}),\quad\text{CR}_g =\frac{1}{n L}\sum_{i=1}^n \sum_{j=1 }^L I(\widetilde{g}_i(\mathbf{x}_{L,j}) -g(\mathbf{x}_{L,j}) \in \text{CI}_{g,ij}),
\end{eqnarray*}
where $\widetilde{\alpha}_i$ and  $\widetilde{g}_i(\cdot) $ stand for the estimates of $\alpha_0$ and $g(\cdot)$, respectively,  at the $i^{th}$ replication. $\text{CR}_{\alpha, ij}$ and $\text{CR}_{g, ij}$ are the 95\% confidence intervals of $\widetilde{\alpha}_i^\ast-\widetilde{\alpha}_i$ and $\widetilde{g}_i^*(\mathbf{x}_{L,j})-\widetilde{g}_i(\mathbf{x}_{L,j})$, respectively,  based on the bootstrap draws from the $i^{th}$ replication. The number of bootstraps is set to be 200. We select $T\in \{800, 1600,2400 \}$, $L=10$ and $n=1000$ without loss of generality. 

RMSEs and coverage rates are reported in Table \ref{Tab1}. As evident in the table, RMSEs of both $\widetilde{\alpha}_i$ and $\widetilde{g}_i(\cdot)$ decrease as $T$ increases from 800 to 2400.  Moreover,  coverage rates are close to 0.95 indicating that the bootstrap procedure behaves reasonably well.  Some additional  facts should also be  mentioned.  The simulation results are not changing significantly with respect to the value of $u_\sigma$, which  confirms our argument in Remark \ref{rmk1}. For comparison, we also report the simulated RMSEs and CRs for the LNN estimation without considering the sparsity structure. As evident in Table \ref{Tab1}, the LNN-based group-LASSO estimators generally outperform the LNN estimation, especially in cases with large parameter sets (e.g., $d=14$ and $q=4$). 

\begin{center}

\fbox{Table 1 near here}

\end{center}

\begin{figure}[htp!]
	\centering
	\subfloat[$T=800$, $q=3$]
	{\includegraphics[width=0.5\textwidth]{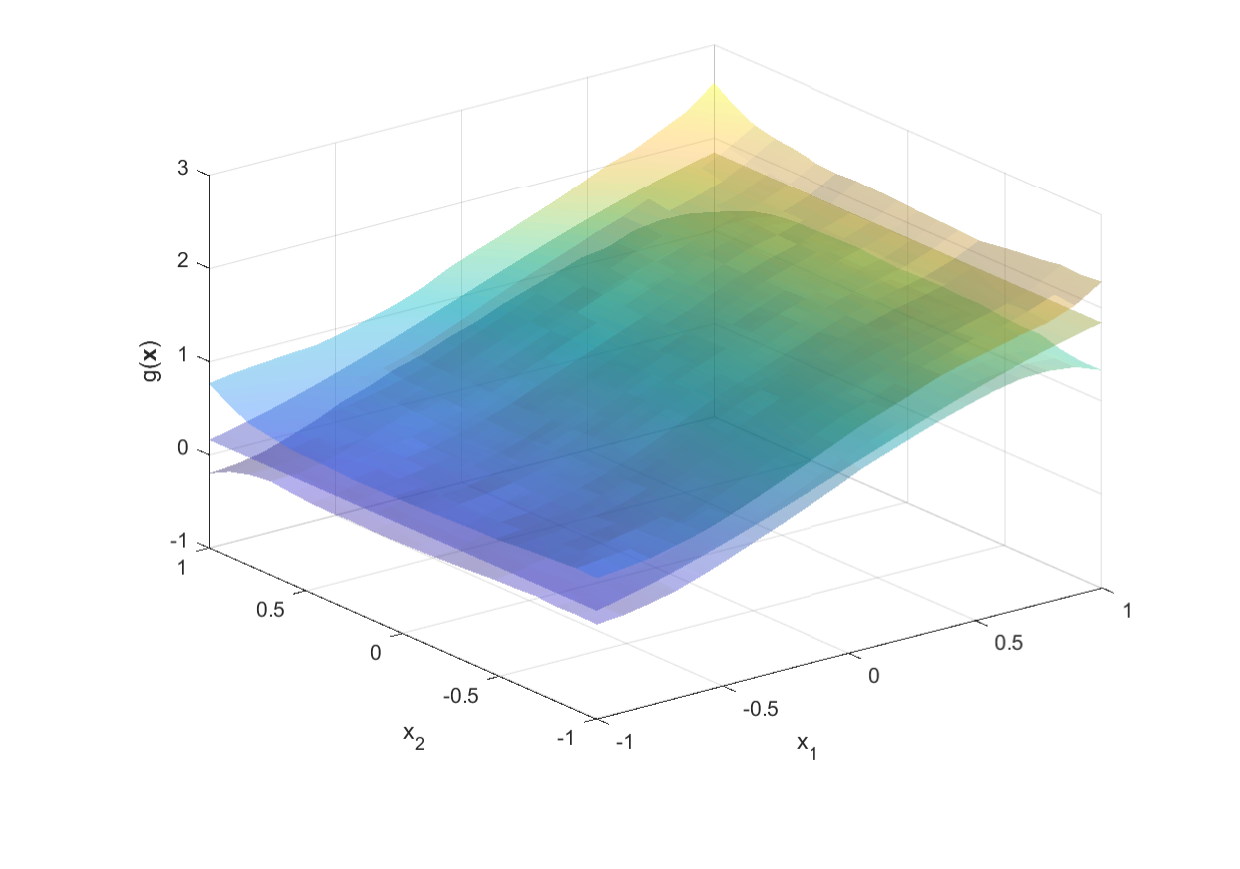}}
	\subfloat[$T=800$, $q=4$]
	{\includegraphics[width=0.5\textwidth]{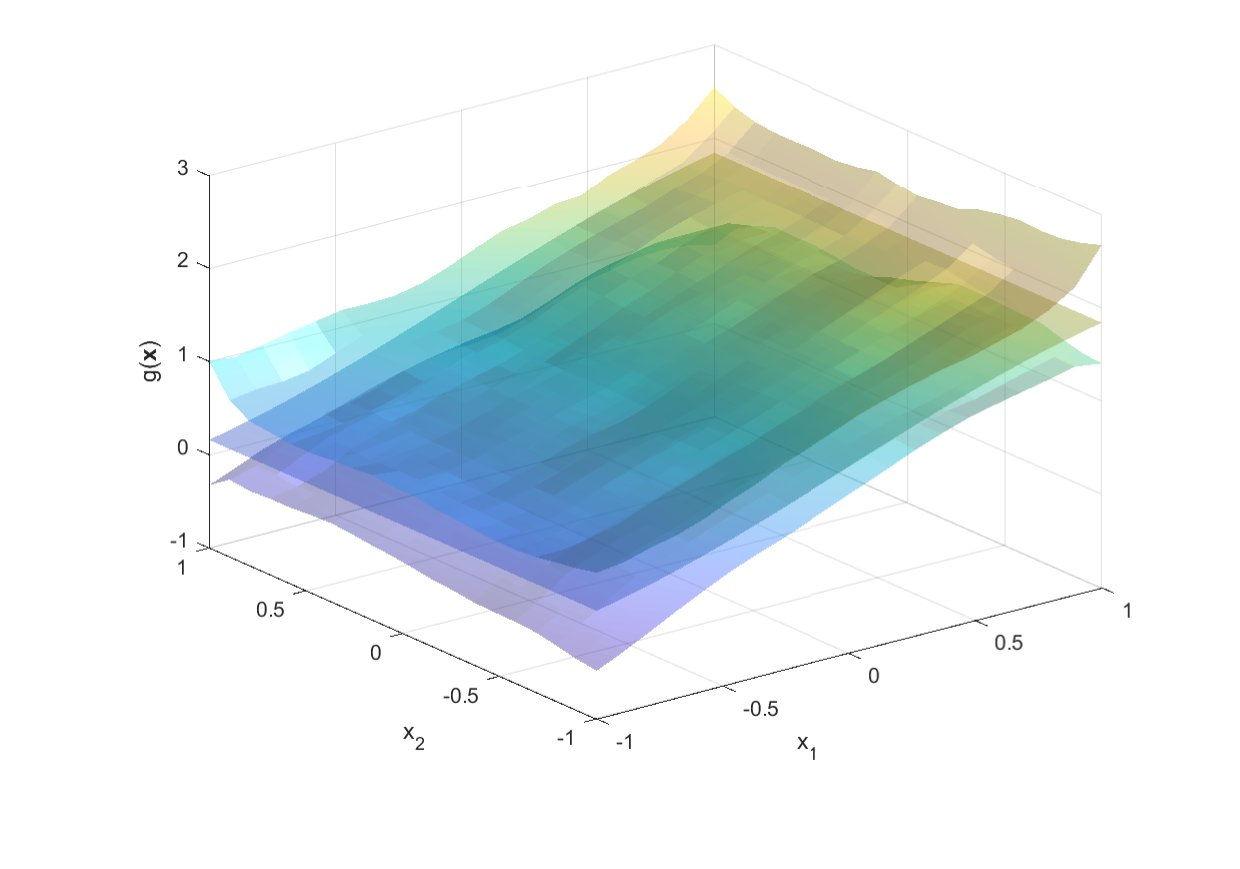}}\\
	\subfloat[$T=1600$, $q=3$]
	{\includegraphics[width=0.5\textwidth]{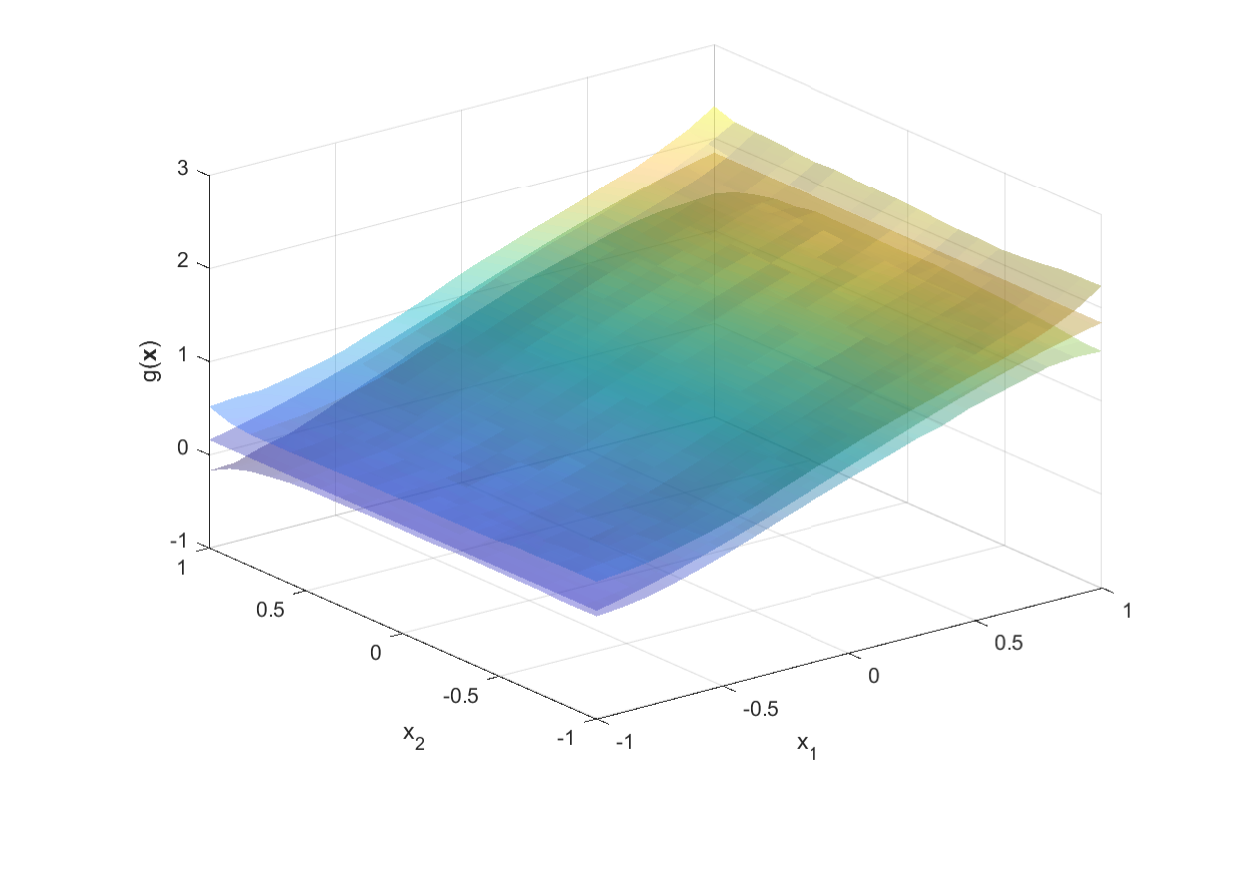}}
	\subfloat[$T=1600$, $q=4$]
	{\includegraphics[width=0.5\textwidth]{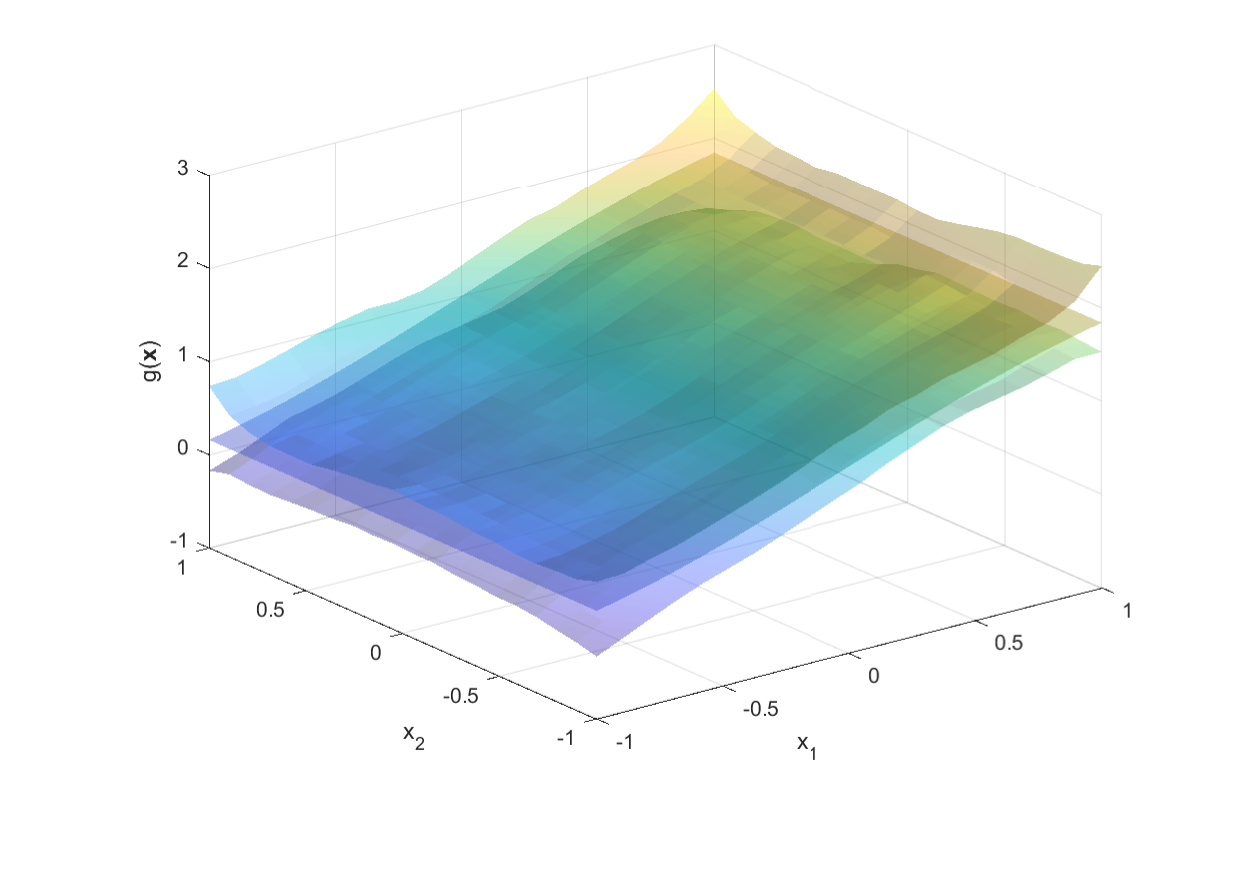}}\\
	\subfloat[$T=2400$, $q=3$]
	{\includegraphics[width=0.5\textwidth]{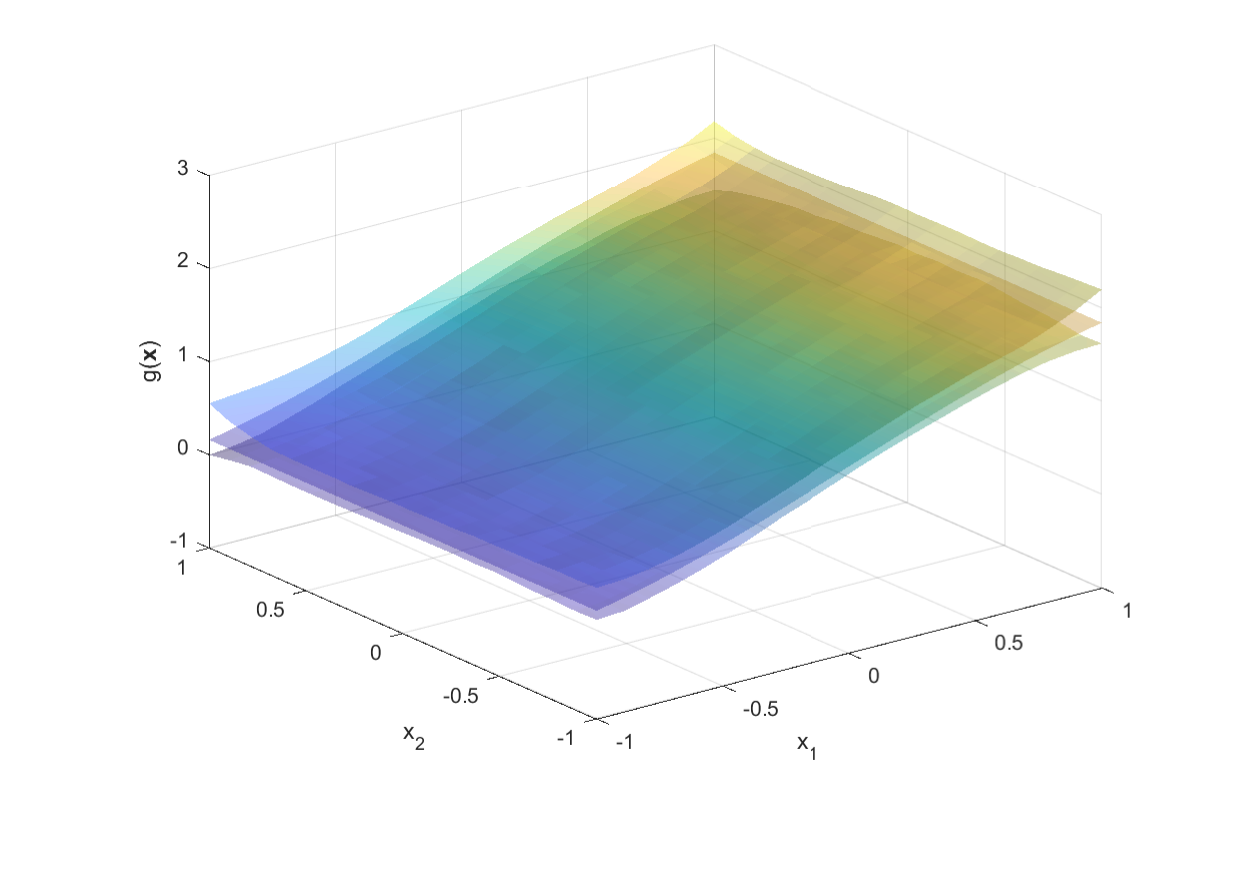}}
	\subfloat[$T=2400$, $q=4$]
	{\includegraphics[width=0.5\textwidth]{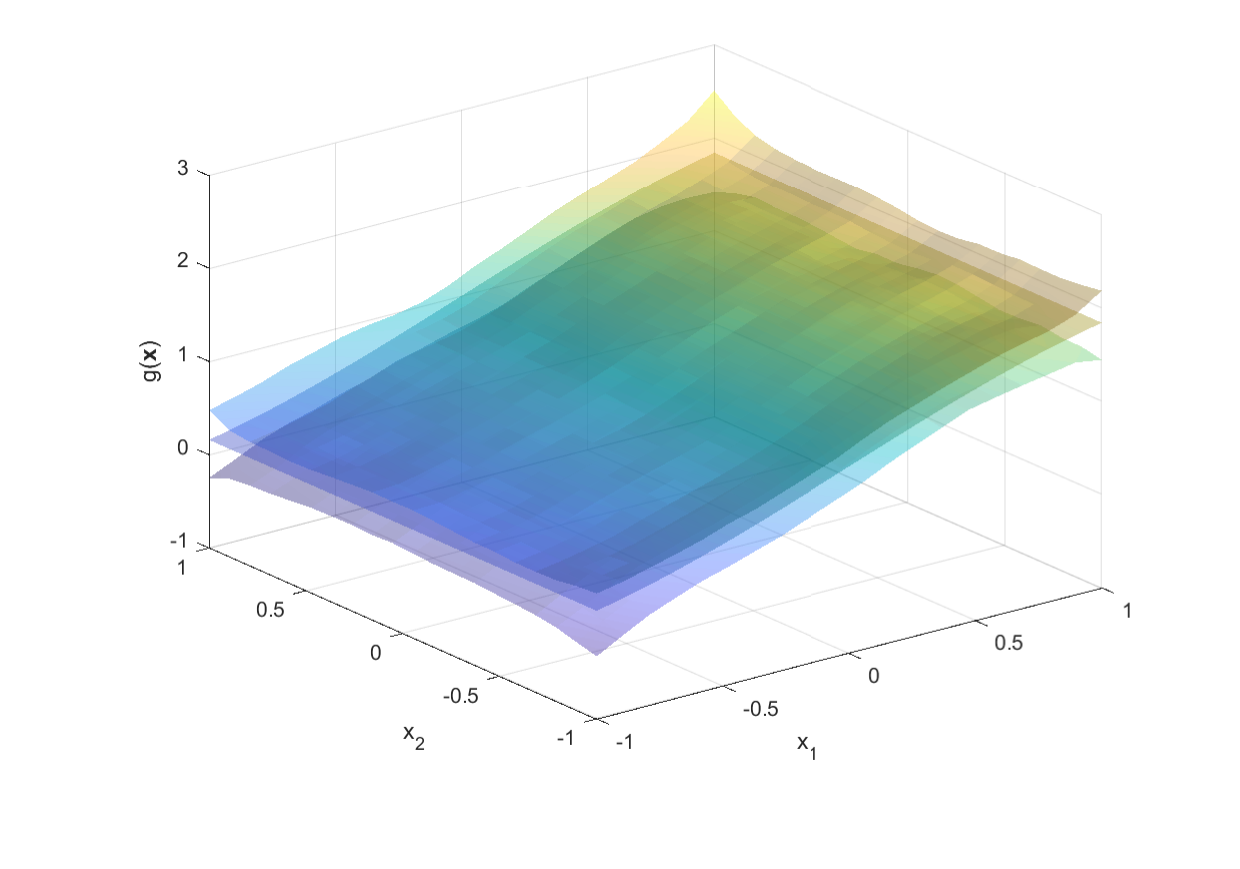}}\\
	\centering
	\caption{\textbf{Confidence intervals  with $d=2$ and $\mu_\sigma=0.5$.} In each plot, the central surface corresponds to the true function, while the upper and lower surfaces depict the averages of the 95\% confidence intervals based on 200 simulations}
	\label{Simplot1}
\end{figure}

\begin{figure}[htp!]
	\centering
	\subfloat[$T=800$, $q=3$]
	{\includegraphics[width=0.5\textwidth]{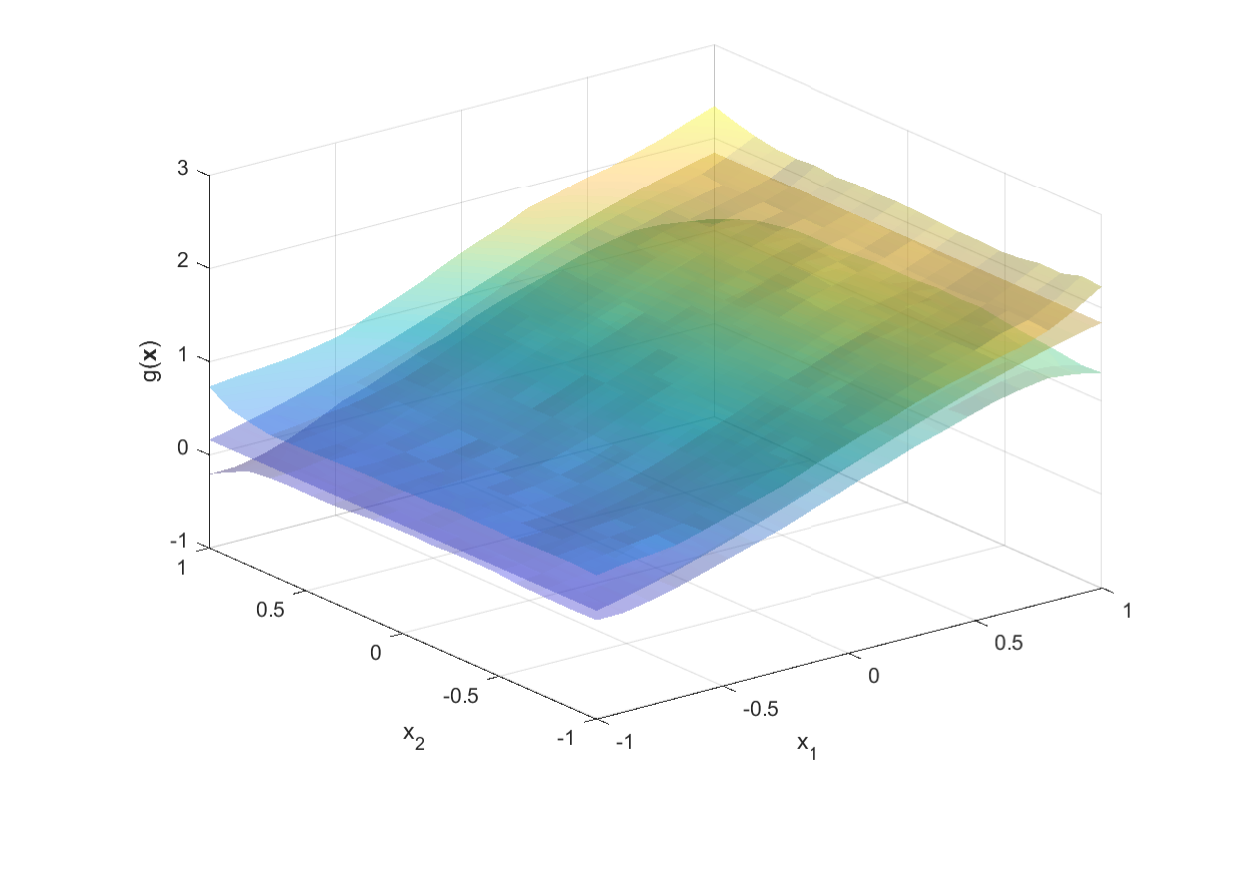}}
	\subfloat[$T=800$, $q=4$]
	{\includegraphics[width=0.5\textwidth]{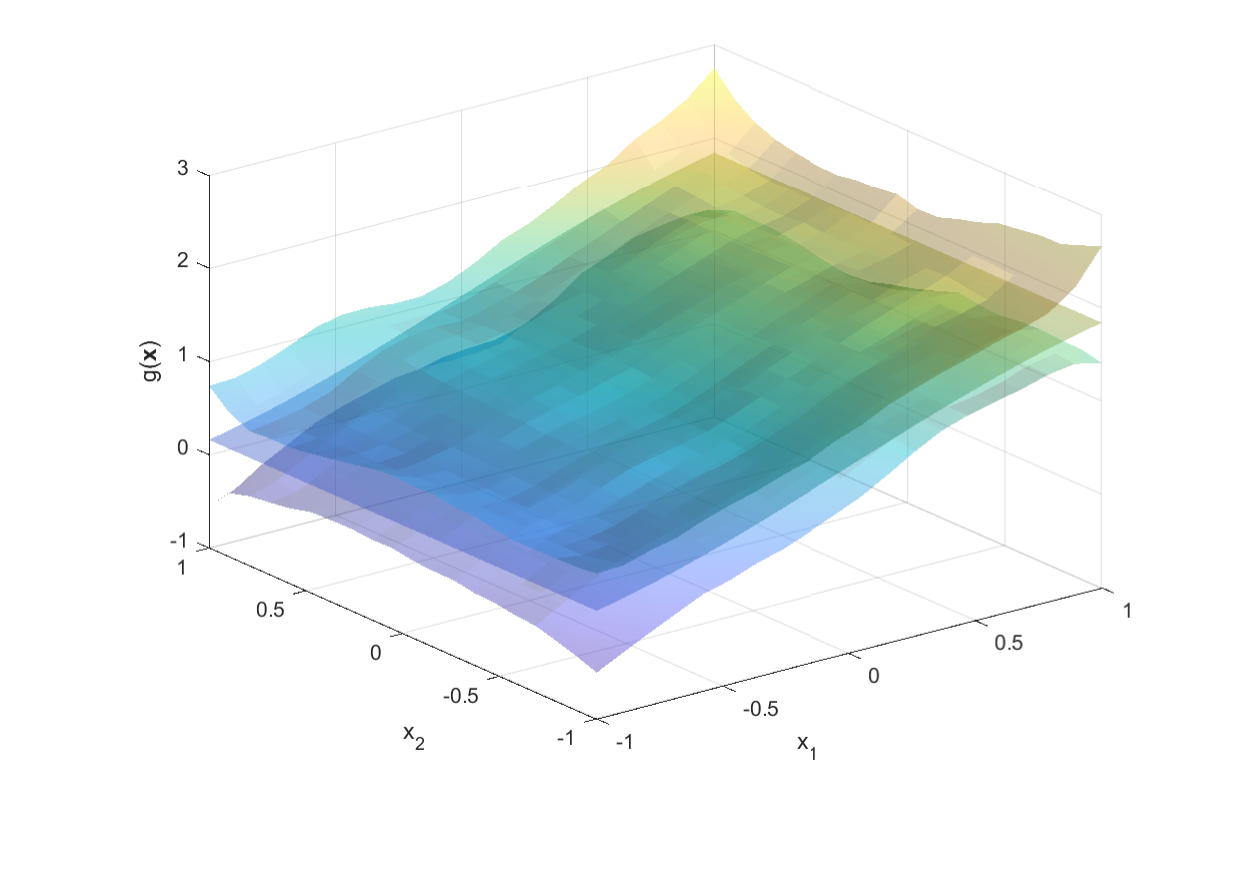}}\\
	\subfloat[$T=1600$, $q=3$]
	{\includegraphics[width=0.5\textwidth]{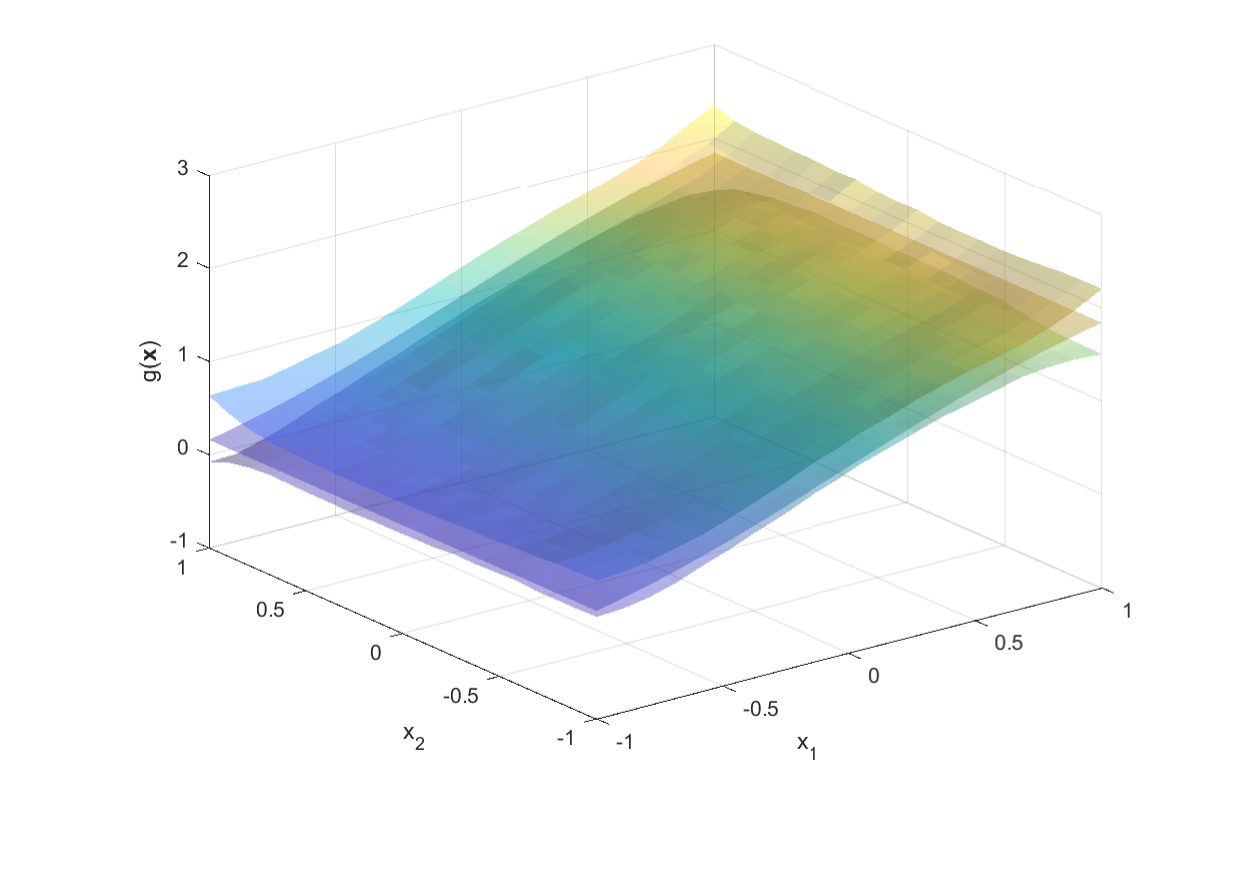}}
	\subfloat[$T=1600$, $q=4$]
	{\includegraphics[width=0.5\textwidth]{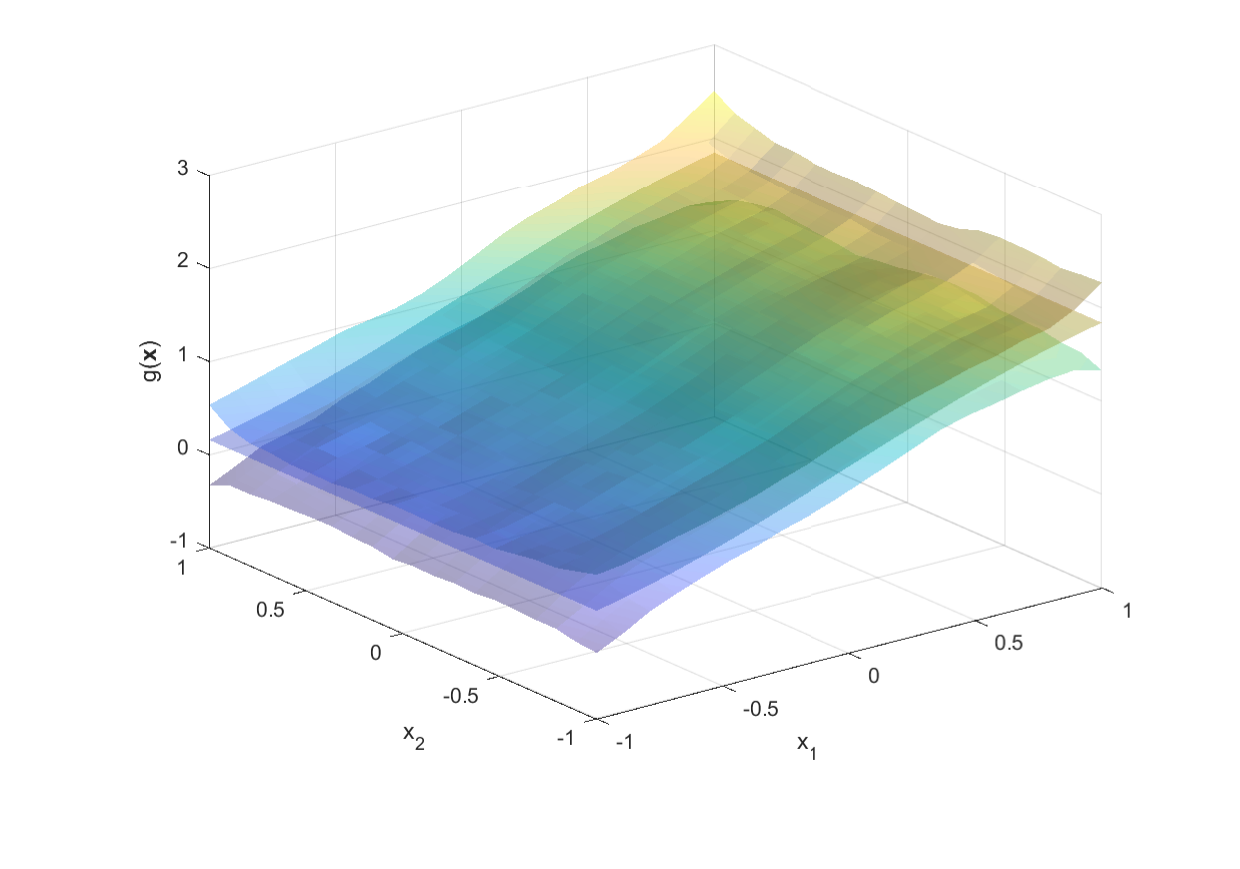}}\\
	\subfloat[$T=2400$, $q=3$]
	{\includegraphics[width=0.5\textwidth]{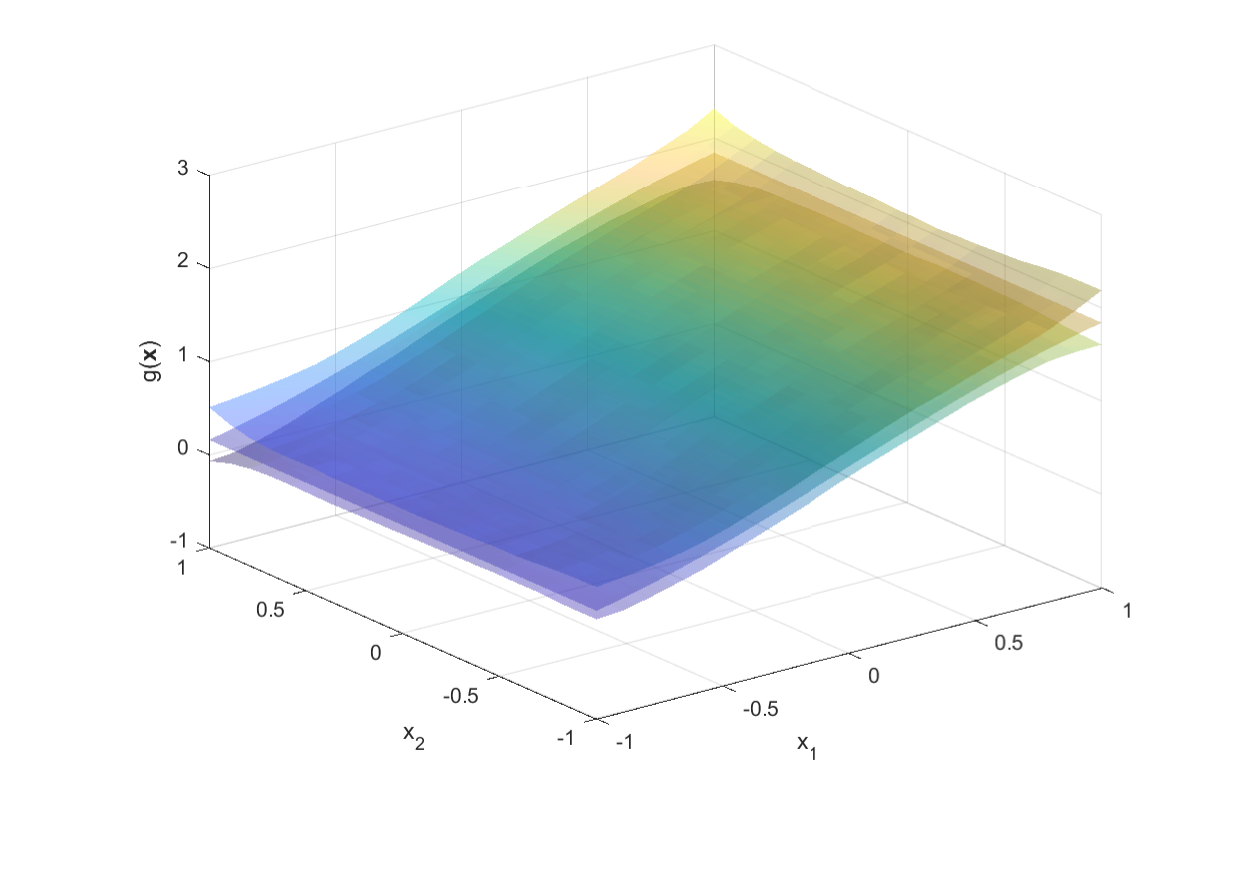}}
	\subfloat[$T=2400$, $q=4$]
	{\includegraphics[width=0.5\textwidth]{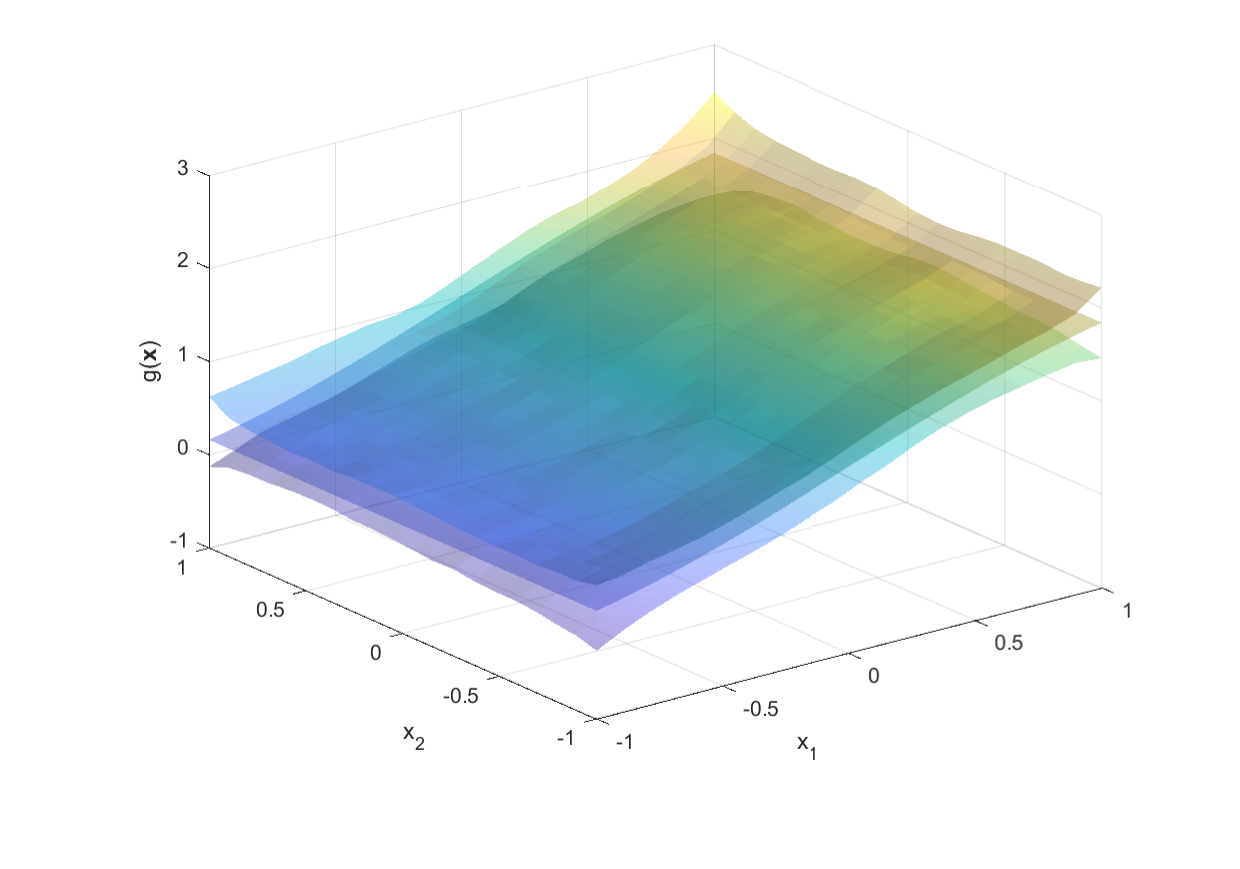}}\\
	\centering
	\caption{\textbf{Confidence intervals  with $d=2$ and $\mu_\sigma=-0.5$.} In each plot, the central surface corresponds to the true function, while the upper and lower surfaces depict the averages of the 95\% confidence intervals based on 200 simulations}
	\label{Simplot2}
\end{figure}

To further illustrate the proposed bootstrap method, we present the plots of 95\% bootstrap confidence intervals of $g(\cdot)$ for the cases $(d,\mu_\sigma)=(2,0.5)$ and $(d,\mu_\sigma)=(2,-0.5)$ in Figures \ref{Simplot1} and \ref{Simplot2}, respectively. For enhanced clarity, we employ a denser grid of evaluation points $\mathbf{x}_{L,ij} = (-a+\frac{2ai}{L+1}, -a+\frac{2aj}{L+1} )$ for $i,j\in[L]$.  At each point, we apply the bootstrap procedure and compute the 95\% confidence interval.  The plots in Figures \ref{Simplot1} and \ref{Simplot2} demonstrate that the true functions are effectively covered across different choices of $T$ and $q$. Moreover, with the increasing sample size, the bootstrap confidence intervals exhibit clear convergence. Furthermore, $q=3$ seems to yield better coverage overall, and the choice of $u_\sigma$ has less impact on the inference.

\begin{center}

\fbox{Figures 2 and 3 near here}

\end{center}

Having demonstrated the finite sample performance of the newly proposed framework, we are now ready to examine the impacts of US monetary policy in the following section.

\section{A Case Study}\label{S4}

It is widely acknowledged that monetary policy shocks may have significant influence over macroeconomic variables. As  noted by \cite{Clarida2000}, ``{\it the difference in policy behaviour could be an important underlying source of the shift in macroeconomic behaviour}". Consequently, over the past few decades, there has been a surge of empirical and theoretical research to study the relationship between monetary policies and various macroeconomic indicators, including interest rate, unemployment rate, economic growth, and asset price. 
For example, \cite{Clarida2000} investigate the role of monetary policy in the macroeconomic stability by establishing connections between the policy interest rate with expected inflation; \cite{Bernanke2005} explore how the stock index responds to the  unanticipated monetary policy actions; \cite{Blanchard2010} study the policy effects on the relationship between inflation and unemployment rate; etc.

\subsection{The Monetary Policy Effect Model}

As widely recognized in the literature, a fundamental question in this field is how to characterize monetary policy shocks or unanticipated policy changes  (see \citealp{Christians1996,Bernanke2005}; among others). A commonly adopted approach involves using the disturbance in a regression of monetary policy indicators on lagged observable macroeconomic variables to capture the information conveyed by policy actions that cannot be anticipated by the market. Then, the response of the macroeconomic variables to the policy shock can be measured through another regression between such variables.
In this section, we adopt a nonparametric specification for the monetary policy evolution, which can be regarded as a generalization of the linear model employed by \cite{Christians1996}. Specifically,
\begin{equation}\label{empbinary}
P_t=I( G_P(\mathbf{x}_{t})-\eta_t \ge 0 ),
\end{equation}
where $P_t$ is an indicator variable for the shift to a more tightening monetary policy,  $\mathbf{x}_{t}$ contains the macroeconomic predictors available when 
$P_t$ is determined, $G_P(\cdot)$ is a nonparametricalyy unknown function, and $\eta_t$ is a disturbance term. We then have
\begin{eqnarray*}
\Pr(P_t=1\, |\, \mathbf{x}_t) =\Phi_\eta(G_P(\mathbf{x}_t)),
\end{eqnarray*}
where $\Phi_\eta(\cdot)$ denotes the CDF of $\eta_t$. Therefore, $\Phi_\eta(G_P(\mathbf{x}_t))$ captures the market anticipation of a tightening policy action by the authority. This concept is in line with the idea of policy propensity scores adopted by \cite{Angrist2018}.
Then, the unanticipated monetary policy shock $S_t$ can be specified as follows:
\begin{equation}\label{Empmodel1}
S_t=P_t-\Phi_\eta(G_P(\mathbf{x}_t)).
\end{equation}
We can then characterize its relationship with macroeconomic or financial outcome variables ($M_t$) through the following semiparametric model: 
\begin{equation}\label{Empmodel2}
M_{t+1}=S_t\alpha_P+G_M(\mathbf{x}_{t})+\varepsilon_t,
\end{equation}
where the parameter $\alpha_P$ captures the effects of unanticipated  policy shocks and $G_M(\cdot)$ is another nonparametric function that controls the influence from the lagged macroeconomic predictors. By substituting $S_t$ from \eqref{Empmodel1} into \eqref{Empmodel2}, we obtain 
\begin{eqnarray}\label{Empmodel3}
M_{t+1}=P_t\alpha_P+g_M(\mathbf{x}_{t})+\varepsilon_t,
\end{eqnarray}
where $g_M(\mathbf{x}_{t})=G_M(\mathbf{x}_{t})-\Phi_\eta(G_P(\mathbf{x}_t))\alpha_P$.   The setup of \eqref{Empmodel3} enables us to estimate the policy effects $\alpha_P$ using the methodology that is proposed in Section \ref{S2.2}.\footnote{In the online supplement, we provide an LNN-based method to estimate the nonparametric binary model  \eqref{empbinary}. Then, using the estimators $\widehat{G}_P(\mathbf{x}_t)$ and $\widehat{g}_M(\mathbf{x}_{t})$ that are obtained by estimating \eqref{empbinary} and \eqref{Empmodel3}, respectively, we can recover  $G_M(\mathbf{x}_{t})$ in \eqref{Empmodel2} by $\widehat{G}_M(\mathbf{x}_{t}) = \widehat{g}_M(\mathbf{x}_{t})+\Phi_\eta(\widehat{G}_P(\mathbf{x}_t))\widehat{\alpha}_P$. 

An alternative approach to recovering the policy effects is to estimate  \eqref{empbinary}  and \eqref{Empmodel2} sequentially.  However, this procedure involves an essential step where we have to replace the unobservable unanticipated policy shock $S_t$ in  \eqref{Empmodel2} with its estimator $\widehat{S}_t=P_t-\Phi_\eta(\widehat{G}_P(\mathbf{x}_t))$   to construct a feasible estimator for $\alpha_P$. This substitution inevitably introduces additional approximation errors compared to the direct estimation of \eqref{Empmodel3}. }
Compared with the traditional parametric models of monetary policy effects,  our framework offers several advantages.   First, the relationship between variables is not constrained to be linear, allowing for more flexible and realistic modelling of complex interactions. Second, our approach is capable to accommodate a large set of control variables and automatically detect the insignificant predictors.

In what follows, we revisit the impacts of a tightening monetary policy action on a variety of economic variables, including short-/long-term interest rate, inflation, unemployment rate, industrial price and equity return. We accomplish this by employing the proposed semiparametric model with treatment effects on a monthly dataset of the US. 

\subsection{Variables and Data}

In the literature  \citep[e.g.,][]{Christians1996,Romer2000,Cochrane2002}, a commonly used indicator of monetary policy shifts is the change in the federal funds target rate which is announced during Federal Reserve Open Market Committee (FOMC) meetings. Accordingly, we construct an indicator variable to capture the tightening or easing monetary policy whenever there is an increase in the announced target rate. We examine the effects of these policy changes on key macroeconomic and financial outcome variables that have been extensively studied in the literature. The sample period considered in this study is from January, 1989 to September, 2015.

Specifically, we first analyze the policy effects on a variety of interest rates: the effective federal funds rate (FFR) and treasury bond yields quoted on an investment basis at 3-month (Yield3m), 1-year (Yield1y), 2-year (Yield2y), 5-year (Yield5y), and 10-year (Yield10y) maturities. For these variables, we obtain their monthly average data  from the FRED, Federal Reserve Bank of St. Louis.
Following the approach of \cite{Angrist2018}, we investigate the effects on  inflation, unemployment rate, and industrial output, which are measured by (change in the log value of) the Personal Consumption Expenditures Price Index (PCE), (change in) the Civilian Unemployment Rate (UNRATE), and (change in the log value of) the Industrial Production Index (IP), respectively. The data for these variables are also collected from FRED.
In addition to these macroeconomic variables, we  follow \cite{Bernanke2005} to study the  value-weighted return of the S$\&$P500 stocks as a representative of equity returns and the data is sourced from the CRSP index series database at WRDS. 

To isolate the effects from the anticipation of the shifts in monetary policy, we incorporate a set of control variables  that are sourced from three categories:
(i) Monetary policy persistence: we measure the persistence of monetary policy using lagged federal funds target rate and real target rate changes (TRC), along with an indicator variable for FOMC meeting occurrences. (ii) Economic conditions: we include the first and second lags of inflation and unemployment rate to reflect the potential influence of underlying economic conditions on monetary policy changes.
(iii) Market expectations: we utilize the federal funds future (FFF) index developed by \cite{Angrist2018} to capture market expectations regarding future monetary policy decisions
\footnote{To measure market expectations of  target rate changes, \cite{Angrist2018} develop the FFF variable using  federal funds rate derivatives with specific adjustments made for data during FOMC meeting months.  For detailed information about the construction of this index, we refer the readers to \cite{Angrist2018}.}.

\begin{center}

\fbox{Table 2 near here}

\end{center}

We present a summary of the aforementioned variables along with their descriptive statistics and unit root test results in Table \ref{Tab_DES} below. As shown in the table, all variables are stationary at the 5\% significance level, except 10-year treasury bond yields which is stationary at the 10\% significance level. Therefore, these variables fit our assumption reasonably well.

\subsection{Estimation Results}

Using the methodology outlined in Section \ref{S2.2}, we investigate the response of macroeconomic and financial variables in the subsequent month following the announcement of a  federal funds target rate increase. To set up the LNN architecture,  we specify $h=T^{-1/(d+2p-0.5)}$, $\mu_\sigma=0.5$ and $q=3$.
The estimated policy effects captured by $\widehat{\alpha}_P$ and their bootstrap confidence intervals are reported in Panel A of Table \ref{Emptable1}.

Our estimation results reveal that an unanticipated increase in the target rate exerts a significant and positive influence on the federal funds rate and on the treasury bond yield with the maturity of 3 months.
However, the effect is statistically insignificant for treasury bond yields with maturities that are more than 1 year, suggesting that tightening monetary policy has a more pronounced impact on short-term interest rates compared to long-term rates. This aligns with previous studies by \cite{Cochrane2002} and \cite{Angrist2018}, who also observe diminishing effects along the yield curve with increasing maturity. Controlling for anticipated changes in the target rate, our estimation indicates no significant effects of monetary policy shocks on changes in inflation, unemployment rate, or industrial price. These findings are consistent with those reported by \cite{Angrist2018}.
Table \ref{Tab_Insig} presents information about the insignificant predictors identified by the LNN-based group-LASSO method. As shown in Table 3, different numbers of significant predictors, ranging from one to five, are detected for FFR, Yield1y, Yield2y, PCE, UNRATE, and IP. This result demonstrates the empirical significance of the proposed method. 

\begin{center}

\fbox{Tables 3 and 4 near here}

\end{center}

We then explore the sensitivity of our results by varying the set of control variables, specifically by excluding the FFF or both FFF and lagged values of PCE and UNRATE. The results, which are also presented in Panel A of Table \ref{Emptable1}, demonstrate that policy effects appear to be more pronounced with fewer control variables, which is likely due to the inclusion of the influence from anticipated monetary policy changes. Notably, the unemployment rate response becomes significantly negative when FFF is not controlled for. Furthermore, without controlling for FFF and the lagged economic indicators, a noticeable short-term ``price puzzle'' in the literature \citep[see, e.g.,][]{Sims1992} emerges --- a temporary increase in price in response to higher interest rate. 

\begin{figure}[htp!]
	\centering
	\subfloat[FFR]
	{\includegraphics[width=0.5\textwidth]{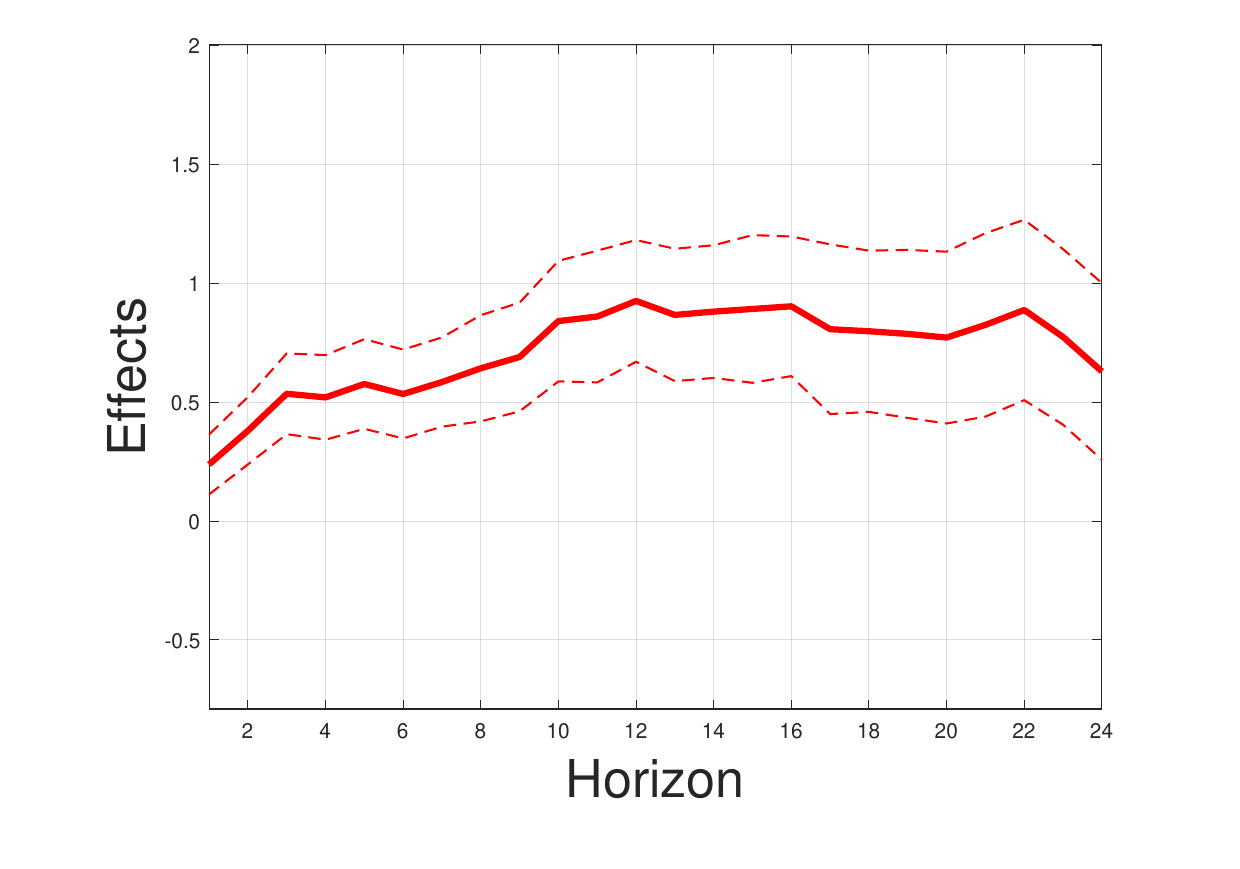}}
	\subfloat[Yield3m]
	{\includegraphics[width=0.5\textwidth]{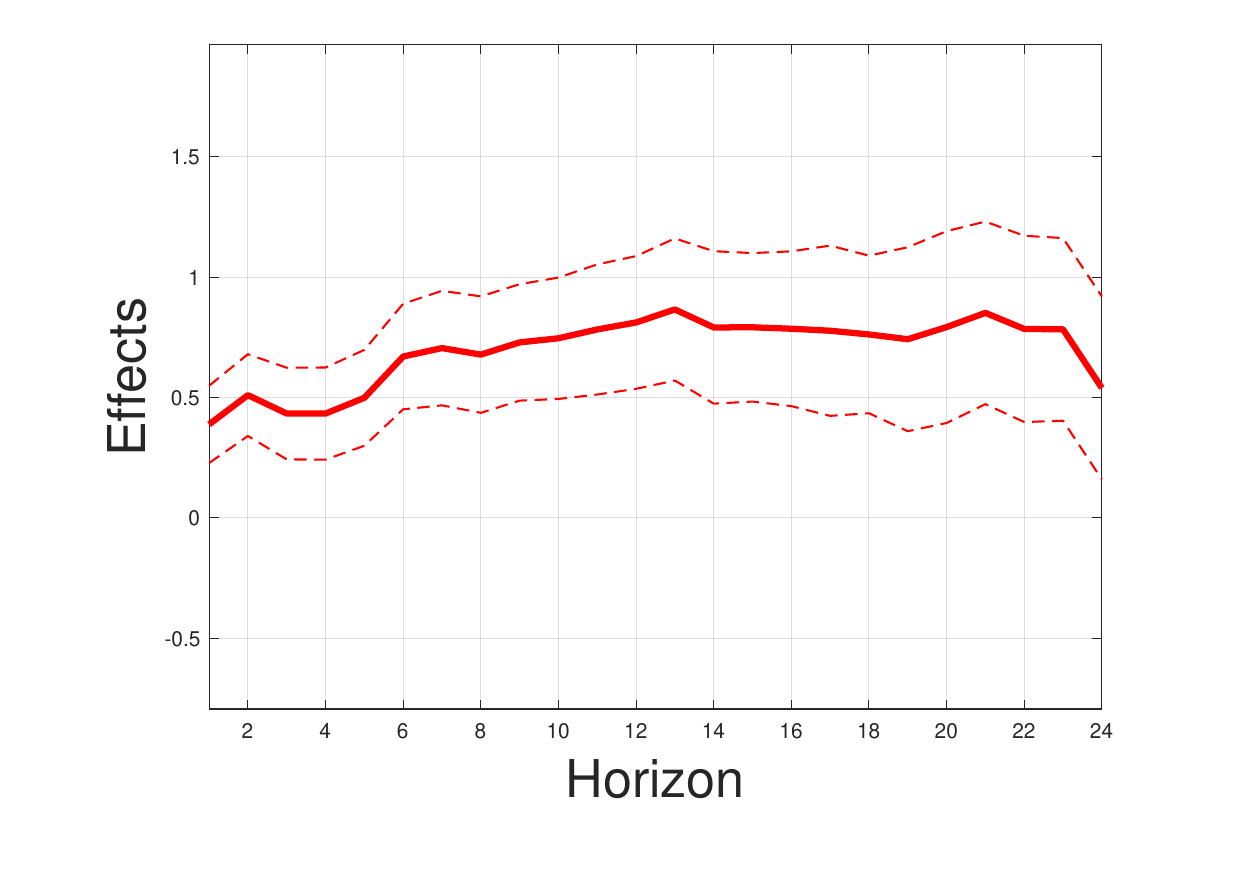}}\\
	\subfloat[Yield1y]
	{\includegraphics[width=0.5\textwidth]{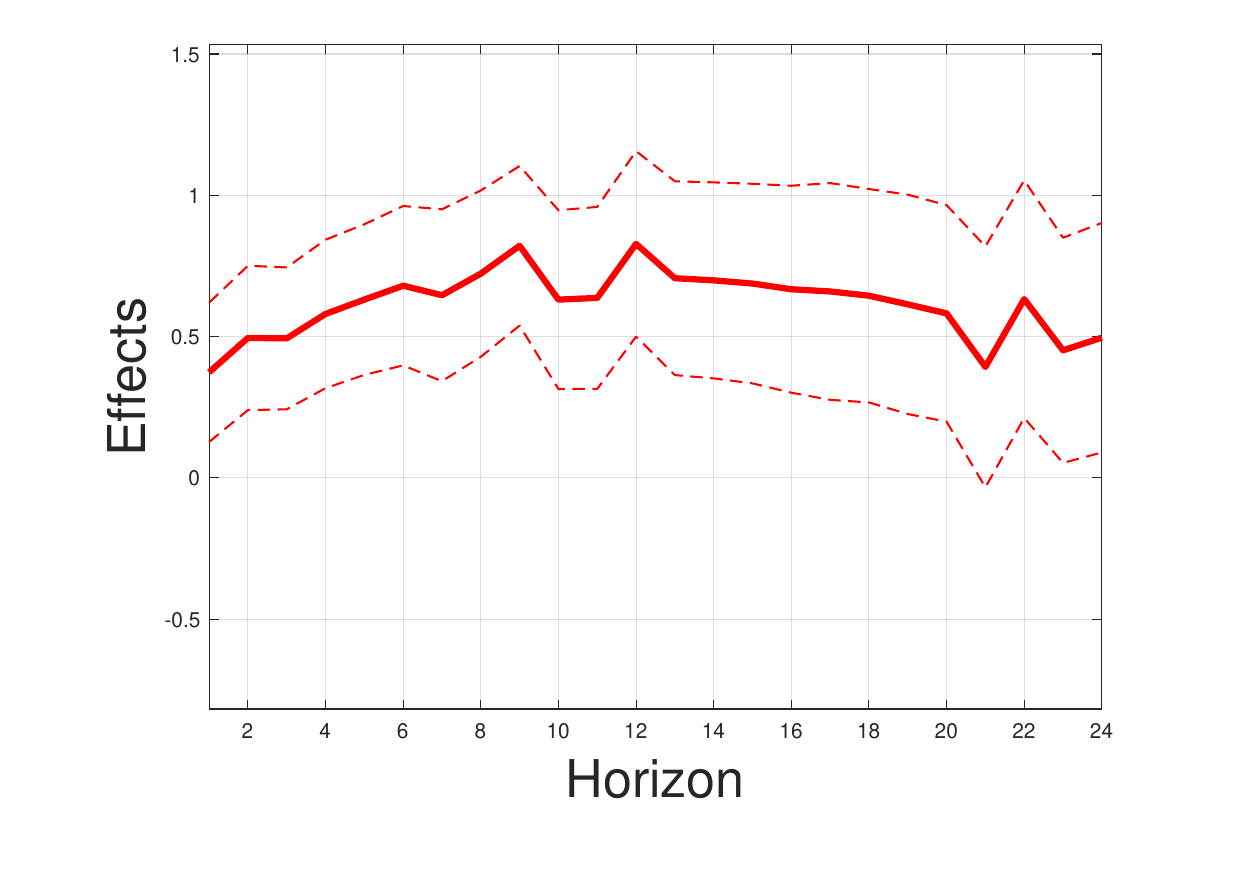}}	
	\subfloat[Yield2y]
	{\includegraphics[width=0.5\textwidth]{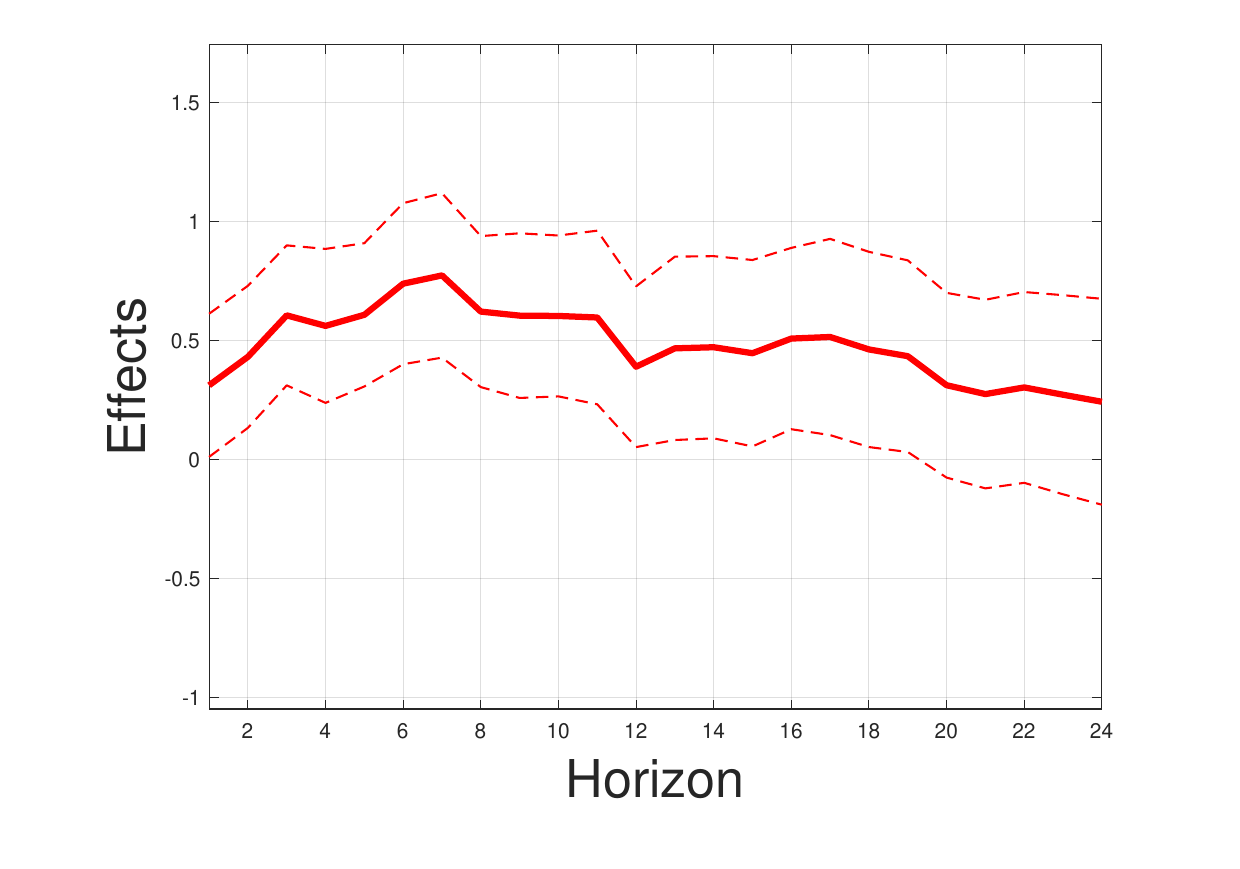}}\\
	\subfloat[Yield5y]
	{\includegraphics[width=0.5\textwidth]{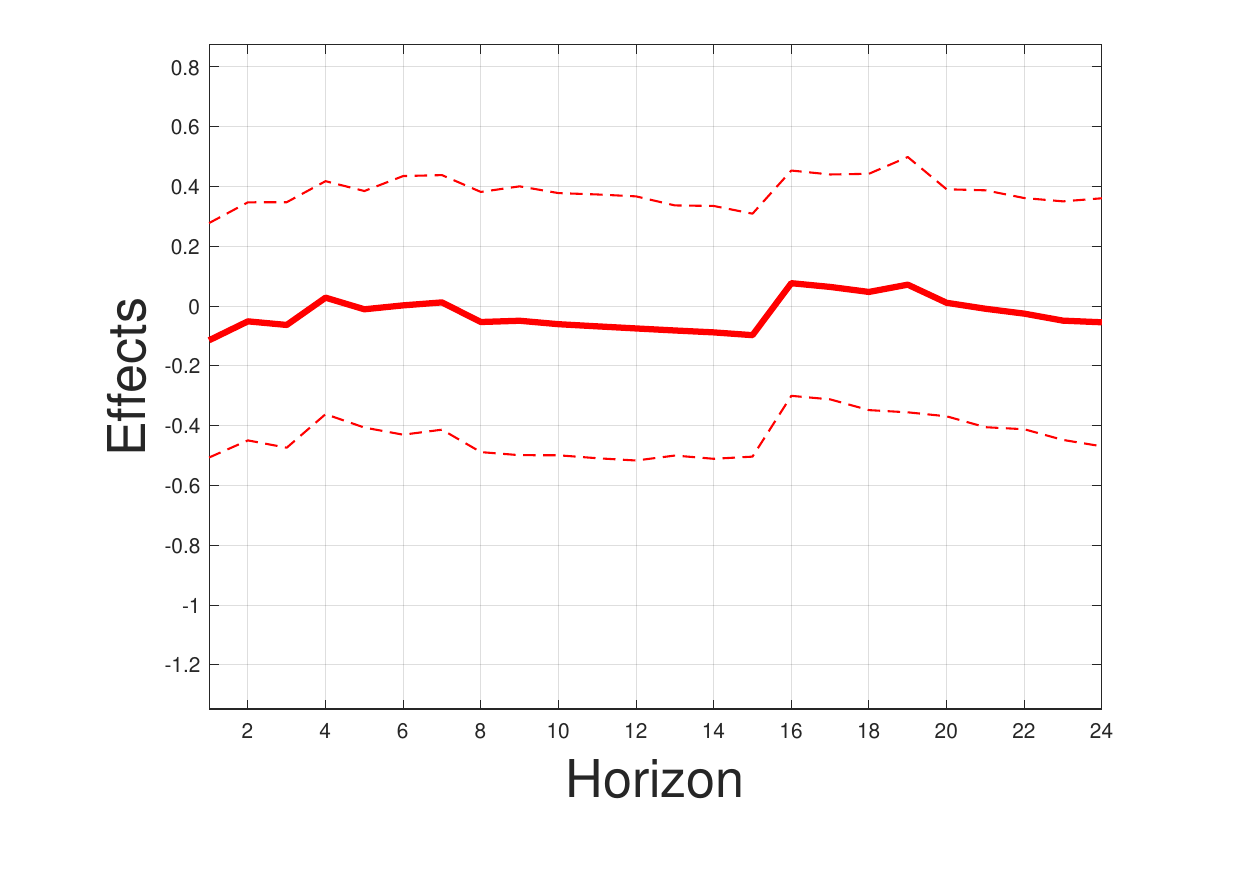}}
	\subfloat[Yield10y]
	{\includegraphics[width=0.5\textwidth]{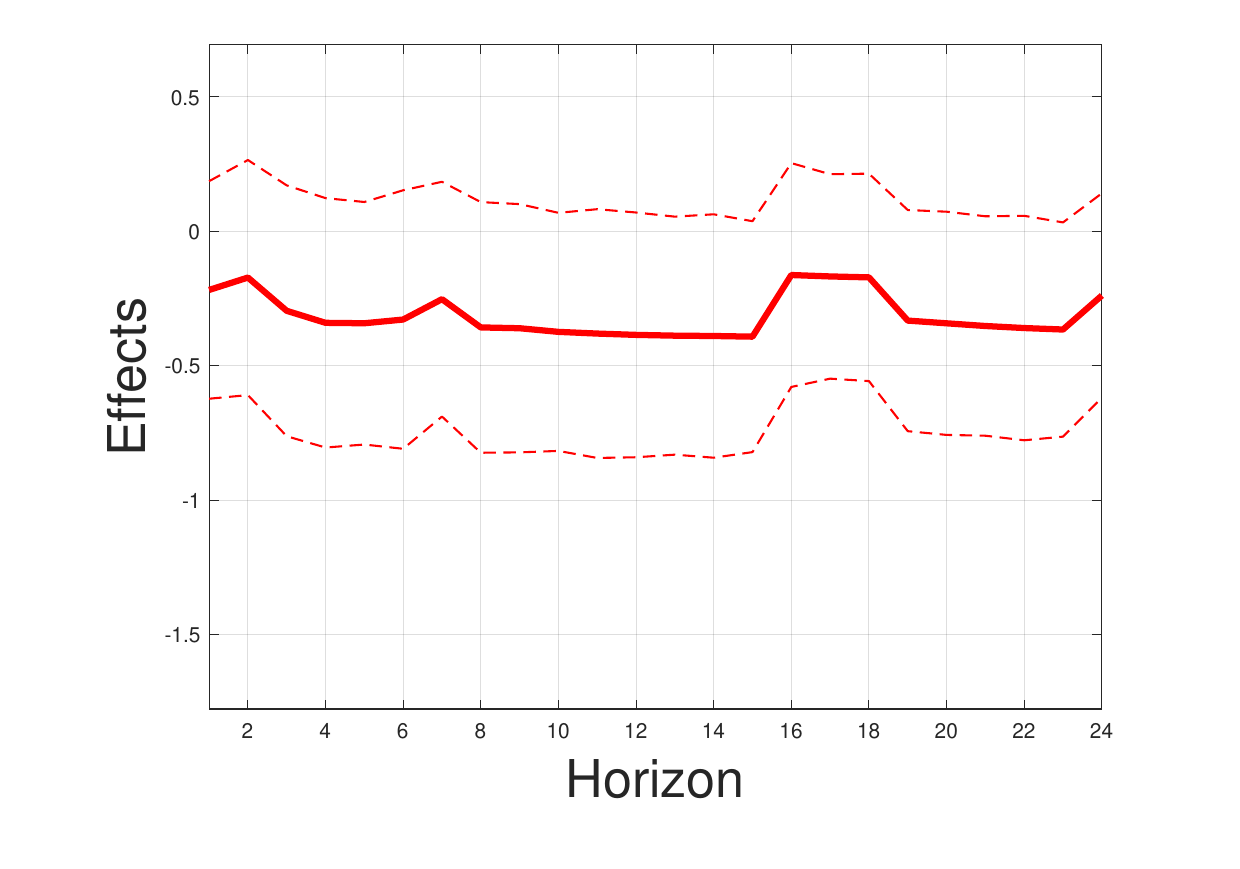}}\\
	\caption{\textbf{Estimated effects of the monetary policy change on federal funds rate and bond yields.} In each plot, the central solid line corresponds to the estimate, while the upper and lower dashed lines represent the the 90\% bootstrap confidence intervals. The sample period is from January, 1989 to September, 2015.}
	\label{Emplot1}
\end{figure}

\begin{figure}[htp!]
	\centering
	\subfloat[PCE]
	{\includegraphics[width=0.5\textwidth]{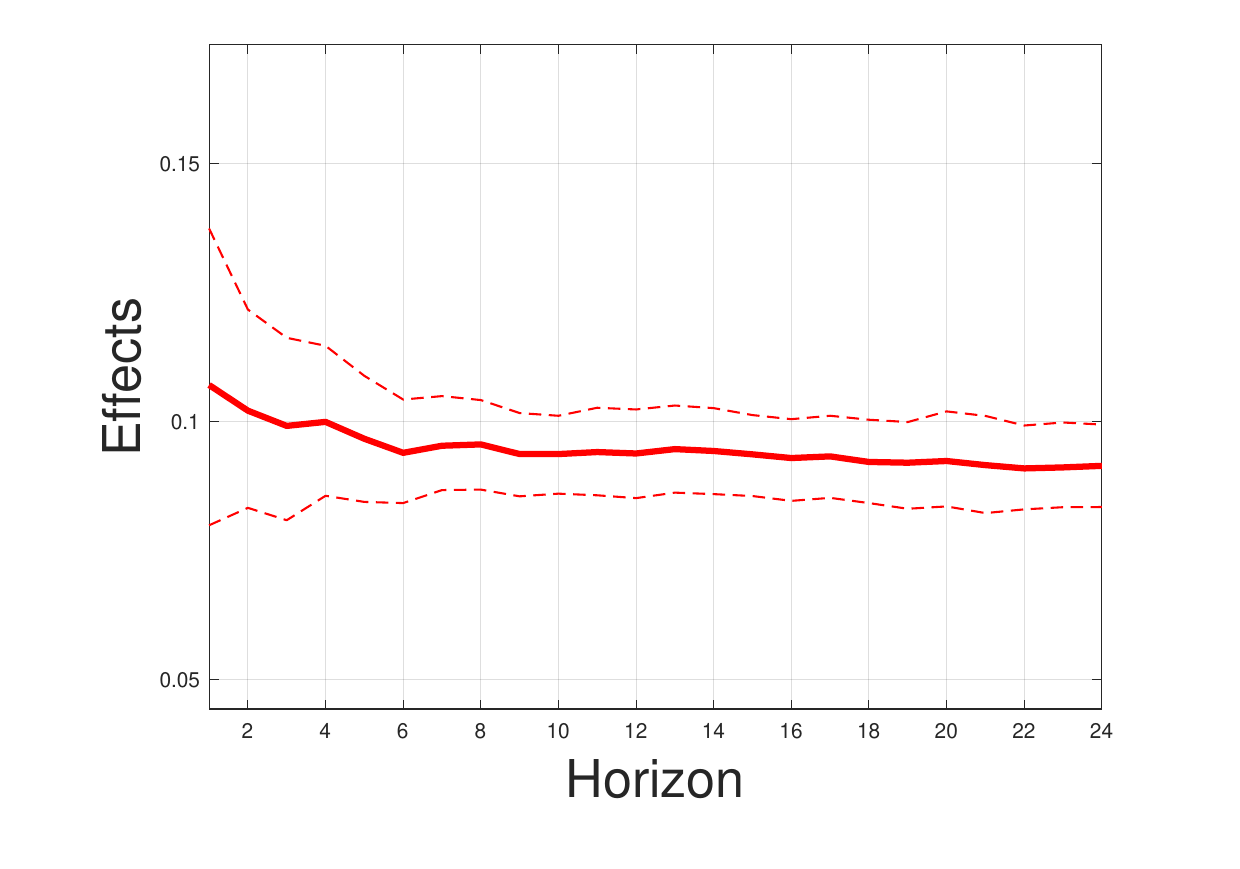}}
	\subfloat[UNRATE]
    {\includegraphics[width=0.5\textwidth]{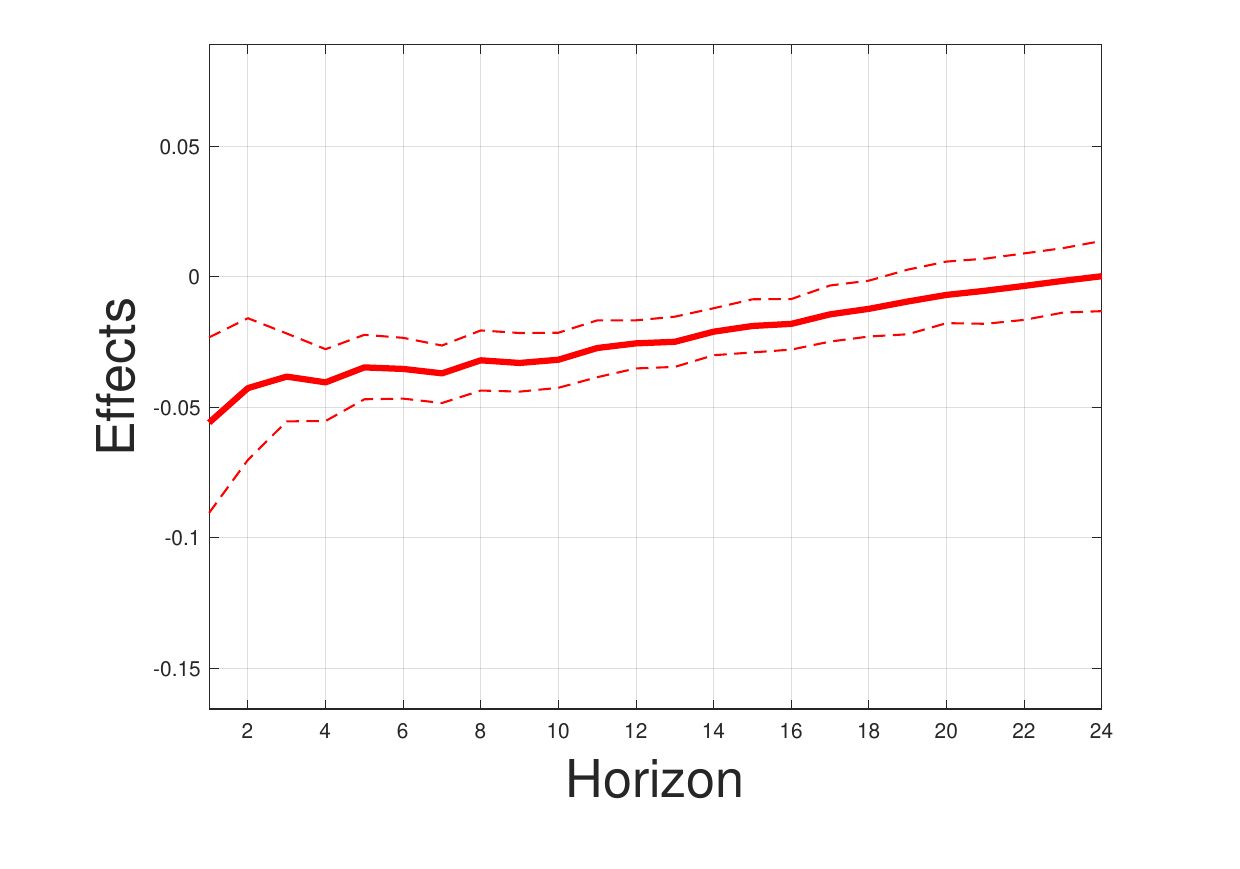}}\\	
	\subfloat[IP]
	{\includegraphics[width=0.5\textwidth]{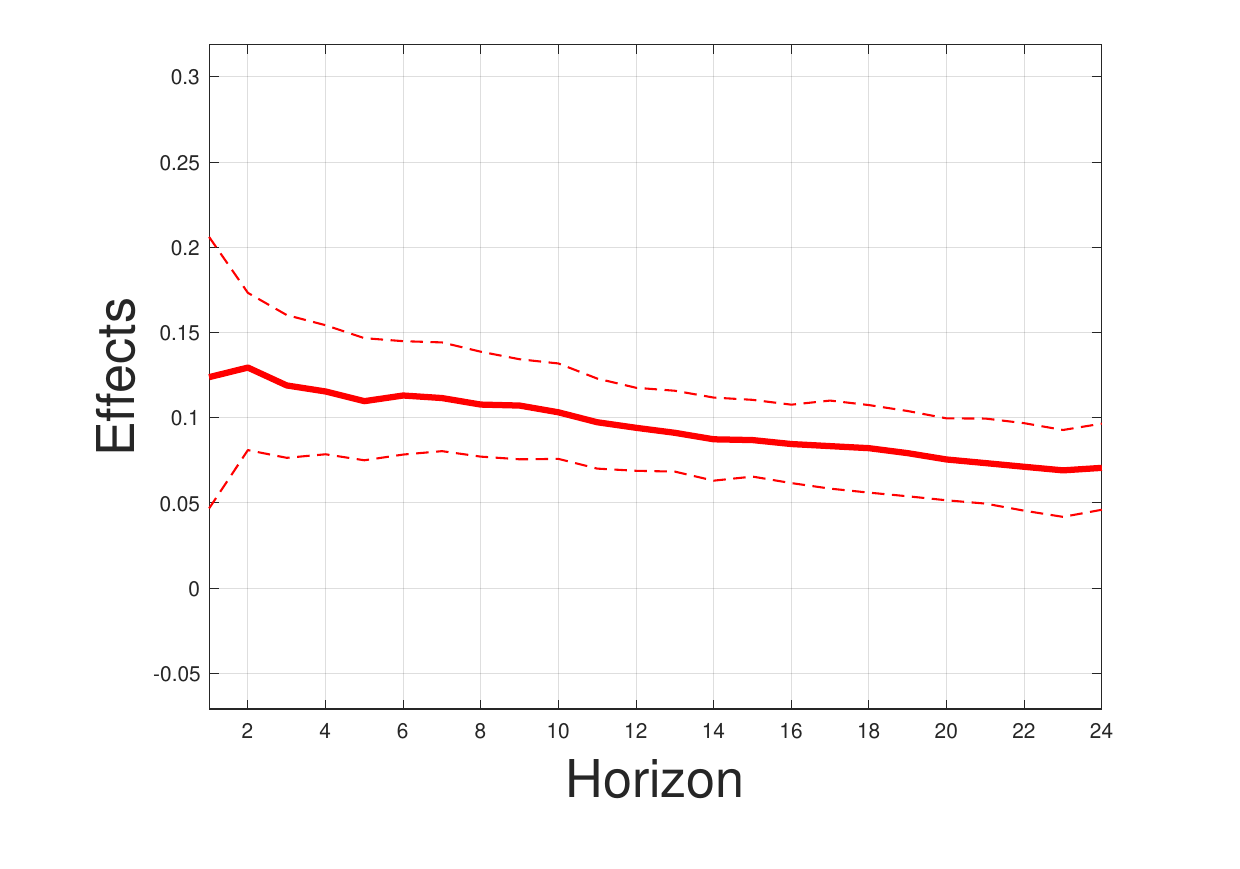}}
	\subfloat[SPR]
	{\includegraphics[width=0.5\textwidth]{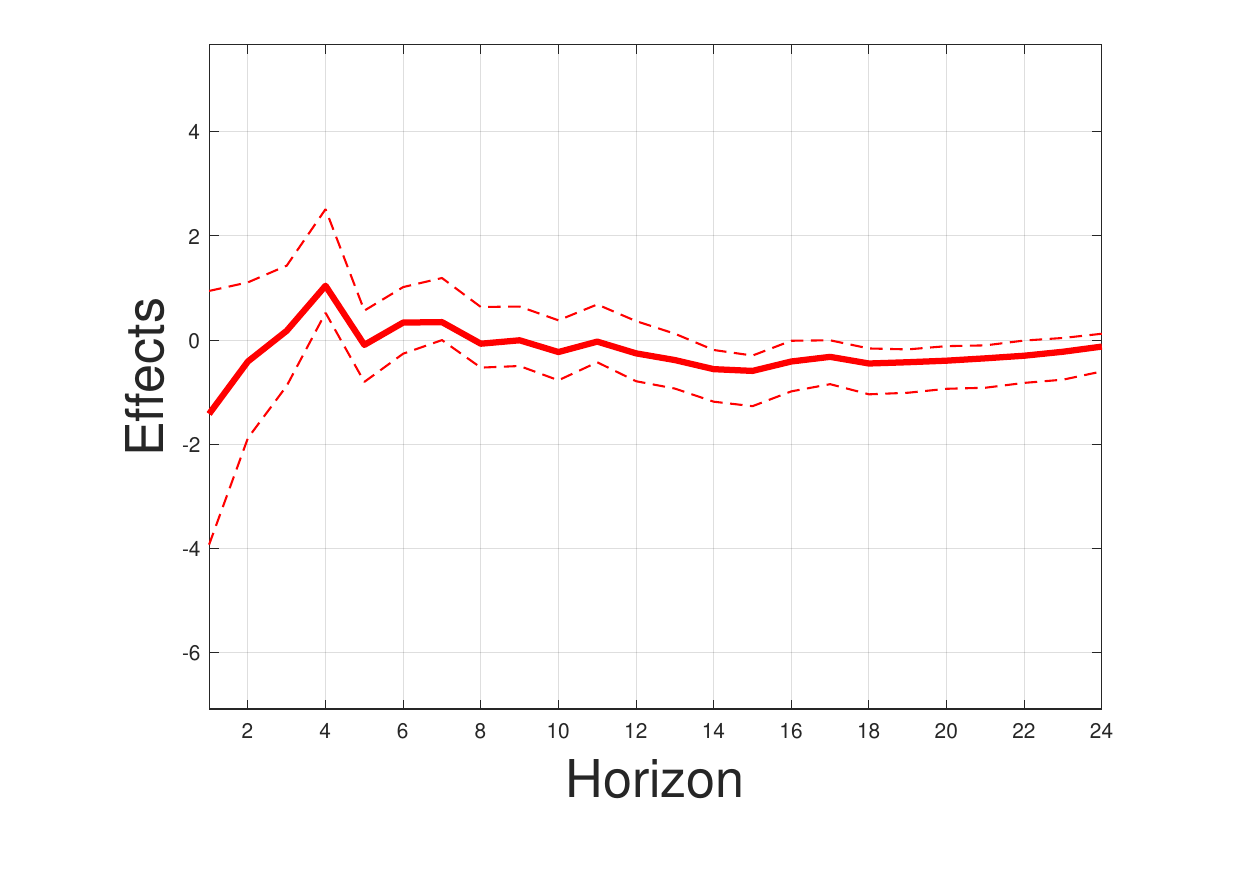}}
	\\
	\caption{\textbf{Estimated effects of the monetary policy change on macroeconomic variables and equity return.} In each plot, the central solid line corresponds to the estimate, while the upper and lower dashed lines represent the the 90\% bootstrap confidence intervals. The sample period is from January, 1989 to September, 2015.}
	\label{Emplot2}
\end{figure}

To examine the longer-term effects of tightening monetary policy, we compute the effects of target rate changes ($P_t$)  on the average changes of future outcome variables: $\overline{M}_{t+L}=\frac{1}{L}\sum_{l=1}^LM_{t+l}$,  across different horizons ($L$) up to 24 months. The estimation results are depicted in Figures \ref{Emplot1} and \ref{Emplot2}, illustrating an accumulation of positive effects on short-term bond yields within the first 12 months, followed by a gradual decline in subsequent months.
As a robustness check, we re-estimate the monetary policy effects using data before August 2008, when the global financial crisis started.  The estimation results are reported in Panel B of Table \ref{Emptable1}. Notably, the impact of tightening monetary policy in curbing the increase in unemployment rate was significant before the global financial crisis, but it became considerably more muted after the crisis. 

\begin{center}

\fbox{Figures 4 and 5 near here}

\end{center}

We follow the relevant literature in treatment effects \citep[e.g.,][]{HIR2003} to estimate the average treatment effect (ATE) using the inverse probability weighting approach. Specifically, we calculate the policy propensity scores $\Phi_\eta(\widehat{G}_P(\mathbf{x}_t))$ based on the LNN estimation of the binary model \eqref{empbinary}, as discussed in Appendix \ref{App.22} of the online supplement, and define the weight as  $w_t=\Phi_\eta(\widehat{G}_P(\mathbf{x}_t))^{-P_t}[1-\Phi_\eta(\widehat{G}_P(\mathbf{x}_t))]^{-(1-P_t)}$. Then, we construct the inverse probability weighting estimator for ATE as follows:
\begin{equation*}
\widehat{\text{ATE}}=\frac{1}{T}\sum_{t=1}^Tw_tP_tM_{t+1}-\frac{1}{T}\sum_{t=1}^Tw_t(1-P_t)M_{t+1}.
\end{equation*}

We omit these results from this study to maintain focus on our main objective.
 
\section{Conclusion}\label{Sec6}

NN has gained considerable attentions over the past few decades. Yet, many questions, as outlined in Section \ref{S1}, have not been satisfactorily addressed in the relevant literature.  In this paper, we bring in identification restrictions to the LNN framework from a semiparametric regression perspective, and consider the LNN based estimation and inference for the unknown parameter and function involved in the modelling of time series data. We then integrate the LNN architecture with the group-LASSO technique to achieve consistent estimation and automatic detection of insignificant regressors. The asymptotic distributions are derived accordingly, demonstrating that the sparsity structure can be effectively identified and the estimators exhibit the distributions that are asymptotically equivalent to those for the infeasible oracle estimators.  

Additionally, we propose a dependent wild bootstrap procedure to obtain valid inferences in practice. Last but not least,  we validate our theoretical findings through extensive numerical studies. In the empirical study, we revisit the impacts of a tightening monetary policy action on a variety of economic variables, including short-/long-term interest rate, inflation, unemployment rate, industrial price and equity return via the newly proposed framework using a monthly dataset of the US.

Several major comments have been made here and there in Section \ref{S2}, and some future research directions have been acknowledged along the way.  Finally, we hope the current article will shed light on how to produce transparent algorithms to ensure that our research findings are useful for practical implementations and applications.

\section{Acknowledgements}

Gao, Peng and Yang acknowledge financial support from the Australian Research Council Discovery Grants Program under Grant  Numbers: {\small DP200102769, DP210100476 and DP230102250}, respectively. Liu's research was financially supported by National Natural Science Foundation of China under Grant Number: 72203114.

{\footnotesize
\bibliography{Refs}

@article{HIR2003,
  title={Efficient estimation of average treatment effects using the estimated propensity score},
  author={Hirano, Keisuke and Imbens, Guido W and Ridder, Geert},
  journal={Econometrica},
  volume={71},
  number={4},
  pages={1161--1189},
  year={2003},
  publisher={Wiley Online Library}
}

@UNPUBLISHED{hhll24,
    author = "Yu-Chin Hsu and Martin Huber and Ying-Ying Lee and Chu-An Liu",
    title  = "{Testing Monotonicity of Mean Potential Outcomes in a Continuous Treatment with High-Dimensional Data. Forthcoming in {\em The Review of Economics and Statistics}}",
    note   = "\url{https://doi.org/10.1162/rest_a_01416}",
    year   = "2024"
}

@article{Chatterjee2011,
  title={Bootstrapping lasso estimators},
  author={Chatterjee, Arindam and Lahiri, Soumendra Nath},
  journal={Journal of the American Statistical Association},
  volume={106},
  number={494},
  pages={608--625},
  year={2011},
  publisher={Taylor \& Francis}
}

@book{hlg2000,
  title={Partially Linear Models},
  author={Wolfgang H\"{a}rdle and Hua Liang and Jiti Gao},
  series={Contributions to Economics and Statistics},
  year={2000},
  publisher={Springer, New York}
}

@book{ttg2010,
  title={Modelling Nonlinear Economic Time Series},
  author={Timo Ter\"{a}svirta and Dag Tj{\o}stheim and Clive W. J. Granger},
  series={Advanced Texts in Econometrics},
  year={2010},
  publisher={Oxford University Press}
}

@article{robinson1988,
  title={Root-N-consistent semiparametric regression},
  author={Peter M. Robinson},
  journal={Econometrica},
  volume={56},
  number={2},
  pages={931--964},
  year={1988},
  publisher={Wiley, New York}
}

@article{Sims1992,
  title={Interpreting the macroeconomic time series facts: The effects of monetary policy},
  author={Sims, Christopher A},
  journal={European Economic Review},
  volume={36},
  number={5},
  pages={975--1000},
  year={1992},
  publisher={Elsevier}
}

@article{Christians1996,
  title={The Effects of Monetary Policy stocks. Evidence from the Flow of Finds},
  author={Christians, LJ and Eichenbaum, M and Evans, C},
  journal={The Review of Economics and Statistics},
  volume={7811},
  pages={16--34},
  year={1996}
}

@article{Cochrane2002,
  title={The Fed and interest rates—a high-frequency identification},
  author={Cochrane, John H and Piazzesi, Monika},
  journal={The American Economic Review},
  volume={92},
  number={2},
  pages={90--95},
  year={2002},
  publisher={American Economic Association}
}

@article{Romer2000,
  title={Federal Reserve information and the behavior of interest rates},
  author={Romer, Christina D and Romer, David H},
  journal={The American Economic Review},
  volume={90},
  number={3},
  pages={429--457},
  year={2000},
  publisher={American Economic Association}
}

@article{Angrist2018,
  title={Semiparametric estimates of monetary policy effects: string theory revisited},
  author={Angrist, Joshua D and Jord{\`a}, {\`O}scar and Kuersteiner, Guido M},
  journal={Journal of Business \& Economic Statistics},
  volume={36},
  number={3},
  pages={371--387},
  year={2018},
  publisher={Taylor \& Francis}
}

@article{Bernanke2005,
  title={What explains the stock market's reaction to Federal Reserve policy?},
  author={Bernanke, Ben S and Kuttner, Kenneth N},
  journal={The Journal of Finance},
  volume={60},
  number={3},
  pages={1221--1257},
  year={2005},
  publisher={Wiley Online Library}
}

@article{Clarida2000,
  title={Monetary policy rules and macroeconomic stability: evidence and some theory},
  author={Clarida, Richard and Gali, Jordi and Gertler, Mark},
  journal={The Quarterly Journal of Economics},
  volume={115},
  number={1},
  pages={147--180},
  year={2000},
  publisher={MIT Press}
}

@article{Blanchard2010,
 author = {Olivier Blanchard and Jordi Galí},
 journal = {American Economic Journal: Macroeconomics},
 number = {2},
 pages = {1--30},
 publisher = {American Economic Association},
 title = {Labor Markets and Monetary Policy: A New Keynesian Model with Unemployment},
 urldate = {2024-04-14},
 volume = {2},
 year = {2010}
}

@article{Shao2010,
  title={The dependent wild bootstrap},
  author={Shao, Xiaofeng},
  journal={Journal of the American Statistical Association},
  volume={105},
  number={489},
  pages={218--235},
  year={2010},
  publisher={Taylor \& Francis}
}

@article{McLeish1975,
  title={A maximal inequality and dependent strong laws},
  author={McLeish, Don L},
  journal={The Annals of Probability},
  volume={3},
  number={5},
  pages={829--839},
  year={1975},
  publisher={Institute of Mathematical Statistics}
}

@article{Hansen1992,
  title={Consistent covariance matrix estimation for dependent heterogeneous processes},
  author={Hansen, Bruce E},
  journal={Econometrica},
  pages={967--972},
  year={1992},
  publisher={JSTOR}
}

@book{Bosq1996,
  title={Nonparametric Statistics for Stochastic Processes: Estimation and Prediction},
  author={Bosq, D.},
  isbn={9781468404890},
  lccn={96013588},
  series={Lecture Notes in Statistics},
  year={2012},
  publisher={Springer New York}
}

@article{FL2001,
  title={Variable selection via nonconcave penalized likelihood and its oracle properties},
  author={Fan, Jianqing and Li, Runze},
  journal={Journal of the American Statistical Association},
  volume={96},
  number={456},
  pages={1348--1360},
  year={2001},
  publisher={Taylor \& Francis}
}

@article{Chernozhukov2018,
    author = {Chernozhukov, Victor and Chetverikov, Denis and Demirer, Mert and Duflo, Esther and Hansen, Christian and Newey, Whitney and Robins, James},
    title = "{Double/debiased machine learning for treatment and structural parameters}",
    journal = {The Econometrics Journal},
    volume = {21},
    number = {1},
    pages = {C1-C68},
    year = {2018}
}

@article{BCH2014,
    author = {Belloni, Alexandre and Chernozhukov, Victor and Hansen, Christian},
    title = "{Inference on treatment effects after selection among high-dimensional controls}",
    journal = {The Review of Economic Studies},
    volume = {81},
    number = {2},
    pages = {608-650},
    year = {2014}
}

@article{HL2005,
  title={Variable selection using MM algorithms},
  author={Hunter, David R and Li, Runze},
  journal={The Annals of Statistics},
  volume={33},
  number={4},
  pages={1617},
  year={2005},
  publisher={NIH Public Access}
}

@article{YL2006,
  title={Model selection and estimation in regression with grouped variables},
  author={Yuan, Ming and Lin, Yi},
  journal={Journal of the Royal Statistical Society Series B: Statistical Methodology},
  volume={68},
  number={1},
  pages={49--67},
  year={2006},
  publisher={Oxford University Press}
}

@article{WX2009,
  title={Shrinkage estimation of the varying coefficient model},
  author={Wang, Hansheng and Xia, Yingcun},
  journal={Journal of the American Statistical Association},
  volume={104},
  number={486},
  pages={747--757},
  year={2009},
  publisher={Taylor \& Francis}
}

@BOOK {AS1972,
    author    = "M. Abramovitz and I. A. Stegun",
    title     = "Handbook of Mathematical Functions",
    publisher = "Dover Publications, New York, U.S.",
    year      = "1972"
}

@article{cs1998,
title={Sieve Extremum Estimates for Weakly Dependent Data},
author={Xiaohong Chen and Xiaotong Shen},
journal={Econometrica},
volume={66},
number={2},
pages={298--314},
year={1998}}

@article{crs2001,
title={Semiparametric {ARX} Neural Network Models with an Application to Forecasting Inflation},
author={Xiaohong Chen and Jeffrey Racine and Norman Swanson},
journal={IEEE Transactions on Neural Networks},
volume={12},
number={6},
pages={674--683},
year={2001}}

@article{chen2007,
title={{Large Sample Sieve Estimation of Semi--Nonparametric Models, Chapter 76 edited by James J. Heckman and Edward E. Leamer}},
author={Xiaohong Chen},
journal={Handbook of Econometrics},
volume={6B},
pages={5549--5632},
year={2007}}

@incollection{NEWEY19942111,
title = {Chapter 36: Large sample estimation and hypothesis testing},
series = {Handbook of Econometrics},
publisher = {Elsevier},
volume = {4},
pages = {2111-2245},
year = {1994},

author = {Whitney K. Newey and Daniel McFadden}
}

@article{MUDELSEE2019310,
title = {Trend analysis of climate time series: {A} review of methods},
journal = {Earth-Science Reviews},
volume = {190},
pages = {310-322},
year = {2019},
author = {Manfred Mudelsee}
}

@ARTICLE {LTG2016,
    author  = "D. Li and D. Tj{\o}stheim and J. Gao",
    title   = "Estimation in nonlinear regression with {H}arris recurrent {M}arkov chains",
    journal = "The Annals of Statistics",
    year    = "2016",
    volume  = "44",
    number  = "5",
    pages   = "1957-1987"
}

@book{fan1996local,
  title={Local Polynomial Modelling and its Applications},
  author={Fan, Jianqing and Gijbels, Irene},
  year={1996},
  publisher={Chapman \& Hall/CRC}
}

@ARTICLE{MYA1994,
  author={Murata, N. and Yoshizawa, S. and Amari, S.},
  journal={IEEE Transactions on Neural Networks}, 
  title={Network information criterion-determining the number of hidden units for an artificial neural network model}, 
  year={1994},
  volume={5},
  number={6},
  pages={865-872}}

@article{MINAI1993845,
title = {On the derivatives of the sigmoid},
journal = {Neural Networks},
volume = {6},
number = {6},
pages = {845-853},
year = {1993},
author = {Ali A. Minai and Ronald D. Williams}
}

@Article{PA2012,
  author={Kline Patrick and Santos Andres},
  title={A Score Based Approach to Wild Bootstrap Inference},
  journal={Journal of Econometric Methods},
  year=2012,
  volume={1},
  number={1},
  pages={23-41}
}

@article{WP2003,
 author = {Whitney K. Newey and James L. Powell},
 journal = {Econometrica},
 number = {5},
 pages = {1565-1578},
 title = {Instrumental Variable Estimation of Nonparametric Models},
 volume = {71},
 year = {2003}
}

@BOOK{Magnus,
        author = {J. R. Magnus and H. Neudecker},
        title = {Matrix Differential Calculus with Applications in Statistics and Econometrics},
        year = {2007},
        publisher = {John Wiley \& Sons Ltd},
        edition = {third}
        }

@article{FG2010,
author = {G{\"u}nther, Frauke and Fritsch, Stefan},
year = {2010},
month = {06},
pages = {30-38},
title = {Neuralnet: Training of Neural Networks},
volume = {2},
journal = {R Journal}
}

@article{FLM2021,
author = {Farrell, Max H. and Liang, Tengyuan and Misra, Sanjog},
title = {Deep Neural Networks for Estimation and Inference},
journal = {Econometrica},
volume = {89},
number = {1},
pages = {181-213},
year = {2021}
}

@unpublished{DFLSP2021,
	title={Dimension-Free Average Treatment Effect Inference with Deep Neural Networks},
	author={Xinze Du and Yingying Fan and Jinchi Lv and Tianshu Sun and Patrick Vossler},
	note={Available at \url{https://doi.org/10.48550/arXiv.2112.01574}},
	year={2021},
}

@unpublished{CLMZ2022,
	title={Casual inference of General Treatment Effects using Neural Networks with a Diverging Number of Confounders},
	author={Xiaohong Chen and Ying Liu and Shujie Ma and Zheng Zhang},
	note={Available at \url{https://arxiv.org/abs/2009.07055v5}},
	year={2022},
}

@article{HOR1996,
 author = {Tim Hill and Marcus {O'C}onnor and William Remus},
 journal = {Management Science},
 number = {7},
 pages = {1082-1092},
 title = {Neural Network Models for Time Series Forecasts},
 volume = {42},
 year = {1996}
}

@article{GU2021429,
title = {Autoencoder asset pricing models},
journal = {Journal of Econometrics},
volume = {222},
number = {1, Part B},
pages = {429-450},
year = {2021},
author = {Shihao Gu and Bryan Kelly and Dacheng Xiu}
}

@article{Gu2020,
    author = {Gu, Shihao and Kelly, Bryan and Xiu, Dacheng},
    title = "Empirical Asset Pricing via Machine Learning",
    journal = {The Review of Financial Studies},
    volume = {33},
    number = {5},
    pages = {2223-2273},
    year = {2020},
}

@INCOLLECTION {SAUER2006191,
    author    = "Tomas Sauer",
    title     = "Polynomial Interpolation in Several Variables: Lattices, Differences, and Ideals",
    booktitle = "Topics in Multivariate Approximation and Interpolation",
    year      = "2006",
    editor    = "Kurt Jetter and Martin D. Buhmann and Werner Haussmann and Robert Schaback and Joachim St{\"o}ckler",
    volume    = "12",
    series    = "Studies in Computational Mathematics",
    pages     = "191-230"
}

@INCOLLECTION {Athey2019,
    author    = "Susan Athey",
    title     = "The Impact of Machine Learning on Economics",
    booktitle = "The Economics of Artificial Intelligence: {A}n Agenda",
    year      = "2019",
    editor    = "Ajay Agrawal and Joshua Gans and Avi Goldfarb",
    pages     = "507-547"
}

@ARTICLE {BK2019,
    author  = "Benedikt Bauer and Michael Kohler",
    title   = "{On deep learning as a remedy for the curse of dimensionality in nonparametric regression}",
    journal = "The Annals of Statistics",
    year    = "2019",
    volume  = "47",
    number  = "4",
    pages   = "2261-2285"
}

@ARTICLE {SH2020,
    author  = "Johannes Schmidt-Hieber",
    title   = "{Nonparametric regression using deep neural networks with ReLU activation function}",
    journal = "The Annals of Statistics",
    year    = "2020",
    volume  = "48",
    number  = "4",
    pages   = "1875-1897"
}

@ARTICLE {Cybenko1989,
    author  = "G. Cybenko",
    title   = "{Approximation by superpositions of a sigmoidal function}",
    journal = "Mathematics of Control, Signals and Systems",
    year    = "1989",
    volume  = "2",
    pages   = "303-314"
}

@BOOK {FanYao,
    author    = "J. Fan and Q. Yao",
    title     = "Nonlinear Time Series: Nonparametric and Parametric Methods",
    publisher = "Springer-Verlag",
    year      = "2003"
}

@BOOK {LR2007,
    author    = "Li, Qi and Racine, Jeffrey",
    title     = "Nonparametric Econometrics Theory and Practice",
    publisher = "Princeton University Press, New Jersey",
    year      = "2007"
}

@article{DUBEY202292,
title = {Activation functions in deep learning: A comprehensive survey and benchmark},
journal = {Neurocomputing},
volume = {503},
pages = {92-108},
year = {2022},
issn = {0925-2312},
author = {Shiv Ram Dubey and Satish Kumar Singh and Bidyut Baran Chaudhuri},
}

@BOOK{Gao,
        author = {Jiti Gao},
        title = {Nonlinear Time Series: Semiparametric and Nonparametric Methods},
        year = {2007},
        publisher = {Chapman \& Hall/CRC},
        edition = {}
        }

@unpublished{WL2021,
	title={Harmless Overparametrization in Two-layer Neural Networks},
	author={Huiyuan Wang and Wei Lin},
	note={Available at \url{
https://doi.org/10.48550/arXiv.2106.04795}},
	year={2021},
}

@article{BMR2021, 
title={Deep learning: {A} statistical viewpoint}, 
volume={30},
journal={Acta Numerica}, 
author={Bartlett, Peter L. and Montanari, Andrea and Rakhlin, Alexander}, 
year={2021}, 
pages={87201}}

@article{FMZ2021,
author = {Jianqing Fan and Cong Ma and Yiqiao Zhong},
title = "A Selective Overview of Deep Learning",
volume = {36},
journal = {Statistical Science},
number = {2},
publisher = {Institute of Mathematical Statistics},
pages = {264-290},
year = {2021}
}
}

\newpage

\begin{table}[htp!]\setlength{\tabcolsep}{3pt}\renewcommand{\arraystretch}{0.8}\centering
\caption{{\bf Simulation Results of RMSEs and CRs.} This table presents the simulation results of RMSEs and CRs for the LNN-based group-LASSO method and LNN estimation (without modelling the sparsity structure), respectively. }\label{Tab1}
\resizebox{\textwidth}{!}{
\begin{tabular}{llcccccccccccccccc}
\hline\hline
                &     &                    & \multicolumn{3}{c}{$\text{RMSE}_\alpha$} &  & \multicolumn{3}{c}{$\text{CR}_\alpha$} &  & \multicolumn{3}{c}{$\text{RMSE}_g$} &  & \multicolumn{3}{c}{$\text{CR}_g$} \\
LNN-GLASSO     &     & $T\setminus d$ & 2            & 8           & 14          &  & 2           & 8           & 14         &  & 2          & 8          & 14        &  & 2         & 8         & 14        \\
$u_\sigma=0.5$  & $q=3$ & 800                & 0.103        & 0.115       & 0.204       &  & 0.966       & 0.937       & 0.921      &  & 0.143      & 0.590      & 0.691     &  & 0.907     & 0.904     & 0.924     \\
                &     & 1600               & 0.074        & 0.083       & 0.091       &  & 0.951       & 0.965       & 0.945      &  & 0.093      & 0.536      & 0.447     &  & 0.915     & 0.915     & 0.916     \\
                &     & 2400               & 0.060        & 0.068       & 0.064       &  & 0.950       & 0.954       & 0.945      &  & 0.074      & 0.408      & 0.430     &  & 0.934     & 0.940     & 0.961     \\
                & $q=4$ & 800                & 0.106        & 0.133       & 0.538       &  & 0.966       & 0.892       & 0.901      &  & 0.134      & 0.543      & 1.580     &  & 0.921     & 0.904     & 0.899     \\
                &     & 1600               & 0.074        & 0.081       & 0.165       &  & 0.951       & 0.938       & 0.924      &  & 0.092      & 0.399      & 0.733     &  & 0.917     & 0.926     & 0.912     \\
                &     & 2400               & 0.058        & 0.068       & 0.063       &  & 0.952       & 0.950       & 0.957      &  & 0.077      & 0.332      & 0.523     &  & 0.936     & 0.957     & 0.941     \\
                &     &                    &              &             &             &  &             &             &            &  &            &            &           &  &           &           &           \\
LNN & $q=3$ & 800                & 0.110        & 0.108       & 0.109       &  & 0.963       & 0.923       & 0.944      &  & 0.126      & 0.562      & 0.455     &  & 0.923     & 0.914     & 0.911     \\
                &     & 1600               & 0.076        & 0.076       & 0.081       &  & 0.960       & 0.936       & 0.957      &  & 0.087      & 0.474      & 0.329     &  & 0.923     & 0.919     & 0.923     \\
                &     & 2400               & 0.060        & 0.064       & 0.065       &  & 0.952       & 0.957       & 0.948      &  & 0.068      & 0.421      & 0.295     &  & 0.955     & 0.933     & 0.955     \\
                & $q=4$ & 800                & 0.110        & 0.120       & 0.225       &  & 0.931       & 0.925       & 0.912      &  & 0.139      & 0.575      & 1.078     &  & 0.920     & 0.901     & 0.911     \\
                &     & 1600               & 0.075        & 0.082       & 0.138       &  & 0.966       & 0.921       & 0.954      &  & 0.098      & 0.419      & 0.872     &  & 0.913     & 0.915     & 0.914     \\
                &     & 2400               & 0.060        & 0.065       & 0.086       &  & 0.942       & 0.940       & 0.947      &  & 0.081      & 0.383      & 0.598     &  & 0.959     & 0.941     & 0.936     \\
                \hline
LNN     &           \\
$u_\sigma=0.5$  & $q=3$ & 800                & 0.101        & 0.125       & 0.181       &  & 0.942       & 0.930       & 0.912      &  & 0.153      & 0.757      & 1.218     &  & 0.902     & 0.911     & 0.918     \\
                &     & 1600               & 0.073        & 0.084       & 0.091       &  & 0.966       & 0.963       & 0.943      &  & 0.101      & 0.565      & 0.821     &  & 0.901     & 0.905     & 0.929     \\
                &     & 2400               & 0.065        & 0.068       & 0.069       &  & 0.926       & 0.955       & 0.951      &  & 0.080      & 0.433      & 0.689     &  & 0.933     & 0.930     & 0.937     \\
                & $q=4$ & 800                & 0.106        & 0.188       & 1.384       &  & 0.966       & 0.907       & 0.854      &  & 0.158      & 1.067      & 6.886     &  & 0.914     & 0.891     & 0.831     \\
                &     & 1600               & 0.075        & 0.094       & 0.567       &  & 0.951       & 0.954       & 0.903      &  & 0.110      & 0.823      & 2.306     &  & 0.915     & 0.917     & 0.873     \\
                &     & 2400               & 0.059        & 0.073       & 0.134       &  & 0.950       & 0.941       & 0.925      &  & 0.091      & 0.789      & 1.517     &  & 0.931     & 0.923     & 0.889     \\
                &     &                    &              &             &             &  &             &             &            &  &            &            &           &  &           &           &           \\
$u_\sigma=-0.5$ & $q=3$ & 800                & 0.112        & 0.118       & 0.122       &  & 0.943       & 0.934       & 0.938      &  & 0.152      & 0.719      & 0.678     &  & 0.913     & 0.898     & 0.914     \\
                &     & 1600               & 0.078        & 0.077       & 0.085       &  & 0.941       & 0.922       & 0.961      &  & 0.107      & 0.556      & 0.477     &  & 0.925     & 0.917     & 0.922     \\
                &     & 2400               & 0.061        & 0.066       & 0.066       &  & 0.955       & 0.961       & 0.949      &  & 0.081      & 0.479      & 0.408     &  & 0.938     & 0.969     & 0.959     \\
                & $q=4$ & 800                & 0.111        & 0.198       & 1.468       &  & 0.922       & 0.904       & 0.872      &  & 0.166      & 1.559      & 3.437     &  & 0.909     & 0.881     & 0.837     \\
                &     & 1600               & 0.072        & 0.096       & 0.863       &  & 0.936       & 0.921       & 0.896      &  & 0.113      & 1.103      & 1.824     &  & 0.923     & 0.901     & 0.860     \\
                &     & 2400               & 0.061        & 0.073       & 0.137       &  & 0.952       & 0.942       & 0.914      &  & 0.090      & 0.880      & 1.277     &  & 0.930     & 0.925     & 0.881  \\
                \hline\hline  
\end{tabular}}
\end{table}

\begin{table}[htp!]
\setlength{\tabcolsep}{2pt}\renewcommand{\arraystretch}{0.8}\small\centering
\caption{Descriptive Statistics and ADF Test Results}
\begin{tabular}{lcccccc}
\hline\hline
Variable                     & Abbreviation & In Difference & Mean   & SD    & ADF $t$-stats & $p$-value        \\
Effective federal funds rate & FFR          & No            & 3.399  & 2.651 & -2.477        & 0.013            \\
3-month treasury bond yields & Yield3m      & No            & 3.138  & 2.446 & -2.480        & 0.013            \\
1-year treasury bond yields  & Yield1y      & No            & 3.490  & 2.545 & -2.382        & 0.017            \\
2-year treasury bond yields  & Yield2y      & No            & 3.810  & 2.524 & -2.240        & 0.025            \\
5-year treasury bond yields  & Yield5y      & No            & 4.449  & 2.241 & -2.014        & 0.042            \\
10-year treasury bond yields & Yield10y     & No            & 5.009  & 1.927 & -1.898        & 0.055            \\
Inflation                    & PCE          & Yes           & 0.075  & 0.086 & -6.575        & \textless{}0.001 \\
Unemployment rate            & UNRATE       & Yes           & -0.001 & 0.158 & -9.113        & \textless{}0.001 \\
Industrial price             & IP           & Yes           & 0.066  & 0.277 & -8.029        & \textless{}0.001 \\
Equity prices                & SPR          & No            & 0.886  & 4.182 & -11.632       & \textless{}0.001 \\
Target rate                  & TR           & Yes           & 3.410  & 2.647 & -2.504        & 0.012            \\
Federal funds future index   & FFF          & No            & 0.008  & 0.136 & -7.660        & \textless{}0.001\\
\hline\hline
\end{tabular}\label{Tab_DES}
\end{table}

\begin{table}[htp!]
\setlength{\tabcolsep}{3pt}\renewcommand{\arraystretch}{0.8}\small \centering
\caption{{\bf Insignificant Predictors}. 
This table provides information about the insignificant predictors identified by the LNN-based group-LASSO method for each outcome variable. For each variable $V$, $V(-p)$ denotes its $p$-lag value.}
\begin{tabular}{ll}
\hline
Outcome variable & Insignificant predictors\\
\hline
FFR       & PCE(-1), UNRATE(-1), PCE(-2), UNRATE(-2)\\  
Yield3m  & \\ 
Yield1y  &  FOMC \\
Yield2y  &  FOMC \\
Yield5y  &         \\
Yield10y  &     \\
PCE      &  TRC(-1), PCE(-1), UNRATE(-1), UNRATE(-2),  FOMC      \\
UNRATE    &  FFF, FOMC       \\
IP       &   FFF  \\
SPR        &  \\
\hline  
\end{tabular}\label{Tab_Insig}
\end{table}

\begin{table}[htp!]\setlength{\tabcolsep}{1pt}\renewcommand{\arraystretch}{0.8}\centering
\begin{center}
\caption{\textbf{Estimated Monetary Policy Effects.} This table contains the estimated effects of a target rate increase, which is captured by $\widehat{\alpha}_P$, along with their 90$\%$ confidence intervals.
Panels A presents the estimates for the models with all control variables, without controlling FFF, and without controlling FFF and the lags of PCE and  UNRATE, respectively. 
Panels B presents the estimates for the models using observations before August 2008, with all control variables, without controlling FFF, and without controlling FFF and the lags of PCE and  UNRATE, respectively.}\label{Emptable1}
\resizebox{\textwidth}{!}{
\begin{tabular}{lccccccccc}
\hline\hline
Panel A (Full   sample)&   & \multicolumn{2}{c}{With all controls} &     & \multicolumn{2}{c}{Without FFF}       &  & \multicolumn{2}{c}{Without FFF,}   \\
&   & &  & & &  &  & \multicolumn{2}{c}{PCE, and UNRATE}   \\
 \cline{3-4}\cline{6-7}\cline{9-10}
Variable & & Estimate  & CI    &  & Estimate & CI   &  & Estimate    & CI          \\
FFR                      &                  & 0.190      & (0.086, 0.283) &  & 0.422     & (0.334, 0.499) &  & 0.450        & (0.358, 0.542) \\
Yield3m                  &                  & 0.221      & (0.091, 0.336) &  & 0.495     & (0.379, 0.622) &  & 0.522        & (0.394, 0.647) \\
Yield1y                  &                  & 0.132      & (-0.204, 0.239) &  & 0.693     & (0.488, 0.912) &  & 0.732        & (0.557, 0.963) \\
Yield2y                  &                  & 0.107      & (-0.153, 0.429) &  & 0.605     & (0.278, 0.860) &  & 0.710        & (0.445, 0.956) \\
Yield5y                  &                  & -0.108     & (-0.460, 0.354) &  & 0.330     & (-0.022, 0.701) &  & 0.412        & (0.102, 0.780) \\
Yield10y                 &                  & -0.245     & (-0.604, 0.226) &  & 0.096     & (-0.366, 0.491) &  & 0.188        & (-0.213, 0.543) \\
PCE                      &                  & 0.029      & (-0.013, 0.053) &  & 0.039     & (0.008, 0.068) &  & 0.038        & (0.006, 0.066) \\
UNRATE                   &                  & -0.036     & (-0.077, 0.018) &  & -0.055    & (-0.099, -0.018) &  & -0.051       & (-0.088, -0.009) \\
IP                       &                  & -0.037     & (-0.150, 0.047) &  & 0.024     & (-0.060, 0.123) &  & 0.055        & (-0.044, 0.159) \\
SPR                      &                  & -1.478     & (-4.548, 1.353) &  & 0.230     & (-1.755, 2.045) &  & 0.321        & (-1.293, 1.793) \\
\hline 
Panel B (Before crisis)&           \\
FFR                      &                  & 0.266      & (0.141, 0.336) &  & 0.416     & (0.299, 0.474) &  & 0.422        & (0.330, 0.513) \\
Yield3m                  &                  & 0.251      & (0.086, 0.386) &  & 0.490     & (0.347, 0.601) &  & 0.483        & (0.345, 0.598) \\
Yield1y                  &                  & 0.175      & (-0.136, 0.326) &  & 0.715     & (0.502, 0.934) &  & 0.674        & (0.464, 0.882) \\
Yield2y                  &                  & 0.200      & (-0.046, 0.517) &  & 0.642     & (0.341, 0.927) &  & 0.695        & (0.436, 1.009) \\
Yield5y                  &                  & 0.301      & (0.222, 1.009) &  & 0.420     & (0.095, 0.780) &  & 0.417        & (0.116, 0.786) \\
Yield10y                 &                  & 0.202      & (-0.081, 0.865) &  & 0.172     & (-0.198, 0.552) &  & 0.217        & (-0.137, 0.632) \\
PCE                      &                  & 0.026      & (-0.005, 0.056) &  & 0.024     & (-0.008, 0.049) &  & 0.025        & (-0.011, 0.052) \\
UNRATE                   &                  & -0.049     & (-0.091, -0.001) &  & -0.058    & (-0.095, -0.013) &  & -0.062       & (-0.095, -0.030) \\
IP                       &                  & -0.029     & (-0.140, 0.057) &  & 0.039     & (-0.069, 0.135) &  & 0.056        & (-0.022, 0.162) \\
SPR                      &                  & -0.523     & (-3.293, 2.923) &  & 0.799     & (-1.101, 2.732) &  & -0.247       & (-1.762, 1.219)\\
\hline\hline
\end{tabular}}
\end{center}
\end{table}

\newpage

{\small

\begin{center}

{\large \bf  Appendix A}

\end{center}

Appendix A includes some discussions on issues associated with practical implementation and also presents a detailed algorithm.  Appendix \ref{App.2} discusses the estimation of a fully nonparametric model which is a special case of \eqref{EQ1.1}, explains how to relax the restriction about $a$, and infers $G(\cdot)$ of $z_t$ via LNN. Appendix \ref{App.3} includes additional simulations.  We finally present the preliminary lemmas and the proofs in Appendix \ref{App.4}.  

 
\renewcommand{\theequation}{A\arabic{section}.\arabic{equation}}
\renewcommand{\thesection}{A.\arabic{section}}
\renewcommand{\thefigure}{A.\arabic{figure}}
\renewcommand{\thetable}{A.\arabic{table}}
\renewcommand{\thetheorem}{A.\arabic{theorem}}
\renewcommand{\thelemma}{A.\arabic{lemma}}
\renewcommand{\theremark}{A.\arabic{remark}}
\renewcommand{\thecorollary}{A.\arabic{corollary}}
\renewcommand{\theassumption}{A.\arabic{assumption}}

\setcounter{equation}{0}
\setcounter{lemma}{0}
\setcounter{section}{0}
\setcounter{table}{0}
\setcounter{figure}{0}
\setcounter{remark}{0}
\setcounter{corollary}{0}
\setcounter{assumption}{0}

\renewcommand{\theequation}{A\arabic{section}.\arabic{equation}}
\renewcommand{\thesection}{A.\arabic{section}}
\renewcommand{\thefigure}{A.\arabic{figure}}
\renewcommand{\thetable}{A.\arabic{table}}
\renewcommand{\thetheorem}{A.\arabic{theorem}}
\renewcommand{\thelemma}{A.\arabic{lemma}}
\renewcommand{\theremark}{A.\arabic{remark}}
\renewcommand{\thecorollary}{A.\arabic{corollary}}
\renewcommand{\theassumption}{A.\arabic{assumption}}

\setcounter{equation}{0}
\setcounter{lemma}{0}
\setcounter{section}{0}
\setcounter{table}{0}
\setcounter{figure}{0}
\setcounter{remark}{0}
\setcounter{corollary}{0}
\setcounter{assumption}{0}

\section{Practical Implementation}\label{App.1}

First, we provide some examples to illustrate our concern about the existing software packages. In the literature, the ``\textsf{neuralnet}" R package (see \citealp{FG2010} for detailed illustration) has been well adopted. When training a NN, a key parameter is called ``\textsf{hidden}" that is \textit{a vector of integers specifying the number of hidden neurons in each layer} by R document, and the package refers to \cite{MYA1994} regarding the choice of the number of neurons. However, it is worth pointing out that \cite{MYA1994} use a modified AIC criterion to investigate the case with the number of activation functions (as well as the number of parameters) being finite which is reflected in their asymptotic development. As a consequence, the arguments of \cite{MYA1994} no longer hold when the number of activation functions is diverging. As we show in this paper, having a diverging number of activation functions is the minimum requirement to achieve asymptotic consistency, and the rate of divergence is also associated with the sample size under a set of minor conditions. Similar issues also apply to ``\textsf{deepnet}" package of {\tt R}, ``\textsf{torch.nn.Linear}" and ``\textsf{tfl.layers.Linear}" of Python, ``\textsf{feedforwardnet}" of Matlab, etc. 


\medskip

Next, we comment on the matrix $\mathbf{D}$, the selection of the tuning parameter, a computational algorithm for the LNN based group-LASSO, and  some properties of Sigmoidal squasher.

\medskip

\noindent \textbf{On the Matrix $\mathbf{D}$} --- By the proof of Lemma \ref{LEM2.2}, we can determine $\mathbf{D}$ through $\mathbf{m}(\mathbf{x}\, |\, \mathbf{x}_0)=\mathbf{D}\mathbf{A}(\mathbf{x} \, | \,\mathbf{x}_0)$, where 
\begin{eqnarray*}
\mathbf{A}(\mathbf{x} \, | \,\mathbf{x}_0)&=&\left([(1,\mathbf{x}^\top-\mathbf{x}_0^\top)\pmb{\alpha}_1 ]^q,\ldots, [(1,\mathbf{x}^\top-\mathbf{x}_0^\top)\pmb{\alpha}_{d_q} ]^q\right)^\top
\end{eqnarray*}
with 
$[\pmb{\alpha}_1 ,\ldots, \pmb{\alpha}_{d_q}]= \frac{1}{d+1}\cdot\diag\{h,\mathbf{I}_d \} \mathbf{W}$. 

\medskip

\noindent \textbf{On Tuning Parameters} --- An essential consideration for the practical application lies in  the selection of tuning parameters: $\psi_1,\cdots,\psi_{d_q}$. We simplify the selection of $d_q$ by utilizing the unpenalized estimators $\widetilde{\pmb{\Theta}}^\ast_j$ along with a common tuning parameter $\psi$:
\begin{equation}\label{A.EQ1}
\widetilde{\psi}_{j} = \frac{\psi }{h^{d}H_j^{-2}\|\widetilde{\pmb{\Theta}}^\ast_{j}\|^2}.
\end{equation}
Using analogous arguments in the proof of Lemma \ref{LEM2.4} and Lemma \ref{Lem_lasso}, we can show that $h^{d}H_j^{-2}\|\widetilde{\pmb{\Theta}}^\ast_{j}\|^2$ converges to a positive constant in probability for $j\in[d_q^0]$, while it has the order of $O_P\bigl(\frac{1}{Th^d}\bigr)$ for $j=d_q^0+1,\ldots,d_q$. With \eqref{A.EQ1}, Assumption \ref{Ass.4}.3 becomes
\begin{eqnarray*}
\psi_0 \max_{j\in[d_q^0]}\{H_j\}T^{-\frac{1}{2}} \rightarrow 0\quad\text{and}\quad
\psi_0 \min_{j=d_q^0+1,\ldots,d_q}\{H_j\}\sqrt{T}h^{d}\rightarrow\infty.
\end{eqnarray*}
By the definition of $H_j$, it is evident that $\max_{j\in[d_q^0]}\{H_j\}=O(h^{-q})$ and $\min_{j=d_q^0+1,\ldots,d_q}\{H_j\}=O(1)$. 
Therefore, a sufficient condition for the tuning parameter will be $\psi_0\sqrt{T}h^{d}\rightarrow\infty$ and $\psi_0T^{-\frac{1}{2}}h^{-q}\rightarrow0$.
The remaining task involves the selection of $\psi_0$, which can be achieved by minimising the following information criterion: 
\begin{eqnarray*}
{\rm IC}(\psi) = \frac{1}{T}\widetilde{Q}(\widetilde{\alpha}_\psi,\widetilde{\pmb{\Theta}}_\psi)+{\rm df}_\psi\frac{\log(Th)}{Th},
\end{eqnarray*}
where $ (\widetilde{\alpha}_\psi,\widetilde{\pmb{\Theta}}_\psi)$ are the group-LASSO estimators using \eqref{A.EQ1}, and ${\rm df}_\psi$ is the number of nonzero coefficients identified by $\widetilde{\pmb{\Theta}}_\psi$. Accordingly, the optimal tuning parameter is obtained by
\begin{equation*}
\widehat{\psi} = \argmin_{\psi}{\rm IC}(\psi).
\end{equation*}

\medskip

\noindent \textbf{Algorithm for the LNN Based Group-LASSO} --- The literature provides well-established computational algorithms for the LASSO estimation \citep[see, for instance,][]{FL2001, HL2005}. Herein, we adopt the local quadratic approximation procedure.

\begin{enumerate}[leftmargin=*, noitemsep] 
\item Obtain the unpenalized estimators $(\widetilde{\alpha}^{(0)},\widetilde{\pmb{\Theta}}^{(0)})$ as the initial estimators. 

\item The  estimators $(\widetilde{\alpha}^{(m)},\widetilde{\pmb{\Theta}}^{(m)})$ in the $m^{th}$ step are constructed as
\begin{eqnarray}\label{A.EQ2}
\left(\widetilde{\alpha}^{(m)},\widetilde{\pmb{\Theta}}^{(m)}\right) = \argmin_{\alpha, \pmb{\Theta}}\left( \widetilde{Q}  (\alpha,\pmb{\Theta})+\sum_{j=1}^{d_q}\psi_j\frac{\|\pmb{\Theta}_{D,j}\|^2}{\|\widetilde{\pmb{\Theta}}^{(m-1)}_{D,j}\|}\right),
\end{eqnarray}
where $\widetilde{\pmb{\Theta}}^{(m-1)}_{D,j}$ denotes the vector containing the $j^{th}$ elements of $\mathbf{D}^{\top,-1}\widetilde{\pmb{\theta}}^{(m-1)}_{\mathbf{i}}$'s in the $(m-1)^{th}$ step. Simple algebra shows that the first-order conditions for $\widetilde{\alpha}^{(m)}$ and $\widetilde{\pmb{\theta}}^{(m)}_{\mathbf{i}}$  are given by
\begin{eqnarray}\label{A.EQ3}
\sum_{t=1}^Tz_t[y_t-z_t\widetilde{\alpha}^{(m)} -\sum_{\mathbf{i}\in [M]^d} \widetilde{\mathbf{x}}_{\mathbf{i}, t}^\top \widetilde{\pmb{\theta}}^{(m)}_{\mathbf{i}}]&=&0,
\nonumber\\
-\sum_{t=1}^T\widetilde{\mathbf{x}}_{\mathbf{i}, t}[y_t-z_t\widetilde{\alpha}^{(m)} - \widetilde{\mathbf{x}}_{\mathbf{i}, t}^\top \widetilde{\pmb{\theta}}^{(m)}_{\mathbf{i}}]+\mathbf{D}^{-1}\pmb{\Phi}^{(m-1)}\mathbf{D}^{\top, -1}\widetilde{\pmb{\theta}}^{(m)}_{\mathbf{i}}&=&\mathbf{0},
\end{eqnarray}
where $\pmb{\Phi}^{(m-1)}$ is a $d_q\times d_q$ diagonal matrix with its $j^{th}$ diagonal element being $\psi_j/(\|\widetilde{\pmb{\Theta}}^{(m-1)}_{D,j}\|)$.  Solving the equations in \eqref{A.EQ3}, we obtain 

\begin{eqnarray*}
\widetilde{\alpha}^{(m)}&=& [\mathbf{Z}^\top \mathbf{M}^{(m-1)}_{x,\phi} \mathbf{Z} ]^{-1}\mathbf{Z}^\top \mathbf{M}^{(m-1)}_{x,\phi} \mathbf{Y},
\nonumber\\
\widetilde{\pmb{\theta}}^{(m)}_{\mathbf{i}}&=& [\widetilde{\mathbf{X}}_{\mathbf{i}}^\top \widetilde{\mathbf{X}}_{\mathbf{i}}+\mathbf{D}^{ -1}\pmb{\Phi}^{(m-1)} \mathbf{D}^{\top, -1}]^{-1}\widetilde{\mathbf{X}}_{\mathbf{i}}^\top [\mathbf{Y}-\mathbf{Z}\widetilde{\alpha}^{(m)} ],
\end{eqnarray*}
where $\mathbf{Z}=(z_1,\cdots,z_T)^\top$, $\mathbf{Y}=(y_1,\cdots,y_T)^\top$,  $\widetilde{\mathbf{X}}_{\mathbf{i}} = (\widetilde{\mathbf{x}}_{\mathbf{i}, 1}, \cdots, \widetilde{\mathbf{x}}_{\mathbf{i}, T})^\top$, and $$\mathbf{M}^{(m)}_{x,\phi}=\mathbf{I}_{T}-\sum_{\mathbf{i}\in [M]^d} \widetilde{\mathbf{X}}_{\mathbf{i}}\left[\widetilde{\mathbf{X}}_{\mathbf{i}}^\top \widetilde{\mathbf{X}}_{\mathbf{i}}+\mathbf{D}^{ -1}\pmb{\Phi}^{(m)}\mathbf{D}^{\top, -1}\right]^{-1}\widetilde{\mathbf{X}}_{\mathbf{i}}^\top.$$

\item Repeat Step 2 until  numerical convergence.
\end{enumerate}

\noindent \textbf{On Sigmoidal Squasher}  --- For Sigmoidal squasher, we have

\begin{eqnarray*}
\sigma^{(n)}(x) = \sum_{k=1}^{n+1} (-1)^{k+1}(k-1)! S(n+1,k) \cdot \sigma(x)^{k},
\end{eqnarray*}
in which 

\begin{eqnarray}\label{B1}
S(n+2,k) =kS(n+1,k) +S(n+1,k-1)
\end{eqnarray}
is the recursion relation for Stirling numbers of the second kind. We refer interested readers to \cite{MINAI1993845} for more details. Sigmoidal squasher is easy to use in the sense that we can arbitrarily choose $u_\sigma$ of Lemma \ref{LEM2.1}. Without loss of generality, we let $u_\sigma \in \{-0.5, 0.5\}$ in the numerical studies. In Figure \ref{FigSig}, we plot $\sigma^{(k)}(x) $ for $k=0,\ldots, 5$ for the purpose of demonstration.

\section{Extra Theoretical Results}\label{App.2}

\subsection{Treatment on $a$}\label{App.21}

Before we explain how to relax the restriction on $a$, we consider a fully nonparametric model by letting $\alpha_0\equiv 0$ and ignoring the sparsity (i.e., $d\equiv d_0$). The rest settings are identical to Section \ref{S2}.

The model to be investigated becomes

\begin{eqnarray*}
y_t = g(\mathbf{x}_t)+\varepsilon_t,
\end{eqnarray*}
and the objective function is a simplified version of that involved in \eqref{EQ2.2}:

\be\label{Eq3.6}
Q (\pmb{\Theta})  = \sum_{t=1}^T[y_t -\widetilde{s}(\mathbf{x}_t \, |\, \pmb{\Theta}) ]^2.
\ee
Accordingly the OLS estimator of $\widetilde{\pmb{\Lambda}}$ is obtained by

\begin{eqnarray}\label{Eq3.7}
\widehat{\pmb{\Theta}} = \argmin_{\widetilde{s}\in \mathcal{S}} Q  (\pmb{\Theta}) \quad\text{with}\quad \widehat{\pmb{\Theta}} =\{ \widehat{\pmb{\theta}}_{\mathbf{i}}\, |\, \mathbf{i}\in [M]^d\}
\end{eqnarray}
and, for $\forall \mathbf{x}_0 \in [-a,a]^d$, the estimator of $g( \mathbf{x}_0)$ is then defined by $\widehat{g}(\mathbf{x}_0) =  \widetilde{s}(\mathbf{x}_0 \, |\, \widehat{\pmb{\Theta}} ) $. Equation \eqref{Eq3.7} admits a closed-form estimator for each $\widehat{\pmb{\theta}}_{\mathbf{i}}$. To see this, we write

\begin{eqnarray*}
\frac{\partial Q  (\pmb{\Theta})  }{\partial \pmb{\theta}_{\mathbf{i}}} 
&=&-2\sum_{t=1}^T [y_t - \widetilde{\mathbf{x}}_{\mathbf{i}, t}^\top \pmb{\theta}_{\mathbf{i}}  ]  \cdot  \widetilde{\mathbf{x}}_{ \mathbf{i}, t},
\end{eqnarray*}
where $\widetilde{\mathbf{x}}_{\mathbf{i},t}  =I_{\mathbf{i},h} (\mathbf{x}_t)(\mathbf{I}_{d_q}\otimes \boldsymbol{\gamma}^\top)  \boldsymbol{\sigma}( \mathbf{x}_t\, |\,\mathbf{x}_{0\mathbf{i}})$ and the equality follows from the fact that $I_{\mathbf{i},h} (\mathbf{x}_t)I_{\mathbf{j},h} (\mathbf{x}_t)=0$ for $\mathbf{i}\ne \mathbf{j}$. Thus, for $\forall\mathbf{i}$, the first order condition yields 

\begin{eqnarray}\label{Eq3.9}
\widehat{\pmb{\theta}}_{\mathbf{i}} &=&\left( \sum_{t=1}^T\widetilde{\mathbf{x}}_{\mathbf{i},t} \widetilde{\mathbf{x}}_{\mathbf{i},t} ^\top \right)^{-1}  \sum_{t=1}^T\widetilde{\mathbf{x}}_{\mathbf{i},t}  y_t,
\end{eqnarray}
where the invertibility of $\sum_{t=1}^T\widetilde{\mathbf{x}}_{\mathbf{i},t} \widetilde{\mathbf{x}}_{\mathbf{i},t} ^\top$ is guaranteed asymptotically in view of \eqref{EqA.6} and \eqref{EqA.7}. 

%
%


After carefully studying \eqref{Eq3.9} for each $\mathbf{i}$ and repeatedly invoking  $I_{\mathbf{i},h} (\mathbf{x}_t)I_{\mathbf{j},h} (\mathbf{x}_t)=0$ for $\mathbf{i}\ne \mathbf{j}$, the following theorem holds.

\begin{theorem}\label{Thm3.1}
Suppose that Assumptions \ref{Ass.1} and \ref{Ass.3} hold. For $\forall \mathbf{x}_0\in [-a,a]^d$, 

\begin{eqnarray*}
\sqrt{Th^d} \widehat{\sigma}_{\mathbf{x}_0}^{ -1}(\widehat{g}(\mathbf{x}_0) - g(\mathbf{x}_0) +O_P(h^p)) \to_D N (0, 1 ),
\end{eqnarray*}
where $\widehat{\sigma}_{\mathbf{x}_0}^2 = \sigma_\varepsilon^2\mathbf{m}(\mathbf{x}_0 \, |\, \mathbf{x}_0)^\top \mathbf{H} \pmb{\Sigma}_{\mathbf{x}_0}^{-1} \mathbf{H} \mathbf{m}(\mathbf{x}_0 \, |\,\mathbf{x}_0)$, and $\pmb{\Sigma}_{\mathbf{x}_0} = f_{\mathbf{x}}( \mathbf{x}_0) \int_{[-1,1]^d}  \mathbf{m}(\mathbf{x}\, |\, \mathbf{0})  \mathbf{m}(\mathbf{x}\, |\, \mathbf{0})^\top  \mathrm{d} \mathbf{x}$. 
\end{theorem}

Obviously, our LNN based estimation method is simple and easy to implement. Accordingly, we propose the following bootstrap procedure to establish inference in practice. 

\begin{enumerate}[leftmargin=*, noitemsep] 
\item We calculate $\widehat{\varepsilon}_t = y_t -\widehat{g}(\mathbf{x}_t)$ for $t\in [T]$.

\item Collect i.i.d. draws of $\{\eta_t\, |\, t\in [T]\}$ from $ N(0,1)$, and construct the bootstrap version dependent variables as follows:  $ y_t^* = \widehat{g}(\mathbf{x}_t) + \widehat{\varepsilon}_t \eta_t$. We re-estimate $g(\cdot)$ using $\{ (y_t^*, \mathbf{x}_T) \, |\, t\in [T]\}$ as under \eqref{Eq3.7}, and denote the estimate as $ \widehat{g}^*(\cdot)$.

\item Repeat Step 2 $R$ times, where $R$ is sufficiently large.
\end{enumerate}

For the bootstrap procedure, the following result holds immediately.

\begin{theorem}\label{Thm3.2}
Let the conditions of Theorem \ref{Thm3.1} hold. Suppose further that $T h^{d+ 2p}\to 0$. For $\forall \mathbf{x}_0\in [-a,a]^d$, we have

\begin{eqnarray*}
\sup_{w}|\text{\normalfont Pr}^* (\sqrt{Th^d} \widehat{\sigma}_{\mathbf{x}_0}^{-1}[\widehat{g}^*(\mathbf{x}_0) - \widehat{g}(\mathbf{x}_0)]\le w ) - \Pr (\sqrt{Th^d} \widehat{\sigma}_{\mathbf{x}_0}^{-1}[\widehat{g}(\mathbf{x}_0) - g(\mathbf{x}_0)]\le w )|=o_P(1).
\end{eqnarray*}
where $\text{\normalfont Pr}^*$ is the probability measure induced by the bootstrap procedure.
\end{theorem}

Note that, we require $g(\mathbf{x})$ to be defined on a compact set, but do not impose restriction on the range of $\{\mathbf{x}_t\}$. In fact, for time series data, it may make more sense to assume that $a$ is diverging, which is indeed achievable.  We now provide two treatments to relax the restriction on $a$.

\medskip

\noindent \textbf{Treatment 1}: Suppose that  $\mathbf{x}_t$ follows a sub-Gaussian distribution, and  we can then require $\sqrt{\log (Th^d)} \cdot a\to \infty.$ A similar treatment has also been discussed in \cite{LTG2016} for example, so we do not further elaborate it here. However, there is a price that we have to pay, i.e., the slow rate of convergence.

\medskip

\noindent  \textbf{Treatment 2}: Alternatively, we can modify the construction of LNN from a nonparametric viewpoint. In the literature of  kernel regression, one normally pre-specifies a point of interest (e.g., $\mathbf{x}_0$), and investigates a small area nearby only which is usually decided by some bandwidth(s) converging to 0. As a result, the parameters obtained from the estimation procedure usually vary with respect to the point of interest. In other words, when evaluating different points from the test set, the number of parameters to be estimated will be proportional to the cardinality of the training set (although estimation is always carried on using the same training set). Provided a large test set, it may create lots of overhead from a computational viewpoint, but the advantage is that in theory it allows us to consider $\forall \mathbf{x}_0\in \mathbb{R}^d$.

That said, for $\forall\mathbf{x}_0\in\mathbb{R}^d$, we consider the following objective function:

\be\label{eqA1}
Q_L (\pmb{\theta})  = \sum_{t=1}^T[y_t - s(\mathbf{x}_t \, |\, \mathbf{x}_0 ,\pmb{\theta}) ]^2\cdot I_{0,h}(\mathbf{x}_t) ,
\ee
where the subscript $L$ infers the local version, $\pmb{\theta}$ is $d_q\times 1$ vector satisfying $\| \pmb{\theta}\|<\infty$, we let $I_{0,h}(\mathbf{x}_t)  :=I(\mathbf{x}_t \in C_{\mathbf{x}_0,h})$ for short, and $C_{\mathbf{x}_0,h}$ is defined in \eqref{EQ1.5} already. Then the OLS estimate of $\widetilde{\pmb{\lambda}}$ defined in Lemma \ref{LEM2.2} is obtained by

\begin{eqnarray*} 
\widehat{\pmb{\theta}} = \argmin_{\pmb{\theta}} Q_L (\pmb{\theta}) 
\end{eqnarray*}
and, accordingly, the estimate of $g(\mathbf{x}_0)$ is defined by $\widehat{g}_L(\mathbf{x}_0) =  s(\mathbf{x}_0 \, |\, \mathbf{x}_0, \widehat{\pmb{\theta}} )  .$  We can then produce the following corollary.

\begin{corollary}\label{CoroA1}
Suppose that Assumptions \ref{Ass.1} and \ref{Ass.3} hold. As $(1/h, Th^d)\to (\infty,  \infty)$, for $\forall \mathbf{x}_0\in \mathbb{R}^d$,

\begin{enumerate}[leftmargin=*, noitemsep] 
\item $\sqrt{Th^d} \, \left(\mathbf{H}^{-1}  \mathbf{D}^{\top,-1}(\widehat{\pmb{\theta}}  -\widetilde{\pmb{\lambda}}) +O_P(h^p))\right) \to_D N\left(\mathbf{0},\sigma_\varepsilon^2\pmb{\Sigma}_{\mathbf{x}_0}^{-1}\right),$

\item $\sqrt{Th^d} \widehat{\sigma}_{\mathbf{x}_0}^{-1}(\widehat{g}_L(\mathbf{x}_0) - g(\mathbf{x}_0) +O_P(h^p)) \to_D N (0, 1 ),$
\end{enumerate}
where $\widetilde{\pmb{\lambda}}$ is uniquely determined by $g(\mathbf{x}_0)$.
\end{corollary}

\subsection{A Nonparametric Binary Model}\label{App.22}

So far, we have not explored the binary structure of $z_t$ much, which is also of great interest widely adopted in a wide range of applications (\citealp{Athey2019}). To close our investigation about the model \eqref{EQ1.1}, we use LNN approach to infer $G(\cdot)$. 

Recall that

\begin{eqnarray*}
z =I(G(\mathbf{x})-\eta \ge 0 ).
\end{eqnarray*}
For simplicity, we suppose that the respective probability density function (PDF) and the cumulative distribution function (CDF) of $\eta$ are known, and denote the PDF and CDF by $\phi_\eta(\cdot)$ and $\Phi_\eta(\cdot)$ respectively. Here, the information about $\phi_\eta(\cdot)$ and $\Phi_\eta(\cdot)$ is necessary for carrying on likelihood estimation.

Direct calculation shows that

\begin{eqnarray*}
\Pr(z=1\, |\, \mathbf{x}) =\Phi_\eta(G(\mathbf{x})) \quad\text{and}\quad\Pr(z=0\, |\, \mathbf{x}) =1-\Phi_\eta(G(\mathbf{x})),
\end{eqnarray*}
which yield $E[z  \, |\, \mathbf{x} ] =\Phi_\eta(G(\mathbf{x}))$. Accordingly, the log-likelihood function is defined below:

\begin{eqnarray}\label{Eq3.10}
\log L(G)&=& \sum_{t=1}^T l_t(G(\mathbf{x}_t))=\sum_{t=1}^T\left\{  (1-z_t)\cdot \log [1-\Phi_\eta(G(\mathbf{x}_t))]+ z_t\cdot \log \Phi_\eta(G(\mathbf{x}_t))\right\},
\end{eqnarray}
where the definition of $l_t(\cdot)$ is obvious. To infer $G(\cdot)$, we consider the following objective function:

\begin{eqnarray*} 
\log L(\widetilde{s}(\cdot\, | \, \pmb{\Theta})) &=& \sum_{t=1}^T\log l_t(\widetilde{s}(\mathbf{x}_t\, | \, \pmb{\Theta}))\nonumber \\
&=&\sum_{t=1}^T\left\{  (1-y_t)\cdot \log [1-\Phi_\eta(\widetilde{s}(\mathbf{x}_t\, | \, \pmb{\Theta}))]+ y_t\cdot \log \Phi_\eta(\widetilde{s}(\mathbf{x}_t\, | \, \pmb{\Theta}))\right\},
\end{eqnarray*}
which yields the following maximum likelihood estimator:

\begin{eqnarray}\label{Eq3.11}
\widehat{\pmb{\Theta}} =\argmax_{\widetilde{s}\in \mathcal{S}} \log L(\widetilde{s}(\cdot\, | \, \pmb{\Theta})) .
\end{eqnarray}

\medskip

Note that our LNN method involves a general unknown function form. As a result, the classical results of likelihood estimation, such as those in \cite{NEWEY19942111}, no longer hold. Therefore, before establishing an asymptotic distribution using $\widehat{\pmb{\Theta}}$, we state a lemma to show the feasibility of LNN architecture when modelling binary outcomes.

\begin{lemma}\label{Lem34}
Under Assumptions \ref{Ass.1} and \ref{Ass.3}, 

\begin{eqnarray*}
\frac{1}{T}\sum_{t=1}^T[\Phi_\eta( g(\mathbf{x}_t))- \Phi_\eta( \widetilde{s}(\mathbf{x}_t\, | \, \widehat{\pmb{\Theta}}) )]^2 =o_P(1) .
\end{eqnarray*}
\end{lemma}

Lemma \ref{Lem34} provides the consistency, and also bridges the likelihood estimation and the nonlinear least squares approach to some extent. To be precise, Lemma \ref{Lem34} does not provide any specific consistency for $\forall\, \widehat{\pmb{\theta}}_{\mathbf{i}}$. Instead, it evaluates the overall performance of LNN. More importantly, it says when modelling a binary outcome, the likelihood estimation using the LNN architecture is approximately equivalent to implementing a nonlinear least squares method provided the distribution of $\eta_t$ is correctly specified. In addition,  Lemma \ref{Lem34} further infers that

\begin{eqnarray*}
 \frac{1}{T}\sum_{t=1}^T [ g(\mathbf{x}_t)-   \widetilde{s}(\mathbf{x}_t\, | \, \widehat{\pmb{\Theta}}) ]^2=o_P(1),
\end{eqnarray*}
so many remarks made previously can be directly applied. Last but not least, Lemma \ref{Lem34} facilitates numerical implementation in practice, which will be further discussed in Appendix \ref{App.3}.

\medskip

Below, we establish the following asymptotic distribution.

\begin{theorem}\label{Thm3.3}
Suppose that Assumptions \ref{Ass.1} and \ref{Ass.3} hold. For $\forall \mathbf{x}_0\in [-a,a]^d$, 

\begin{eqnarray*}
\sqrt{Th^d} \widetilde{\sigma}_{\mathbf{x}_0}^{-1} (\widehat{g}(\mathbf{x}_0) - g(\mathbf{x}_0) +O_P(h^p)) \to_D N (0,1),
\end{eqnarray*}
where $ \widehat{g}(\mathbf{x}_0) =  \widetilde{s}(\mathbf{x}_0 \, |\, \widehat{\pmb{\Theta}} )$, $\widetilde{\sigma}_{\mathbf{x}_0}^2 =\mathbf{m}(\mathbf{x}_0 \, |\, \mathbf{x}_0)^\top \mathbf{H} \widetilde{\pmb{\Sigma}}_{\mathbf{x}_0}^{-1}\mathbf{H} \mathbf{m}(\mathbf{x}_0 \, |\, \mathbf{x}_0)$, and 

\begin{eqnarray*}
\widetilde{\pmb{\Sigma}}_{\mathbf{x}_0}=\frac{f_{\mathbf{x}}(\mathbf{x}_0) \phi_\eta(g(\mathbf{x}_0))^2 }{[1- \Phi_\eta(g(\mathbf{x}_0))] \Phi_\eta(g(\mathbf{x}_0))} \int_{[-1,1]^d}  \mathbf{m}(\mathbf{x}\, |\, \mathbf{0})  \mathbf{m}(\mathbf{x}\, |\, \mathbf{0})^\top  \mathrm{d} \mathbf{x} .
\end{eqnarray*}
\end{theorem}

 In light of \cite{PA2012}, we propose a score based wild bootstrap approach for inferential purposes as follows.

\begin{enumerate}[leftmargin=*, noitemsep] 
\item For each bootstrap replication, we collect i.i.d. draws of $\{\eta_t \, |\, t\in [T] \}$ from $N(0,1)$, and calculate 

\begin{eqnarray}\label{Eq3.12}
\widehat{\pmb{\theta}}_{\mathbf{i}}^* = \widehat{\pmb{\theta}}_{\mathbf{i}} + \left(\sum_{t=1}^T\frac{\partial^2  \log l_t(\widetilde{s}(\mathbf{x}_t\, | \, \widehat{\pmb{\Theta}}))  }{\partial \pmb{\theta}_{\mathbf{i}} \partial \pmb{\theta}_{\mathbf{i}}^\top } \right)^{-1} \sum_{t=1}^T\frac{\partial \log l_t(\widetilde{s}(\mathbf{x}_t\, | \, \widehat{\pmb{\Theta}})) }{\partial \pmb{\theta}_{\mathbf{i}}} \eta_t,
\end{eqnarray}
where $l_t(\cdot)$ is defined in \eqref{Eq3.10}.

\item Repeat Step 1 $R$ times, where $R$ is sufficiently large.
\end{enumerate}

It is worth pointing out that the above procedure is computationally efficient in the sense that the right hand side of \eqref{Eq3.12} enjoys a closed-form expression, which is given in \eqref{EqA.240} and \eqref{EqA.24} specifically. Practically, the bootstrap procedure may require much less time compared with the estimation of \eqref{Eq3.11} itself. 
 
The following theorem holds for the above bootstrap procedure.  
 
\begin{theorem}\label{Thm3.4}
Let the conditions of Theorem \ref{Thm3.3} hold. Suppose further that $T h^{d+ 2p}\to 0$. For $\forall \mathbf{x}_0\in [-a,a]^d$, we have

\begin{eqnarray*}
\sup_{w}|\text{\normalfont Pr}^* (\sqrt{Th^d} \widetilde{\sigma}_{\mathbf{x}_0}^{-1}[\widehat{g}^*(\mathbf{x}_0) - \widehat{g}(\mathbf{x}_0)]\le w ) - \Pr (\sqrt{Th^d} \widetilde{\sigma}_{\mathbf{x}_0}^{-1}[\widehat{g}(\mathbf{x}_0) - g(\mathbf{x}_0)]\le w )|=o_P(1),
\end{eqnarray*}
where  $\text{\normalfont Pr}^*$ is the probability measure induced by the bootstrap procedure, and $\widehat{g}^*(\cdot)$ is yielded by the bootstrap draws in an obvious manner.
\end{theorem}

\section{Extra Simulation}\label{App.3}

We now provide extra simulations, which have two focuses: (1). examining the theoretical results  in Appendix \ref{App.2}, and (2). demonstrating the newly proposed method works reasonably well even without involving sparsity. 

\medskip

\textbf{On the Fully Nonparametric Model} --- Consider the following regression model: 
\begin{equation}
y_t =g(\mathbf{x}_t) +\varepsilon_t,
\end{equation}
where the variables are generated in the same manner as in Section \ref{S3}. In what follows, we only vary the values of $(q, d, u_\sigma)$. Specifically, we consider the cases $d=2, 8$. The rest parameters are identical to those in the main text.

To measure the finite sample performance,  we select the points as follows:

\begin{eqnarray*}
\mathbf{x}_{L,\mathbf{j}}&=&\left( -a+\frac{2a}{L-1} (j_1-1), -a+\frac{2a}{L-1} (j_2-1) \right)^\top\quad\text{for}\quad \mathbf{j} \in [20]^2 \ (\text{i.e., } d=2 ), \nonumber \\
\mathbf{x}_{L,\mathbf{j}} &=& -a \mathbf{1}_d+(j-1)\cdot 0.05 \mathbf{1}_d \quad\text{with} \quad\mathbf{j} = j \cdot \mathbf{1}_d \quad\text{for}\quad d= 8.
\end{eqnarray*}
With each dataset, we first estimate all $g(\mathbf{x}_{L,\mathbf{j}})$ using the approach of Appendix \ref{App.21}, and then construct the corresponding 95\% confidence interval using the bootstrap procedure documented in Theorem \ref{Thm3.2} for each point. We report $\text{RMSE}_g$ and $\text{CR}_g$ which are defined in the main text already.

\medskip

First, we draw some plots for the case with $d=2$. In both Figures \ref{FigSim1a} and \ref{FigSim2a}, the first sub-plot is always the true $g(\mathbf{x})$. For the rest of sub-plots, each has three layers. The middle one is the average of estimates  over $n$ replications. The top and bottom layers are the averages of the bootstrap draws corresponding to the 97.5\% and 2.5\% quantiles respectively. A few facts emerge. Overall, the LNN approach can recover the unknown function reasonably well.  Also, both figures are very similar, so the results are not sensitive to the choice of $u_\sigma$ as explained in Remark \ref{rmk1}. 

More detailed numbers are summarized Table \ref{Tab1a}. As expected, when $T$ goes up, RMSE$_g$ converges to 0 and CR$_g$ converges to 0.95.  Also, we note that when $d$ increases, RMSE$_g$ increases but still has reasonable performance expect the case with $(d=8,q=4)$.  Therefore, it seems that for large $d$, smaller $q$ yields better finite sample performance.  

\medskip

\textbf{On the Binary Model} --- We consider the following data generating process:

\begin{eqnarray*}
 z_t = \left\{\begin{array}{ll}
1 & G(\mathbf{x}_t)- \eta_t \ge 0\\
0 & \text{otherwise}
\end{array} \right.  ,
\end{eqnarray*}
in which $\eta_t = 0.5\eta_{t-1} + N(0, 0.75)$, the $j^{th}$ element of $\mathbf{x}_t$ is generated as $x_{t,j}\sim U(-a, a)$, and $G(\mathbf{x}) = 1+\sin(\mathbf{x}^\top \mathbf{1}_d/d)$.  The rest parameters are identical to the simulation design of the fully nonparametric model.

We first note a computational issue. We reply on ``fminunc" function of Matlab to solve

\begin{eqnarray*}
\argmin_{\widetilde{s}\in \mathcal{S}} [-\log L(\widetilde{s}(\cdot\, | \, \pmb{\Theta})) ], 
\end{eqnarray*}
which is  the same as that in \eqref{Eq3.11}. In order to invoke the minimization process in any statistical software (including R, Matlab, etc.), one needs to provide initial values to the parameters under estimation. As a consequence, the numbers reported below are affected by the initial values more or less. Although it is not our intention to tackle this complicated computational issue in this paper, Lemma \ref{Lem34} does become useful in this case. Recall that Lemma \ref{Lem34} bridges the log likelihood estimation and the nonlinear least squares estimation. Therefore, we first conduct an OLS estimation using the approach of Appendix \ref{App.21} as the initial value of $\pmb{\Theta}$ for each generated $\{( z_t, \mathbf{x}_t)\, |\, t\in [T]\}$. We then invoke log likelihood estimation as our final estimate of $\pmb{\Theta}$ for each dataset. Even in this case, the computation is rather slow, and the computational time increases dramatically when the number of parameters goes up. 

We draw a few plots  in Figures \ref{FigSim3} and \ref{FigSim4}. The first sub-plot is always the true $G(\mathbf{x})$. For the rest of sub-plots, each has three layers. The middle one is the average of estimates  over $n$ replications. The top and bottom layers are the averages of the bootstrap draws corresponding to the 97.5\% and 2.5\% quantiles respectively. Overall, the LNN approach can recover the unknown function reasonably well.  Also, both figures are very similar, so the results are not sensitive to the choice of $u_\sigma$. 

We further summarize the detailed numbers in Table \ref{Tab2}. A few facts should be mentioned. First, the coverage rates are reasonably well. As expected, when $T$ goes up, RMSE converges to 0 and CR converges to 0.95.  Also, we note that when $d$ increases, RMSE increases but still has reasonable performance. The results are not changing much with respect to the value of $u_\sigma$. Again, it seems that for large $d$, smaller $q$ yields better finite sample performance.

 {\small
 
\section{Preliminary Lemmas \& Proofs}\label{App.4}

Before proving the theoretical results in Appendices \ref{App.42}-\ref{App.45}, we first present all preliminary lemmas in Appendix \ref{App.41}. 

\subsection{Preliminary Lemmas}\label{App.41}

For $\mathbf{a}= (a_0, a_1,\ldots, a_d)^\top \in \mathbb{R}^{d+1}$, $\mathbf{r} = (r_0,r_1,\ldots, r_d)^\top\in \mathbb{N}_0^{d+1}$, and $ \mathbf{x}\in \mathbb{R}^{d}$, we let

\begin{eqnarray*}
f_{\mathbf{a}} (\mathbf{x}) =[(1, \mathbf{x}^\top)\mathbf{a} ]^q =\sum_{|\mathbf{r}|=q}\binom{q}{\mathbf{r}} \cdot a_0^{r_0}\prod_{k=1}^d a_k^{r_k}   x_k^{r_k}.
\end{eqnarray*}
Obviously, we have $f_{\mathbf{a}}\in \mathscr{P}_q$, where $ \mathscr{P}_q$ is defined in \eqref{EQ1.3}.

\medskip

\begin{lemma}\label{LemA.1}
Let $f\,:\, \mathbb{R}^d\to \mathbb{R}$ be a $(p,\mathscr{C})$-smooth function. For $\forall \mathbf{x}_0\in \mathbb{R}^d$, let
\begin{eqnarray*}
p_q(\mathbf{x}\, |\, \mathbf{x}_0) =\sum_{0\le |\mathbf{J}|\le q} \frac{1}{\mathbf{J}!}\cdot \frac{\partial^{|\mathbf{J}|} f(\mathbf{x}_0) }{\partial \mathbf{x}^{\mathbf{J}}} (\mathbf{x}-\mathbf{x}_0)^{\mathbf{J}}.
\end{eqnarray*}
Then 
\begin{eqnarray*}
\|f(\mathbf{x})-p_q(\mathbf{x}\, |\, \mathbf{x}_0)\|_{\infty}\le O(1)\|\mathbf{x}-\mathbf{x}_0 \|^p,
\end{eqnarray*}
where $p=q+s$, $O(1)$ depends on $d$ and $q$ only.
\end{lemma}

\medskip

\begin{lemma}\label{LemA.2}
For almost all $\mathbf{a}_1,\ldots,\mathbf{a}_{d_q}\in \mathbb{R}^{d+1}$ (with respect to the Lebesgue measure in $\mathbb{R}^{(d+1)\times d_q}$), we have that $\{f_{\mathbf{a}_1} (\mathbf{x}),\ldots,f_{\mathbf{a}_{d_q}} (\mathbf{x})\}$ is a basis of the linear vector space $ \mathscr{P}_q$.
\end{lemma}

\medskip

\begin{lemma}\label{Lem_lasso}
Under Assumptions \ref{Ass.1}, \ref{Ass.3}, and \ref{Ass.4}, we have 
\begin{eqnarray*}
\|\widetilde{\alpha}-\alpha_0\|^2+ h^d\sum_{\mathbf{i}\in [M]^d}\|\mathbf{H}^{-1}\mathbf{D}^{\top,-1}(\widetilde{\pmb{\theta}}_{\mathbf{i}}-\widetilde{\pmb{\lambda}}_{\mathbf{i}})\|^2=O_P\left( \frac{1}{Th^{d}}\right).
\end{eqnarray*}
\end{lemma}

\medskip

\begin{lemma}\label{Lem_Ora}
Suppose  Assumptions \ref{Ass.1}, \ref{Ass.3}, and \ref{Ass.4} hold. As $(1/h, Th^{d_0})\to (\infty,  \infty)$, 
\begin{enumerate}
\item $\sqrt{T}(\widetilde{\alpha}_{c}-\alpha_0+O_P(h^p))\to_D N\left(\mathbf{0}, M_{c,z}^{-2}\sigma_{c,z}^2\right)$;
\item $\sigma_\varepsilon^{-1}\pmb{\Sigma}_{c,\mathbf{i}}^{1/2}\sqrt{Th^{d_0}}\left(\mathbf{H}_c^{-1}  \mathbf{D}_c^{\top,-1}(\widetilde{\pmb{\theta}}_{c,\mathbf{i}} -\widetilde{\pmb{\lambda}}_{c,\mathbf{i}})+O_P(h^p)\right) \to_D N\left(\mathbf{0},\mathbf{I}_{d_q^0}\right)$, for each $\mathbf{i}\in [M]^{d_0}$,
\end{enumerate}
where $\mathbf{H}_c$, $\mathbf{D}_c$, and $\widetilde{\pmb{\lambda}}_{c,\mathbf{i}}$ are counterpart matrices of $\mathbf{H}$, $\mathbf{D}$, and $\widetilde{\pmb{\lambda}}_{\mathbf{i}}$ for the true model, and
\begin{eqnarray*}
M_{c,z}&=&\int_{\mathbf{x}\in [-a,a]^d}    \Phi_\eta(G(\mathbf{x}))  f_{\mathbf{x}}(\mathbf{x})\mathrm{d} \mathbf{x}-\lim_{T\rightarrow\infty}h^{d_0}\sum_{\mathbf{i}\in [M]^{d_0}}\mathbf{M}_{c,\mathbf{i}}^\top \pmb{\Sigma}_{c,\mathbf{i}}^{-1} \mathbf{M}_{c,\mathbf{i}}
\nonumber\\
\pmb{\Sigma}_{c,\mathbf{i}}&=&f_{\mathbf{x}_c}(\mathbf{x}_{c,\mathbf{i}}) \int_{[-1,1]^{d_0}}  \mathbf{m}_c(\mathbf{x}_c\, |\, \mathbf{0})  \mathbf{m}_c(\mathbf{x}_c\, |\, \mathbf{0})^\top  \mathrm{d} \mathbf{x}_c,
\end{eqnarray*}
with
\begin{equation*}
\mathbf{M}_{c,\mathbf{i}}=\int_{[-a,a]^{d-d_0}}\Phi_\eta(G(\mathbf{x}_{c,0\mathbf{i}},\mathbf{z}))    f_{\mathbf{x}}(\mathbf{x}_{c,0\mathbf{i}},\mathbf{z})\mathrm{d}\mathbf{z} \int_{[-1,1]^{d_0}}    \mathbf{m}_c(\mathbf{x}_c\, |\, \mathbf{0}) \mathrm{d} \mathbf{x}_c.
\end{equation*}
\end{lemma}

\medskip

\begin{lemma}\label{LemA.3}
Suppose  Assumptions \ref{Ass.1} and \ref{Ass.3} hold. As $(1/h, Th^d)\to (\infty,  \infty)$, for each $\mathbf{i}\in [M]^d$,

\begin{eqnarray*}
\sigma_\varepsilon^{-1}\pmb{\Sigma}_{\mathbf{i}}^{1/2}\sqrt{Th^d}\left(\mathbf{H}^{-1}  \mathbf{D}^{\top,-1}(\widehat{\pmb{\theta}}_{\mathbf{i}} -\widetilde{\pmb{\lambda}}_{\mathbf{i}})+O_P(h^p)\right) \to_D N\left(\mathbf{0},\mathbf{I}_{d_q}\right),
\end{eqnarray*}
where $ \pmb{\Sigma}_{\mathbf{i}} =f_{\mathbf{x}}(\mathbf{x}_{\mathbf{i},0}) \int_{[-1,1]^d}  \mathbf{m}(\mathbf{x}\, |\, \mathbf{0})  \mathbf{m}(\mathbf{x}\, |\, \mathbf{0})^\top  \mathrm{d} \mathbf{x}$. 
\end{lemma}

\medskip

Before presenting the next lemma, we  calculate the partial derivatives of $\log L(\widetilde{s}(\cdot\, | \, \pmb{\Theta}))$ with respect to each $\pmb{\theta}_{\mathbf{i}}$.

\begin{eqnarray}\label{EqA.240}
\frac{\partial \log L(\widetilde{s}(\cdot\, | \, \pmb{\Theta})) }{\partial \pmb{\theta}_{\mathbf{i}}} &=&\sum_{t=1}^T\left\{ -\frac{(1-z_t) \cdot \phi_\eta(\widetilde{s}(\mathbf{x}_t\, | \, \pmb{\Theta}))}{1-\Phi_\eta(\widetilde{s}(\mathbf{x}_t\, | \, \pmb{\Theta}) )} +\frac{z_t\cdot \phi_\eta(\widetilde{s}(\mathbf{x}_t\, | \, \pmb{\Theta})) }{\Phi_\eta(\widetilde{s}(\mathbf{x}_t\, | \, \pmb{\Theta}))} \right\}\widetilde{\mathbf{x}}_{\mathbf{i}, t} \nonumber \\
&=&\sum_{t=1}^T  \frac{[ z_t - \Phi_\eta(\widetilde{s}(\mathbf{x}_t\, | \, \pmb{\Theta}))]\cdot \phi_\eta(\widetilde{s}(\mathbf{x}_t\, | \, \pmb{\Theta}))}{\Phi_\eta(\widetilde{s}(\mathbf{x}_t\, | \, \pmb{\Theta})) [1-\Phi_\eta(\widetilde{s}(\mathbf{x}_t\, | \, \pmb{\Theta}))]}  \widetilde{\mathbf{x}}_{\mathbf{i}, t} \nonumber \\
&=&\sum_{t=1}^T  \frac{[ z_t - \Phi_\eta(s(\mathbf{x}_t\, | \, \mathbf{x}_{\mathbf{i},0},\pmb{\theta}_{\mathbf{i}} ))]\cdot \phi_\eta(s(\mathbf{x}_t\, | \, \mathbf{x}_{\mathbf{i},0},\pmb{\theta}_{\mathbf{i}} ))}{\Phi_\eta(s(\mathbf{x}_t\, | \, \mathbf{x}_{\mathbf{i},0},\pmb{\theta}_{\mathbf{i}} )) [1-\Phi_\eta(s(\mathbf{x}_t\, | \, \mathbf{x}_{\mathbf{i},0},\pmb{\theta}_{\mathbf{i}} ))]}  \widetilde{\mathbf{x}}_{\mathbf{i}, t} ,
\end{eqnarray}
where the third equality follows from the fact that $\mathbf{x}_t$ can not simultaneous belong to $C_{\mathbf{x}_{0\mathbf{i}},h}$ and $C_{\mathbf{x}_{0\mathbf{j}},h}$ for $\mathbf{i}\ne \mathbf{j}$ by the construction of $\widetilde{s}(\cdot\, | \, \pmb{\Theta})$, and $\widetilde{\mathbf{x}}_{\mathbf{i}, t}$ is the same as that defined in Section \ref{S2}.

Based on $\frac{\partial \log L(\widetilde{s}(\cdot\, | \, \pmb{\Theta})) }{\partial \pmb{\theta}_{\mathbf{i}}}$ and some tedious calculation, the second order derivative is

\begin{eqnarray}\label{EqA.24}
&&\frac{\partial^2 \log L(\widetilde{s}(\cdot\, | \, \pmb{\Theta})) }{\partial \pmb{\theta}_{\mathbf{i}} \partial \pmb{\theta}_{\mathbf{i}}^\top }  \nonumber \\
&=&-\sum_{t=1}^T \frac{\phi_\eta(s(\mathbf{x}_t\, | \, \mathbf{x}_{\mathbf{i},0},\pmb{\theta}_{\mathbf{i}} ) )^2 }{[1- \Phi_\eta(s(\mathbf{x}_t\, | \, \mathbf{x}_{\mathbf{i},0},\pmb{\theta}_{\mathbf{i}} ))] \Phi_\eta(s(\mathbf{x}_t\, | \, \mathbf{x}_{\mathbf{i},0},\pmb{\theta}_{\mathbf{i}} ))}\widetilde{\mathbf{x}}_{\mathbf{i}, t} \widetilde{\mathbf{x}}_{\mathbf{i}, t} ^\top \nonumber \\
&&+\sum_{t=1}^T[z_t -\Phi_\eta(s(\mathbf{x}_t\, | \, \mathbf{x}_{\mathbf{i},0},\pmb{\theta}_{\mathbf{i}} ))] \frac{\phi_\eta^{(1)}(s(\mathbf{x}_t\, | \, \mathbf{x}_{\mathbf{i},0},\pmb{\theta}_{\mathbf{i}} ))}{[1-\Phi_\eta(s(\mathbf{x}_t\, | \, \mathbf{x}_{\mathbf{i},0},\pmb{\theta}_{\mathbf{i}} )) ]\Phi_\eta(s(\mathbf{x}_t\, | \, \mathbf{x}_{\mathbf{i},0},\pmb{\theta}_{\mathbf{i}} ))}  \widetilde{\mathbf{x}}_{\mathbf{i}, t} \widetilde{\mathbf{x}}_{\mathbf{i}, t} ^\top \nonumber \\ 
&&-\sum_{t=1}^T [z_t -\Phi_\eta(s(\mathbf{x}_t\, | \, \mathbf{x}_{\mathbf{i},0},\pmb{\theta}_{\mathbf{i}} ))]\frac{ \phi_\eta(s(\mathbf{x}_t\, | \, \mathbf{x}_{\mathbf{i},0},\pmb{\theta}_{\mathbf{i}} ))^2 [1-2\Phi_\eta(s(\mathbf{x}_t\, | \, \mathbf{x}_{\mathbf{i},0},\pmb{\theta}_{\mathbf{i}} ))] }{[1-\Phi_\eta(s(\mathbf{x}_t\, | \, \mathbf{x}_{\mathbf{i},0},\pmb{\theta}_{\mathbf{i}} ))]^2 \Phi_\eta(s(\mathbf{x}_t\, | \, \mathbf{x}_{\mathbf{i},0},\pmb{\theta}_{\mathbf{i}} ))^2}  \widetilde{\mathbf{x}}_{\mathbf{i}, t} \widetilde{\mathbf{x}}_{\mathbf{i}, t} ^\top. \quad 
\end{eqnarray}

\begin{lemma}\label{LemA.4}
Under Assumptions \ref{Ass.1} and \ref{Ass.3}, for each $\mathbf{i}\in [M]^d$,

\begin{enumerate}
\item $\frac{1}{T}\sum_{t=1}^TI_{\mathbf{i}, h}(\mathbf{x}_t) [g(\mathbf{x}_t) - s(\mathbf{x}_t \, | \, \widetilde{\mathbf{x}}_{\mathbf{i}}, \widehat{\pmb{\theta}}_{\mathbf{i}} )]^2 =o_P(1)$,

\item $\left\|\frac{1}{T}\mathbf{H}\mathbf{D}\frac{\partial^2 \log L(\widetilde{s}(\cdot\, | \, \widehat{\pmb{\Theta}})) }{\partial \pmb{\theta}_{\mathbf{i}} \partial \pmb{\theta}_{\mathbf{i}}^\top }\mathbf{D}^\top \mathbf{H} -\widetilde{\pmb{\Sigma}}_{\mathbf{i}}\right\| =o_P(1)$, 

\item $\left\|\frac{1}{T}\mathbf{H}\mathbf{D}\frac{\partial^2 \log L(s(\cdot\, | \,  \widetilde{\mathbf{x}}_{\mathbf{i}}, \pmb{\theta}_{\mathbf{i}}^*) ) }{\partial \pmb{\theta}_{\mathbf{i}} \partial \pmb{\theta}_{\mathbf{i}}^\top }\mathbf{D}^\top \mathbf{H} -\widetilde{\pmb{\Sigma}}_{\mathbf{i}}\right\| =o_P(1)$,  

\item $\widetilde{\pmb{\Sigma}}_{\mathbf{i}}^{-1/2}\frac{1}{\sqrt{Th^d}} \mathbf{H} \mathbf{D}\frac{\partial \log L(g(\cdot) ) }{\partial \pmb{\theta}_{\mathbf{i}}  }\to_D N(\mathbf{0}, \mathbf{I}_{d_q})$,
\end{enumerate}
where $\pmb{\theta}_{\mathbf{i}}^*$ lies between $\widehat{\pmb{\theta}}_{\mathbf{i}}$ and $ \widetilde{\pmb{\lambda}}_{\mathbf{i}}$, and

\begin{eqnarray*}
\widetilde{\pmb{\Sigma}}_{\mathbf{i}}=\frac{f_{\mathbf{x}}(\widetilde{\mathbf{x}}_{\mathbf{i}}) \phi_\varepsilon(g(\widetilde{\mathbf{x}}_{\mathbf{i}}))^2 }{[1- \Phi_\varepsilon(g(\widetilde{\mathbf{x}}_{\mathbf{i}}))] \Phi_\varepsilon(g(\widetilde{\mathbf{x}}_{\mathbf{i}}))} \int_{[-1,1]^d}  \mathbf{m}(\mathbf{x}\, |\, \mathbf{0})  \mathbf{m}(\mathbf{x}\, |\, \mathbf{0})^\top  \mathrm{d} \mathbf{x} .
\end{eqnarray*}
\end{lemma}

\subsection{Proofs for the LNN Architecture}\label{App.42}

\noindent \textbf{Proof of Lemma \ref{LemA.1}:}

This is Lemma 8 of \cite{BK2019}, so the derivation is omitted.  \hspace*{\fill}{$\blacksquare$}

\bigskip

\noindent \textbf{Proof of Lemma \ref{LemA.2}:}

In what follows, let

\begin{eqnarray*}
\mathbf{a}_k =(a_{k,0}, a_{k,1},\ldots, a_{k,d})^\top \ \text{for}\ k\in [d_q],\quad\text{and}\quad \widetilde{\mathbf{r}} =\{\mathbf{r} \in \mathbb{N}_0^{d+1}\, |\, |\mathbf{r}|=q\}.
\end{eqnarray*}

It suffices to show that $f_{\mathbf{a}_1} (\mathbf{x}),\ldots,f_{\mathbf{a}_{d_q}} (\mathbf{x})$ are linearly independent. To do this, let $b_1,\ldots, b_{d_q}\in \mathbb{R}$ be such that

\begin{eqnarray}\label{EqA.1}
\sum_{k=1}^{d_q}b_kf_{\mathbf{a}_k}(\mathbf{x})=0.
\end{eqnarray}
As we explained under \eqref{EQ1.3}, the monomials involved in \eqref{EQ1.3} are linearly independent. Thus, \eqref{EqA.1} implies that

\begin{eqnarray*}
\sum_{k=1}^{d_q}b_k \prod_{j=0}^d a_{k,j}^{r_j} =0 \ \text{ for }\ \forall\mathbf{r}  \in \widetilde{\mathbf{r}} .
\end{eqnarray*}

Note that $\sharp\widetilde{\mathbf{r}}=d_q$ by design, so we can construct a one-to-one relationship between $k\in [d_q]$ and $\mathbf{r}\in \widetilde{\mathbf{r}}$. Using this relationship, we can construct $p(\mathbf{x})\in  \mathscr{P}_q$ as follows:

\begin{eqnarray*}
p(\mathbf{x}) =\sum_{k=1}^{d_q} b_k  \prod_{j=1}^d x_j^{r_j}=\sum_{\mathbf{r}\in \widetilde{\mathbf{r}}} b_\mathbf{r} \prod_{j=1}^d x_j^{r_j},
\end{eqnarray*}
which satisfies 

\begin{eqnarray}\label{EqA.2}
p(\mathbf{a}_k)=0  \quad \text{for}\quad \forall k\in [d_q].
\end{eqnarray}
Position 4 in \cite{SAUER2006191} implies that \eqref{EqA.2} has the only solution $p(\cdot)=0$ in $\mathscr{P}_q$ for Lebesgue almost all $\mathbf{a}_1,\ldots,\mathbf{a}_{d_q}\in \mathbb{R}^{d+1}$, which in turn implies $b_1=\cdots =b_{d_q}=0$. The proof is now completed. \hspace*{\fill}{$\blacksquare$}

\bigskip

\noindent \textbf{Proof of Lemma \ref{LEM2.1}:}

Before proceeding further, we would like to point out that in what follows,  $\mathtt{C}$ is a constant for the purpose of rescaling only.

\medskip

By Assumption \ref{Ass.1}.2, there is a point $u_\sigma\in \mathbb{R}$ such that none of the derivatives up to the order $q$ is 0 at $u_\sigma$. Thus, we construct the following one-layer NN:

\begin{eqnarray}\label{EqA.3}
&&\sum_{k=1}^{q+1}(-1)^{q+k-1}\cdot \frac{\mathtt{C}^q}{\sigma^{(q)} (u_\sigma)} \binom{q}{k-1} \cdot\sigma\left( \frac{k-1}{\mathtt{C}}\cdot (x-x_0) + u_\sigma\right) \nonumber \\
&=& \sum_{k=0}^q (-1)^{q+k}\cdot \frac{\mathtt{C}^q}{\sigma^{(q)} (u_\sigma)} \binom{q}{k} \cdot\sigma\left( \frac{k}{\mathtt{C}}\cdot (x-x_0)  + u_\sigma\right) \nonumber \\
&=& (-1)^{q} \frac{\mathtt{C}^q}{\sigma^{(q)} (u_\sigma)} \sum_{k=0}^q (-1)^{k}\cdot \binom{q}{k} \cdot\sigma\left( \frac{k}{\mathtt{C}}\cdot (x-x_0)  + u_\sigma\right),
\end{eqnarray}
in which the definitions of $\gamma_k$'s and $\beta_k$'s are obvious. 

By Assumption \ref{Ass.1}.2 again, $\sigma(\cdot)$ is $q+1$ times continuously differentiable. Thus, it can be expanded in a Taylor series with Lagrange remainder around $u_\sigma$ up to order $q$:

\begin{eqnarray}\label{EqA.4}
&&\sum_{k=0}^q (-1)^{k}\cdot \binom{q}{k} \cdot\sigma\left( \frac{k}{\mathtt{C}}\cdot (x-x_0)  + u_\sigma\right)\nonumber \\
&=&\sum_{k=0}^q (-1)^{k}\cdot \binom{q}{k} \cdot\left(\sum_{j=0}^q \frac{\sigma^{(j)} (u_\sigma) \cdot ((x-x_0) k)^j}{\mathtt{C}^j\cdot j!} +\frac{\sigma^{(q+1)} (\xi_k) \cdot ((x-x_0) k)^{q+1}}{\mathtt{C}^{q+1} \cdot (q+1)! }\right)\nonumber \\
&=& \sum_{j=0}^q\frac{\sigma^{(j)} (u_\sigma) \cdot (x-x_0) ^j}{\mathtt{C}^j\cdot j!}  \sum_{k=0}^q (-1)^{k}\cdot k^j\cdot \binom{q}{k} \nonumber \\
&&+ \frac{ (x-x_0) ^{q+1}}{\mathtt{C}^{q+1} \cdot (q+1)!  }\sum_{k=0}^q (-1)^{k} \cdot k^{q+1} \cdot\sigma^{(q+1)} (\xi_k) \cdot \binom{q}{k},
\end{eqnarray}
where $\xi_k\in [u_\sigma - \frac{k}{\mathtt{C}}\cdot |x-x_0|, u_\sigma + \frac{k}{\mathtt{C}}\cdot |x-x_0|]$ for all $0\le k\le q$.

Note that

\begin{eqnarray*}
\sum_{k=0}^q (-1)^{k}\cdot k^j\cdot \binom{q}{k} &=& q!(-1)^{q}\cdot \frac{1}{q!}\sum_{k=0}^q (-1)^{k-q}\cdot (q-(q-k))^j\cdot \binom{q}{q-k} \nonumber \\
&=& q!(-1)^{q} \cdot\frac{1}{q!}\sum_{k=0}^q (-1)^{q-k}\cdot (q-(q-k))^j\cdot \binom{q}{q-k} \nonumber \\
&=&q!(-1)^{q}\cdot \Big\{\begin{array}{c}
j\\ q
\end{array} \Big\},
\end{eqnarray*}
where $ \Big\{\begin{array}{c}
j\\ q
\end{array} \Big\}$ is the Stirling number of the second kind. The Stirling number of the second kind describes the number of options to split a set of $j$ elements into $n$ non-empty subsets, which is equal to 0 for $0\le j<n$, and is equal to 1 for $j=n$. The result holds true for all $j, n\in \mathbb{N}$ (\citealp[p. 825]{AS1972}). 

Thus, we can further simplify the right hand side of \eqref{EqA.4}, and write

\begin{eqnarray*}
&&\sum_{k=0}^q (-1)^{k}\cdot \binom{q}{k} \cdot\sigma\left( \frac{k}{\mathtt{C}}\cdot (x-x_0)  + u_\sigma\right)\nonumber \\
&=& \frac{\sigma^{(q)} (u_\sigma) \cdot (x-x_0) ^q}{\mathtt{C}^q} \cdot (-1)^q \nonumber \\
&&+ \frac{ (x-x_0) ^{q+1}}{\mathtt{C}^{q+1} \cdot (q+1)!  }\sum_{k=0}^q (-1)^{k} \cdot k^{q+1} \cdot\sigma^{(q+1)} (\xi_k) \cdot \binom{q}{k},
\end{eqnarray*}
which in connection with \eqref{EqA.3} yields that

\begin{eqnarray*}
&&(-1)^{q} \frac{\mathtt{C}^q}{\sigma^{(q)} (u_\sigma)} \sum_{k=0}^q (-1)^{k}\cdot \binom{q}{k} \cdot\sigma\left( \frac{k}{\mathtt{C}}\cdot (x-x_0)  + u_\sigma\right) \nonumber \\
&=&(-1)^{q} \frac{\mathtt{C}^q}{\sigma^{(q)} (u_\sigma)}\Big\{  \frac{\sigma^{(q)} (u_\sigma) \cdot (x-x_0) ^q}{\mathtt{C}^q} \cdot (-1)^q \nonumber \\
&&+ \frac{ (x-x_0) ^{q+1}}{\mathtt{C}^{q+1} \cdot (q+1)!  }\sum_{k=0}^q (-1)^{k} \cdot k^{q+1} \cdot\sigma^{(q+1)} (\xi_k) \cdot \binom{q}{k}\Big\}\nonumber \\
&=&(x-x_0) ^q +  \frac{  (-1)^{q}(x-x_0) ^{q+1}}{\mathtt{C} \cdot \sigma^{(q)} (u_\sigma)\cdot (q+1)!  }\sum_{k=0}^q (-1)^{k} \cdot k^{q+1} \cdot\sigma^{(q+1)} (\xi_k) \cdot \binom{q}{k}.
\end{eqnarray*}
In view of Assumption \ref{Ass.1}.2 and $\mathtt{C}$ being a fixed value, the proof is now completed.  \hspace*{\fill}{$\blacksquare$}

\bigskip

\noindent \textbf{Proof of Lemma \ref{LEM2.2}:}

By Lemma \ref{LemA.2}, we can reconstruct all of $\{m_i(\mathbf{x}\, |\, \mathbf{x}_0)\}$ as follows:

\begin{eqnarray*}
\mathbf{m}(\mathbf{x}\, |\, \mathbf{x}_0)=\mathbf{D} \mathbf{A}(\mathbf{x} \, | \,\mathbf{x}_0)
\end{eqnarray*}
where 

\begin{eqnarray*}
\mathbf{D}&=&\{d_{ij}\}_{d_q\times d_q}=(\mathbf{d}_1,\ldots, \mathbf{d}_{d_q}) ,\nonumber \\
\mathbf{A}(\mathbf{x} \, | \,\mathbf{x}_0)&=&\left([(1,\mathbf{x}^\top-\mathbf{x}_0^\top)\pmb{\alpha}_1 ]^q,\ldots, [(1,\mathbf{x}^\top-\mathbf{x}_0^\top)\pmb{\alpha}_{d_q} ]^q\right)^\top.
\end{eqnarray*}
Note that the rotation matrix $\mathbf{D}$ is determined by $\pmb{\alpha}_{j}$'s only, so they are fixed. 

Apparently, we have an issue of identification here, because for example we can arbitrarily rescale $\pmb{\alpha}_{j}$'s, and modify $\mathbf{D}$ accordingly without changing $m_i(\mathbf{x}\, |\, \mathbf{x}_0)$ as follows:

\begin{eqnarray*}
\mathbf{D} \mathbf{A}(\mathbf{x} \, | \,\mathbf{x}_0) =\mathbf{D} \mathbf{B}\mathbf{B}^{-1}\mathbf{A}(\mathbf{x} \, | \,\mathbf{x}_0),
\end{eqnarray*}
in which $ \mathbf{B}$ is full rank. Therefore, for the purpose of identification, we regulate $\pmb{\alpha}_j$'s as follows:

\begin{eqnarray}\label{EqA.5}
[\pmb{\alpha}_1 ,\ldots, \pmb{\alpha}_{d_q}]= \frac{1}{d+1}\cdot\mathbf{I}_h \mathbf{W}
\end{eqnarray}
in which $\mathbf{I}_h=\diag\{h,\mathbf{I}_d \}$, and $\mathbf{W}$ is defined in the body of this lemma. As a result, for  $\forall j\in [d_q]$,
 
\begin{eqnarray*}
\sup_{\mathbf{x}\in C_{\mathbf{x}_0,h}} |(1,\mathbf{x}^\top-\mathbf{x}_0^\top) \pmb{\alpha}_j |&=&\sup_{\mathbf{x}\in C_{\mathbf{x}_0,h}}\frac{1}{d+1} | (h,\mathbf{x}^\top-\mathbf{x}_0^\top)\mathbf{w}_j| \\
&\le &\frac{h}{d+1}\sqrt{d+1}\cdot \max_j\| \mathbf{w}_j\| =h,
\end{eqnarray*}
so we can invoke Lemma \ref{LEM2.1} later on. 

Treating $(1,\mathbf{x}^\top-\mathbf{x}_0^\top)\pmb{\alpha}_{j}$ as a whole and using Lemma \ref{LEM2.1}, we write

\begin{eqnarray}\label{EqA.6}
&&\sup_{\mathbf{x}\in C_{\mathbf{x}_0,h}}\left| m_i(\mathbf{x}\, |\, \mathbf{x}_0)-\sum_{j=1}^{d_q} d_{ij} \cdot \sum_{k=1}^{q+1} \gamma_k\cdot \sigma\left(\beta_k  (1,\mathbf{x}^\top-\mathbf{x}_0^\top)\pmb{\alpha}_{j}  +u_\sigma \right)\right| \nonumber \\
&\le &\sum_{j=1}^{d_q}|d_{ij}|\cdot\sup_{\mathbf{x}\in C_{\mathbf{x}_0,h}}  \left|[(1,\mathbf{x}^\top-\mathbf{x}_0^\top)\pmb{\alpha}_{j}]^q- \sum_{k=1}^{q+1} \gamma_k\cdot \sigma\left(\beta_k  (1,\mathbf{x}^\top-\mathbf{x}_0^\top) \pmb{\alpha}_{j}  +u_\sigma  \right)\right| \nonumber \\
&= &O\left(h^{q+1}\right),
\end{eqnarray}
where the last line follows from Lemma \ref{LEM2.1}. Also, $\pmb{\beta}$, $\pmb{\gamma}$ and $u_\sigma$ are known as discussed in Remark \ref{rmk1}. Thus, we can further write

\begin{eqnarray}\label{EqA.7}
&&\sup_{\mathbf{x}\in C_{\mathbf{x}_0,h}}  \left| p(\mathbf{x}\, |\, \mathbf{x}_0, \pmb{\lambda})-   \sum_{j=1}^{d_q}\pmb{\lambda}^\top \mathbf{d}_j \cdot \sum_{k=1}^{q+1} \gamma_k\cdot \sigma\left(\beta_k  (1,\mathbf{x}^\top-\mathbf{x}_0^\top)  \pmb{\alpha}_{j}  +u_\sigma  \right)\right| \nonumber\\
&=&\sup_{\mathbf{x}\in C_{\mathbf{x}_0,h}}  \left| \pmb{\lambda}^\top \mathbf{m}(\mathbf{x}\, |\, \mathbf{x}_0)-   \sum_{j=1}^{d_q}\pmb{\lambda}^\top \mathbf{d}_j \cdot \sum_{k=1}^{q+1} \gamma_k\cdot \sigma\left(\beta_k  (1,\mathbf{x}^\top-\mathbf{x}_0^\top)  \pmb{\alpha}_{j}  +u_\sigma  \right)\right| \nonumber\\
&\le &\| \pmb{\lambda}\|\cdot \sup_{\mathbf{x}\in C_{\mathbf{x}_0,h}}  \left\| \mathbf{m}(\mathbf{x}\, |\, \mathbf{x}_0)-\sum_{j=1}^{d_q} \mathbf{d}_j \cdot \sum_{k=1}^{q+1} \gamma_k\cdot \sigma\left(\beta_k  (1,\mathbf{x}^\top-\mathbf{x}_0^\top)\pmb{\alpha}_{j}  +u_\sigma  \right)\right\| \nonumber \\
&= &O\left( h^{q+1}\right),
\end{eqnarray}
where the last line follows from the facts that $d_q$ is fixed and $\|\pmb{\lambda} \|=O(1)$.

Finally, let $\widetilde{\pmb{\lambda}} =(\widetilde{\lambda}_1,\ldots, \widetilde{\lambda}_{d_q})^\top$ with $\widetilde{\lambda}_j = \pmb{\lambda}^\top \mathbf{d}_j$. Further, in view of the definitions of $\pmb{\sigma}(\mathbf{x}\, |\, \mathbf{x}_0 ) $ and $(\pmb{\pi}_1,\ldots, \pmb{\pi}_{d_q (q+1)})$ in the body of this lemma and \eqref{EqA.5}, the proof is then completed. \hspace*{\fill}{$\blacksquare$}

\bigskip

\noindent \textbf{Proof of Lemma \ref{LEM2.3}:}

Before starting the proof, we introduce a few notations to facilitate the development. First, recall that in the body of this theorem, we have defined

\begin{eqnarray*}
\widetilde{s}(\mathbf{x} \, |\, \widetilde{\pmb{\Lambda}}) = \sum_{\mathbf{i}\in [M]^d}I_{\mathbf{i},h}(\mathbf{x})\cdot s(\mathbf{x} \,|\, \widetilde{\mathbf{x}}_{\mathbf{i}},  \widetilde{\pmb{\lambda}}_{\mathbf{i}}) ,
\end{eqnarray*}
where $s (\mathbf{x}\, |\, \widetilde{\mathbf{x}}_{\mathbf{i}}, \widetilde{\pmb{\lambda}}_{\mathbf{i}}) = (\widetilde{\pmb{\lambda}}_{\mathbf{i}}\otimes \pmb{\gamma})^\top\pmb{\sigma} (\mathbf{x}\, |\, \widetilde{\mathbf{x}}_{\mathbf{i}} )$ by the definition of Lemma \ref{LEM2.2}. Second, note that the leading terms of the $q^{th}$ order Taylor expansion of $g(\mathbf{x}) $ at each $ \widetilde{\mathbf{x}}_{\mathbf{i}}\in (-a,a)^d$ can be written as follows:

\begin{eqnarray*}
 \sum_{0\le |\mathbf{J}|\le q} \frac{1}{\mathbf{J}!}\cdot \frac{\partial^{|\mathbf{J}|} g( \widetilde{\mathbf{x}}_{\mathbf{i}}) }{\partial \mathbf{x}^{\mathbf{J}}} (\mathbf{x}- \widetilde{\mathbf{x}}_{\mathbf{i}})^{\mathbf{J}}= \pmb{\lambda}_{\mathbf{i}}^\top \mathbf{m}(\mathbf{x}\, |\,  \widetilde{\mathbf{x}}_{\mathbf{i}})\coloneqq p(\mathbf{x}\, |\,  \widetilde{\mathbf{x}}_{\mathbf{i}}, \pmb{\lambda}_{\mathbf{i}}) ,
\end{eqnarray*}
where the definition of $\pmb{\lambda}_{\mathbf{i}}$ should be obvious in view of the definition of $\mathbf{m}(\mathbf{x}\, |\, \widetilde{\mathbf{x}}_{\mathbf{i}})$ according to \eqref{EQ1.4}.

We are now ready to start the proof, and write

\begin{eqnarray*}
g(\mathbf{x}) -\widetilde{s}(\mathbf{x} \, |\, \widetilde{\pmb{\Lambda}}) &=& \sum_{\mathbf{i}\in [M]^d} I_{\mathbf{i},h}(\mathbf{x})\cdot p (\mathbf{x} \,|\,  \widetilde{\mathbf{x}}_{\mathbf{i}}, \pmb{\lambda}_{\mathbf{i}})  -\sum_{\mathbf{i}\in [M]^d} I_{\mathbf{i},h}(\mathbf{x})\cdot s (\mathbf{x} \,|\, \widetilde{\mathbf{x}}_{\mathbf{i}}, \widetilde{\pmb{\lambda}}_{\mathbf{i}})  \nonumber \\
& &+g(\mathbf{x}) -\sum_{\mathbf{i}\in [M]^d}I_{\mathbf{i},h}(\mathbf{x})\cdot p (\mathbf{x} \,|\,  \widetilde{\mathbf{x}}_{\mathbf{i}}, \pmb{\lambda}_{\mathbf{i}}) .
\end{eqnarray*}
Note that by Lemma \ref{LEM2.2} we choose $\widetilde{\pmb{\lambda}}_{\mathbf{i}}$ which fulfils the relationship: $\widetilde{\pmb{\lambda}}_{\mathbf{i}} =\mathbf{D}^\top \pmb{\lambda}_{\mathbf{i}}$. It is worth mentioning that although $(\widetilde{\mathbf{x}}_{\mathbf{i}},  \pmb{\lambda}_{\mathbf{i}})$ vary with respect to $\mathbf{i}$, the rotation matrix $\mathbf{D}$ in facts is solely determined by $\mathbf{W}$ of Lemma \ref{LEM2.2}. Therefore, without loss of generality, we can fix $\mathbf{W}$ over $\mathbf{i}$, as it is user chosen. Then $\mathbf{D}$ remains the same in view of the proof of Lemma \ref{LEM2.2}.

Next, we write

\begin{eqnarray} \label{EqA.8}
&&\left\|\sum_{\mathbf{i}\in [M]^d} I_{\mathbf{i},h}(\mathbf{x})\cdot p (\mathbf{x} \,|\,  \widetilde{\mathbf{x}}_{\mathbf{i}}, \pmb{\lambda}_{\mathbf{i}}) -g(\mathbf{x})\right\|_{\infty}  \nonumber \\
&=& \left\|\sum_{\mathbf{i}\in [M]^d} [ p (\mathbf{x}\, |\,  \widetilde{\mathbf{x}}_{\mathbf{i}},  \pmb{\lambda}_{\mathbf{i}}) -g(\mathbf{x}) ] \cdot I_{\mathbf{i},h}(\mathbf{x}) \right\|_{\infty} \nonumber \\
&\le & \sum_{\mathbf{i}\in [M]^d}I_{\mathbf{i},h}(\mathbf{x}) \cdot \sup_{\mathbf{x}\in C_{\mathbf{x}_{0\mathbf{i}},h}}   | p (\mathbf{x}\, |\,  \widetilde{\mathbf{x}}_{\mathbf{i}}, \pmb{\lambda}_{\mathbf{i}}) -g(\mathbf{x}) |  =O(h^p),
\end{eqnarray}
where the inequality follows from the definition of $I_{\mathbf{i},h}(\mathbf{x})$, and the last step follows from Lemma \ref{LemA.1}. Also, we can obtain that

\begin{eqnarray}\label{EqA.9}
&& \left\| \sum_{\mathbf{i}\in [M]^d}I_{\mathbf{i},h}(\mathbf{x})\cdot s (\mathbf{x}  \,|\,  \widetilde{\mathbf{x}}_{\mathbf{i}}, \widetilde{\pmb{\lambda}}_{\mathbf{i}})  -\sum_{\mathbf{i}\in [M]^d}I_{\mathbf{i},h}(\mathbf{x})\cdot p (\mathbf{x} \,|\,  \widetilde{\mathbf{x}}_{\mathbf{i}}, \pmb{\lambda}_{\mathbf{i}})   \right\|_{\infty}\nonumber \\
&\le &  \sum_{\mathbf{i}\in [M]^d} I_{\mathbf{i},h}(\mathbf{x})\cdot \sup_{\mathbf{x}\in C_{\mathbf{x}_{0\mathbf{i}},h}} | s (\mathbf{x}  \,|\,  \widetilde{\mathbf{x}}_{\mathbf{i}}, \widetilde{\pmb{\lambda}}_{\mathbf{i}})- p (\mathbf{x} \,|\,  \widetilde{\mathbf{x}}_{\mathbf{i}}, \pmb{\lambda}_{\mathbf{i}}) |  =O(h^{q+1}),
\end{eqnarray}
where the inequality follows from the definition of $I_{\mathbf{i},h}(\mathbf{x})$, and the last step follows from Lemma \ref{LEM2.2} by letting  $\widetilde{\pmb{\lambda}}_{\mathbf{i}} =\mathbf{D}^\top \pmb{\lambda}_{\mathbf{i}}$.

Therefore, based on \eqref{EqA.8} and \eqref{EqA.9}, we obtain

\begin{eqnarray*}
\| g(\mathbf{x}) - \widetilde{s}(\mathbf{x} \, |\, \widetilde{\pmb{\Lambda}})  \|_{\infty}=O(h^p),
\end{eqnarray*}
where $p=q+s$. The proof is now completed. \hspace*{\fill}{$\blacksquare$}

\subsection{Proofs for the Main Results}\label{App.43}

\noindent \textbf{Proof of Lemma \ref{Lem_lasso}:}

We adopt a similar strategy with that for Lemma A.1 of \cite{WX2009} to establish the results in Lemma \ref{Lem_lasso}. For notational simplicity, let $\widetilde{Q}_\psi  (\alpha,\pmb{\Theta})=\widetilde{Q}  (\alpha,\pmb{\Theta})+\sum_{l=1}^{d_q}\psi_l\|\pmb{\Theta}_{l}\|$, where $\widetilde{Q}  (\alpha,\pmb{\Theta})= \sum_{t=1}^T[y_t-z_t\alpha -\widetilde{s}(\mathbf{x}_t \, |\, \pmb{\Theta}) ]^2$. Let $\mathbf{B} =\{   \mathbf{b}_{\mathbf{i}}\, |\, \mathbf{i}\in [M]^d\}$ and $\|\mathbf{B}\|_H^2=\sum_{\mathbf{i}\in [M]^d}\|\mathbf{H}^{-1}\mathbf{D}^{\top,-1}\mathbf{b}_{\mathbf{i}}\|^2$, where $\mathbf{b}_{\mathbf{i}}$ is a $d_q\times 1$ vector of constants. Additionally, denote $d_T=\frac{1}{\sqrt{T}h^d}$.

By \cite{FL2001}, it suffices to show that for any $\epsilon>0$, there exists a constant $C>0$ such that
\begin{eqnarray}\label{int0}
\liminf_{T\rightarrow\infty} P\left(\inf_{b_0^2+h^d\|\mathbf{B}\|_H^2=C^2h^d} \widetilde{Q}_\psi  (\alpha_0+d_Tb_0,\widetilde{\pmb{\Lambda}}+d_T\mathbf{B})>  \widetilde{Q}_\psi  (\alpha_0, \widetilde{\pmb{\Lambda}})\right)=1-\epsilon.
\end{eqnarray}

We write
\begin{eqnarray}\label{int10}
\frac{1}{Th^d}\widetilde{Q}_\psi  (\alpha_0+d_Tb_0,\widetilde{\pmb{\Lambda}}+d_T\mathbf{B})-  \frac{1}{Th^d}\widetilde{Q}_\psi  (\alpha_0, \widetilde{\pmb{\Lambda}})&=& \frac{1}{Th^d}\left(\widetilde{Q}  (\alpha_0+d_Tb_0,\widetilde{\pmb{\Lambda}}+d_T\mathbf{B})-  \widetilde{Q}  (\alpha_0, \widetilde{\pmb{\Lambda}})\right)
\nonumber\\
&&+\frac{1}{Th^d}\sum_{l=1}^{d_q}\psi_l\left(\|\pmb{\Lambda}_l+d_T\mathbf{B}_{D,l}\| -\|\pmb{\Lambda}_l\|\right)
\nonumber\\
&\coloneqq&R_1+R_2,
\end{eqnarray}
where $\pmb{\Lambda}_l$ and $\mathbf{B}_{D,l}$ are vectors that contain the $l$-th elements of $\pmb{\lambda}_{\mathbf{i}}$ and $\mathbf{D}^{\top,-1}\mathbf{b}_{\mathbf{i}}$, respectively.

For $R_1$, directly using the definition of $\widetilde{s}(\mathbf{x} \, |\, \pmb{\Theta})$ and $s(\mathbf{x} \,|\, \mathbf{x}_{0\mathbf{i}},  \pmb{\theta}_{\mathbf{i}})$ gives
\begin{eqnarray*}
\widetilde{s}(\mathbf{x} \, |\, \widetilde{\pmb{\Lambda}}+d_T\mathbf{B}) &=& \sum_{\mathbf{i}\in [M]^d} I_{\mathbf{i},h}(\mathbf{x})\cdot s(\mathbf{x} \,|\, \mathbf{x}_{0\mathbf{i}},  \widetilde{\pmb{\lambda}}_{\mathbf{i}}+d_T \mathbf{b}_{\mathbf{i}})
\nonumber\\
&=&  \sum_{\mathbf{i}\in [M]^d} I_{\mathbf{i},h}(\mathbf{x})\cdot  ((\widetilde{\pmb{\lambda}}_{\mathbf{i}}+d_T \mathbf{b}_{\mathbf{i}})\otimes \pmb{\gamma})^\top\pmb{\sigma}(\mathbf{x}\, |\, \mathbf{x}_{0\mathbf{i}} )
\nonumber\\
&=&\widetilde{s}(\mathbf{x} \, |\, \widetilde{\pmb{\Lambda}})+d_T\widetilde{s}(\mathbf{x} \, |\, \mathbf{B}).
\end{eqnarray*}
We can further write
\begin{eqnarray*}
R_1&=&
\frac{1}{Th^d}\sum_{t=1}^T\left[g(\mathbf{x}_t)-\widetilde{s}(\mathbf{x}_t \, |\, \widetilde{\pmb{\Lambda}})+\varepsilon_t-d_T b_0z_t -d_T\widetilde{s}(\mathbf{x}_t \, |\, \mathbf{B}) \right]^2
-\frac{1}{Th^d}\sum_{t=1}^T\left[g(\mathbf{x}_t) -\widetilde{s}(\mathbf{x}_t \, |\, \widetilde{\pmb{\Lambda}}) +\varepsilon_t\right]^2
\nonumber\\
&=&\frac{d_T^2}{Th^d}\sum_{t=1}^T \left[b_0z_t +\widetilde{s}(\mathbf{x}_t \, |\, \mathbf{B})\right]^2
-\frac{2d_T}{Th^d}\sum_{t=1}^T\left[g(\mathbf{x}_t)-\widetilde{s}(\mathbf{x}_t \, |\, \widetilde{\pmb{\Lambda}})\right]\left[b_0z_t +\widetilde{s}(\mathbf{x}_t \, |\, \mathbf{B}) \right]
\nonumber\\
&&-\frac{2d_T}{Th^d}\sum_{t=1}^T\left[b_0z_t +\widetilde{s}(\mathbf{x}_t \, |\, \mathbf{B}) \right]\varepsilon_t
\nonumber\\
&\coloneqq&R_{1,1}+R_{1,2}+R_{1,3}.
\end{eqnarray*}
We can further write
\begin{eqnarray}\label{int4}
b_0z_t +\widetilde{s}(\mathbf{x}_t \, |\, \mathbf{B})&=&b_0z_t+\sum_{\mathbf{i}\in [M]^d} I_{\mathbf{i},h} (\mathbf{x}_t)  \pmb{\sigma}( \mathbf{x}_t\, |\,\mathbf{x}_{0\mathbf{i}})^\top (\mathbf{I}_{d_q}\otimes \pmb{\gamma}^\top)^\top\mathbf{b}_{\mathbf{i}}
\nonumber\\
&=&\sum_{\mathbf{i}\in [M]^d} I_{\mathbf{i},h} (\mathbf{x}_t) (z_t,\pmb{\sigma}( \mathbf{x}_t\, |\,\mathbf{x}_{0\mathbf{i}})^\top (\mathbf{I}_{d_q}\otimes \pmb{\gamma}^\top)^\top)(a,\mathbf{b}_{\mathbf{i}}^\top)^\top
\nonumber\\
&\coloneqq&\sum_{\mathbf{i}\in [M]^d} \widetilde{\mathbf{x}}^{\ast\top}_{\mathbf{i},t}\mathbf{b}_{\mathbf{i}}^\ast,
\end{eqnarray}
where $\widetilde{\mathbf{x}}^\ast_{\mathbf{i},t}  =I_{\mathbf{i},h} (\mathbf{x}_t)(z_t,\pmb{\sigma}( \mathbf{x}_t\, |\,\mathbf{x}_{0\mathbf{i}})^\top (\mathbf{I}_{d_q}\otimes \pmb{\gamma}^\top)^\top)^\top$ and $\mathbf{b}_{\mathbf{i}}^\ast=(a,\mathbf{b}_{\mathbf{i}}^\top)^\top$. With this notation, we can rewrite $R_{1,1}$:
\begin{eqnarray}\label{int5}
R_{1,1}&=&\frac{d_T^2}{Th^d}\sum_{t=1}^T \left[\sum_{\mathbf{i}\in [M]^d} \widetilde{\mathbf{x}}^{\ast\top}_{\mathbf{i},t}\mathbf{b}_{\mathbf{i}}^\ast\right]^2
\nonumber\\
&=&\frac{d_T^2}{Th^d}\sum_{t=1}^T \sum_{\mathbf{i}\in [M]^d}\mathbf{b}_{\mathbf{i}}^{\ast\top}\widetilde{\mathbf{x}}^{\ast}_{\mathbf{i},t} \widetilde{\mathbf{x}}^{\ast\top}_{\mathbf{i},t}\mathbf{b}_{\mathbf{i}}^\ast
\nonumber\\
&\geq& d_T^2\sum_{\mathbf{i}\in [M]^d} \widetilde{\lambda}_{\mathbf{i},\min}\|\widetilde{\mathbf{H}}^{\top,-1}\mathbf{b}_{\mathbf{i}}^\ast\|^2
\nonumber\\
&\geq& d_T^2\widetilde{\lambda}_{\min}\sum_{\mathbf{i}\in [M]^d} \|\widetilde{\mathbf{H}}^{\top,-1}\mathbf{b}_{\mathbf{i}}^\ast\|^2
\nonumber\\
&=& d_T^2\widetilde{\lambda}_{\min} (\frac{1}{h^d}b_0^2+\|\mathbf{B}\|_H^2)
\nonumber\\
&=&d_T^2\widetilde{\lambda}_{\min} C^2
,
\end{eqnarray}
where $\widetilde{\mathbf{H}}=\text{diag}(1, \mathbf{H}\mathbf{D})$, $\widetilde{\lambda}_{\mathbf{i},\min}$ denotes the smallest eigenvalue of $\frac{1}{Th^d}\sum_{t=1}^T \widetilde{\mathbf{H}} \widetilde{\mathbf{x}}^{\ast}_{\mathbf{i},t} \widetilde{\mathbf{x}}^{\ast\top}_{\mathbf{i},t}\widetilde{\mathbf{H}}^\top$, $\widetilde{\lambda}_{\min}=\min\{\widetilde{\lambda}_{\mathbf{i},\min},|, \mathbf{i}\in [M]^d\}$, and the second equality holds by the fact that $I_{\mathbf{i},h} (\mathbf{x}_t)I_{\mathbf{j},h} (\mathbf{x}_t)=0$ for $\mathbf{i}\ne \mathbf{j}$.

It suffices to explore  $\frac{1}{Th^d}\sum_{t=1}^T\mathbf{H}\mathbf{D}\widetilde{\mathbf{x}}_{\mathbf{i},t} \widetilde{\mathbf{x}}^{\top}_{\mathbf{i},t}\mathbf{D}^\top\mathbf{H}$, $\frac{1}{Th^d}\sum_{t=1}^TI_{\mathbf{i},h} (\mathbf{x}_t)z_t^2$, and $\frac{1}{Th^d}\sum_{t=1}^Tz_t\widetilde{\mathbf{x}}^{\top}_{\mathbf{i},t}\mathbf{D}^\top\mathbf{H}$ before we can obtain the probability limit of $\widetilde{\lambda}_{\min}$. 

We proceed with $\frac{1}{Th^d}\sum_{t=1}^TI_{\mathbf{i},h} (\mathbf{x}_t)z_t^2$. Note that
\begin{eqnarray}\label{int1}
E\left[\frac{1}{Th^d}\sum_{t=1}^T I_{\mathbf{i},h} (\mathbf{x}_t)z_t^2\right]&=&\frac{1}{h^d}E\left[ I_{\mathbf{i},h} (\mathbf{x}_t)E\left[z_t^2|\mathbf{x}_t\right]\right]
\nonumber\\
&=&\frac{1}{h^d}E\left[ I_{\mathbf{i},h} (\mathbf{x}_t)\Phi_\eta(G(\mathbf{x}_t))\right]
\nonumber\\
&=& \frac{1}{h^d}\int_{\mathbf{x}\in C_{\mathbf{x}_{0\mathbf{i}},h}}    \Phi_\eta(G(\mathbf{x}))  f_{\mathbf{x}}(\mathbf{x})\mathrm{d} \mathbf{x}
\nonumber\\
&=& \Phi_\eta(G(\mathbf{x}_{0\mathbf{i}}))  f_{\mathbf{x}}(\mathbf{x}_{0\mathbf{i}})\int_{[-1,1]^d}    1\mathrm{d} \mathbf{x}\cdot(1+O(h))
\nonumber\\
&=&  2^d\Phi_\eta(G(\mathbf{x}_{0\mathbf{i}}))  f_{\mathbf{x}}(\mathbf{x}_{0\mathbf{i}})(1+O(h))
,
\end{eqnarray}
where the fourth equality can be proved by directly applying the Lipschitz continuity of $G(\mathbf{x})$ and $f_{\mathbf{x}}(\mathbf{x})$ in Assumption \ref{Ass.3}. 

Analogously, we can compute the second moment:
\begin{eqnarray}\label{int2}
E\left[\left(\frac{1}{Th^d}\sum_{t=1}^T I_{\mathbf{i},h} (\mathbf{x}_t)z_t^2\right)^2\right]&=&\frac{1}{T^2h^{2d}}\sum_{s,t=1}^TE\left[I_{\mathbf{i},h} (\mathbf{x}_t)I_{\mathbf{i},h} (\mathbf{x}_s)E\left[z_t^2z_s^2|\mathbf{x}_t,\mathbf{x}_s\right]\right]
\nonumber\\
&=&\frac{1}{T^2h^{2d}}\sum_{t=1}^TE\left[I_{\mathbf{i},h} (\mathbf{x}_t)E\left[z_t^4|\mathbf{x}_t\right]\right]
\nonumber\\
&&+\frac{1}{T^2h^{2d}}\sum_{t=1}^T\sum_{s=t+1}^TE\left[I_{\mathbf{i},h} (\mathbf{x}_t)I_{\mathbf{i},h} (\mathbf{x}_s)E\left[z_t^2z_s^2|\mathbf{x}_t,\mathbf{x}_s\right]\right]
\nonumber\\
&&+\frac{1}{T^2h^{2d}}\sum_{t=1}^T\sum_{s=1}^{t-1}E\left[I_{\mathbf{i},h} (\mathbf{x}_t)I_{\mathbf{i},h} (\mathbf{x}_s)E\left[z_t^2z_s^2|\mathbf{x}_t,\mathbf{x}_s\right]\right]
.
\end{eqnarray}
For the first term on the right-hand side of \eqref{int2}, by \eqref{int1},
\begin{eqnarray}\label{int2_0}
\frac{1}{T^2h^{2d}}\sum_{t=1}^TE\left[I_{\mathbf{i},h} (\mathbf{x}_t)E\left[z_t^4|\mathbf{x}_t\right]\right]&=&\frac{1}{Th^{2d}}E\left[I_{\mathbf{i},h} (\mathbf{x}_t)\Phi_{\eta}(G(\mathbf{x}_t))\right]
\nonumber\\
&=&\frac{1}{Th^{d}}2^d\Phi_\eta(G(\mathbf{x}_{0\mathbf{i}}))  f_{\mathbf{x}}(\mathbf{x}_{0\mathbf{i}})(1+O(h)).
\end{eqnarray}
For the second term on the right-hand side of \eqref{int2}, 
\begin{eqnarray}\label{int2_1}
&&\frac{1}{T^2h^{2d}}\sum_{t=1}^T\sum_{s=t+1}^TE\left[I_{\mathbf{i},h} (\mathbf{x}_t)I_{\mathbf{i},h} (\mathbf{x}_s)E\left[z_t^2z_s^2|\mathbf{x}_t,\mathbf{x}_s\right]\right]
\nonumber\\
&=&\frac{1}{T^2h^{2d}}\sum_{t=1}^T\sum_{s=1}^{T-t}E\left[I_{\mathbf{i},h} (\mathbf{x}_1)I_{\mathbf{i},h} (\mathbf{x}_{1+s})\widetilde{\Phi}_{\eta,s}(G(\mathbf{x}_1), G(\mathbf{x}_{1+s}))\right]
\nonumber\\
&=&\frac{1}{T^2h^{2d}}\sum_{t=1}^T\sum_{s=1}^{T-t}\int_{\mathbf{x},\mathbf{z}\in C_{\mathbf{x}_{0\mathbf{i}},h}}\widetilde{\Phi}_{\eta,s}(G(\mathbf{x}), G(\mathbf{z}))f_{\mathbf{x},s}(\mathbf{x},\mathbf{z})
\mathrm{d} \mathbf{x}\mathrm{d} \mathbf{z}
\nonumber\\
&=&\frac{2^{2d}}{T^2}\sum_{t=1}^T\sum_{s=1}^{T-t}\widetilde{\Phi}_{\eta,s}(G(\mathbf{x}_{0\mathbf{i}}), G(\mathbf{x}_{0\mathbf{i}}))f_{\mathbf{x},s}(\mathbf{x}_{0\mathbf{i}},\mathbf{x}_{0\mathbf{i}})(1+O(h))
\nonumber\\
&=&2^{2d-1}\Phi_{\eta}(G(\mathbf{x}_{0\mathbf{i}}))^2f_{\mathbf{x}}(\mathbf{x}_{0\mathbf{i}})^2(1+o(1)),
\end{eqnarray}
where $\widetilde{\Phi}_{\eta,s}(\cdot,\cdot)$ denotes the joint CDF of $(\eta_1,\eta_{1+s})$, $f_{\mathbf{x},s}(\mathbf{x}, \mathbf{z})$ is the joint PDF of $(\mathbf{x}_1, \mathbf{x}_{1+s})$, and the last equality holds by the following results which are implied by the $\alpha$-mixing conditions in Assumption \ref{Ass.3}, 
\begin{equation}\label{dk20}
\sum_{s=1}^T\sup_{\mathbf{x}, \mathbf{z}\in[-a,a]^d}\left|f_{\mathbf{x},s}(\mathbf{x}, \mathbf{z})-f_{\mathbf{x}}(\mathbf{x})f_{\mathbf{x}}(\mathbf{z})\right|=O(1),\quad \sum_{s=1}^T\sup_{x, z\in \mathbb{R}}\left|\widetilde{\Phi}_{\eta,s}(x, z)-\Phi_\eta(x)\Phi_\eta(z)\right|=O(1).
\end{equation}
Analogously, for the third term on the right-hand side of \eqref{int2}, we have
\begin{eqnarray}\label{int2_3}
\frac{1}{T^2h^{2d}}\sum_{t=1}^T\sum_{s=1}^{t-1}E\left[I_{\mathbf{i},h} (\mathbf{x}_t)I_{\mathbf{i},h} (\mathbf{x}_s)E\left[z_t^2z_s^2|\mathbf{x}_t,\mathbf{x}_s\right]\right]
=2^{2d-1}\Phi_{\eta}(G(\mathbf{x}_{0\mathbf{i}}))^2f_{\mathbf{x}}(\mathbf{x}_{0\mathbf{i}})^2(1+o(1)).
\end{eqnarray}
Combining \eqref{int2}, \eqref{int2_0}, \eqref{int2_1}, and \eqref{int2_3}, we obtain
\begin{equation}\label{int2_4}
E\left[\left(\frac{1}{Th^d}\sum_{t=1}^T I_{\mathbf{i},h} (\mathbf{x}_t)z_t^2\right)^2\right]=2^{2d}\Phi_{\eta}(G(\mathbf{x}_{0\mathbf{i}}))^2f_{\mathbf{x}}(\mathbf{x}_{0\mathbf{i}})^2(1+o(1)).
\end{equation}
Together with \eqref{int1},  it yields that
\begin{equation}\label{int3}
\frac{1}{Th^d}\sum_{t=1}^T I_{\mathbf{i},h} (\mathbf{x}_t)z_t^2=2^d\Phi_\eta(G(\mathbf{x}_{0\mathbf{i}}))  f_{\mathbf{x}}(\mathbf{x}_{0\mathbf{i}})(1+o_P(1)).
\end{equation}

Then, we consider $\frac{1}{Th^d}\sum_{t=1}^Tz_t\widetilde{\mathbf{x}}^{\top}_{\mathbf{i},t}\mathbf{D}^\top\mathbf{H}$. Drawing upon \eqref{EqA.6} and \eqref{EqA.7}, it follows from the definition of $\mathbf{H}$ and $\max_j|\mathbf{n}_j|=q$ that 

\begin{eqnarray}\label{EqA.10}
\| \widetilde{\mathbf{x}}_{\mathbf{i},t} ^\top \mathbf{D}^\top\mathbf{H}  -I_{\mathbf{i},h}(\mathbf{x}_t)\mathbf{m}(\mathbf{x}_t\, |\, \mathbf{x}_{\mathbf{i},0})^\top \cdot \mathbf{H} \| =O_p(h).
\end{eqnarray}

Therefore, we only need to study the convergence of $\frac{1}{Th^d}\sum_{t=1}^TI_{\mathbf{i},h} (\mathbf{x}_t)z_t\mathbf{m}(\mathbf{x}_t\, |\, \mathbf{x}_{0\mathbf{i}})^\top   \mathbf{H}$. We have
\begin{eqnarray*}
E\left[\frac{1}{Th^d}\sum_{t=1}^TI_{\mathbf{i},h} (\mathbf{x}_t)z_t\mathbf{m}(\mathbf{x}_t\, |\, \mathbf{x}_{0\mathbf{i}})^\top   \mathbf{H}\right]&=&\frac{1}{h^d}E\left[I_{\mathbf{i},h} (\mathbf{x}_t)\mathbf{m}(\mathbf{x}_t\, |\, \mathbf{x}_{0\mathbf{i}})^\top   E\left[z_t|\mathbf{x}_t\right]\right]\mathbf{H}
\nonumber\\
&=&\frac{1}{h^d}E\left[I_{\mathbf{i},h} (\mathbf{x}_t)\Phi_\eta(G(\mathbf{x}_t))\mathbf{m}(\mathbf{x}_t\, |\, \mathbf{x}_{0\mathbf{i}})^\top   \right]\mathbf{H}
\nonumber\\
&=&\frac{1}{h^d}\int_{\mathbf{x}\in C_{\mathbf{x}_{0\mathbf{i}},h}}    \Phi_\eta(G(\mathbf{x})) \mathbf{m}(\mathbf{x}\, |\, \mathbf{x}_{0\mathbf{i}})^\top   f_{\mathbf{x}}(\mathbf{x})\mathrm{d} \mathbf{x}\mathbf{H}
\nonumber\\
&=&\Phi_\eta(G(\mathbf{x}_{0\mathbf{i}}))    f_{\mathbf{x}}(\mathbf{x}_{0\mathbf{i}}) \int_{[-1,1]^d}    \mathbf{m}(\mathbf{x}\, |\, \mathbf{0})^\top  \mathrm{d} \mathbf{x}(1+O(h)).
\end{eqnarray*}
Using analogous arguments to those in \eqref{int2_4}, we can show the convergence of its second moment and obtain that 
\begin{eqnarray}\label{dk2}
\frac{1}{Th^d}\sum_{t=1}^TI_{\mathbf{i},h} (\mathbf{x}_t)z_t\widetilde{\mathbf{x}}^{\top}_{\mathbf{i},t}\mathbf{D}^\top\mathbf{H}&=&\Phi_\eta(G(\mathbf{x}_{0\mathbf{i}}))    f_{\mathbf{x}}(\mathbf{x}_{0\mathbf{i}}) \int_{[-1,1]^d}    \mathbf{m}(\mathbf{x}\, |\, \mathbf{0})^\top  \mathrm{d} \mathbf{x}(1+o_P(1)).
\end{eqnarray}

Finally, we consider $\frac{1}{Th^d}\sum_{t=1}^T\mathbf{H}\mathbf{D}\widetilde{\mathbf{x}}_{\mathbf{i},t} \widetilde{\mathbf{x}}^{\top}_{\mathbf{i},t}\mathbf{D}^\top\mathbf{H}$ and write

\begin{eqnarray} \label{EqA.12}
&&\frac{1}{Th^d}\sum_{t=1}^T\mathbf{H} \mathbf{D} E\left[\widetilde{\mathbf{x}}_{\mathbf{i},t}  \widetilde{\mathbf{x}}_{\mathbf{i},t}  ^\top\right]\mathbf{D}^\top \mathbf{H} \nonumber \\
&=&\frac{1}{h^d}\mathbf{H} \mathbf{D}E[\widetilde{\mathbf{x}}_{\mathbf{i},1} \widetilde{\mathbf{x}}_{\mathbf{i},1} ^\top]\mathbf{D}^\top\mathbf{H} \nonumber \\
&=&\frac{1}{h^d} E[I_{\mathbf{i}, h}(\mathbf{x}_1)\mathbf{H} \cdot \mathbf{m}(\mathbf{x}_1\, |\, \mathbf{x}_{\mathbf{i},0}) \mathbf{m}(\mathbf{x}_1\, |\, \mathbf{x}_{\mathbf{i},0})^\top \cdot \mathbf{H}  ]\cdot (1+o(1))\nonumber \\
&=&\frac{1}{h^d} \int_{\mathbf{x}\in C_{\mathbf{x}_{0\mathbf{i}},h}} \mathbf{H} \cdot \mathbf{m}(\mathbf{x}\, |\, \mathbf{x}_{\mathbf{i},0}) \mathbf{m}(\mathbf{x}\, |\, \mathbf{x}_{\mathbf{i},0})^\top \cdot \mathbf{H} \cdot f_{\mathbf{x}}(\mathbf{x})\mathrm{d} \mathbf{x}\cdot (1+o(1))\nonumber \\
&=& f_{\mathbf{x}}(\mathbf{x}_{\mathbf{i},0}) \int_{[-1,1]^d}  \mathbf{m}(\mathbf{x}\, |\, \mathbf{0})  \mathbf{m}(\mathbf{x}\, |\, \mathbf{0})^\top  \mathrm{d} \mathbf{x}\cdot (1+o(1))>0
\end{eqnarray}
where the first equality follows from Assumption \ref{Ass.3}.1, the second equality follows from \eqref{EqA.10}, and the last step follows from Assumption \ref{Ass.3}.2 and the definition of $\mathbf{m}(\mathbf{x}\, |\, \mathbf{0})$ given in \eqref{EQ1.4}.  
 
For the second moment, using the arguments that are analogous to the proof of \eqref{int2_4}, we obtain 

\begin{eqnarray}\label{EqA.13}
E\left\|  \frac{1}{Th^d} \sum_{t=1}^T\mathbf{H} \mathbf{D}\bigl(\widetilde{\mathbf{x}}_{\mathbf{i},t} \widetilde{\mathbf{x}}_{\mathbf{i},t}^\top-E\bigl[\widetilde{\mathbf{x}}_{\mathbf{i},t} \widetilde{\mathbf{x}}_{\mathbf{i},t}^\top\bigr]\bigr)\mathbf{D}^\top \mathbf{H}\right\|^2=o(1),
\end{eqnarray}
so we omit the details for now.

Together with  \eqref{int3} and \eqref{dk2}, it proves that 
\begin{equation}\label{int6}
P\left(\lim_{T\rightarrow\infty}\widetilde{\lambda}_{\min}=\lambda_{\min,0}\right)\rightarrow1,
\end{equation}
where $\lambda_{\min,0}=\inf_{\mathbf{x}\in[-a,a]^d}\lambda_{\min}\left\{\pmb{\Omega}^\ast_0(\mathbf{x})\right\}$ with  
\begin{eqnarray*}
\pmb{\Omega}^\ast_0(\mathbf{x})=\left(
\begin{array}{c c}
2^d\Phi_\eta(G(\mathbf{x}))  f_{\mathbf{x}}(\mathbf{x})& \Phi_\eta(G(\mathbf{x}))    f_{\mathbf{x}}(\mathbf{x}) \int_{[-1,1]^d}    \mathbf{m}(\mathbf{x}\, |\, \mathbf{0})^\top  \mathrm{d} \mathbf{x}\\
\Phi_\eta(G(\mathbf{x}))    f_{\mathbf{x}}(\mathbf{x}) \int_{[-1,1]^d}    \mathbf{m}(\mathbf{x}\, |\, \mathbf{0}) \mathrm{d} \mathbf{x} &
f_{\mathbf{x}}(\mathbf{x}) \int_{[-1,1]^d}  \mathbf{m}(\mathbf{x}\, |\, \mathbf{0})  \mathbf{m}(\mathbf{x}\, |\, \mathbf{0})^\top  \mathrm{d} \mathbf{x}
\end{array}.
\right)
\end{eqnarray*}
Additionally, Assumption \ref{Ass.4} ensures that $\lambda_{\min,0}>0$.

For $R_{1,2}$, by Lemma \ref{LEM2.3}, Cauchy–Schwarz inequality, and using the analogous arguments to the derivations of $R_{1,1}$, we can obtain 
\begin{eqnarray}\label{int7}
\|R_{1,2}\|&\leq&
\frac{2d_T}{Th^d}\left[\sum_{t=1}^T\left(g(\mathbf{x}_t)-\widetilde{s}(\mathbf{x}_t \, |\, \widetilde{\pmb{\Lambda}})\right)^2\right]^{\frac{1}{2}}\left[\sum_{t=1}^T\left(b_0z_t +\widetilde{s}(\mathbf{x}_t \, |\, \mathbf{B})\right)^2 \right]^{\frac{1}{2}}
\nonumber\\
&=&O_P\left(h^{p-\frac{d}{2}} d_T \left(\sum_{\mathbf{i}\in [M]^d} \|\widetilde{\mathbf{H}}^{\top,-1}\mathbf{b}_{\mathbf{i}}^\ast\|^2\right)^{1/2}\right)
\nonumber\\
&=&O_P\left(h^{p-\frac{d}{2}} d_T C\right).
\end{eqnarray}

For $R_{1,3}$, by \eqref{EqA.17} and \eqref{int4}, we can further write
\begin{eqnarray}\label{int8}
R_{1,3}&=&-\frac{2d_T}{Th^d}\sum_{t=1}^T\left[b_0z_t +\widetilde{s}(\mathbf{x}_t \, |\, \mathbf{B}) \right]\varepsilon_t
\nonumber\\
&=&-\frac{2d_T}{Th^d}\sum_{t=1}^T \varepsilon_t\left[ \sum_{\mathbf{i}\in [M]^d} \widetilde{\mathbf{x}}^{\ast\top}_{\mathbf{i},t}\mathbf{b}_{\mathbf{i}}^\ast\right]
\nonumber\\
&=&O_P\left(\frac{d_T}{\sqrt{Th^d}}\left(\sum_{\mathbf{i}\in [M]^d} \|\widetilde{\mathbf{H}}^{\top,-1}\mathbf{b}_{\mathbf{i}}^\ast\|^2\right)^{1/2}\right)
\nonumber\\
&=&O_P\left(\frac{d_T}{\sqrt{Th^d}} C\right).
\end{eqnarray}

Under Assumption \ref{Ass.4}, it is clear to see that $\|R_{1,2}\|=o_P(d_T^2)$ and $\|R_{1,3}\|=o_P(d_T^2)$. Up to now, we have finished the investigation of $R_{1}$ and then proceed with  $R_{2}$. Using simple algebra, we obtain
\begin{eqnarray}\label{int9}
R_2&=&\frac{1}{Th^d}\sum_{l=1}^{d^0_q}\psi_l\left(\|\pmb{\Lambda}_l+d_T\mathbf{B}_{D,l}\| -\|\pmb{\Lambda}_l\|\right)+\frac{d_T}{Th^d}\sum_{l=d^0_q+1}^{d_q}\psi_l\left(\|\mathbf{B}_{D,l}\| \right)
\nonumber\\
&\geq& -\frac{d_T}{Th^d}\sum_{l=1}^{d^0_q}\psi^\ast_lH_l^{-1}\|\mathbf{B}_{D,l}\|
\nonumber\\
&\geq&  -\frac{d_T}{Th^d}\psi^\ast_{\max}\sum_{l=1}^{d^0_q}H_l^{-1}\|\mathbf{B}_{D,l}\|
\nonumber\\
&\geq&  -\frac{d_T}{Th^d}\sqrt{d^0_q}\psi^\ast_{\max}\|\mathbf{B}\|_H,
\end{eqnarray}
where $\psi_l^\ast = \psi_lH_l$, $\psi^\ast_{\max}=\max\{\psi^\ast_{l},l\in[d^0_q]\}$ and the third inequality holds by the fact $\sum_{l=1}^{d^0_q}H_l^{-1}\|\mathbf{B}_{D,l}\|\leq \sqrt{d^0_q\sum_{l=1}^{d^0_q}H_l^{-2}\|\mathbf{B}_{D,l}\|^2}=\sqrt{d^0_q}\|\mathbf{B}\|_H $. By Assumption \ref{Ass.4}, it is straightforward to see that $\|R_2\|=o_P(d_T^2)$.

In summary of the results that are established in \eqref{int10}, \eqref{int5}, \eqref{int6}, \eqref{int7}, \eqref{int8}, and \eqref{int9}, it is implied by Assumption \ref{Ass.4} that the leading order term in $\frac{1}{d_T^2Th^d}\widetilde{Q}_\psi  (\alpha_0+d_Tb_0,\widetilde{\pmb{\Lambda}}+d_T\mathbf{B})-  \frac{1}{d_T^2Th^d}\widetilde{Q}_\psi  (\alpha_0, \widetilde{\pmb{\Lambda}})$ is a quadratic function of $C$ with the coefficient for the quadratic term having the probability limit $\lambda_{\min,0}>0$ and the coefficients for the linear terms being bounded in probability by $O_P(1)$. Consequently, with a sufficiently large $C$, the left-hand side of  \eqref{int10} is guaranteed to be positive with probability one. This completes the proof of Lemma \ref{Lem_lasso}. \hspace*{\fill}{$\blacksquare$}

\bigskip

\noindent \textbf{Proof of Lemma \ref{Lem_Ora}:}

(1). With  knowledge of the true model ($g(\mathbf{x})=g_c(\mathbf{x}_c)$), we can  define the following oracle LNN candidates:
\begin{eqnarray*}
\mathcal{S}_{c} =\left\{ \widetilde{s}_{c}(\mathbf{x}_c \, |\, \pmb{\Theta}_c) =\sum_{\mathbf{i}\in [M]^{d_0}} I_{\mathbf{i},h}(\mathbf{x}_c) \cdot s_c(\mathbf{x}_c\, |\, \mathbf{x}_{c,0\mathbf{i}},  \pmb{\theta}_{c,\mathbf{i}})\, \Big|\, \| \pmb{\theta}_{c,\mathbf{i}}\|<\infty \right\},
\end{eqnarray*}
where $\mathbf{x}_c$, $\mathbf{x}_{c,0\mathbf{i}}$, and $\pmb{\theta}_{c,\mathbf{i}}$ are oracle counterparts of $\mathbf{x}$, $\mathbf{x}_{0\mathbf{i}}$, and  $\pmb{\theta}_{\mathbf{i}}$, respectively, and
\begin{eqnarray*}
s_c(\mathbf{x}_c\, |\, \mathbf{x}_{c,0}, \widetilde{\pmb{\lambda}}_c) =(\widetilde{\pmb{\lambda}}_c\otimes \pmb{\gamma}_c)^\top\pmb{\sigma}_c(\mathbf{x}_c\, |\, \mathbf{x}_{c,0} ) \quad\text{with}\quad \pmb{\sigma}_c(\mathbf{x}_c\, |\, \mathbf{x}_{c,0} ) =\{\sigma\left([1,\mathbf{x}_c^\top- \mathbf{x}_{c,0}^\top]\pmb{\pi}_{c,j}\right) \}_{d^0_q(q+1)\times 1} 
\end{eqnarray*}
where $\widetilde{\pmb{\lambda}}_c$, $\pmb{\gamma}_c$, and $\pmb{\pi}_{c,j}$ are the counterparts of $\widetilde{\pmb{\lambda}}$, $\pmb{\gamma}$, and $\pmb{\pi}_{j}$ in the true model. Noteworthily, we assume $\mathbf{m}_c(\mathbf{x}_c|\mathbf{x}_{c,0}) = (m_{c,1}(\mathbf{x}_c|\mathbf{x}_{c,0}),\ldots, m_{c,d_q^0}(\mathbf{x}_c|\mathbf{x}_{c,0}))^\top$ constitute a basis for the true space $\mathscr{P}^0_q$ without loss of generality, and  
\begin{equation*}
m_{c,j}(\mathbf{x}_c|\mathbf{x}_{c,0})=m_{j}(\mathbf{x}|\mathbf{x}_{0}),
\end{equation*}
for $j=1,\ldots,d_q^0$. 

Using arguments that are analogous to those in the proof of Lemma \ref{LEM2.3}, we can obtain
\begin{eqnarray}\label{dk15}
\| g_c(\mathbf{x}_c) -  \widetilde{s}_c(\mathbf{x}_c \, |\, \widetilde{\pmb{\Lambda}}_c)  \|_{\infty}=O(h^p) ,
\end{eqnarray}
where $\widetilde{\pmb{\Lambda}}_c =\{\widetilde{\pmb{\lambda}}_{c,\mathbf{i}}\, |\, \mathbf{i}\in [M]^{d_0}\}$, $\widetilde{\pmb{\lambda}}_{c,\mathbf{i}}$ corresponds to $\pmb{\lambda}_{c,\mathbf{i}}$ up to a rotation matrix $\mathbf{D}_c$, and $\pmb{\lambda}_{c,\mathbf{i}}$ is decided by the Taylor expansion of $g_c(\mathbf{x}_c)$ at the point $\mathbf{x}_{c,0\mathbf{i}}$.

For the oracle estimators $\widetilde{\alpha}_{c}$ and $\widetilde{\pmb{\theta}}_{c,\mathbf{i}}$, it is easy to see that they satisfy the following first-order conditions:
\begin{eqnarray}\label{dk6}
\sum_{t=1}^Tz_t\Bigl[y_t-z_t\widetilde{\alpha}_{c} -\sum_{\mathbf{i}\in [M]^{d_0}} \widetilde{\mathbf{x}}_{c,\mathbf{i}, t}^\top \widetilde{\pmb{\theta}}_{c,\mathbf{i}}\Bigr]&=&0,
\nonumber\\
\sum_{t=1}^T\widetilde{\mathbf{x}}_{c,\mathbf{i}, t}[y_t-z_t\widetilde{\alpha}_{c} - \widetilde{\mathbf{x}}_{c,\mathbf{i}, t}^\top \widetilde{\pmb{\theta}}_{c,\mathbf{i}}]&=&\mathbf{0}.
\end{eqnarray}
By solving the equations in \eqref{dk6}, we obtain the expressions for $\widetilde{\pmb{\theta}}_{c,\mathbf{i}}$ and $\widetilde{\alpha}_{c}$:  
\begin{eqnarray}\label{dk12}
\widetilde{\pmb{\theta}}_{c,\mathbf{i}}&=&\left(\widetilde{\mathbf{X}}_{c,\mathbf{i}}^\top \widetilde{\mathbf{X}}_{c,\mathbf{i}}\right)^{-1}\widetilde{\mathbf{X}}_{c,\mathbf{i}}^\top[\mathbf{Y}-\mathbf{Z}\widetilde{\alpha}_{c}],
\nonumber\\
\widetilde{\alpha}_{c}&=&\left(\mathbf{Z}^\top \mathbf{M}_{c,x} \mathbf{Z}\right)^{-1}\mathbf{Z}^\top \mathbf{M}_{c,x} \mathbf{Y},
\end{eqnarray}
where  $\mathbf{Z}=(z_1,\cdots,z_T)^\top$, $\mathbf{Y}=(y_1,\cdots,y_T)^\top$, and $\mathbf{M}_{c,x}=\mathbf{I}_{T}-\sum_{\mathbf{i}\in [M]^{d_0}} \widetilde{\mathbf{X}}_{c,\mathbf{i}}\bigl[\widetilde{\mathbf{X}}_{c,\mathbf{i}}^\top \widetilde{\mathbf{X}}_{c,\mathbf{i}}\bigr]^{-1}\widetilde{\mathbf{X}}_{c,\mathbf{i}}^\top$ with $\widetilde{\mathbf{X}}_{c,\mathbf{i}} = (\widetilde{\mathbf{x}}_{c,\mathbf{i}, 1}, \cdots, \widetilde{\mathbf{x}}_{c,\mathbf{i}, T})^\top$.

Using \eqref{dk15} and the expression  in \eqref{dk12}, we can further expand $\widetilde{\alpha}_{c}$:
\begin{eqnarray*}
\widetilde{\alpha}_{c}-\alpha_0&=&\left(\mathbf{Z}^\top \mathbf{M}_{c,x} \mathbf{Z}\right)^{-1}\mathbf{Z}^\top \mathbf{M}_{c,x}\sum_{\mathbf{i}\in [M]^{d_0}} \widetilde{\mathbf{X}}_{c,\mathbf{i}}\widetilde{\pmb{\lambda}}_{c, \mathbf{i}}+\left(\mathbf{Z}^\top \mathbf{M}_{c,x} \mathbf{Z}\right)^{-1}\mathbf{Z}^\top \mathbf{M}_{c,x}\pmb{\varepsilon}+O_P(h^{p})
\nonumber\\
&=&\left(\mathbf{Z}^\top \mathbf{M}_{c,x} \mathbf{Z}\right)^{-1}\mathbf{Z}^\top \mathbf{M}_{c,x}\pmb{\varepsilon}+O_P(h^{p}).
\end{eqnarray*}
where $\pmb{\varepsilon}=(\varepsilon_1,\cdots,\varepsilon_T)^\top$. We then proceed with the convergence of $\frac{1}{T}\mathbf{Z}^\top \mathbf{M}_{c,x} \mathbf{Z}$. We write
\begin{eqnarray}\label{dk30}
\frac{1}{T}\mathbf{Z}^\top \mathbf{M}_{c,x} \mathbf{Z}&=&\frac{1}{T}\mathbf{Z}^\top\mathbf{Z}-\frac{1}{T}\sum_{\mathbf{i}\in [M]^{d_0}} \mathbf{Z}^\top\widetilde{\mathbf{X}}_{c,\mathbf{i}}\bigl[\widetilde{\mathbf{X}}_{c,\mathbf{i}}^\top \widetilde{\mathbf{X}}_{c,\mathbf{i}}\bigr]^{-1}\widetilde{\mathbf{X}}_{c,\mathbf{i}}^\top\mathbf{Z}
\nonumber\\
&\coloneqq&Q_1+Q_2.
\end{eqnarray}
For $Q_1$, it is clear to see that
\begin{eqnarray}\label{dk21}
E[Q_1]&=&E[E[z_t^2|\mathbf{x}_t]]
\nonumber\\
&=&E[\Phi_\eta(G(\mathbf{x}_t))]
\nonumber\\
&=&\int_{\mathbf{x}\in [-a,a]^d}    \Phi_\eta(G(\mathbf{x}))  f_{\mathbf{x}}(\mathbf{x})\mathrm{d} \mathbf{x}.
\end{eqnarray}

For the second moment,
\begin{eqnarray}\label{dk22}
E[Q_1^2]&=&\frac{1}{T^2}\sum_{t=1}^T\sum_{s=1}^TE[z_tz_s]
\nonumber\\
&=&\frac{1}{T^2}\sum_{t=1}^TE[E[z_t^2|\mathbf{x}_t]]+\frac{1}{T^2}\sum_{t=1}^T\sum_{s=t+1}^TE[E[z_tz_s|\mathbf{x}_t,\mathbf{x}_s]]+\frac{1}{T^2}\sum_{t=1}^T\sum_{s=1}^{t-1}E[E[z_tz_s|\mathbf{x}_t,\mathbf{x}_s]]
\nonumber\\
&=&\frac{1}{T^2}\sum_{t=1}^TE[\Phi_\eta(G(\mathbf{x}_t))]+\frac{1}{T^2}\sum_{t=1}^T\sum_{s=t+1}^TE[\widetilde{\Phi}_{\eta,s}(G(\mathbf{x}_1), G(\mathbf{x}_{1+s}))]\nonumber \\
&&+\frac{1}{T^2}\sum_{t=1}^T\sum_{s=1}^{t-1}E[\widetilde{\Phi}_{\eta,s}(G(\mathbf{x}_1), G(\mathbf{x}_{1+s}))]
\nonumber\\
&=&\frac{1}{T^2}\sum_{t=1}^T\sum_{s=1,\neq t}^T\int_{\mathbf{x},\mathbf{z}\in [-a,a]^d}   \widetilde{\Phi}_{\eta,s}(G(\mathbf{x}), G(\mathbf{z}))f_{\mathbf{x},s}(\mathbf{x},\mathbf{z})\mathrm{d}\mathbf{x}\mathrm{d}\mathbf{z}+O(T^{-1})
\nonumber\\
&=&\Bigl[\int_{\mathbf{x}\in [-a,a]^d}    \Phi_\eta(G(\mathbf{x}))  f_{\mathbf{x}}(\mathbf{x})\mathrm{d} \mathbf{x}\Bigr]^2+O\bigl(T^{-1}\bigr).
\end{eqnarray}
where $\widetilde{\Phi}_{\eta,s}(\cdot,\cdot)$ denotes the joint CDF of $(\eta_1,\eta_{1+s})$, $f_{\mathbf{x},s}(\mathbf{x}, \mathbf{z})$ is the joint PDF of $(\mathbf{x}_1, \mathbf{x}_{1+s})$, and the last equality holds by \eqref{dk20}, which is a standard result under the $\alpha$-mixing conditions in Assumption \ref{Ass.4}. Combing \eqref{dk21} and \eqref{dk22},  we obtain
\begin{equation}\label{dk31}
Q_1 = \int_{\mathbf{x}\in [-a,a]^d}    \Phi_\eta(G(\mathbf{x}))  f_{\mathbf{x}}(\mathbf{x})\mathrm{d} \mathbf{x} +O_P\left(\frac{1}{\sqrt{T}}\right).
\end{equation}

For $Q_2$, using arguments that are analogous to those for \eqref{EqA.12}, \eqref{int3}, and \eqref{dk2}, we can readily obtain
\begin{eqnarray}\label{dk32}
Q_2&=&h^{d_0}\sum_{\mathbf{i}\in [M]^{d_0}} \left[\frac{1}{Th^{d_0}}\mathbf{Z}^\top\widetilde{\mathbf{X}}_{c,\mathbf{i}}\mathbf{D}_c^\top\mathbf{H}_c\right]\left[\frac{1}{Th^{d_0}}\mathbf{H}_c\mathbf{D}_c\widetilde{\mathbf{X}}_{c,\mathbf{i}}^\top \widetilde{\mathbf{X}}_{c,\mathbf{i}}\mathbf{D}_c^\top\mathbf{H}_c\right]^{-1}\left[\frac{1}{Th^{d_0}}\mathbf{H}_c\mathbf{D}_c\widetilde{\mathbf{X}}_{c,\mathbf{i}}^\top\mathbf{Z}\right]
\nonumber\\
&=& h^{d_0}\sum_{\mathbf{i}\in [M]^{d_0}}\mathbf{M}_{c,\mathbf{i}}^\top \pmb{\Sigma}_{c,\mathbf{i}}^{-1} \mathbf{M}_{c,\mathbf{i}}(1+o_P(1)),
\end{eqnarray}
where $\mathbf{H}_c$ and $\mathbf{D}_c$ are counterpart matrices of $\mathbf{H}$ and $\mathbf{D}$ for the true model, 
\begin{eqnarray}\label{dk41}
\mathbf{M}_{c,\mathbf{i}}&=&\int_{[-a,a]^{d-d_0}}\Phi_\eta(G(\mathbf{x}_{c,0\mathbf{i}},\mathbf{z}))    f_{\mathbf{x}}(\mathbf{x}_{c,0\mathbf{i}},\mathbf{z})\mathrm{d}\mathbf{z} \int_{[-1,1]^{d_0}}    \mathbf{m}_c(\mathbf{x}_c\, |\, \mathbf{0}) \mathrm{d} \mathbf{x}_c,
\nonumber\\
\pmb{\Sigma}_{c,\mathbf{i}}&=&f_{\mathbf{x}_c}(\mathbf{x}_{c,\mathbf{i}}) \int_{[-1,1]^{d_0}}  \mathbf{m}_c(\mathbf{x}_c\, |\, \mathbf{0})  \mathbf{m}_c(\mathbf{x}_c\, |\, \mathbf{0})^\top  \mathrm{d} \mathbf{x}_c.
\end{eqnarray}

In summary of \eqref{dk30}, \eqref{dk31}, and \eqref{dk32}, 
\begin{eqnarray}\label{dk35}
\frac{1}{T}\mathbf{Z}^\top \mathbf{M}_{c,x} \mathbf{Z}&\xrightarrow{~P~}& \int_{\mathbf{x}\in [-a,a]^d}    \Phi_\eta(G(\mathbf{x}))  f_{\mathbf{x}}(\mathbf{x})\mathrm{d} \mathbf{x}-\lim_{T\rightarrow\infty}h^{d_0}\sum_{\mathbf{i}\in [M]^{d_0}}\mathbf{M}_{c,\mathbf{i}}^\top \pmb{\Sigma}_{c,\mathbf{i}}^{-1} \mathbf{M}_{c,\mathbf{i}}
\nonumber\\
&\coloneqq&M_{c,z}
.
\end{eqnarray}

We then proceed with $\mathbf{Z}^\top \mathbf{M}_{c,x}\pmb{\varepsilon}$. Using arguments that are analogous to those in the proof of \eqref{dk35},  we can write
\begin{eqnarray}\label{dk100}
&&\frac{1}{T}\mathbf{Z}^\top \mathbf{M}_{c,x}\pmb{\varepsilon} \nonumber \\
&=&\frac{1}{T}\Bigl[\mathbf{Z}^\top-\sum_{\mathbf{i}\in [M]^{d_0}} \mathbf{Z}^\top\widetilde{\mathbf{X}}_{c,\mathbf{i}}\bigl(\widetilde{\mathbf{X}}_{c,\mathbf{i}}^\top \widetilde{\mathbf{X}}_{c,\mathbf{i}}\bigr)^{-1}\widetilde{\mathbf{X}}_{c,\mathbf{i}}^\top\Bigr]\pmb{\varepsilon}
\nonumber\\
&=&\frac{1}{T}\sum_{t=1}^T\Bigl[z_t-\sum_{\mathbf{i}\in [M]^{d_0}} \mathbf{Z}^\top\widetilde{\mathbf{X}}_{c,\mathbf{i}}\bigl(\widetilde{\mathbf{X}}_{c,\mathbf{i}}^\top \widetilde{\mathbf{X}}_{c,\mathbf{i}}\bigr)^{-1}\widetilde{\mathbf{x}}_{c,\mathbf{i},t}\Bigr]\varepsilon_t
\nonumber\\
&=&-\frac{1}{T}\sum_{t=1}^T\sum_{\mathbf{i}\in [M]^{d_0}} I_{\mathbf{i},h} (\mathbf{x}_{c,t}) \Bigl[\frac{1}{Th^{d_0}}\mathbf{Z}^\top\widetilde{\mathbf{X}}_{c,\mathbf{i}}\mathbf{D}_c^\top\mathbf{H}_c-\mathbf{M}_{c,\mathbf{i}}^\top\Bigr] \pmb{\Sigma}_{c,\mathbf{i}}^{-1}\mathbf{H}_c\mathbf{m}_c(\mathbf{x}_{c,t}\, |\, \mathbf{x}_{c,0\mathbf{i}})\varepsilon_t
\nonumber\\
&&-\frac{1}{T}\sum_{t=1}^T\sum_{\mathbf{i}\in [M]^{d_0}} I_{\mathbf{i},h} (\mathbf{x}_{c,t}) \mathbf{M}_{c,\mathbf{i}}^\top \Bigl[\Bigl(\frac{1}{Th^{d_0}}\mathbf{H}_c\mathbf{D}_c\widetilde{\mathbf{X}}_{c,\mathbf{i}}^\top \widetilde{\mathbf{X}}_{c,\mathbf{i}}\mathbf{D}_c^\top\mathbf{H}_c\Bigr)^{-1}-\pmb{\Sigma}_{c,\mathbf{i}}^{-1}\Bigr]\mathbf{H}_c\mathbf{m}_c(\mathbf{x}_{c,t}\, |\, \mathbf{x}_{c,0\mathbf{i}})\varepsilon_t
\nonumber\\
&&-\frac{1}{T}\sum_{t=1}^T\sum_{\mathbf{i}\in [M]^{d_0}} I_{\mathbf{i},h} (\mathbf{x}_{c,t}) \Bigl[\frac{1}{Th^{d_0}}\mathbf{Z}^\top\widetilde{\mathbf{X}}_{c,\mathbf{i}}\mathbf{D}_c^\top\mathbf{H}_c-\mathbf{M}_{c,\mathbf{i}}^\top\Bigr] \Bigl[\Bigl(\frac{1}{Th^{d_0}}\mathbf{H}_c\mathbf{D}_c\widetilde{\mathbf{X}}_{c,\mathbf{i}}^\top \widetilde{\mathbf{X}}_{c,\mathbf{i}}\mathbf{D}_c^\top\mathbf{H}_c\Bigr)^{-1}-\pmb{\Sigma}_{c,\mathbf{i}}^{-1}\Bigr]
\nonumber\\
&&\times 
\mathbf{H}_c\mathbf{m}_c(\mathbf{x}_{c,t}\, |\, \mathbf{x}_{c,0\mathbf{i}})\varepsilon_t
\nonumber\\
&&+\frac{1}{T}\sum_{t=1}^T\left(z_t-\sum_{\mathbf{i}\in [M]^{d_0}} I_{\mathbf{i},h} (\mathbf{x}_{c,t}) \mathbf{M}_{c,\mathbf{i}}^\top \pmb{\Sigma}_{c,\mathbf{i}}^{-1}\mathbf{H}_c\mathbf{m}_c(\mathbf{x}_{c,t}\, |\, \mathbf{x}_{c,0\mathbf{i}})\right)\varepsilon_t+O_P(h^p)
\nonumber\\
&=&\widetilde{Q}_1+\cdots+\widetilde{Q}_4+O_P(h^p),
\end{eqnarray}
where the definitions of $\widetilde{Q}_1$, $\cdots$, $\widetilde{Q}_4$ are obvious.

For notational simplicity, let $\pmb{\xi}_{M,\mathbf{i}}=\frac{1}{Th^{d_0}}\sum_{t=1}^TI_{\mathbf{i},h} (\mathbf{x}_{c,t})z_t\mathbf{H}_c\mathbf{m}_c(\mathbf{x}_{c,t}\, |\, \mathbf{x}_{c,0\mathbf{i}})   -\mathbf{M}_{c,\mathbf{i}}$.  For the first term, using the oracle counterpart of \eqref{EqA.10}, we can readily obtain that
\begin{eqnarray*}
\widetilde{Q}_1&=&-\frac{1}{T}\sum_{t=1}^T\sum_{\mathbf{i}\in [M]^{d_0}} I_{\mathbf{i},h} (\mathbf{x}_{c,t}) \pmb{\xi}_{M,\mathbf{i}}^\top \pmb{\Sigma}_{c,\mathbf{i}}^{-1}\mathbf{H}_c\mathbf{m}_c(\mathbf{x}_{c,t}\, |\, \mathbf{x}_{c,0\mathbf{i}})\varepsilon_t(1+o_P(1))
\nonumber\\
&\coloneqq&\widetilde{Q}_1^\ast(1+o_P(1)).
\end{eqnarray*}
To show the convergence of $\widetilde{Q}_1$,  it suffices only to study $\widetilde{Q}_1^\ast$. It is clear to see that $E[\widetilde{Q}^\ast_1]=0$ under Assumption \ref{Ass.3} and 
{\footnotesize
\begin{eqnarray}\label{dk101}
E[\widetilde{Q}_1^{\ast 2}]&=&\frac{1}{T^2}\sum_{t=1}^T\sum_{\mathbf{i}\in [M]^{d_0}}E\Bigl[I_{\mathbf{i},h} (\mathbf{x}_{c,t}) \pmb{\xi}_{M,\mathbf{i}}^\top \pmb{\Sigma}_{c,\mathbf{i}}^{-1}\mathbf{H}_c\mathbf{m}_c(\mathbf{x}_{c,t}\, |\, \mathbf{x}_{c,0\mathbf{i}})\mathbf{m}_c(\mathbf{x}_{c,t}\, |\, \mathbf{x}_{c,0\mathbf{i}})^\top\mathbf{H}_c\pmb{\Sigma}_{c,\mathbf{i}}^{-1}
\pmb{\xi}_{M,\mathbf{i}}\Bigr]E[\varepsilon_t^2]
\nonumber\\
&&+\frac{1}{T^2}\sum_{t=1}^T\sum_{s=t+1}^T\sum_{\mathbf{i},\mathbf{j}\in [M]^{d_0}}E\Bigl[I_{\mathbf{i},h} (\mathbf{x}_{c,t})I_{\mathbf{j},h} (\mathbf{x}_{c,s})\pmb{\xi}_{M,\mathbf{i}}^\top \pmb{\Sigma}_{c,\mathbf{i}}^{-1}\mathbf{H}_c\mathbf{m}_c(\mathbf{x}_{c,t}\, |\, \mathbf{x}_{c,0\mathbf{i}})\mathbf{m}_c(\mathbf{x}_{c,s}\, |\, \mathbf{x}_{c,0\mathbf{j}})^\top\mathbf{H}_c\pmb{\Sigma}_{c,\mathbf{j}}^{-1}
\pmb{\xi}_{M,\mathbf{j}}\Bigr]E[\varepsilon_t\varepsilon_s]
\nonumber\\
&&+\frac{1}{T^2}\sum_{t=1}^T\sum_{s=1}^{t-1}\sum_{\mathbf{i},\mathbf{j}\in [M]^{d_0}}E\Bigl[I_{\mathbf{i},h} (\mathbf{x}_{c,t})I_{\mathbf{j},h} (\mathbf{x}_{c,s})\pmb{\xi}_{M,\mathbf{i}}^\top \pmb{\Sigma}_{c,\mathbf{i}}^{-1}\mathbf{H}_c\mathbf{m}_c(\mathbf{x}_{c,t}\, |\, \mathbf{x}_{c,0\mathbf{i}})\mathbf{m}_c(\mathbf{x}_{c,s}\, |\, \mathbf{x}_{c,0\mathbf{j}})^\top\mathbf{H}_c\pmb{\Sigma}_{c,\mathbf{j}}^{-1}
\pmb{\xi}_{M,\mathbf{j}}\Bigr]E[\varepsilon_t\varepsilon_s]
\nonumber\\
&\coloneqq&\widetilde{Q}_{1,1}^\ast +\widetilde{Q}_{1,2}^\ast +\widetilde{Q}_{1,3}^\ast .
\end{eqnarray}
}For $\widetilde{Q}_{1,1}^\ast$, using the arguments that are closely related to those in the proof of \eqref{int2_4}, we can obtain
\begin{eqnarray*}
E\Bigl[I_{\mathbf{i},h} (\mathbf{x}_{c,t}) \|\pmb{\xi}_{M,\mathbf{i}}\|^2 \bigl\|\pmb{\Sigma}_{c,\mathbf{i}}^{-1}\mathbf{H}_c\mathbf{m}_c(\mathbf{x}_{c,t}\, |\, \mathbf{x}_{c,0\mathbf{i}})\bigr\|^2\Bigr]&=&o\bigl(h^{d_0}\bigr).
\end{eqnarray*}
Therefore,  $\widetilde{Q}_{1,1}^\ast=o\bigl(\frac{1}{T} \bigr)$. Analogously, for the second term
\begin{eqnarray}
\widetilde{Q}_{1,2}^\ast
&\leq&o(1)\frac{1}{T^2}\sum_{t=1}^T\sum_{s=t+1}^TE[\varepsilon_t\varepsilon_s]
\nonumber\\
&=&o(1)\frac{1}{T^2}\sum_{t=1}^{T-1}\Bigl(1-\frac{t}{T}\Bigr)E[\varepsilon_1\varepsilon_{1+t}]
\nonumber\\
&\leq&o(1)\frac{1}{T^2}\sum_{t=1}^{T-1}\bigl|\text{cov}(\varepsilon_1,\varepsilon_{1+t})\bigr|
\nonumber\\
&\leq&o(1)\frac{1}{T^2}\sum_{t=1}^{T-1} \alpha(t)^{\nu/(2+\nu)} E[|\varepsilon_1|^{2+\nu}]^{\frac{2}{2+\nu}}
\nonumber\\
&=&o\Bigl(\frac{1}{T}\Bigr),
\end{eqnarray}
where the third inequality holds by Assumption \ref{Ass.3} and the Davydov's inequality for $\alpha$-mixing processes   \citep[see pages 19-20 of][]{Bosq1996}.

It immediately implies that the second and third terms on the right-hand side of \eqref{dk101} are also asymptotically negligible. In summary of these results, we have $E[\widetilde{Q}_1^{\ast 2}]=o\bigl(\frac{1}{T} \bigr)$ and 
\begin{equation}\label{dk102}
\widetilde{Q}_1=o_p\Bigl(\frac{1}{\sqrt{T}} \Bigr).
\end{equation}
Using similar arguments, we can obtain 
\begin{equation}\label{dk103}
\widetilde{Q}_2=o_p\bigl(\frac{1}{\sqrt{T}} \bigr),\quad \widetilde{Q}_3=o_p\bigl(\frac{1}{\sqrt{T}} \bigr).
\end{equation}

Then, it suffices only to study $\widetilde{Q}_4$.
For notational simplicity, we define 
\begin{equation*}
\widetilde{z}_t=z_t-\sum_{\mathbf{i}\in [M]^{d_0}} I_{\mathbf{i},h} (\mathbf{x}_{c,t}) \mathbf{M}_{c,\mathbf{i}}^\top \pmb{\Sigma}_{c,\mathbf{i}}^{-1}\mathbf{H}_c\mathbf{m}(\mathbf{x}_{c,t}\, |\, \mathbf{x}_{c,0\mathbf{i}}).
\end{equation*}
In what follows, we first compute the asymptotic covariance of $\frac{1}{\sqrt{T}}\sum_{t=1}^T\widetilde{z}_t\varepsilon_t$ and then employ the small-block and large-block technique for $\alpha$-mixing processes to establish its asymptotic normality.

Write
\begin{eqnarray}\label{revn1}
\widetilde{Q}_4&=&E\Bigl[\Bigl(\frac{1}{\sqrt{T}}\sum_{t=1}^T\widetilde{z}_t\varepsilon_t\Bigr)^2\Bigr]
\nonumber\\
&=&\frac{1}{T}\sum_{t=1}^T\sum_{s=1}^TE\Bigl[\widetilde{z}_t\widetilde{z}_s\varepsilon_t\varepsilon_s\Bigr]
\nonumber\\
&=&\frac{1}{T}\sum_{t=1}^TE\Bigl[\widetilde{z}^2_t\varepsilon^2_t\Bigr]+\frac{1}{T}\sum_{t=1}^T\sum_{s=t+1}^{T}E\Bigl[\widetilde{z}_t\widetilde{z}_s\varepsilon_t\varepsilon_s\Bigr]+\frac{1}{T}\sum_{t=1}^T\sum_{s=1}^{t-1}E\Bigl[\widetilde{z}_t\widetilde{z}_s\varepsilon_t\varepsilon_s\Bigr]
\nonumber\\
&=&\widetilde{Q}_{4,1}+\widetilde{Q}_{4,2}+\widetilde{Q}_{4,3}.
\end{eqnarray}
For the first term, we can use analogous arguments in the proof of \eqref{int1} and Assumption \ref{Ass.3} to show that
 
\begin{eqnarray}\label{fi2}
\widetilde{Q}_{4,1}&=&\frac{1}{T}\sum_{t=1}^TE\Bigl[\widetilde{z}^2_t\Bigr]\sigma_\varepsilon^2
\nonumber\\
&=& M_{c,z}\sigma_\varepsilon^2+o(1).
\end{eqnarray}
For $\widetilde{Q}_{4,2}$ and $\widetilde{Q}_{4,3}$, it is clear to see that $\widetilde{Q}_{4,2}=\widetilde{Q}_{4,3}$. It suffices only to study $\widetilde{Q}_{4,2}$. Write

{\footnotesize
\begin{eqnarray}\label{fi1}
\widetilde{Q}_{4,2}&=&\frac{1}{T}\sum_{t=1}^T\sum_{s=t+1}^{T}E\Bigl[\widetilde{z}_t\widetilde{z}_{s}\Bigr]E[\varepsilon_t\varepsilon_s]
\nonumber\\
&=&\frac{1}{T}\sum_{t=1}^T\sum_{s=t+1}^{T}E\bigl[E[\widetilde{z}_t\widetilde{z}_{s}|\mathbf{x}_t,\mathbf{x}_s]\bigr]E[\varepsilon_t\varepsilon_s]
\nonumber\\
&=&\frac{1}{T}\sum_{t=1}^T\sum_{s=t+1}^{T}E[z_tz_{s}]\sigma^2_{t,s}+\frac{1}{T}\sum_{t=1}^T\sum_{s=t+1}^{T}\sum_{\mathbf{i},\mathbf{j}\in [M]^{d_0}}\mathbf{M}_{c,\mathbf{i}}^\top \pmb{\Sigma}_{c,\mathbf{i}}^{-1}\mathbf{H}_c\mathbf{D}_cE\bigl[\widetilde{\mathbf{x}}_{c,\mathbf{i},t} \widetilde{\mathbf{x}}_{c,\mathbf{j},s}^\top\bigr] \mathbf{D}_c^\top \mathbf{H}_c \pmb{\Sigma}_{c,\mathbf{j}}^{-1}\mathbf{M}_{c,\mathbf{j}}
E[\varepsilon_t\varepsilon_s]
\nonumber\\
&&-\frac{1}{T}\sum_{t=1}^T\sum_{s=t+1}^{T}\sum_{\mathbf{i}\in [M]^{d_0}} \mathbf{M}_{c,\mathbf{i}}^\top \pmb{\Sigma}_{c,\mathbf{i}}^{-1}\mathbf{H}_c\mathbf{D}_cE\bigl[\widetilde{\mathbf{x}}_{c,\mathbf{i},t}z_s\bigr]E[\varepsilon_t\varepsilon_s]
\nonumber\\
&&-\frac{1}{T}\sum_{t=1}^T\sum_{s=t+1}^{T}\sum_{\mathbf{i}\in [M]^{d_0}} E\bigl[\widetilde{\mathbf{x}}^\top_{c,\mathbf{i},s}z_t\bigr]E[\varepsilon_t\varepsilon_s]\mathbf{D}^\top_c  \mathbf{H}_c \pmb{\Sigma}_{c,\mathbf{i}}^{-1}\mathbf{M}_{c,\mathbf{i}}
\nonumber\\
&=& \sum_{s=1}^{T-1}\Bigl(1-\frac{s}{T}\Bigr) \sigma^2_{\varepsilon,s}\int_{\mathbf{x},\mathbf{z}\in [-a,a]^d}   \widetilde{\Phi}_{\eta,s}(G(\mathbf{x}), G(\mathbf{z}))f_{\mathbf{x},s}(\mathbf{x},\mathbf{z})\mathrm{d}\mathbf{x}\mathrm{d}\mathbf{z}
\nonumber\\
&&+ h^{2d_0}\sum_{s=1}^{T-1}\Bigl(1-\frac{s}{T}\Bigr) \sigma^2_{\varepsilon,s}\sum_{\mathbf{i},\mathbf{j}\in [M]^{d_0}}\mathbf{M}_{c,\mathbf{i}}^\top \pmb{\Sigma}_{c,\mathbf{i}}^{-1}
f_{\mathbf{x}_c,s}(\mathbf{x}_{c,0\mathbf{i}},\mathbf{x}_{c,0\mathbf{j}}) \int_{\mathbf{x}_c,\mathbf{z}_c\in[-1,1]^{d_0}}  \mathbf{m}_c(\mathbf{x}_c\, |\, \mathbf{0})  \mathbf{m}_c(\mathbf{z}_c\, |\, \mathbf{0})^\top  \mathrm{d} \mathbf{x}_c\mathrm{d} \mathbf{z}_c
\pmb{\Sigma}_{c,\mathbf{j}}^{-1}\mathbf{M}_{c,\mathbf{j}}
\nonumber\\
&&-h^{d_0}\sum_{s=1}^{T-1}\Bigl(1-\frac{s}{T}\Bigr)\sigma^2_{\varepsilon,s}\sum_{\mathbf{i}\in [M]^{d_0}} \mathbf{M}_{c,\mathbf{i}}^\top\pmb{\Sigma}_{c,\mathbf{i}}^{-1}\int_{\mathbf{x}_1\in[-a,a]^{d-d_0},\mathbf{z}\in[-a,a]^{d}}\Phi_{\eta}(G(\mathbf{z}))    f_{\mathbf{x},s}(\mathbf{x}_{\mathbf{i}}, \mathbf{x}_{c,0\mathbf{i}},\mathbf{z})\mathrm{d}\mathbf{x}_1\mathrm{d}\mathbf{z} 
\int_{[-1,1]^{d_0}}   \mathbf{m}_c(\mathbf{x}_c\, |\, \mathbf{0}) \mathrm{d} \mathbf{x}_c
\nonumber\\
&&-h^{d_0}\sum_{s=1}^{T-1}\Bigl(1-\frac{s}{T}\Bigr)\sigma^2_{\varepsilon,s}\sum_{\mathbf{i}\in [M]^{d_0}} \int_{\mathbf{x}_1\in[-a,a]^{d-d_0},\mathbf{z}\in[-a,a]^{d}}\Phi_{\eta}(G(\mathbf{z}))    f_{\mathbf{x},s}(\mathbf{z},\mathbf{x}_{\mathbf{i}}, \mathbf{x}_{c,0\mathbf{i}})\mathrm{d}\mathbf{x}_1\mathrm{d}\mathbf{z} 
\int_{[-1,1]^{d_0}}   \mathbf{m}^\top_c(\mathbf{x}_c\, |\, \mathbf{0}) \mathrm{d} \mathbf{x}_c   \pmb{\Sigma}_{c,\mathbf{i}}^{-1}      \mathbf{M}_{c,\mathbf{i}}
,
\nonumber\\
\end{eqnarray}
}where $\sigma_{\varepsilon,s}^2=E[\varepsilon_1\varepsilon_{1+s}]$. Let $Q_0$ represent the limit of the terms on the right-hand side of \eqref{fi1}, as $T\rightarrow\infty$. Together with \eqref{fi2} and Assumption \ref{Ass.4}, it yields that 
\begin{eqnarray*}
\sigma_{c,z}^2 = \lim_{T\rightarrow \infty}\widetilde{Q}_{4}=M_{c,z}\sigma_\varepsilon^2+2Q_0.
\end{eqnarray*}

Below, we further use small-block and large-block to prove the normality. To employ the small-block and large-block arguments, we partition the set $\{1,\ldots, T \}$ into $2k_T+1$ subsets with large blocks of size $l_T$ and small blocks of size $s_T$ and the last remaining set of size $T-k_T(l_T+s_T)$, where $l_T$ and $s_T$ are selected such that

\begin{eqnarray*}
s_T\to \infty,\quad  \frac{s_T}{l_T}\to 0,\quad \frac{l_T^{1+\nu} }{  T^{\frac{\nu}{2}}}\to 0,\quad\text{and}\quad k_T\equiv \left\lfloor \frac{T}{l_T+s_T} \right\rfloor,
\end{eqnarray*}
and $\nu$ is defined in Assumption \ref{Ass.3}.1.

For $j=1,\ldots, k_T$, define

\begin{eqnarray*}
&&\pmb{\xi}_{j,1} = \sum_{t=(j-1)(l_T+s_T)+1}^{jl_T+(j-1)s_T} \widetilde{z}_t \varepsilon_t,\quad \pmb{\xi}_{j,2} = \sum_{t=jl_T+ (j-1)s_T+1}^{j(l_T+s_T)} \widetilde{z}_t  \varepsilon_t,\quad \pmb{\xi}_0 =\sum_{t=k_T(l_T+s_T)+1}^T \widetilde{z}_t  \varepsilon_t.
\end{eqnarray*}
Note that $\alpha(T) = o(1/T)$ and $k_Ts_T/T\to 0$. By direct calculation, we immediately obtain that

\begin{eqnarray*}
\frac{1}{T}E\left\|\sum_{j=1}^{k_T} \pmb{\xi}_{j,2}  \right\|^2\to 0\quad \text{and}\quad \frac{1}{T}E\left\| \pmb{\xi}_{0} \right\|^2\to 0.
\end{eqnarray*}
Therefore,

\begin{eqnarray*}
\frac{1}{\sqrt{T}}\sum_{t=1}^T\widetilde{z}_t   \varepsilon_t = \frac{1}{\sqrt{T}}\sum_{j=1}^{k_T}\pmb{\xi}_{j,1} +o_P(1).
\end{eqnarray*}
By Proposition 2.6 of \cite{FanYao}, we have as $T\to 0$

\begin{eqnarray*}
&&\left| E\left[\exp\left( \frac{iw}{\sqrt{T}}\sum_{j=1}^{k_T}\pmb{\xi}_{j,1}\right) \right] -\prod_{j=1}^{k_T}E\left[\exp\left( \frac{iw \pmb{\xi}_{j,1}}{\sqrt{T}} \right) \right]\right| \nonumber \\
&\le & 16(k_T-1) \alpha(s_T)\to 0,
\end{eqnarray*}
where $i$ is the imaginary unit.

In connection with \eqref{EqA.14}-\eqref{EqA.17}, the Feller condition is fulfilled as follows:

\begin{eqnarray*}
\frac{1}{T}\sum_{j=1}^{k_T}E[\pmb{\xi}_{j,1}\pmb{\xi}_{j,1}^\top]\to \sigma_{c,z}^2.
\end{eqnarray*}
 
Also, we note that

\begin{eqnarray}\label{dk151}
E[\|\pmb{\xi}_{1,1}\|^2 \cdot I(\|\pmb{\xi}_{1,1}\| \ge \epsilon \sqrt{T})] &\le &\{E\|\pmb{\xi}_{1,1}\|^{2\cdot \frac{2+\nu}{2}}\}^{\frac{2}{2+\nu}} \left\{ E[ I(\|\pmb{\xi}_{1,1}\| \ge \epsilon \sqrt{T})] \right\}^{\frac{\nu}{2+\nu}}\nonumber \\
&\le  &\{E\|\pmb{\xi}_{1,1}\|^{2+\nu}\}^{\frac{2}{2+\nu}} \left\{ \frac{E\|\pmb{\xi}_{1,1}\|^{2+\nu}}{\epsilon^{2+\nu} T^{\frac{2+\nu}{2}}}\right\}^{\frac{\nu}{2+\nu}}\nonumber \\
&= &  \frac{1}{\epsilon^{\nu} T^{\frac{\nu}{2}}}\left\{ E\|\pmb{\xi}_{1,1}\|^{2+\nu}\right\} \nonumber \\
&=&  O(1) \frac{l_T^{2+\nu} }{ T^{\frac{\nu}{2}}} ,
\end{eqnarray}
where the first inequality follows from H\"older inequality, the second inequality follows from  Chebyshev's inequality, and the second equality follows from Minkowski inequality. Consequently,

\begin{eqnarray*}
\frac{1}{T}\sum_{j=1}^{k_T}E[\|\pmb{\xi}_{j,1}\|^2 \cdot I(\|\pmb{\xi}_{j,1}\| \ge \epsilon \sqrt{T})] =O\left( \frac{k_Tl_T^{2+\nu} }{ T ^{1+\frac{\nu}{2}}}\right) =o(1),
\end{eqnarray*}
where the last step follows from the choice of $l_T$ as specified above. Therefore, the Lindberg condition is justified. Using a Cram{\'e}r-Wold device, the CLT follows immediately by the standard argument:
\begin{eqnarray}\label{dk36}
\frac{1}{\sqrt{T}}\mathbf{Z}^\top \mathbf{M}_{c,x}\pmb{\varepsilon}\to_D N\left(\mathbf{0}, M^{-2}_{c,z}\sigma_{c,z}^2\right).
\end{eqnarray}

Combing \eqref{dk35} and \eqref{dk36} leads to the desired result in Lemma \ref{Lem_Ora}.1.

\medskip

(2). We now study the asymptotic behaviour of $\widetilde{\pmb{\theta}}_{c,\mathbf{i}}$. Using its expression in \eqref{dk12}, we can further write 
\begin{eqnarray}\label{dk40}
\widetilde{\pmb{\theta}}_{c,\mathbf{i}}-\widetilde{\pmb{\lambda}}_{c,\mathbf{i}}&=&\left(\widetilde{\mathbf{X}}_{c,\mathbf{i}}^\top \widetilde{\mathbf{X}}_{c,\mathbf{i}}\right)^{-1}\widetilde{\mathbf{X}}_{c,\mathbf{i}}^\top[\mathbf{Y}-\mathbf{Z}\widetilde{\alpha}_{c}]-\widetilde{\pmb{\lambda}}_{c,\mathbf{i}}
\nonumber\\
&=& \left( \sum_{t=1}^T\widetilde{\mathbf{x}}_{c,\mathbf{i},t} \widetilde{\mathbf{x}}_{c,\mathbf{i},t} ^\top \right)^{-1}   \sum_{t=1}^T\widetilde{\mathbf{x}}_{c,\mathbf{i},t} [g_c(\mathbf{x}_{c,t})-\widetilde{s}_c(\mathbf{x}_{c,t} \, |\, \widetilde{\pmb{\Lambda}}_a) ]
\nonumber \\
&&+\left( \sum_{t=1}^T\widetilde{\mathbf{x}}_{c,\mathbf{i},t} \widetilde{\mathbf{x}}_{c,\mathbf{i},t} ^\top \right)^{-1}   \sum_{t=1}^T\widetilde{\mathbf{x}}_{c,\mathbf{i},t} z_t(\alpha_0-\widetilde{\alpha}_{c})
\nonumber \\
&&+\left( \sum_{t=1}^T\widetilde{\mathbf{x}}_{c,\mathbf{i},t} \widetilde{\mathbf{x}}_{c,\mathbf{i},t} ^\top \right)^{-1}   \sum_{t=1}^T\widetilde{\mathbf{x}}_{c,\mathbf{i},t}  \varepsilon_t.
\end{eqnarray}
Using arguments that are analogous to those for \eqref{EqA.12} and \eqref{dk2}, we can readily obtain
\begin{eqnarray}\label{dk50}
\frac{1}{Th^{d_0}}\sum_{t=1}^T\mathbf{H}_c\mathbf{D}_c\widetilde{\mathbf{x}}_{c,\mathbf{i},t} \widetilde{\mathbf{x}}_{c,\mathbf{i},t} ^\top\mathbf{D}_c^\top\mathbf{H}_c&=&\pmb{\Sigma}_{c,\mathbf{i}}(1+o_P(1)),
\nonumber\\
\frac{1}{Th^{d_0}}\sum_{t=1}^T\mathbf{H}_c\mathbf{D}_c\widetilde{\mathbf{x}}_{c,\mathbf{i},t} z_t&=&\mathbf{M}_{c,\mathbf{i}}(1+o_P(1)),
\end{eqnarray}
where $\pmb{\Sigma}_{c,\mathbf{i}}$ and $\mathbf{M}_{c,\mathbf{i}}$ are defined in \eqref{dk41}. Together with Lemma \ref{Lem_Ora}.1 and \eqref{dk15}, these results yield
\begin{equation}\label{dk42}
\sqrt{Th^{d_0}}\mathbf{H}_c^{-1}  \mathbf{D}_c^{\top,-1}(\widetilde{\pmb{\theta}}_{c,\mathbf{i}} -\widetilde{\pmb{\lambda}}_{c,\mathbf{i}}+O_P(h^p))=\frac{1}{\sqrt{Th^{d_0}}}\pmb{\Sigma}_{c,\mathbf{i}}^{-1}\sum_{t=1}^T\mathbf{H}_c\mathbf{D}_c\widetilde{\mathbf{x}}_{c,\mathbf{i},t}  \varepsilon_t+o_P(1).
\end{equation} 

Drawing upon the $\alpha$-mixing conditions in Assumption \ref{Ass.3},  we can use the arguments that are closely related to the small-block and large-block technique employed in the proof of Lemma \ref{LemA.3} to show that
\begin{equation*}
\frac{1}{\sqrt{Th^{d_0}}}\sum_{t=1}^T\mathbf{H}_c\mathbf{D}_c\widetilde{\mathbf{x}}_{c,\mathbf{i},t}  \varepsilon_t\to_D N\left(\mathbf{0},\sigma^2_\varepsilon \pmb{\Sigma}_{c,\mathbf{i}}\right).
\end{equation*}
Together with \eqref{dk42}, it completes the proof of Lemma \ref{Lem_Ora}.2. \hspace*{\fill}{$\blacksquare$}

\bigskip

\noindent \textbf{Proof of Lemma \ref{LEM2.4}:}

(1). For any $l=d_q^0+1,\ldots,d_q$, if $\|\widetilde{\pmb{\Theta}}_{l}\|\neq 0$, we have the following first-order condition for the minimization problem in \eqref{EQ2.2}:
\begin{eqnarray}\label{dk4}
\mathbf{0}&=&\Bigl\{\left.\frac{\partial\widetilde{Q}  (\alpha,\pmb{\Theta})}{\partial \pmb{\Theta}_{l}}+\frac{\psi_l}{2\|\pmb{\Theta}_{D,l}\|}\frac{\partial \pmb{\Theta}_{D,l}^\top\pmb{\Theta}_{D,l}}{\partial \pmb{\Theta}_l}\Bigr\}\right|_{(\alpha,\pmb{\Theta})=(\widetilde{\alpha},\widetilde{\pmb{\Theta}})}
\nonumber\\
&\coloneqq&\mathbf{Q}_{l,1}+\mathbf{Q}_{l,2},
\end{eqnarray}
where $\widetilde{Q}  (\alpha,\pmb{\Theta})= \sum_{t=1}^T[y_t-z_t\alpha -\widetilde{s}(\mathbf{x}_t \, |\, \pmb{\Theta}) ]^2$.

We proceed with the derivations of $\mathbf{Q}_{l,1}$. In light of \eqref{int4},  by taking first-order partial derivative of $\widetilde{Q}  (\alpha,\pmb{\Theta})$ with respect to $\pmb{\theta}_{\mathbf{i},l}$, we obtain
\begin{eqnarray*}
\mathbf{Q}_{\mathbf{i},l1}&=& \left.\frac{\partial\widetilde{Q}  (\alpha,\pmb{\Theta})}{\partial \pmb{\theta}_{l}}\right|_{(\alpha,\pmb{\Theta})=(\widetilde{\alpha},\widetilde{\pmb{\Theta}})}
\nonumber\\
&=&-2\sum_{t=1}^T \widetilde{x}_{\mathbf{i},tl}\Bigl[y_t-z_t\widetilde{\alpha}-\sum_{\mathbf{i}\in [M]^d} \widetilde{\mathbf{x}}_{\mathbf{i},t}^\top \widetilde{\pmb{\theta}}_{\mathbf{i}} \Bigr]
\nonumber\\
&=&-2\sum_{t=1}^T \widetilde{x}_{\mathbf{i},tl}\Bigl[y_t-\sum_{\mathbf{i}\in [M]^d} \widetilde{\mathbf{x}}^{\ast\top}_{\mathbf{i},t}\widetilde{\pmb{\theta}}_{\mathbf{i}}^\ast \Bigr],
\end{eqnarray*}
where $\widetilde{x}_{\mathbf{i},tl}$ is the $l$-th element of $\widetilde{\mathbf{x}}_{\mathbf{i},t}$ and $\widetilde{\mathbf{x}}^\ast_{\mathbf{i},t}  =I_{\mathbf{i},h} (\mathbf{x}_t)(z_t,\pmb{\sigma}( \mathbf{x}_t\, |\,\mathbf{x}_{0\mathbf{i}})^\top (\mathbf{I}_{d_q}\otimes \pmb{\gamma}^\top)^\top)^\top$, and $\widetilde{\pmb{\theta}}_{\mathbf{i}}^\ast=(\widetilde{a},\widetilde{\pmb{\theta}}_{\mathbf{i}}^ \top)^\top$.

Then, simple algebra gives
\begin{eqnarray}\label{dk1}
\|\mathbf{Q}_{l,1}\|^2&=&\sum_{\mathbf{i}\in [M]^d} \|\mathbf{Q}_{\mathbf{i},l1}\|^2
\nonumber\\
&=&4\sum_{\mathbf{i}\in [M]^d} \left\|\sum_{t=1}^T \widetilde{x}_{\mathbf{i},tl}\Bigl[y_t-\sum_{\mathbf{i}\in [M]^d} \widetilde{\mathbf{x}}^{\ast\top}_{\mathbf{i},t}\widetilde{\pmb{\theta}}_{\mathbf{i}}^\ast \Bigr]\right\|^2
\nonumber\\
&=&4\sum_{\mathbf{i}\in [M]^d} \left\|\sum_{t=1}^T \widetilde{x}_{\mathbf{i},tl}\widetilde{\mathbf{x}}^{\ast\top}_{\mathbf{i},t}(\widetilde{\pmb{\theta}}_{\mathbf{i}}^\ast-\widetilde{\pmb{\lambda}}_{\mathbf{i}}^\ast)\right\|^2+4\sum_{\mathbf{i}\in [M]^d} \left\|\sum_{t=1}^T \widetilde{x}_{\mathbf{i},tl}\left(g(\mathbf{x}_t)-\widetilde{s}(\mathbf{x}_t \, |\, \widetilde{\pmb{\Lambda}})\right) \right\|^2
+4\sum_{\mathbf{i}\in [M]^d} \left\|\sum_{t=1}^T \widetilde{x}_{\mathbf{i},tl}\varepsilon_t\right\|^2
\nonumber\\
&&+\text{interaction terms},
\end{eqnarray}
where  $\widetilde{\pmb{\lambda}}_{\mathbf{i}}^\ast=(a_0,\widetilde{\pmb{\lambda}}_{\mathbf{i}}^ \top)^\top$.

It suffices only to study the first three terms on the right-hand side of \eqref{dk1} to derive the convergence rate of $\mathbf{Q}_{l,1}$. Using \eqref{EqA.12}, \eqref{dk2} and Lemma \ref{Lem_lasso}, we can readily obtain that the first term has the probability order of  $O_P(TH_l^{-2})$. For the second term, by Lemma \ref{LEM2.3} and \eqref{EqA.12}, it has the probability order of  $O_P(T^2h^{d+2p}H_l^{-2})$. Additionally, we can use analogous arguments in \eqref{EqA.17} and show that the third term in \eqref{dk1} is bounded in probability by  $O_P(TH_l^{-2})$. In summary, we can establish the following result for $\mathbf{Q}_{l,1}$:
\begin{eqnarray}\label{dk3}
\|\mathbf{Q}_{l,1}\|=O_P(T^{1/2}H^{-1}_l).
\end{eqnarray}
For $\mathbf{Q}_{l,2}$, it is straightforward to see $\|\mathbf{Q}_{l,2}\|=\psi_l\mathtt{c}$. Under the condition  $\min_{d_q^0+1\leq l\leq d_q}\{\psi_lH_l\}T^{-\frac{1}{2}}\rightarrow \infty$, \eqref{dk3} yields that
\begin{equation*}
P(\|\mathbf{Q}_{l,2}\|>\|\mathbf{Q}_{l,1}\|)\rightarrow 1,
\end{equation*}
which leads to a contradictory result to that in \eqref{dk4}. Therefore, we must have $P\left(\|\widetilde{\pmb{\Theta}}_{D,l}\|= 0\right)\rightarrow 1$. It completes the proof of Lemma  \ref{LEM2.4}.1.

(2).  Together with Lemma \ref{LEM2.3} and the fact that $g(\mathbf{x})=g_c(\mathbf{x}_c)$, \eqref{dk15} yields
\begin{eqnarray}\label{dk8}
\|\widetilde{s}(\mathbf{x} \, |\, \widetilde{\pmb{\Lambda}}) -  \widetilde{s}_c(\mathbf{x}_c \, |\, \widetilde{\pmb{\Lambda}}_c)  \|_{\infty}=O(h^p) ,
\end{eqnarray}

Recall that $\widetilde{Q}  (\alpha,\pmb{\Theta})= \sum_{t=1}^T[y_t-z_t\alpha -\widetilde{s}(\mathbf{x}_t \, |\, \pmb{\Theta}) ]^2$.
Simple algebra further gives $ \widetilde{s}_{c}(\mathbf{x}_c \, |\, \pmb{\Theta}_c)=\sum_{\mathbf{i}\in [M]^{d_0}} \widetilde{\mathbf{x}}^\top_{c,\mathbf{i},t}\pmb{\theta}_{c,\mathbf{i}}$,
where $\widetilde{\mathbf{x}}_{c,\mathbf{i},t}  =I_{\mathbf{i},h} (\mathbf{x}_{c,t})(\mathbf{I}_{d^0_q}\otimes \pmb{\gamma}_c^\top)  \pmb{\sigma}_c( \mathbf{x}_{c,t}\, |\,\mathbf{x}_{c, 0\mathbf{i}})$.

For the group-LASSO estimators, we can formulate the following first-order conditions:
\begin{eqnarray}\label{dk7}
\sum_{t=1}^Tz_t\Bigl[y_t-z_t\widetilde{\alpha} -\sum_{\mathbf{i}\in [M]^{d}} \widetilde{\mathbf{x}}_{\mathbf{i}, t}^\top \widetilde{\pmb{\theta}}_{\mathbf{i}}\Bigl]&=&0,
\nonumber\\
-\sum_{t=1}^T\widetilde{\mathbf{x}}_{\mathbf{i}, t}\Bigl[y_t-z_t\widetilde{\alpha} - \sum_{\mathbf{j}\in [M]^{d}}\widetilde{\mathbf{x}}_{\mathbf{j}, t}^\top \widetilde{\pmb{\theta}}_{\mathbf{j}}\Bigr]+\pmb{\Phi}_c\widetilde{\pmb{\theta}}_{\mathbf{i}}&=&\mathbf{0},
\end{eqnarray}
where $\pmb{\Phi}_c=\mathbf{D}^{-1}\pmb{\Phi}_D\mathbf{D}^{\top,-1}$ with $\pmb{\Phi}_D$ being a $d_q\times d_q$ diagonal matrix with its $l$-th diagonal element being $\psi_l/(2\|\widetilde{\pmb{\Theta}}_{D,l}\|)$, for $l=1,\ldots,d_q$.

Recall that $\widetilde{\pmb{\theta}}_{\mathbf{i},\star}$ contains the first $d_q^0$ elements in $\mathbf{D}^{\top,-1}\widetilde{\pmb{\theta}}_{\mathbf{i}}$.
Let $\mathbf{D}_\star$ be the matrix that contains the first $d_q^0$ rows in $\mathbf{D}$ and $\widetilde{\mathbf{x}}_{\mathbf{i}, t,\star} = \mathbf{D}_\star\widetilde{\mathbf{x}}_{\mathbf{i}, t}$. Then, using the sparsity in $g(\mathbf{x})$, Lemma \ref{LEM2.4}.1, \eqref{dk2}, and \eqref{dk8}, we can rewrite  \eqref{dk7} as 
\begin{eqnarray}\label{dk9}
0&=&\frac{1}{T}\sum_{t=1}^Tz_t\Bigl[y_t-z_t\widetilde{\alpha} -\sum_{\mathbf{i}\in [M]^{d_0}} \widetilde{\mathbf{x}}_{\mathbf{i}, t,\star}^\top \widetilde{\pmb{\theta}}_{\mathbf{i},\star}\Bigl]+O_P(h^p),
\nonumber\\
\mathbf{0}&=&-\frac{1}{Th^{d_0}}\sum_{t=1}^T\mathbf{H}_c\widetilde{\mathbf{x}}_{\mathbf{i}, t,\star}\Bigl[y_t-z_t\widetilde{\alpha} - \widetilde{\mathbf{x}}_{\mathbf{i}, t,\star}^\top \widetilde{\pmb{\theta}}_{\mathbf{i},\star}\Bigr]+\frac{1}{Th^{d_0}}\mathbf{H}_c\pmb{\Phi}_{D,\star}\widetilde{\pmb{\theta}}_{\mathbf{i},\star}+O_P(h^p),
\end{eqnarray}
where $\pmb{\Phi}_{D,\star}$ is a $d_q^0\times d_q^0$ diagonal matrix that contains the first $d_q^0$ diagonal elements of $\pmb{\Phi}_D$ and $\mathbf{H}_c$, as defined at an early stage, consists of the first $d_q^0$ elements in $\mathbf{H}$. 

Solving  \eqref{dk9},  we can establish the following results for $(\widetilde{\alpha}, \widetilde{\pmb{\theta}}_{\mathbf{i},\star})$:

\begin{eqnarray}\label{dk11}
\widetilde{\alpha}&=&\left[\mathbf{Z}^\top \mathbf{M}_{x,\phi,\star} \mathbf{Z}\right]^{-1}\mathbf{Z}^\top \mathbf{M}_{x,\phi,\star} \mathbf{Y}+O_P(h^p),
\nonumber\\
\mathbf{H}_c^{-1}\widetilde{\pmb{\theta}}_{\mathbf{i},\star}&=&\mathbf{H}_c^{-1}\left[\widetilde{\mathbf{X}}_{\mathbf{i},\star}^\top \widetilde{\mathbf{X}}_{\mathbf{i},\star}+\pmb{\Phi}_{D,\star}\right]^{-1}\widetilde{\mathbf{X}}_{\mathbf{i},\star}^\top\left[\mathbf{Y}-\mathbf{Z}\widetilde{\alpha}\right]+O_P(h^p),
\end{eqnarray}
where
$\mathbf{M}_{x,\phi,\star}=\mathbf{I}_{T}-\sum_{\mathbf{i}\in [M]^{d_0}} \widetilde{\mathbf{X}}_{\mathbf{i},\star}\left[\widetilde{\mathbf{X}}_{\mathbf{i},\star}^\top \widetilde{\mathbf{X}}_{\mathbf{i},\star}+\pmb{\Phi}_{D,\star}\right]^{-1}\widetilde{\mathbf{X}}_{\mathbf{i},\star}^\top$ with $\widetilde{\mathbf{X}}_{\mathbf{i},\star} = (\widetilde{\mathbf{x}}_{\mathbf{i}, 1,\star}, \cdots, \widetilde{\mathbf{x}}_{\mathbf{i}, T,\star})^\top$.

In light of \eqref{dk12} and \eqref{dk11}, it suffices only to study the asymptotic behaviour of $\mathbf{M}_{x,\phi,\star}$ before we can establish  Lemma  \ref{LEM2.4}.2. Define $\mathbf{M}_{c,x,\phi}=\mathbf{I}_{T}-\sum_{\mathbf{i}\in [M]^{d_0}} \widetilde{\mathbf{X}}_{c,\mathbf{i}}\left[\widetilde{\mathbf{X}}_{c,\mathbf{i}}^\top \widetilde{\mathbf{X}}_{c,\mathbf{i}}+\pmb{\Phi}_{D,\star}\right]^{-1}\widetilde{\mathbf{X}}_{c,\mathbf{i}}^\top$.

We write
\begin{eqnarray*}
\mathbf{M}_{c,x,\phi}-\mathbf{M}_{c,x}&=&\sum_{\mathbf{i}\in [M]^{d_0}} \widetilde{\mathbf{X}}_{c,\mathbf{i}}\left[\widetilde{\mathbf{X}}_{c,\mathbf{i}}^\top \widetilde{\mathbf{X}}_{c,\mathbf{i}}\right]^{-1}\widetilde{\mathbf{X}}_{c,\mathbf{i}}^\top-\sum_{\mathbf{i}\in [M]^{d_0}} \widetilde{\mathbf{X}}_{c,\mathbf{i}}\left[\widetilde{\mathbf{X}}_{c,\mathbf{i}}^\top \widetilde{\mathbf{X}}_{c,\mathbf{i}}+\pmb{\Phi}_{D,\star}\right]^{-1}\widetilde{\mathbf{X}}_{c,\mathbf{i}}^\top
\nonumber\\
&=&\sum_{\mathbf{i}\in [M]^{d_0}} \widetilde{\mathbf{X}}_{c,\mathbf{i}}\left[\widetilde{\mathbf{X}}_{c,\mathbf{i}}^\top \widetilde{\mathbf{X}}_{c,\mathbf{i}}\right]^{-1}\pmb{\Phi}_{D,\star} \left[\widetilde{\mathbf{X}}_{c,\mathbf{i}}^\top \widetilde{\mathbf{X}}_{c,\mathbf{i}}+\pmb{\Phi}_{D,\star}\right]^{-1}\widetilde{\mathbf{X}}_{c,\mathbf{i}}^\top.
\end{eqnarray*}

Together with \eqref{dk50},  it yields that
\begin{eqnarray}\label{dkd}
\frac{1}{T}\mathbf{Z}^\top\left(\mathbf{M}_{c, x,\phi}-\mathbf{M}_{c,x}\right)\mathbf{Z}&=& \frac{1}{T}\sum_{\mathbf{i}\in [M]^{d_0}} \mathbf{Z}^\top\widetilde{\mathbf{X}}_{c,\mathbf{i}}\bigl[\widetilde{\mathbf{X}}_{c,\mathbf{i}}^\top \widetilde{\mathbf{X}}_{c,\mathbf{i}}\bigr]^{-1}\pmb{\Phi}_{D,\star} \bigl[\widetilde{\mathbf{X}}_{c,\mathbf{i}}^\top \widetilde{\mathbf{X}}_{c,\mathbf{i}}+\pmb{\Phi}_{D,\star}\bigr]^{-1}\widetilde{\mathbf{X}}_{c,\mathbf{i}}^\top\mathbf{Z}
\nonumber\\
&=&h^{d_0}\sum_{\mathbf{i}\in [M]^{d_0}}\left[\frac{\mathbf{Z}^\top\widetilde{\mathbf{X}}_{c,\mathbf{i}}}{Th^{d_0}}\right]
\left[\frac{\widetilde{\mathbf{X}}_{c,\mathbf{i}}^\top \widetilde{\mathbf{X}}_{c,\mathbf{i}}}{Th^{d_0}}\right]^{-1}\left[\frac{\pmb{\Phi}_{D,\star}}{Th^{d_0}}\right] \left[\frac{\widetilde{\mathbf{X}}_{c,\mathbf{i}}^\top \widetilde{\mathbf{X}}_{c,\mathbf{i}}+\pmb{\Phi}_{D,\star}}{Th^{d_0}}\right]^{-1}\left[\frac{\widetilde{\mathbf{X}}_{c,\mathbf{i}}^\top\mathbf{Z}}{Th^{d_0}}\right]
\nonumber\\
&=&O_P\left(\frac{1}{Th^{d_0}}\|\mathbf{H}_c\mathbf{D}_c\pmb{\Phi}_{D,\star}\mathbf{D}_c^\top\mathbf{H}_c\|\right)
\nonumber\\
&=&o_P\left(\frac{1}{\sqrt{T}}\right),
\end{eqnarray}
where the last equality is ensured by the fact  that $\{\psi_lH_l\}T^{-\frac{1}{2}}\rightarrow0$ and $h^{d_0}H^{-1}_l\|\widetilde{\pmb{\Theta}}_l\|\geq c$ hold uniformly for $l\in[d^0_q]$ and a positive constant $c$ under Lemma \ref{Lem_lasso} and the conditions in Assumption \ref{Ass.4}. 

In light of the definition of $\widetilde{\mathbf{x}}_{\mathbf{i}, t,\star} $ and $\widetilde{\mathbf{x}}_{c,\mathbf{i}, t}$, analogously to \eqref{EqA.6} and \eqref{EqA.7}, we can obtain
\begin{eqnarray}\label{dke}
\max_t\left\| \mathbf{D}_c\widetilde{\mathbf{x}}_{c,\mathbf{i}, t} -\widetilde{\mathbf{x}}_{\mathbf{i}, t,\star} \right\| =O_P(h^{q+1}).
\end{eqnarray}
Therefore, it is clear to see that $\frac{1}{T}\mathbf{Z}^\top\left(\mathbf{M}_{x,\phi,\star}-\mathbf{M}_{c,x,\phi}\right)\mathbf{Z}=O_P(h^{q+1})$. Together with \eqref{dkd}, it implies that $\frac{1}{T}\mathbf{Z}^\top\left(\mathbf{M}_{x,\phi,\star}-\mathbf{M}_{c,x}\right)\mathbf{Z}$ has the probability order of $o_P\left(\frac{1}{\sqrt{T}}+h^p\right)$.
Analogously, we can show that $\frac{1}{T}\mathbf{Z}^\top\left(\mathbf{M}_{x,\phi,\star}-\mathbf{M}_{c,x}\right)\mathbf{Y}$ is also negligible. Therefore, we have
\begin{equation}\label{dk53}
\widetilde{\alpha}-\widetilde{\alpha}_{c}=o_P\left(\frac{1}{\sqrt{T}}\right)+O_P(h^p).
\end{equation} 
Therefore, the proof of Lemma \ref{LEM2.4}.2 is complete. \hspace*{\fill}{$\blacksquare$}

\bigskip

\noindent \textbf{Proof of Theorem \ref{THM2.1}:}

(1). Directly applying Lemma \ref{LEM2.4} and Lemma \ref{Lem_Ora}, we can establish the desired result in Theorem \ref{THM2.1}.1.

(2).  Similarly to \eqref{EqA.6} and \eqref{EqA.7}, we have
\begin{eqnarray}\label{dka}
\left\| \mathbf{D}_c^{-1}I_{\mathbf{i},h}(\mathbf{x}_{c,0})\cdot \mathbf{m}_c(\mathbf{x}_{c,0}\, |\, \mathbf{x}_{c,0\mathbf{i}})-\widetilde{\mathbf{x}}_{c,\mathbf{i}, 0} \right\| =O_P(h^{q+1}),
\end{eqnarray}
where $\widetilde{\mathbf{x}}_{c,\mathbf{i},0}  =I_{\mathbf{i},h} (\mathbf{x}_{c,0})(\mathbf{I}_{d^0_q}\otimes \pmb{\gamma}_c^\top)  \pmb{\sigma}_c( \mathbf{x}_{c,0}\, |\,\mathbf{x}_{c,0\mathbf{i}})$.

Together with  Lemma \ref{LEM2.4}.1, \eqref{dk15}, \eqref{dkd}, and \eqref{dke}, it gives
\begin{eqnarray}\label{dkb}
\widetilde{g}(\mathbf{x}_{0})-g_c(\mathbf{x}_{c,0})&=&\sum_{\mathbf{i}\in [M]^{d_0}} \widetilde{\mathbf{x}}_{c, 0,\mathbf{i}}^\top\mathbf{D}_c^{\top} \mathbf{H}_c\bigl[\mathbf{H}_c^{-1}(\widetilde{\pmb{\theta}}_{\mathbf{i},\star}-\mathbf{D}_c^{\top,-1}\widetilde{\pmb{\lambda}}_{c,\mathbf{i}})\bigr]+O_P(h^p),
\end{eqnarray}

Combining \eqref{dka} and \eqref{dkb}, we obtain
\begin{eqnarray*}
\widetilde{g}(\mathbf{x}_{0})-g_c(\mathbf{x}_{c,0})
&=&\sum_{\mathbf{i}\in [M]^{d_0}} I_{\mathbf{i},h}(\mathbf{x}_{c,0}) \mathbf{m}_c(\mathbf{x}_{c,0}\, |\, \mathbf{x}_{c,0\mathbf{i}})^\top \mathbf{H}_c\bigl[\mathbf{H}_c^{-1} (\widetilde{\pmb{\theta}}_{\mathbf{i},\star}-\mathbf{D}_c^{\top, -1}\widetilde{\pmb{\lambda}}_{c,\mathbf{i}})\bigr]+O_P(h^p)+o_P\left(\frac{1}{\sqrt{Th^{d_0}}}\right).
\end{eqnarray*}

We now study $\widetilde{\pmb{\theta}}_{\mathbf{i},\star}$. With \eqref{dk50}, similarly to \eqref{dkd},  we can further write
\begin{eqnarray}\label{dk54}
&&\frac{1}{Th^{d_0}}\left\|\mathbf{H}_c^{-1}\mathbf{D}_c^{\top,-1}\left[\bigl(\widetilde{\mathbf{X}}_{c,\mathbf{i}}^\top \widetilde{\mathbf{X}}_{c,\mathbf{i}}\bigr)^{-1}-\bigl(\widetilde{\mathbf{X}}_{c,\mathbf{i}}^\top \widetilde{\mathbf{X}}_{c,\mathbf{i}}+\pmb{\Phi}_{D,\star}\bigr)^{-1}\right]\mathbf{D}_c^{-1}\mathbf{H}_c^{-1}\right\|
\nonumber\\
&=&\frac{1}{Th^{d_0}}\left\|\mathbf{H}_c^{-1}\mathbf{D}_c^{\top,-1}\bigl[\widetilde{\mathbf{X}}_{c,\mathbf{i}}^\top \widetilde{\mathbf{X}}_{c,\mathbf{i}}\bigr]^{-1}
\pmb{\Phi}_{D,\star}\bigl[\widetilde{\mathbf{X}}_{c,\mathbf{i}}^\top \widetilde{\mathbf{X}}_{c,\mathbf{i}}+\pmb{\Phi}_{D,\star}\bigr]^{-1}\mathbf{D}_c^{-1}\mathbf{H}_c^{-1}\right\|
\nonumber\\
&=&O_P\left(\frac{1}{Th^{d_0}}\|\mathbf{H}_c\mathbf{D}_c\pmb{\Phi}_{D,\star}\mathbf{D}_c^\top\mathbf{H}_c\|\right)
\nonumber\\
&=&o_P\left(\frac{1}{\sqrt{T}}\right).
\end{eqnarray}
Drawing upon \eqref{dke} and the expressions for $\widetilde{\pmb{\theta}}_{c,\mathbf{i}}$ and $\widetilde{\pmb{\theta}}_{\mathbf{i},\star}$ that we have derived in \eqref{dk12} and \eqref{dk11}, respectively,  
\begin{eqnarray}\label{dk55}
\mathbf{H}_c^{-1}\bigl(\widetilde{\pmb{\theta}}_{\mathbf{i},\star}-\mathbf{D}_c^{\top,-1}\widetilde{\pmb{\theta}}_{c,\mathbf{i}}\bigr)&=&\mathbf{H}_c^{-1}\mathbf{D}_c^{\top,-1}\left[\bigl(\widetilde{\mathbf{X}}_{c,\mathbf{i}}^\top \widetilde{\mathbf{X}}_{c,\mathbf{i}}+\pmb{\Phi}_{D,\star}\bigr)^{-1}-\bigl(\widetilde{\mathbf{X}}_{c,\mathbf{i}}^\top \widetilde{\mathbf{X}}_{c,\mathbf{i}}\bigr)^{-1}\right]\widetilde{\mathbf{X}}_{c,\mathbf{i}}^\top\left[\mathbf{Y}-\mathbf{Z}\widetilde{\alpha}_{c}\right]
\nonumber\\
&&
+\mathbf{H}_c^{-1}\mathbf{D}_c^{\top,-1}\left[\bigl(\widetilde{\mathbf{X}}_{c,\mathbf{i}}^\top \widetilde{\mathbf{X}}_{c,\mathbf{i}}+\pmb{\Phi}_{D,\star}\bigr)^{-1}-\bigl(\widetilde{\mathbf{X}}_{c,\mathbf{i}}^\top \widetilde{\mathbf{X}}_{c,\mathbf{i}}\bigr)^{-1}\right]\widetilde{\mathbf{X}}_{c,\mathbf{i}}^\top\mathbf{Z}\bigl[\widetilde{\alpha}_{c}-\widetilde{\alpha}\bigr]
\nonumber\\
&&+\mathbf{H}_c^{-1}\mathbf{D}_c^{\top,-1}\bigl[\widetilde{\mathbf{X}}_{c,\mathbf{i}}^\top \widetilde{\mathbf{X}}_{c,\mathbf{i}}\bigr]^{-1}\widetilde{\mathbf{X}}_{c,\mathbf{i}}^\top\mathbf{Z}\bigl[\widetilde{\alpha}_{c}-\widetilde{\alpha}\bigr]
+O_P(h^p).
\end{eqnarray}
Using the results that are established in \eqref{dk53} and \eqref{dk54}, we can readily obtain the orders of the first three terms on the right-hand side of \eqref{dk55} as $o_P\left(\frac{1}{\sqrt{T}}\right)$, $o_P\left(\frac{1}{T}\right)$ and $o_P\left(\frac{1}{\sqrt{T}}\right)$. 
Therefore, we have
\begin{equation}\label{Lem_Ora_dif}
\mathbf{H}_c^{-1} (\widetilde{\pmb{\theta}}_{\mathbf{i},\star} - \mathbf{D}_c^{\top,-1}\widetilde{\pmb{\theta}}_{c,\mathbf{i}})=O_P(h^p)+o_P\left(\frac{1}{\sqrt{Th^{d_0}}}\right).
\end{equation}

Using the central limit theorem for $\widetilde{\pmb{\theta}}_{c,\mathbf{i}}$ in Lemma \ref{Lem_Ora}, we can establish the asymptotic normality of $\sqrt{Th^{d_0}}(\widetilde{g}(\mathbf{x}_{0})-g_c(\mathbf{x}_{c,0}))$, which has the asymptotic covariance as the limit of 
\begin{eqnarray*}
\widetilde{\sigma}_{\mathbf{x}_{c,0}}^2&=&\sigma_\varepsilon^2\sum_{\mathbf{i}\in [M]^{d_0}} I_{\mathbf{i},h}(\mathbf{x}_{c,0}) \mathbf{m}_c(\mathbf{x}_{c,0}\, |\, \mathbf{x}_{c,0\mathbf{i}})^\top\mathbf{H}_c \pmb{\Sigma}_{c,\mathbf{i}}^{-1}\mathbf{H}_c\mathbf{m}_c(\mathbf{x}_{c,0}\, |\, \mathbf{x}_{c,0\mathbf{i}}),
\end{eqnarray*}
where $\pmb{\Sigma}_{c,\mathbf{i}}$ is defined in Lemma \ref{Lem_Ora}. 
Then, the proof of Theorem \ref{THM2.1}.2 is complete. \hspace*{\fill}{$\blacksquare$}

\bigskip

\noindent \textbf{Proof of Theorem \ref{Thm_Bootstrap}:}

(1) Using arguments that are analogous to those in the proof of Lemma \ref{LEM2.4}, we can show that the bootstrap Group-LASSO estimators can approximate the bootstrap oracle estimators up to some asymptotically negligible bias terms. Therefore, it suffices only to study the asymptotic behaviour of the  bootstrap oracle estimators. 
Define the bootstrap oracle estimators $ (\widetilde{\alpha}^\ast_c,\widetilde{\pmb{\Theta}}^\ast_c)$ as follows:
\begin{equation*}
(\widetilde{\alpha}^\ast_c,\widetilde{\pmb{\Theta}}^\ast_c)=\argmin_{\alpha\in \mathbb{R}, \widetilde{s}_c\in \mathcal{S}_{c}}  \widetilde{Q}^\ast_c  (\alpha,\pmb{\Theta}_c).
\end{equation*}
where $\widetilde{Q}^\ast_c  (\alpha,\pmb{\Theta}_c)= \sum_{t=1}^T[y^\ast_t-z_t\alpha -\widetilde{s}_c(\mathbf{x}_{c,t} \, |\, \pmb{\Theta}_c) ]^2$. 

By solving the first-order conditions, we obtain the following expressions for $\widetilde{\pmb{\theta}}^\ast_{c,\mathbf{i}}$ and $\widetilde{\alpha}^\ast_{c}$:  
\begin{eqnarray}\label{dk12B}
\widetilde{\pmb{\theta}}^\ast_{c,\mathbf{i}}&=&\left(\widetilde{\mathbf{X}}_{c,\mathbf{i}}^\top \widetilde{\mathbf{X}}_{c,\mathbf{i}}\right)^{-1}\widetilde{\mathbf{X}}_{c,\mathbf{i}}^\top[\mathbf{Y}^\ast-\mathbf{Z}\widetilde{\alpha}^\ast_{c}],
\nonumber\\
\widetilde{\alpha}^\ast_{c}&=&\left(\mathbf{Z}^\top \mathbf{M}_{c,x} \mathbf{Z}\right)^{-1}\mathbf{Z}^\top \mathbf{M}_{c,x} \mathbf{Y}^\ast,
\end{eqnarray}
where  $\mathbf{Z}=(z_1,\cdots,z_T)^\top$, $\mathbf{Y}^\ast=(y^\ast_1,\cdots,y^\ast_T)^\top$, and $\mathbf{M}_{c,x}=\mathbf{I}_{T}-\sum_{\mathbf{i}\in [M]^{d_0}} \widetilde{\mathbf{X}}_{c,\mathbf{i}}\bigl[\widetilde{\mathbf{X}}_{c,\mathbf{i}}^\top \widetilde{\mathbf{X}}_{c,\mathbf{i}}\bigr]^{-1}\widetilde{\mathbf{X}}_{c,\mathbf{i}}^\top$ with $\widetilde{\mathbf{X}}_{c,\mathbf{i}} = (\widetilde{\mathbf{x}}_{c,\mathbf{i}, 1}, \cdots, \widetilde{\mathbf{x}}_{c,\mathbf{i}, T})^\top$.

Drawing upon the DGP of $y_t^\ast$ and the second expression  in \eqref{dk12B}, we can further expand $\widetilde{\alpha}^\ast_{c}$ as follows: 
\begin{eqnarray}\label{boot1}
\widetilde{\alpha}^\ast_{c}-\widetilde{\alpha}&=&\left(\mathbf{Z}^\top \mathbf{M}_{c,x} \mathbf{Z}\right)^{-1}\mathbf{Z}^\top \mathbf{M}_{c,x}\sum_{\mathbf{i}\in [M]^{d}} \widetilde{\mathbf{X}}_{\mathbf{i}}\widetilde{\pmb{\theta}}_{\mathbf{i}}+\left(\mathbf{Z}^\top \mathbf{M}_{c,x} \mathbf{Z}\right)^{-1}\mathbf{Z}^\top \mathbf{M}_{c,x}\pmb{\varepsilon}^\ast,
\end{eqnarray}
where $\widetilde{\mathbf{X}}_{\mathbf{i}} = (\widetilde{\mathbf{x}}_{\mathbf{i}, 1}, \cdots, \widetilde{\mathbf{x}}_{\mathbf{i}, T})^\top$ and $\pmb{\varepsilon}^\ast=(\varepsilon_1^\ast,\cdots,\varepsilon_T^\ast)^\top$. 

 For the first term, it is clear to see that
\begin{eqnarray*}
\mathbf{Z}^\top \mathbf{M}_{c,x} \sum_{\mathbf{i}\in [M]^{d}} \widetilde{\mathbf{X}}_{\mathbf{i}}\widetilde{\pmb{\theta}}_{\mathbf{i}}&=&\mathbf{Z}^\top \mathbf{M}_{c,x}\Bigl( \sum_{\mathbf{i}\in [M]^{d}} \widetilde{\mathbf{X}}_{\mathbf{i}}\widetilde{\pmb{\theta}}_{\mathbf{i}}-\sum_{\mathbf{i}\in [M]^{d_0}} \widetilde{\mathbf{X}}_{c,\mathbf{i}}\widetilde{\pmb{\theta}}_{\mathbf{i},\star}\Bigr).
\end{eqnarray*}
Together with Lemma \ref{LEM2.4}.1 and \eqref{dk35}, it immediately yields 
that the first term in \eqref{boot1} is asymptotically negligible. 

Recall that $\widetilde{z}_t=z_t-\sum_{\mathbf{i}\in [M]^{d_0}} I_{\mathbf{i},h} (\mathbf{x}_{c,t}) \mathbf{M}_{c,\mathbf{i}}^\top \pmb{\Sigma}_{c,\mathbf{i}}^{-1}\mathbf{H}_c\mathbf{m}_c(\mathbf{x}_{c,t}\, |\, \mathbf{x}_{c,0\mathbf{i}})$. For notational simplicity,  we further define $\widetilde{z}_t^\ast = z_t-\sum_{\mathbf{i}\in [M]^{d_0}} \mathbf{Z}^\top\widetilde{\mathbf{X}}_{c,\mathbf{i}}\bigl[\widetilde{\mathbf{X}}_{c,\mathbf{i}}^\top \widetilde{\mathbf{X}}_{c,\mathbf{i}}\bigr]^{-1}\widetilde{\mathbf{x}}_{c,\mathbf{i},t}$.
For the second term  on the right-hand side of \eqref{boot1},  we write
\begin{eqnarray}\label{dk155}
\frac{1}{\sqrt{T}}\mathbf{Z}^\top \mathbf{M}_{c,x}\pmb{\varepsilon}^\ast
&=&\frac{1}{\sqrt{T}}\sum_{t=1}^T\widetilde{z}^\ast_t\widehat{\varepsilon}_t\varsigma_t
\nonumber\\
&=&\frac{1}{\sqrt{T}}\sum_{t=1}^T(\widetilde{z}^\ast_t-\widetilde{z}_t)\varepsilon_t\varsigma_t+\frac{1}{\sqrt{T}}\sum_{t=1}^T\widetilde{z}_t(\widehat{\varepsilon}_t-\varepsilon_t)\varsigma_t
+\frac{1}{\sqrt{T}}\sum_{t=1}^T(\widetilde{z}^\ast_t-\widetilde{z}_t)(\widehat{\varepsilon}_t-\varepsilon_t)\varsigma_t
\nonumber\\
&&+\frac{1}{\sqrt{T}}\sum_{t=1}^T\widetilde{z}_t\varepsilon_t\varsigma_t
\nonumber\\
&\coloneqq&\mathcal{J}_1+\cdots+\mathcal{J}_4.
\end{eqnarray}
Using arguments that are analogous to those in the proofs of \eqref{dk102} and \eqref{dk103}, we can show that $\mathcal{J}_1=o_P(1)$. For $\mathcal{J}_2$, write
\begin{eqnarray}
\mathcal{J}_2&=&-\frac{1}{\sqrt{T}}\sum_{t=1}^T\widetilde{z}_tz_t(\widetilde{\alpha}-\alpha) \varsigma_t-\frac{1}{\sqrt{T}}\sum_{t=1}^T\widetilde{z}_t(\widetilde{g}(\mathbf{x}_t)-g_c(\mathbf{x}_{c,t})) \varsigma_t
\nonumber\\
&\coloneqq&\mathcal{J}_{2,1}+\mathcal{J}_{2,2}.
\end{eqnarray}
For $\mathcal{J}_{2,1}$, it is straightforward to see that $\frac{1}{\sqrt{T}}\sum_{t=1}^TE[\widetilde{z}_tz_t\varsigma_t]=0$ and 
\begin{eqnarray*}
E\left[\Bigl(\frac{1}{\sqrt{T}}\sum_{t=1}^T\widetilde{z}_tz_t\varsigma_t\Bigr)^2\right]&=& \frac{1}{T}\sum_{t,s=1}^TE[\widetilde{z}_t\widetilde{z}_sz_tz_s]E[\varsigma_t\varsigma_s]
\nonumber\\
&=&\frac{1}{T}\sum_{t=1}^TE[\widetilde{z}_t^2z_t^2]E[\varsigma_t^2]
\nonumber\\
&&+ \frac{1}{T}\sum_{t=1}^T\sum_{s=t+1}^TE[\widetilde{z}_t\widetilde{z}_sz_tz_s]E[\varsigma_t\varsigma_s]
\nonumber\\
&&+ \frac{1}{T}\sum_{t=1}^T\sum_{s=1}^{t-1}E[\widetilde{z}_t\widetilde{z}_sz_tz_s]E[\varsigma_t\varsigma_s].
\end{eqnarray*}
It is obvious that the first term has the order $O(1)$ and the second and third terms have the same order. Therefore, it suffices only to study the second term. We have
\begin{eqnarray*}
\frac{1}{T}\sum_{t=1}^T\sum_{s=t+1}^T\bigl|E[\widetilde{z}_t\widetilde{z}_sz_tz_s]\bigr|\bigl|E[\varsigma_t\varsigma_s]\bigr|
&=&\frac{1}{T}\sum_{t=1}^T\sum_{s=t+1}^T K\Bigl(\frac{t-s}{\ell}\Bigr)\bigl|E[\widetilde{z}_t\widetilde{z}_sz_tz_s]\bigr|
\nonumber\\
&=&\sum_{t=1}^{T-1}\Bigl(1-\frac{t}{T}\Bigr)K\Bigl(\frac{t}{\ell}\Bigr)\bigl|E[\widetilde{z}_1\widetilde{z}_{1+t}z_1z_{1+t}]\bigr|
\nonumber\\
&=& O(1)\sum_{t=1}^{\ell}\Bigl(1-\frac{t}{T}\Bigr)K\Bigl(\frac{t}{\ell}\Bigr)
\nonumber\\
&=& O(\ell).
\end{eqnarray*}

In summary of these results, we have
\begin{eqnarray}\label{dk105}
E\left[\Bigl(\frac{1}{\sqrt{T}}\sum_{t=1}^T\widetilde{z}_tz_t\varsigma_t\Bigr)^2\right]=O(\ell).
\end{eqnarray}
with implies that $\frac{1}{\sqrt{T}}\sum_{t=1}^T\widetilde{z}_tz_t\varsigma_t=O_P(\sqrt{\ell})$. Together with Theorem \ref{THM2.1}.1, it yields that
\begin{eqnarray}\label{dk106}
\mathcal{J}_{2,1}&=&O_P(\sqrt{\ell}(\widetilde{\alpha}-\alpha))
\nonumber\\
&=&O_P\left(\sqrt{\frac{\ell}{T}}\right)+O_P(\sqrt{\ell}h^p)
\nonumber\\
&=&o_P(1).
\end{eqnarray}

We now proceed with the derivations of $\mathcal{J}_{2,2}$.
By  Lemma \ref{LEM2.4}.1 and \eqref{dk15}, we obtain
\begin{eqnarray*}
\widetilde{g}(\mathbf{x}_{t})-g_c(\mathbf{x}_{c,t})&=&\sum_{\mathbf{i}\in [M]^{d_0}} \widetilde{\mathbf{x}}_{c, \mathbf{i}, t}^\top\mathbf{D}_c^{\top} \mathbf{H}_c \bigl[\mathbf{H}_c^{-1}(\widetilde{\pmb{\theta}}_{\mathbf{i},\star}-\mathbf{D}_c^{\top,-1}\widetilde{\pmb{\lambda}}_{c,\mathbf{i}})\bigr]+O_P(h^p).
\end{eqnarray*}
Similarly to \eqref{dk105}, we can show that 
\begin{equation*}
E\Bigl[\Bigl\|\sum_{t=1}^T\widetilde{z}_t\varsigma_t \widetilde{\mathbf{x}}_{c, \mathbf{i},t}^\top\mathbf{D}_c^{\top} \mathbf{H}_c\Bigr\|^2\Bigr]=O(Th^{d_0}).
\end{equation*}
Additionally, directly applying Lemma \ref{Lem_Ora}.2,  \eqref{Lem_Ora_dif}, 
and Cauchy-Schwarz inequality yields
\begin{eqnarray*}
\frac{1}{\sqrt{T}}\Bigl\|\sum_{t=1}^T\widetilde{z}_t\varsigma_t\sum_{\mathbf{i}\in [M]^{d_0}} \widetilde{\mathbf{x}}_{c, \mathbf{i}, t}^\top \mathbf{H}_c \bigl[\mathbf{H}_c^{-1}(\widetilde{\pmb{\theta}}_{\mathbf{i},\star}-\mathbf{D}_c^{\top,-1}\widetilde{\pmb{\lambda}}_{c,\mathbf{i}})\bigr] \Bigr\|&\leq&  \frac{1}{\sqrt{T}} \left(\sum_{\mathbf{i}\in [M]^{d_0}} \Bigl\|
\sum_{t=1}^T\widetilde{z}_t\varsigma_t \widetilde{\mathbf{x}}_{c, \mathbf{i},t}^\top\mathbf{D}_c^{\top} \mathbf{H}_c
\Bigr\|^2\right)^{\frac{1}{2}}
\nonumber\\
&&\times\left(\sum_{\mathbf{i}\in [M]^{d_0}} \Bigl\| \mathbf{H}_c^{-1}(\widetilde{\pmb{\theta}}_{\mathbf{i},\star}-\mathbf{D}_c^{\top,-1}\widetilde{\pmb{\lambda}}_{c,\mathbf{i}})\Bigr\|^2\right)^{\frac{1}{2}}
\nonumber\\
&=&O_P\Bigl(\frac{1}{\sqrt{Th^{d_0}}} \Bigr).
\end{eqnarray*}
Therefore, we have
\begin{equation}\label{dk107}
\mathcal{J}_{2,2}=O_P\left(\sqrt{\frac{\ell}{Th^{d_0}}} \right)+O_P(\sqrt{\ell}h^p).
\end{equation}
Combining \eqref{dk106} and \eqref{dk107}, we obtain
\begin{eqnarray*}
\mathcal{J}_{2}=o_P(1).
\end{eqnarray*}
Analogously, we can show that $\mathcal{J}_{3}$ is also asymptotically negligible.

In what follows, we proceed to explore $\mathcal{J}_{4}$ which generates the bootstrap distribution. Let $E^\ast[\cdot]$ and $\text{Var}^\ast(\cdot)$ denote the expectation  and variance conditional on the observed sample. We first show that $\text{Var}^\ast( \mathcal{J}_{4})=\sigma_{c,z}^2+o_P(1)$.

 It is clear to see that $ E^\ast\bigl[\mathcal{J}_{4}\bigr]=0$. Moreover, we write
\begin{eqnarray*}
\text{Var}^\ast( \mathcal{J}_{4})&=&\frac{1}{T}\sum_{t,s=1}^T \widetilde{z}_t\widetilde{z}_s \varepsilon_t\varepsilon_s E^\ast[ \varsigma_t\varsigma_s ]
\nonumber\\
&=&\frac{1}{T}\sum_{t,s=1}^T K\Bigl(\frac{t-s}{\ell}\Bigr)\widetilde{z}_t\widetilde{z}_s  \varepsilon_t\varepsilon_s
\nonumber\\
&=& 
\frac{1}{T}\sum_{t=1}^T \widetilde{z}_t^2 \varepsilon_t^2+\frac{1}{T}\sum_{t=1}^T\sum_{s=t+1}^T K\Bigl(\frac{t-s}{\ell}\Bigr)\widetilde{z}_t\widetilde{z}_s  \varepsilon_t\varepsilon_s
+\frac{1}{T}\sum_{t=1}^T\sum_{s=1}^{t-1} K\Bigl(\frac{t-s}{\ell}\Bigr)\widetilde{z}_t\widetilde{z}_s  \varepsilon_t\varepsilon_s
\nonumber\\
&\coloneqq& \mathcal{Q}_1+\mathcal{Q}_2+\mathcal{Q}_3.
\end{eqnarray*}
Using a decomposition that is similar to \eqref{revn1}, we can easily show that  $\mathcal{Q}_1=E[\widetilde{z}^2_1  \varepsilon^2_1]+o_P(1)$. Let $s_T$ satisfy that $s_T\rightarrow\infty$ and  $s_T^2/\ell\rightarrow0$. For $\mathcal{Q}_2$, 
we write 
\begin{eqnarray}\label{dk111}
E[\mathcal{Q}_2]-\frac{1}{T}\sum_{t=1}^T\sum_{s=t+1}^T E[\widetilde{z}_t\widetilde{z}_s  \varepsilon_t\varepsilon_s]&=&\frac{1}{T}\sum_{t=1}^T\sum_{s=t+1}^T\Bigl[ K\Bigl(\frac{t-s}{\ell}\Bigr)-1\Bigr]E[\widetilde{z}_t\widetilde{z}_s  \varepsilon_t\varepsilon_s]
\nonumber\\
&=&\sum_{t=1}^{T-1}\Bigl(1-\frac{t}{T}\Bigr)\Bigl[ K\Bigl(\frac{t}{\ell}\Bigr)-1\Bigr]E[\widetilde{z}_1\widetilde{z}_{1+t}  \varepsilon_1\varepsilon_{1+t}]
\nonumber\\
&=&\sum_{t=1}^{s_T}\Bigl(1-\frac{t}{T}\Bigr)\Bigl[K\Bigl(\frac{t}{\ell}\Bigr)-1\Bigr]E[\widetilde{z}_1\widetilde{z}_{1+t}  \varepsilon_1\varepsilon_{1+t}]
\nonumber\\
&&+\sum_{t=s_T+1}^{T-1}\Bigl(1-\frac{t}{T}\Bigr)\Bigl[K\Bigl(\frac{t}{\ell}\Bigr)-1\Bigr]E[\widetilde{z}_1\widetilde{z}_{1+t}  \varepsilon_1\varepsilon_{1+t}]
\nonumber\\
&\coloneqq&\mathcal{Q}_{2,1}+\mathcal{Q}_{2,2}.
\end{eqnarray}
For $\mathcal{Q}_{2,1}$,  by Davydov's inequality for $\alpha$-mixing processes, 
\begin{eqnarray}\label{dk110}
|E[\widetilde{z}_1\widetilde{z}_{1+t}  \varepsilon_1\varepsilon_{1+t}]|&=&\bigl|\text{cov}(\widetilde{z}_1\varepsilon_1,\widetilde{z}_{1+t}\varepsilon_{1+t})\bigr|
\nonumber\\
&\leq&O(1) \alpha(t)^{\nu/(2+\nu)}E[|\widetilde{z}_1|^{2+\nu}]^{\frac{2}{2+\nu}} E[|\varepsilon_1|^{2+\nu}]^{\frac{2}{2+\nu}}
\nonumber\\
&=&O(1) \alpha(t)^{\nu/(2+\nu)}
.
\end{eqnarray}
Together with Lipschitz continuity of the kernel function, it yields that
\begin{eqnarray}\label{dk112}
|\mathcal{Q}_{2,1}|&\leq &O(1)\sum_{t=1}^{s_T}\frac{t}{\ell} \alpha(t)^{\nu/(2+\nu)}
\nonumber\\
&=&O\Bigl(\frac{s_T^2}{\ell}\Bigr)=o(1).
\end{eqnarray}
For $\mathcal{Q}_{2,2}$, by \eqref{dk110}, we have
\begin{eqnarray}\label{dk113}
|\mathcal{Q}_{2,2}|&\leq& O(1)\sum_{t=s_T+1}^{T-1}|\text{cov}(\widetilde{z}_1\varepsilon_1, \widetilde{z}_{1+t}  \varepsilon_{1+t})|
\nonumber\\
&=&O(1)\sum_{t=s_T+1}^{T-1} \alpha(t)^{\nu/(2+\nu)}
=o(1).
\end{eqnarray}
The second equality  holds by the fact that $\sum_{t=1}^{T-1} \alpha(t)^{\nu/(2+\nu)}$ and $\sum_{t=1}^{s_T} \alpha(t)^{\nu/(2+\nu)}$ have the same limit as $T\rightarrow\infty$ and $s_T\rightarrow\infty$, which is ensured by the order of $\alpha$-mixing coefficients.

In summary of the results that are established in \eqref{dk111}, \eqref{dk112}, and \eqref{dk113}, we can readily obtain
\begin{equation}\label{dk122}
E[\mathcal{Q}_2]-\frac{1}{T}\sum_{t=1}^T\sum_{s=t+1}^T E[\widetilde{z}_t\widetilde{z}_s  \varepsilon_t\varepsilon_s]=o(1). 
\end{equation}

We then study $\mathcal{Q}_2-E[\mathcal{Q}_2]$. With $\nu$ and $\nu^\ast$ that are defined in Assumption \ref{Ass.3} and Theorem \ref{Thm_Bootstrap}, respectively, we can always define a positive number $r$ through the following equation:
\begin{equation*}
\frac{1}{r}=\frac{1}{2+\nu^\ast}+\frac{\nu}{2(2+\nu)}.
\end{equation*}
Since $\nu^\ast>\nu$, it is clear to see that
\begin{eqnarray*}
\frac{1}{r}&<& \frac{1}{2+\nu}+\frac{\nu}{2(2+\nu)}
=\frac{1}{2}.
\end{eqnarray*}
Thus, we have $r>2$. For notational simplicity, we define a norm $\|\zeta\|_{n}=E\bigl[\|\zeta\|^{n}\bigr]^{1/n}$ for any random variable $\zeta$ and any positive number $n\geq 1$. We have
\begin{eqnarray}\label{dk120}
\|\mathcal{Q}_2-E[\mathcal{Q}_2]\|_{r/2}
&=&\frac{1}{T}\biggl\|\sum_{s=1}^{T-1}\sum_{t=1}^{T-s} K\Bigl(\frac{s}{\ell}\Bigr)\bigl(\widetilde{z}_t\widetilde{z}_{t+s}  \varepsilon_t\varepsilon_{t+s}-E[\widetilde{z}_t\widetilde{z}_{t+s}  \varepsilon_t\varepsilon_{t+s}]\bigr)\biggr\|_{r/2}
\nonumber\\
&\leq&\frac{1}{T}\sum_{s=1}^{T-1} K\Bigl(\frac{s}{\ell}\Bigr)\biggl\|\sum_{t=1}^{T-s}\bigl(\widetilde{z}_t\widetilde{z}_{t+s}  \varepsilon_t\varepsilon_{t+s}-E[\widetilde{z}_t\widetilde{z}_{t+s}  \varepsilon_t\varepsilon_{t+s}]\bigr)\biggr\|_{r/2}
\nonumber\\
&=&\frac{1}{T}\sum_{s=1}^{\ell} K\Bigl(\frac{s}{\ell}\Bigr)\biggl\|\sum_{t=1}^{T-s}\bigl(\widetilde{z}_t\widetilde{z}_{t+s}  \varepsilon_t\varepsilon_{t+s}-E[\widetilde{z}_t\widetilde{z}_{t+s}  \varepsilon_t\varepsilon_{t+s}]\bigr)\biggr\|_{r/2}
.
\end{eqnarray}
Additionally, let $\mathcal{F}_t$ and $E_{t}[\cdot]$ be the sigma field generated by $\{\widetilde{z}_s, \varepsilon_s\}_{s=t,t-1,\cdots}$ and the expectation conditional on $\mathcal{F}_t$, respectively. By McLeish's inequality for $\alpha$-mixing processes \citep[see Lemma 2.1 of][]{McLeish1975} and Assumption \ref{Ass.3}, 
\begin{eqnarray*}
\left\|E_{t-t_0}\bigl[\bigl(\widetilde{z}_t\widetilde{z}_{t+s}  \varepsilon_t\varepsilon_{t+s}-E[\widetilde{z}_t\widetilde{z}_{t+s}  \varepsilon_t\varepsilon_{t+s}]\bigr)\bigr]\right\|_{r/2}&\leq&6\alpha^{\frac{\nu}{2+\nu}}(t_0)\Bigl\|\widetilde{z}_t\widetilde{z}_{t+s}  \varepsilon_t\varepsilon_{t+s}-E[\widetilde{z}_t\widetilde{z}_{t+s}  \varepsilon_t\varepsilon_{t+s}]\Bigr\|_{(2+\nu^\ast)/2}
\nonumber\\
&\leq&12\alpha^{\frac{\nu}{2+\nu}}(t_0)\bigl\|\widetilde{z}_1\widetilde{z}_{1+s} \bigr\|_{(2+\nu^\ast)/2} \bigl\|\varepsilon_1\varepsilon_{1+s}\bigr\|_{(2+\nu^\ast)/2}
\nonumber\\
&\leq&12\alpha^{\frac{\nu}{2+\nu}}(t_0)\bigl\|\widetilde{z}_1 \bigr\|_{2+\nu^\ast} \bigl\|\varepsilon_1\bigr\|_{2+\nu^\ast}
,
\end{eqnarray*}
for a positive integer $t_0$. Using this result and the Lemma A of \cite{Hansen1992}, we can readily obtain
\begin{eqnarray}\label{dk121}
\left\|\sum_{t=1}^{T-s}\bigl(\widetilde{z}_t\widetilde{z}_{t+s}  \varepsilon_t\varepsilon_{t+s}-E[\widetilde{z}_t\widetilde{z}_{t+s}  \varepsilon_t\varepsilon_{t+s}]\bigr)\right\|_{r/2}&\leq& 36c_{\nu^\ast} \Bigl(r/(r-2)\Bigr)^{3/2}\sum_{t_0=1}^\infty \alpha^{\frac{\nu}{2+\nu}}(t_0)\left(T-s \right)^{2/r^\ast},
\end{eqnarray}
where $c_{\nu^\ast}=12\bigl\|\widetilde{z}_1 \bigr\|_{2+\nu^\ast} \bigl\|\varepsilon_1\bigr\|_{2+\nu^\ast}$ and $r^\ast=\min(r,4)$.

Combing \eqref{dk120} and \eqref{dk121} gives
\begin{eqnarray*}
\|\mathcal{Q}_2-E[\mathcal{Q}_2]\|_{r/2}&\leq& O(1)T^{2/r^\ast-1}\sum_{s=1}^{\ell} K\Bigl(\frac{s}{\ell}\Bigr)
\nonumber\\
&=&O(\ell T^{2/r^\ast-1}).
\end{eqnarray*}
Under the condition $\ell T^{2/r^\ast-1}\rightarrow 0$, we have
\begin{eqnarray}\label{dk123}
\mathcal{Q}_2-E[\mathcal{Q}_2]=o_P(1).
\end{eqnarray}
By \eqref{dk122} and \eqref{dk123}, we can finish the investigation of $\mathcal{Q}_2$ and obtain
\begin{eqnarray}\label{dk124}
\mathcal{Q}_2-\frac{1}{T}\sum_{t=1}^T\sum_{s=t+1}^T E[\widetilde{z}_t\widetilde{z}_s  \varepsilon_t\varepsilon_s]=o_P(1).
\end{eqnarray}

A similar argument applies with the index $s$ replacing $t$ for $\mathcal{Q}_3$, so it follows that 
\begin{eqnarray}\label{dk125}
\mathcal{Q}_3-\frac{1}{T}\sum_{t=1}^T\sum_{s=1}^{t-1} E[\widetilde{z}_t\widetilde{z}_s  \varepsilon_t\varepsilon_s]=o_P(1).
\end{eqnarray}

Drawing upon the definition of $\sigma_{c,z}^2$ in Assumption \ref{Ass.4} and the convergence of $\mathcal{Q}_1$, the results  that are established in \eqref{dk124} and \eqref{dk125} can immediately yield
\begin{equation}\label{dk150}
\text{Var}^\ast( \mathcal{J}_{4})=\sigma_{c,z}^2+o_P(1).
\end{equation}

In light of the $\ell$-dependent $\varsigma_t$,  we follow the Theorem 3.1 of \cite{Shao2010} and adopt the large-block and small-block argument to prove the central limit theorem for $\mathcal{J}_{4}$ conditional on the observed sample. Define $l_T$ and $s_T$ as the lengths for the large and small blocks and $k_T=\lfloor T/(l_T+s_T)\rfloor$ such that $l_T,s_T\rightarrow\infty$ and
\begin{equation*}
\frac{\ell}{s_T},\,\frac{k_Ts_T}{T},\Bigl(\frac{\ell l_T}{T}\Bigr)^{1+\frac{\nu}{2}}k_T\rightarrow 0.
\end{equation*}
For $j=1,\ldots, k_T$, define

\begin{eqnarray*}
&&\xi^\ast_{j,1} = \sum_{t=(j-1)(l_T+s_T)+1}^{jl_T+(j-1)s_T} \widetilde{z}_t\varepsilon_t\varsigma_t,\quad \xi^\ast_{j,2} = \sum_{t=jl_T+ (j-1)s_T+1}^{j(l_T+s_T)} \widetilde{z}_t\varepsilon_t\varsigma_t,\quad \xi^\ast_0 =\sum_{t=k_T(l_T+s_T)+1}^T \widetilde{z}_t\varepsilon_t\varsigma_t.
\end{eqnarray*}

We first show that $\frac{1}{\sqrt{T}}\sum_{j=1}^{k_T} \xi^\ast_{j,2}=o_P(1)$. Since $\frac{\ell}{s_T}\rightarrow 0$,  we assume $l_T,s_T>\ell$ without loss of generality.
By the definition of $\varsigma_t$, we can observe that $\{\xi^\ast_{1,1},\cdots,\xi^\ast_{k_T,1}\}$ are independent conditional on the observed data, as are $\{\xi^\ast_{1,2},\cdots,\xi^\ast_{k_T,2}\}$. Using \eqref{dk110} and Assumption \ref{Ass.3},  we have
\begin{eqnarray*}
\frac{1}{T}E\Biggl[E^\ast\Biggl[\Biggl(\sum_{j=1}^{k_T} \xi^\ast_{j,2}\Biggr)^2\Biggr]\Biggr]&=&\frac{1}{T}\sum_{j=1}^{k_T}E\left[E^\ast\left[ \xi^{\ast 2}_{j,2}\right]\right]
\nonumber\\
&=&\frac{1}{T}\sum_{j=1}^{k_T}\sum_{t,s=jl_T+ (j-1)s_T+1}^{j(l_T+s_T)}  E\left[E^\ast\left[ \widetilde{z}_t\widetilde{z}_s\varepsilon_t\varepsilon_s\varsigma_t  \varsigma_s\right]\right]
\nonumber\\
&\leq&\frac{1}{T}\sum_{j=1}^{k_T}\sum_{t=jl_T+ (j-1)s_T+1}^{j(l_T+s_T)}\sum_{s=-s_T+1}^{s_T-1}K\Bigl(\frac{s}{\ell}\Bigr)  |E\left[\widetilde{z}_t\widetilde{z}_{t+s}\varepsilon_t\varepsilon_{t+s}\right]|
\nonumber\\
&=& O\left(\frac{k_Ts_T}{T}\right)=o(1).
\end{eqnarray*}
Therefore, $\frac{1}{\sqrt{T}}\sum_{j=1}^{k_T} \xi^\ast_{j,2}=o_P(1)$. Analogously, we also have $\frac{1}{\sqrt{T}} \xi^\ast_{0}=o_P(1)$. Next, we establish the asymptotic normality of $\frac{1}{\sqrt{T}}\sum_{j=1}^{k_T} \xi^\ast_{j,1}$ by verifying the Lindeberg condition. Using analogous arguments to those in the proof of \eqref{dk150}, we can first show that  $\frac{1}{T}E^\ast\bigl[\bigl(\sum_{j=1}^{k_T} \xi^\ast_{j,1}\bigr)^2 \bigr]=\sigma_{c,z}^2+o_P(1)$. Moreover, for any $\epsilon>0$, we can use the same argument as in \eqref{dk151} to obtain
\begin{eqnarray}\label{dk153}
\frac{1}{T}\sum_{j=1}^{k_T}E^\ast\bigl[\xi^{\ast 2}_{j,1}\cdot I(|\xi^\ast_{j,1}|\geq \epsilon \sqrt{T}) \bigr]&\leq&\frac{1}{\epsilon^\nu T^{1+\frac{\nu}{2}}}\sum_{j=1}^{k_T}E^\ast\bigl[\xi^{\ast 2+\nu}_{j,1}\bigr].
\end{eqnarray}

For notational simplicity, we define a norm (conditional on the observed sample) $\|\zeta\|^\ast_{n}=E^\ast\bigl[\|\zeta\|^{n}\bigr]^{1/n}$ for any random variable $\zeta$ and any positive number $n\geq 1$. In what follows, we use the Rosenthal inequality to study the order of $\|\xi^\ast_{1,1}\|^\ast_{2+\nu}$, without loss of generality.  Noteworthily, Rosenthal inequality is designed for the independent random variables and it is not directly applicable for $\widetilde{z}_t\varepsilon_t\varsigma_t$. Therefore, we further decompose $\xi^\ast_{1,1}$ as $\xi^\ast_{1,1}=\sum_{k=1}^{\ell+1} \xi^\ast_{1,1,k}$,
where 
\begin{equation*}
\xi^\ast_{1,1,k}=\sum_{s=1}^{\lfloor (l_T-k)/(\ell+1) \rfloor}\widetilde{z}_{k+(s-1)(\ell+1)}\varepsilon_{k+(s-1)(\ell+1)}\varsigma_{k+(s-1)(\ell+1)}.
\end{equation*}
Then, by the definition of $\varsigma$, it is clear to see that for each $k$, 
all elements that are involved in the summation in $\xi^\ast_{1,1,k}$ are independent conditional on the observed sample. Using the triangle inequality and Rosenthal inequality sequentially, we obtain
\begin{eqnarray}\label{dk154}
\|\xi^\ast_{1,1}\|^\ast_{2+\nu}&\leq& \sum_{k=1}^{\ell+1} \|\xi^\ast_{1,1,k}\|^\ast_{2+\nu}
\nonumber\\
&\leq&O(1)\sum_{k=1}^{\ell+1} \Biggl[\Biggl\|\sum_{s=1}^{\lfloor (l_T-k)/(\ell+1) \rfloor}\widetilde{z}^2_{k+(s-1)(\ell+1)}\varepsilon_{k+(s-1)(\ell+1)}^2\varsigma^2_{k+(s-1)(\ell+1)}\Biggr\|^\ast_{1+\nu/2}\Biggr]^{1/2}
\nonumber\\
&\leq&O(1)\sum_{k=1}^{\ell+1} \Biggl[\sum_{s=1}^{\lfloor (l_T-k)/(\ell+1) \rfloor}\widetilde{z}^2_{k+(s-1)(\ell+1)}\varepsilon_{k+(s-1)(\ell+1)}^2\Biggr]^{1/2}
\nonumber\\
&\leq&O(\sqrt{\ell})\Biggl[\sum_{k=1}^{\ell+1} \sum_{s=1}^{\lfloor (l_T-k)/(\ell+1) \rfloor}\widetilde{z}^2_{k+(s-1)(\ell+1)}\varepsilon_{k+(s-1)(\ell+1)}^2\Biggr]^{1/2}
\nonumber\\
&=&O(\sqrt{\ell})\Biggl[\sum_{t=1}^{l_T}\widetilde{z}_t^2\varepsilon_t^2\Biggr]^{1/2}. 
\end{eqnarray}
Combining \eqref{dk153} and  \eqref{dk154} gives
\begin{eqnarray*}
\frac{1}{T}\sum_{j=1}^{k_T}E \bigl[E^\ast\bigl[\xi^{\ast 2}_{j,1}\cdot I(|\xi^\ast_{j,1}|\geq \epsilon \sqrt{T}) \bigr]\bigr]&\leq&\frac{1}{\epsilon^\nu T^{1+\frac{\nu}{2}}}\sum_{j=1}^{k_T} E\Bigl[\|\xi^\ast_{j,1}\|^{\ast 2+\nu}_{2+\nu}\Bigr]
\nonumber\\
&\leq&O(1)\left(\frac{\ell}{T}\right)^{1+\frac{\nu}{2}}\sum_{j=1}^{k_T}E\Biggl[\Biggl(\sum_{t=(j-1)(l_T+s_T)+1}^{jl_T+(j-1)s_T} \widetilde{z}_t^2\varepsilon_t^2\Biggr)^{1+\frac{\nu}{2}}\Biggr]
\nonumber\\
&=&O\Bigl(\Bigl(\frac{\ell l_T}{T}\Bigr)^{1+\frac{\nu}{2}}k_T\Bigr)=o(1).
\end{eqnarray*}
We therefore have $\frac{1}{T}\sum_{j=1}^{k_T}E^\ast\bigl[\xi^{\ast 2}_{j,1}\cdot I(|\xi^\ast_{j,1}|\geq \epsilon \sqrt{T}) \bigr]=o_P(1)$, which is the last step of the large-block and small-block technique and it immediately yields the following CLT:
$\mathcal{J}_{4}\rightarrow_{D^\ast} N\left(0,\sigma_{c,z}^2\right)$,
where $\rightarrow_{D^\ast}$ denotes the convergence in distribution conditional on the observed sample. Recall that we have shown that $\mathcal{J}_{1},\mathcal{J}_{2},$ and $\mathcal{J}_{3}$ are all asymptotically negligible at an earlier stage. Combining these results with \eqref{dk35}, \eqref{dk155}, and Theorem \ref{THM2.1}.1 leads to the assertion in Theorem \ref{Thm_Bootstrap}.1.

(2) Let $\Omega_T$ denote the event in which the group-LASSO estimation has correctly identified the sparsity structure of the $g(\mathbf{x})$ function. That is $\|\widetilde{\pmb{\Theta}}_{D,j}\|=0$, 
for $j=d_q^0+1,\ldots,d_q$.
By Lemma \ref{LEM2.4}, we have $P(\Omega_T)\rightarrow1$, as $T\rightarrow\infty$. Hence, it suffices only to study the asymptotic distribution of $\widetilde{g}^*(\mathbf{x}_0)$ conditional on  the observed sample and $\Omega_T$. 

On $\Omega_T$ and using analogous arguments to those in the proof of Theorem \ref{THM2.1},  we can obtain the following result for the bootstrap estimator $\widetilde{g}^\ast(\mathbf{x}_{0})$:
\begin{eqnarray*}
\widetilde{g}^\ast(\mathbf{x}_{0})-\widetilde{g}(\mathbf{x}_{0})&=&\sum_{\mathbf{i}\in [M]^{d_0}} \widetilde{\mathbf{x}}_{c, 0,\mathbf{i}}^\top\mathbf{D}_c^{\top} \mathbf{H}_c\bigl[\mathbf{H}_c^{-1}\mathbf{D}_c^{\top,-1}(\widetilde{\pmb{\theta}}^\ast_{c,\mathbf{i}}-\widetilde{\pmb{\theta}}_{c,\mathbf{i}})\bigr]+O_P(h^p),
\end{eqnarray*}
where $\widetilde{\pmb{\theta}}^\ast_{c,\mathbf{i}}$ denotes the oracle bootstrap estimator.

Then, we can use the arguments that are closely related to those in the proof of \eqref{dk42} to obtain

\begin{equation}\label{dkh}
\sqrt{Th^{d_0}}\mathbf{H}_c^{-1}  \mathbf{D}_c^{\top,-1}(\widetilde{\pmb{\theta}}^\ast_{c,\mathbf{i}} -\widetilde{\pmb{\theta}}_{c,\mathbf{i}}+O_P(h^p))=\frac{1}{\sqrt{Th^{d_0}}}\pmb{\Sigma}_{c,\mathbf{i}}^{-1}\sum_{t=1}^T\mathbf{H}_c\mathbf{D}_c\widetilde{\mathbf{x}}_{c,\mathbf{i},t}  \widehat{\varepsilon}_t\varsigma_t+o_P(1).
\end{equation}
Then, similar steps to those in the derivation of $\frac{1}{\sqrt{T}}\mathbf{Z}^\top \mathbf{M}_{c,x}\pmb{\varepsilon}^\ast$'s bootstrap distribution in \eqref{dk155}  can be applied here to establish the bootstrap  behaviour of the first term on the right-hand side of \eqref{dkh}. Specifically, we can obtain that it converges to $N\left(\mathbf{0},\sigma^2_\varepsilon \pmb{\Sigma}_{c,\mathbf{i}}^{-1}\right)$ conditional on the observed sample, up to some asymptotically negligible terms. In connection with \eqref{dkh}, it leads to the desired result in Theorem \ref{Thm_Bootstrap}.2. \hspace*{\fill}{$\blacksquare$}

\subsection{Proofs for the Fully Nonparametric Model}\label{App.44}

\noindent \textbf{Proof of Lemma \ref{LemA.3}:}

First, we expand the expression of $\widehat{\pmb{\theta}}_{\mathbf{i}}$ as follows:

\begin{eqnarray*}
\widehat{\pmb{\theta}}_{\mathbf{i}} -\widetilde{\pmb{\lambda}}_{\mathbf{i}}  &=& \left(  \sum_{t=1}^T\widetilde{\mathbf{x}}_{\mathbf{i},t} \widetilde{\mathbf{x}}_{\mathbf{i},t} ^\top \right)^{-1} \sum_{t=1}^T\widetilde{\mathbf{x}}_{\mathbf{i},t} y_t -\widetilde{\pmb{\lambda}}_{\mathbf{i}}  \nonumber \\
&=&\left(  \sum_{t=1}^T\widetilde{\mathbf{x}}_{\mathbf{i},t} \widetilde{\mathbf{x}}_{\mathbf{i},t} ^\top \right)^{-1}   \sum_{t=1}^T\widetilde{\mathbf{x}}_{\mathbf{i},t} \widetilde{s}(\mathbf{x}_t \, |\, \widetilde{\pmb{\Lambda}})-\widetilde{\pmb{\lambda}}_{\mathbf{i}}   \nonumber \\
&&+\left(  \sum_{t=1}^T\widetilde{\mathbf{x}}_{\mathbf{i},t} \widetilde{\mathbf{x}}_{\mathbf{i},t} ^\top \right)^{-1}   \sum_{t=1}^T\widetilde{\mathbf{x}}_{\mathbf{i},t} [g(\mathbf{x}_t)-\widetilde{s}(\mathbf{x}_t \, |\, \widetilde{\pmb{\Lambda}}) ]\nonumber \\
&&+\left( \sum_{t=1}^T\widetilde{\mathbf{x}}_{\mathbf{i},t} \widetilde{\mathbf{x}}_{\mathbf{i},t} ^\top \right)^{-1}   \sum_{t=1}^T\widetilde{\mathbf{x}}_{\mathbf{i},t}  \varepsilon_t\nonumber \\
&=& \left( \sum_{t=1}^T\widetilde{\mathbf{x}}_{\mathbf{i},t} \widetilde{\mathbf{x}}_{\mathbf{i},t} ^\top \right)^{-1}   \sum_{t=1}^T\widetilde{\mathbf{x}}_{\mathbf{i},t} [g(\mathbf{x}_t)-\widetilde{s}(\mathbf{x}_t \, |\, \widetilde{\pmb{\Lambda}}) ]\nonumber \\
&&+\left( \sum_{t=1}^T\widetilde{\mathbf{x}}_{\mathbf{i},t} \widetilde{\mathbf{x}}_{\mathbf{i},t} ^\top \right)^{-1}   \sum_{t=1}^T\widetilde{\mathbf{x}}_{\mathbf{i},t}  \varepsilon_t,
\end{eqnarray*}
where the third equality follows from the fact that $I_{\mathbf{i}, h}(\mathbf{x}_t)I_{\mathbf{j}, h}(\mathbf{x}_t)=0$ for $\mathbf{i}\ne \mathbf{j}$, and the definition of $\widetilde{s}(\mathbf{x}_t \, |\, \widetilde{\pmb{\Lambda}})$. Below, we consider the terms on the right-hand side one by one.

We further define
\begin{eqnarray}\label{EqA.11}
I_{\mathbf{i},h}(\mathbf{x}_t)\mathbf{H} \cdot \mathbf{m}(\mathbf{x}_t\, |\, \mathbf{x}_{\mathbf{i},0}) \coloneqq\widetilde{\mathbf{m}}(\mathbf{x}_t\, |\, \mathbf{x}_{\mathbf{i},0})=(\widetilde{m}_1(\mathbf{x}_t\, |\, \mathbf{x}_{\mathbf{i},0}),\ldots, \widetilde{m}_{d_q}(\mathbf{x}_t\, |\, \mathbf{x}_{\mathbf{i},0}) )^\top .
\end{eqnarray}

\medskip

First, we consider $\frac{1}{Th^d} \sum_{t=1}^T\widetilde{\mathbf{x}}_{\mathbf{i},t}  \widetilde{\mathbf{x}}_{\mathbf{i},t}  ^\top$. By \eqref{EqA.12} and  \eqref{EqA.13}, we obtain 

\begin{eqnarray*} 
&&\frac{1}{Th^d}\sum_{t=1}^T\mathbf{H} \mathbf{D} \widetilde{\mathbf{x}}_{\mathbf{i},t}  \widetilde{\mathbf{x}}_{\mathbf{i},t}  ^\top\mathbf{D}^\top \mathbf{H} 
= f_{\mathbf{x}}(\mathbf{x}_{\mathbf{i},0}) \int_{[-1,1]^d}  \mathbf{m}(\mathbf{x}\, |\, \mathbf{0})  \mathbf{m}(\mathbf{x}\, |\, \mathbf{0})^\top  \mathrm{d} \mathbf{x}\cdot (1+o_p(1)).
\end{eqnarray*}

\medskip

Also, we note

\begin{eqnarray*}
&& \left( \sum_{t=1}^T\widetilde{\mathbf{x}}_{\mathbf{i},t} \widetilde{\mathbf{x}}_{\mathbf{i},t} ^\top \right)^{-1}   \sum_{t=1}^T\widetilde{\mathbf{x}}_{\mathbf{i},t} [g(\mathbf{x}_t)-\widetilde{s}(\mathbf{x}_t \, |\, \widetilde{\pmb{\Lambda}}) ]\nonumber \\
 &=&\mathbf{D}^{\top}\mathbf{H} \left( \mathbf{H}\mathbf{D}\sum_{t=1}^T\widetilde{\mathbf{x}}_{\mathbf{i},t} \widetilde{\mathbf{x}}_{\mathbf{i},t} ^\top\mathbf{D}^\top  \mathbf{H} \right)^{-1}   \mathbf{H}\mathbf{D}\sum_{t=1}^T\widetilde{\mathbf{x}}_{\mathbf{i},t} [g(\mathbf{x}_t)-\widetilde{s}(\mathbf{x}_t \, |\, \widetilde{\pmb{\Lambda}}) ] \nonumber \\
 &=&\mathbf{D}^{\top}\mathbf{H} (\mathbf{H}\mathbf{D}\widetilde{\mathbf{X}}_{\mathbf{i}}^\top \widetilde{\mathbf{X}}_{\mathbf{i}} \mathbf{D}^\top  \mathbf{H} )^{-1} \mathbf{H}\mathbf{D}\widetilde{\mathbf{X}}_{\mathbf{i}}^\top \Delta \mathbf{G} ,
\end{eqnarray*}
where 

\begin{eqnarray*}
&& \widetilde{\mathbf{X}}_{\mathbf{i}} = (\widetilde{\mathbf{x}}_{\mathbf{i},1},\ldots, \widetilde{\mathbf{x}}_{\mathbf{i},T})^\top ,\\
&&\Delta \mathbf{G} =(I_{\mathbf{i},h}(\mathbf{x}_1)[g(\mathbf{x}_1)-\widetilde{s}(\mathbf{x}_1 \, |\, \widetilde{\pmb{\Lambda}}) ],\ldots,I_{\mathbf{i},h}(\mathbf{x}_T)[ g(\mathbf{x}_T)-\widetilde{s}(\mathbf{x}_T \, |\, \widetilde{\pmb{\Lambda}})] )^\top .
\end{eqnarray*}

By Lemma \ref{LEM2.3}, it is easy to see that

\begin{eqnarray*}
&&\frac{1}{Th^d}E\|\Delta \mathbf{G}\|^2 = \frac{1}{h^d}E\left[I_{\mathbf{i},h}(\mathbf{x}_1)[g(\mathbf{x}_1)-\widetilde{s}(\mathbf{x}_1 \, |\, \widetilde{\pmb{\Lambda}}) ]^2\right] =O(h^{2p}).
\end{eqnarray*}
Then we can write

\begin{eqnarray*}
& & \|( \mathbf{H}\mathbf{D}\widetilde{\mathbf{X}}_{\mathbf{i}}^\top \widetilde{\mathbf{X}}_{\mathbf{i}} \mathbf{D}^\top  \mathbf{H} )^{-1} \mathbf{H}\mathbf{D}\widetilde{\mathbf{X}}_{\mathbf{i}}^\top \Delta \mathbf{G}  \|^2 \nonumber \\
&=& \Delta \mathbf{G}^\top \widetilde{\mathbf{X}}_{\mathbf{i}} \mathbf{D}^\top \mathbf{H}  ( \mathbf{H}\mathbf{D}\widetilde{\mathbf{X}}_{\mathbf{i}}^\top \widetilde{\mathbf{X}}_{\mathbf{i}} \mathbf{D}^\top  \mathbf{H} )^{-1} ( \mathbf{H}\mathbf{D}\widetilde{\mathbf{X}}_{\mathbf{i}}^\top \widetilde{\mathbf{X}}_{\mathbf{i}} \mathbf{D}^\top  \mathbf{H} )^{-1} \mathbf{H}\mathbf{D}\widetilde{\mathbf{X}}_{\mathbf{i}}^\top \Delta \mathbf{G}  \nonumber \nonumber \\
&\le&  \frac{1}{Th^d} \lambda_{\text{\normalfont max}}\left\{  \left(\frac{1}{Th^d} \mathbf{H}\mathbf{D}\widetilde{\mathbf{X}}_{\mathbf{i}}^\top \widetilde{\mathbf{X}}_{\mathbf{i}} \mathbf{D}^\top  \mathbf{H} \right)^{-1}  \right\} \Delta \mathbf{G}^\top \widetilde{\mathbf{X}}_{\mathbf{i}} \mathbf{D}^\top \mathbf{H}   ( \mathbf{H}\mathbf{D}\widetilde{\mathbf{X}}_{\mathbf{i}}^\top \widetilde{\mathbf{X}}_{\mathbf{i}} \mathbf{D}^\top  \mathbf{H} )^{-1} \mathbf{H}\mathbf{D}\widetilde{\mathbf{X}}_{\mathbf{i}}^\top \Delta \mathbf{G} \nonumber \\
&\le&O_P(1) \lambda_{\text{\normalfont max}}\{ \widetilde{\mathbf{X}}_{\mathbf{i}}\mathbf{D}^\top \mathbf{H}   ( \mathbf{H}\mathbf{D}\widetilde{\mathbf{X}}_{\mathbf{i}}^\top \widetilde{\mathbf{X}}_{\mathbf{i}} \mathbf{D}^\top  \mathbf{H} )^{-1} \mathbf{H}\mathbf{D} \widetilde{\mathbf{X}}_{\mathbf{i}}^\top \} \cdot \| \Delta \mathbf{G}  \|^2/(Th^d)\nonumber \\
&=&O_P(h^{2p} ) ,
\end{eqnarray*}
where the first inequality follows from the exercise 5 on page 267 of \cite{Magnus}, and the second inequality follows from \eqref{EqA.12} and \eqref{EqA.13}.

Based on the above development, we can conclude that

\begin{eqnarray*}
\left\| \mathbf{H}^{-1} \mathbf{D}^{\top,-1}\left( \sum_{t=1}^T\widetilde{\mathbf{x}}_{\mathbf{i},t} \widetilde{\mathbf{x}}_{\mathbf{i},t} ^\top \right)^{-1}   \sum_{t=1}^T\widetilde{\mathbf{x}}_{\mathbf{i},t} [g(\mathbf{x}_t)-\widetilde{s}(\mathbf{x}_t \, |\, \widetilde{\pmb{\Lambda}}) ]\right\|=O_P(h^{p}) .
\end{eqnarray*}

\medskip

Finally, in order to establish the asymptotic distribution, we just need to focus on $\frac{1}{\sqrt{Th^d}}\sum_{t=1}^T\widetilde{\mathbf{m}}(\mathbf{x}_t\, |\, \mathbf{x}_{\mathbf{i},0})  \varepsilon_t$  in view of \eqref{EqA.10}. Write

\begin{eqnarray}\label{EqA.14}
&&E\left[ \left(\frac{1}{\sqrt{Th^d}}\sum_{t=1}^T\widetilde{\mathbf{m}}(\mathbf{x}_t\, |\, \mathbf{x}_{\mathbf{i},0})  \varepsilon_t\right) \left(\frac{1}{\sqrt{Th^d}}\sum_{t=1}^T\widetilde{\mathbf{m}}(\mathbf{x}_t\, |\, \mathbf{x}_{\mathbf{i},0}) \varepsilon_t\right)^\top \right]\nonumber \\
&=&\frac{1}{Th^d}\sum_{t=1}^T \sum_{s=1}^T E [\widetilde{\mathbf{m}}(\mathbf{x}_t\, |\, \mathbf{x}_{\mathbf{i},0})  \widetilde{\mathbf{m}}(\mathbf{x}_s\, |\, \mathbf{x}_{\mathbf{i},0})^\top \varepsilon_t\varepsilon_s ] \nonumber \\
&=&\frac{1}{Th^d}\sum_{t=1}^T\sigma_\varepsilon^2 E [\widetilde{\mathbf{m}}(\mathbf{x}_t\, |\, \mathbf{x}_{\mathbf{i},0})  \widetilde{\mathbf{m}}(\mathbf{x}_t\, |\, \mathbf{x}_{\mathbf{i},0})  ^\top  ] \nonumber \\
&&+ \frac{1}{ h^d}\sum_{t=1}^{T-1} (1-t/T)E [\widetilde{\mathbf{m}}(\mathbf{x}_1\, |\, \mathbf{x}_{\mathbf{i},0})  \widetilde{\mathbf{m}}(\mathbf{x}_{1+t}\, |\, \mathbf{x}_{\mathbf{i},0})  ^\top \varepsilon_1\varepsilon_{1+t} ]  \nonumber \\
&&+\frac{1}{ h^d}\sum_{t=1}^{T-1} (1-t/T)E [\widetilde{\mathbf{m}}(\mathbf{x}_{1+t}\, |\, \mathbf{x}_{\mathbf{i},0}) \widetilde{\mathbf{m}}(\mathbf{x}_1\, |\, \mathbf{x}_{\mathbf{i},0})   ^\top \varepsilon_1\varepsilon_{1+t} ],
\end{eqnarray}
where the last two terms are the same up to a transpose operation.

For the first term, we can use similar argument to that in \eqref{EqA.12} and obtain 
\begin{eqnarray}
\frac{1}{Th^d}\sum_{t=1}^T\sigma_\varepsilon^2   E [\widetilde{\mathbf{m}}(\mathbf{x}_t\, |\, \mathbf{x}_{\mathbf{i},0})  \widetilde{\mathbf{m}}(\mathbf{x}_t\, |\, \mathbf{x}_{\mathbf{i},0})  ^\top  ] 
&\to &\sigma_\varepsilon^2  f_{\mathbf{x}}(\mathbf{x}_{\mathbf{i},0}) \int_{[-1,1]^d}  \mathbf{m}(\mathbf{x}\, |\, \mathbf{0})  \mathbf{m}(\mathbf{x}\, |\, \mathbf{0})^\top  \mathrm{d} \mathbf{x}.
\end{eqnarray}

Then, we use Assumption \ref{Ass.3} and the Davydov's inequality for $\alpha$-mixing processes  (see pages 19-20 in \cite{Bosq1996}) to show the convergence of the second term on the right-hand side of \eqref{EqA.14}. Specifically, we have
\begin{eqnarray}\label{dk160}
|E[\varepsilon_1\varepsilon_{1+t}]|&\leq& O(1) \alpha(t)^{\nu/(2+\nu)}\left\{E[  \varepsilon_1^{2+\nu} ]\right\}^{2/(2+\nu)}
\nonumber\\
&=&O(1) \alpha(t)^{\nu/(2+\nu)}.
\end{eqnarray}
Moreover, it is clear to see that
\begin{eqnarray}\label{dk161}
\bigl\|E [\widetilde{\mathbf{m}}(\mathbf{x}_1\, |\, \mathbf{x}_{\mathbf{i},0})  \widetilde{\mathbf{m}}(\mathbf{x}_{1+t}\, |\, \mathbf{x}_{\mathbf{i},0})  ^\top ]\bigr\|&=& 
\Bigl\|
\int_{\mathbf{x},\mathbf{z}\in C_{\mathbf{x}_{0\mathbf{i}},h}} \mathbf{H}  \mathbf{m}(\mathbf{x}\, |\, \mathbf{x}_{\mathbf{i},0}) \mathbf{m}(\mathbf{z}\, |\, \mathbf{x}_{\mathbf{i},0})^\top  \mathbf{H}  f_{\mathbf{x},s}(\mathbf{x},\mathbf{z})\mathrm{d} \mathbf{x}\mathrm{d} \mathbf{z}
\Bigr\|
\nonumber\\
&=&h^{2d}\Bigl\|\int_{\mathbf{x},\mathbf{z}\in [-1,1]^d} \mathbf{m}(\mathbf{x}\, |\,\mathbf{0}) \mathbf{m}(\mathbf{z}\, |\, \mathbf{0})^\top \mathrm{d} \mathbf{x}\mathrm{d} \mathbf{z} f_{\mathbf{x},s}(\mathbf{x}_{\mathbf{i},0},\mathbf{x}_{\mathbf{i},0})\Bigr\|
\nonumber\\
&=&O(h^{2d}).
\end{eqnarray}
By \eqref{dk160} and \eqref{dk161}, 
\begin{eqnarray}\label{jitith21}
\frac{1}{ h^d}\sum_{t=1}^{T-1} (1-t/T)\Bigl\|E [\widetilde{\mathbf{m}}(\mathbf{x}_1\, |\, \mathbf{x}_{\mathbf{i},0})  \widetilde{\mathbf{m}}(\mathbf{x}_{1+t}\, |\, \mathbf{x}_{\mathbf{i},0})  ^\top \varepsilon_1\varepsilon_{1+t} ]\Bigr\|&\leq&O(h^d)\sum_{t=1}^{T-1}\alpha(t)^{\nu/(2+\nu)}
\nonumber\\
&=&o(1).
\end{eqnarray}
Analogously, the third term on the right-hand side of \eqref{EqA.14} is also negligible.

Thus, we can conclude that

\begin{eqnarray}\label{EqA.17}
&&E\left[ \left(\frac{1}{\sqrt{Th^d}}\sum_{t=1}^T\widetilde{\mathbf{m}}(\mathbf{x}_t\, |\, \mathbf{x}_{\mathbf{i},0})  \varepsilon_t\right) \left(\frac{1}{\sqrt{Th^d}}\sum_{t=1}^T\widetilde{\mathbf{m}}(\mathbf{x}_t\, |\, \mathbf{x}_{\mathbf{i},0}) \varepsilon_t\right)^\top \right]\nonumber \\
&=&\frac{1}{Th^d}\sum_{t=1}^T\sigma_\varepsilon^2   E [\widetilde{\mathbf{m}}(\mathbf{x}_t\, |\, \mathbf{x}_{\mathbf{i},0})  \widetilde{\mathbf{m}}(\mathbf{x}_t\, |\, \mathbf{x}_{\mathbf{i},0})  ^\top  ] +o(1)\nonumber \\
&\to &\sigma_\varepsilon^2  f_{\mathbf{x}}(\mathbf{x}_{\mathbf{i},0}) \int_{[-1,1]^d}  \mathbf{m}(\mathbf{x}\, |\, \mathbf{0})  \mathbf{m}(\mathbf{x}\, |\, \mathbf{0})^\top  \mathrm{d} \mathbf{x}.
\end{eqnarray}

Below, we further use small-block and large-block to prove the normality. To employ the small-block and large-block arguments, we partition the set $\{1,\ldots, T \}$ into $2k_T+1$ subsets with large blocks of size $l_T$ and small blocks of size $s_T$ and the last remaining set of size $T-k_T(l_T+s_T)$, where $l_T$ and $s_T$ are selected such that

\begin{eqnarray*}
s_T\to \infty,\quad  \frac{s_T}{l_T}\to 0,\quad \frac{l_T^{1+\nu} }{  (Th^d)^{\frac{\nu}{2}}}\to 0,\quad\text{and}\quad k_T\equiv \left\lfloor \frac{T}{l_T+s_T} \right\rfloor,
\end{eqnarray*}
and $\nu$ is defined in Assumption \ref{Ass.3}.1.

For $j=1,\ldots, k_T$, define

\begin{eqnarray*}
&&\pmb{\xi}_{j,1} = \sum_{t=(j-1)(l_T+s_T)+1}^{jl_T+(j-1)s_T} \frac{1}{\sqrt{h^d}}\widetilde{\mathbf{m}}(\mathbf{x}_t\, |\, \mathbf{x}_{\mathbf{i},0}) \varepsilon_t,\quad \pmb{\xi}_{j,2} = \sum_{t=jl_T+ (j-1)s_T+1}^{j(l_T+s_T)} \frac{1}{\sqrt{h^d}}\widetilde{\mathbf{m}}(\mathbf{x}_t\, |\, \mathbf{x}_{\mathbf{i},0})  \varepsilon_t ,\nonumber \\
&&\pmb{\xi}_0 =\sum_{t=k_T(l_T+s_T)+1}^T \frac{1}{\sqrt{h^d}}\widetilde{\mathbf{m}}(\mathbf{x}_t\, |\, \mathbf{x}_{\mathbf{i},0})  \varepsilon_t.
\end{eqnarray*}
Note that $\alpha(T) = o(1/T)$ and $k_Ts_T/T\to 0$. By direct calculation, we immediately obtain that

\begin{eqnarray*}
\frac{1}{T}E\left\|\sum_{j=1}^{k_T} \pmb{\xi}_{j,2}  \right\|^2\to 0\quad \text{and}\quad \frac{1}{T}E\left\| \pmb{\xi}_{0} \right\|^2\to 0.
\end{eqnarray*}
Therefore,

\begin{eqnarray*}
\frac{1}{\sqrt{Th^d}}\sum_{t=1}^T\widetilde{\mathbf{m}}(\mathbf{x}_t\, |\, \mathbf{x}_{\mathbf{i},0})   \varepsilon_t = \frac{1}{\sqrt{T}}\sum_{j=1}^{k_T}\pmb{\xi}_{j,1} +o_P(1).
\end{eqnarray*}
By Proposition 2.6 of \cite{FanYao}, we have as $T\to 0$

\begin{eqnarray*}
&&\left| E\left[\exp\left( \frac{iw}{\sqrt{T}}\sum_{j=1}^{k_T}\pmb{\xi}_{j,1}\right) \right] -\prod_{j=1}^{k_T}E\left[\exp\left( \frac{iw \pmb{\xi}_{j,1}}{\sqrt{T}} \right) \right]\right| \nonumber \\
&\le & 16(k_T-1) \alpha(s_T)\to 0,
\end{eqnarray*}
where $i$ is the imaginary unit.

In connection with \eqref{EqA.14}-\eqref{EqA.17}, the Feller condition is fulfilled as follows:

\begin{eqnarray*}
\frac{1}{T}\sum_{j=1}^{k_T}E[\pmb{\xi}_{j,1}\pmb{\xi}_{j,1}^\top]\to \sigma_\varepsilon^2  f_{\mathbf{x}}(\mathbf{x}_{\mathbf{i},0}) \int_{[-1,1]^d}  \mathbf{m}(\mathbf{x}\, |\, \mathbf{0})  \mathbf{m}(\mathbf{x}\, |\, \mathbf{0})^\top  \mathrm{d} \mathbf{x}.
\end{eqnarray*}

Also, we note that

\begin{eqnarray*}
E[\|\pmb{\xi}_{1,1}\|^2 \cdot I(\|\pmb{\xi}_{1,1}\| \ge \epsilon \sqrt{T})] &\le &\{E\|\pmb{\xi}_1\|^{2\cdot \frac{2+\nu}{2}}\}^{\frac{2}{2+\nu}} \left\{ E[ I(\|\pmb{\xi}_1\| \ge \epsilon \sqrt{T})] \right\}^{\frac{\nu}{2+\nu}}\nonumber \\
&\le  &\{E\|\pmb{\xi}_{1,1}\|^{2+\nu}\}^{\frac{2}{2+\nu}} \left\{ \frac{E\|\pmb{\xi}_{1,1}\|^{2+\nu}}{\epsilon^{2+\nu} T^{\frac{2+\nu}{2}}}\right\}^{\frac{\nu}{2+\nu}}\nonumber \\
&= &  \frac{1}{\epsilon^{\nu} T^{\frac{\nu}{2}}}\left\{ E\|\pmb{\xi}_{1,1}\|^{2+\nu}\right\}^{\frac{1}{2+\nu} \cdot (2+\nu)} \nonumber \\
&=&  O(1) \frac{l_T^{2+\nu} }{\epsilon^{\nu} T^{\frac{\nu}{2}}}  E\| \widetilde{\mathbf{m}}(\mathbf{x}_1\, |\, \mathbf{x}_{\mathbf{i},0}) \varepsilon_1 \|^{2+\nu} \cdot \frac{1}{h^{\frac{d}{2} \cdot (2+\nu) }} \nonumber \\
&=&  O(1) \frac{l_T^{2+\nu} }{\epsilon^{\nu} (Th^d)^{\frac{\nu}{2}}}  \cdot  \frac{1}{h^d} E\| \widetilde{\mathbf{m}}(\mathbf{x}_1\, |\, \mathbf{x}_{\mathbf{i},0}) \varepsilon_1 \|^{2+\nu} \nonumber \\
&=&  O(1) \frac{l_T^{2+\nu} }{ (Th^d)^{\frac{\nu}{2}}} ,
\end{eqnarray*}
where the first inequality follows from H\"older inequality, the second inequality follows from  Chebyshev's inequality, and the second equality follows from Minkowski inequality. Consequently,

\begin{eqnarray*}
\frac{1}{T}\sum_{j=1}^{k_T}E[\|\pmb{\xi}_{j,1}\|^2 \cdot I(\|\pmb{\xi}_{j,1}\| \ge \epsilon \sqrt{T})] =O\left( \frac{k_Tl_T^{2+\nu} }{ T (Th^d)^{\frac{\nu}{2}}}\right) = O\left( \frac{l_T^{1+\nu} }{  (Th^d)^{\frac{\nu}{2}}}\right)=o(1),
\end{eqnarray*}
where the last step follows from the choice of $l_T$ as specified above. Therefore, the Lindberg condition is justified. Using a Cram{\'e}r-Wold device, the CLT follows immediately by the standard argument. \hspace*{\fill}{$\blacksquare$}

\bigskip

\noindent \textbf{Proof of Theorem \ref{Thm3.1}:}

By  Lemma \ref{LEM2.3}, we can write
\begin{eqnarray}\label{newpr}
\widehat{g}(\mathbf{x}_0)-g(\mathbf{x}_0)&=&  \widetilde{s}(\mathbf{x}_0 \, |\, \widehat{\pmb{\Theta}} ) -\widetilde{s}(\mathbf{x}_0 \, |\, \widetilde{\pmb{\Lambda}} )+O_P(h^p) 
\nonumber\\
&=& \sum_{\mathbf{i}\in [M]^d} I_{\mathbf{i},h}(\mathbf{x}_0) \cdot\left( s(\mathbf{x}_0\, |\, \mathbf{x}_{0\mathbf{i}},  \widehat{\pmb{\theta}}_{\mathbf{i}})-s(\mathbf{x}_0\, |\, \mathbf{x}_{0\mathbf{i}},  \widetilde{\pmb{\lambda}}_{\mathbf{i}})\right)+O_P(h^p) 
\nonumber\\
&=&\sum_{\mathbf{i}\in [M]^d} I_{\mathbf{i},h}(\mathbf{x}_0) \cdot ((\widehat{\pmb{\theta}}_{\mathbf{i}}-\widetilde{\pmb{\lambda}}_{\mathbf{i}})\otimes \pmb{\gamma})^\top\pmb{\sigma}(\mathbf{x}_0\, |\, \mathbf{x}_{0\mathbf{i}})+O_P(h^p) 
\nonumber\\
&=&\sum_{\mathbf{i}\in [M]^d} I_{\mathbf{i},h}(\mathbf{x}_0) \cdot \pmb{\sigma}(\mathbf{x}_0\, |\, \mathbf{x}_{0\mathbf{i}})^\top (\mathbf{I}_{d_q}\otimes \pmb{\gamma}^\top)^\top (\widehat{\pmb{\theta}}_{\mathbf{i}}-\widetilde{\pmb{\lambda}}_{\mathbf{i}})+O_P(h^p) 
\nonumber\\
&=&\sum_{\mathbf{i}\in [M]^d}\widetilde{\mathbf{x}}_{\mathbf{i},0}^\top (\widehat{\pmb{\theta}}_{\mathbf{i}}-\widetilde{\pmb{\lambda}}_{\mathbf{i}})+O_P(h^p) . 
\end{eqnarray}
where $\widetilde{\mathbf{x}}_{\mathbf{i},0}  =I_{\mathbf{i},h} (\mathbf{x}_0)(\mathbf{I}_{d_q}\otimes \pmb{\gamma}^\top)  \pmb{\sigma}( \mathbf{x}_0\, |\,\mathbf{x}_{0\mathbf{i}})$.

By \eqref{EqA.6} and \eqref{EqA.7}, we obtain

\begin{eqnarray*}
\sup_{\mathbf{x} \in C_{\mathbf{x}_{0\mathbf{i}},h}}\left\|\mathbf{D}^{-1}\mathbf{m}(\mathbf{x} \, |\, \mathbf{x}_{0\mathbf{i}})-(\mathbf{I}_{d_q}\otimes \pmb{\gamma}^\top)  \pmb{\sigma}( \mathbf{x} \, |\, \mathbf{x}_{0\mathbf{i}}) \right\| =O(h^{q+1}),
\end{eqnarray*}
which immediately yields

\begin{eqnarray*}
\left\| \mathbf{D}^{-1}I_{\mathbf{i},h}(\mathbf{x}_0)\cdot \mathbf{m}(\mathbf{x}_0\, |\, \mathbf{x}_{0\mathbf{i}})-\widetilde{\mathbf{x}}_{\mathbf{i}, 0} \right\| =O_P(h^{q+1}).
\end{eqnarray*}
Together with Lemma \ref{LemA.3} and \eqref{newpr}, it implies that
\begin{eqnarray}
\widehat{g}(\mathbf{x}_0)-g(\mathbf{x}_0)&=&\sum_{\mathbf{i}\in [M]^d}\widetilde{\mathbf{x}}_{\mathbf{i},0}^\top (\widehat{\pmb{\theta}}_{\mathbf{i}}-\widetilde{\pmb{\lambda}}_{\mathbf{i}})+O_P(h^p)
\nonumber\\
&=&\sum_{\mathbf{i}\in [M]^d} I_{\mathbf{i},h}(\mathbf{x}_0) \mathbf{m}(\mathbf{x}_0\, |\, \mathbf{x}_{0\mathbf{i}})^\top \mathbf{H}\bigl[\mathbf{H}^{-1} \mathbf{D}^{\top, -1}(\widehat{\pmb{\theta}}_{\mathbf{i}}-\widetilde{\pmb{\lambda}}_{\mathbf{i}})\bigr]+O_P\left(h^p\right)+o_P(1).
\end{eqnarray}

Therefore,  Lemma\ref{LemA.3} implies that $\sqrt{Th^d}(\widehat{g}(\mathbf{x}_0)-g(\mathbf{x}_0))$ is asymptotically normal with the asymptotic covariance being the limit of 
\begin{eqnarray*}
\widehat{\sigma}_{\mathbf{x}_0}^2&=&\sigma_\varepsilon^2\sum_{\mathbf{i}\in [M]^d} I_{\mathbf{i},h}(\mathbf{x}_0) \mathbf{m}(\mathbf{x}_0\, |\, \mathbf{x}_{0\mathbf{i}})^\top\mathbf{H} \pmb{\Sigma}_{\mathbf{i}}^{-1}\mathbf{H}\mathbf{m}(\mathbf{x}_0\, |\, \mathbf{x}_{0\mathbf{i}}),
\end{eqnarray*}
where $ \pmb{\Sigma}_{\mathbf{i}} =f_{\mathbf{x}}(\mathbf{x}_{0\mathbf{i}}) \int_{[-1,1]^d}  \mathbf{m}(\mathbf{x}\, |\, \mathbf{0})  \mathbf{m}(\mathbf{x}\, |\, \mathbf{0})^\top  \mathrm{d} \mathbf{x}$.\hspace*{\fill}{$\blacksquare$}

\bigskip

\noindent \textbf{Proof of Theorem \ref{Thm3.2}:}

In what follows, we label the quantities associated with the bootstrap procedure by the superscript $^*$, which will not be further explained unless misunderstanding may arise. 

By design, we have for $\forall \mathbf{i}$

\begin{eqnarray*}
\widehat{\pmb{\theta}}_{\mathbf{i}}^* - \widehat{\pmb{\theta}}_{\mathbf{i}}&=& \left(  \sum_{t=1}^T\widetilde{\mathbf{x}}_{\mathbf{i},t} \widetilde{\mathbf{x}}_{\mathbf{i},t} ^\top \right)^{-1} \sum_{t=1}^T\widetilde{\mathbf{x}}_{\mathbf{i},t} y_t^* - \widehat{\pmb{\theta}}_{\mathbf{i}}  \nonumber \\
&=&\left(  \sum_{t=1}^T\widetilde{\mathbf{x}}_{\mathbf{i},t} \widetilde{\mathbf{x}}_{\mathbf{i},t} ^\top \right)^{-1}   \sum_{t=1}^T\widetilde{\mathbf{x}}_{\mathbf{i},t} \widetilde{s}(\mathbf{x}_t \, |\, \widehat{\pmb{\Theta}})  - \widehat{\pmb{\theta}}_{\mathbf{i}}   +\left( \sum_{t=1}^T\widetilde{\mathbf{x}}_{\mathbf{i},t} \widetilde{\mathbf{x}}_{\mathbf{i},t} ^\top \right)^{-1}   \sum_{t=1}^T\widetilde{\mathbf{x}}_{\mathbf{i},t}  \widehat{\varepsilon}_t \eta_t\nonumber \\
&=&\left( \sum_{t=1}^T\widetilde{\mathbf{x}}_{\mathbf{i},t} \widetilde{\mathbf{x}}_{\mathbf{i},t} ^\top \right)^{-1}   \sum_{t=1}^T\widetilde{\mathbf{x}}_{\mathbf{i},t} \varepsilon_t \eta_t +\left( \sum_{t=1}^T\widetilde{\mathbf{x}}_{\mathbf{i},t} \widetilde{\mathbf{x}}_{\mathbf{i},t} ^\top \right)^{-1}   \sum_{t=1}^T\widetilde{\mathbf{x}}_{\mathbf{i},t}  (g(\mathbf{x}_t) - \widetilde{s}(\mathbf{x}_t \, |\, \widetilde{\pmb{\Lambda}})  )  \eta_t\nonumber \\
&&+\left( \sum_{t=1}^T\widetilde{\mathbf{x}}_{\mathbf{i},t} \widetilde{\mathbf{x}}_{\mathbf{i},t} ^\top \right)^{-1}   \sum_{t=1}^T\widetilde{\mathbf{x}}_{\mathbf{i},t}  (  \widetilde{s}(\mathbf{x}_t \, |\, \widetilde{\pmb{\Lambda}}) - \widetilde{s}(\mathbf{x}_t \, |\, \widehat{\pmb{\Theta}}) )  \eta_t \coloneqq \mathbf{A}_{\mathbf{i},1} + \mathbf{A}_{\mathbf{i},2} + \mathbf{A}_{\mathbf{i},3},
\end{eqnarray*}
where the second equality follows from the definition of $y_t^*$. 

Note that the term $\mathbf{A}_{\mathbf{i,2}}$ has been investigated in the proof of Lemma \ref{LemA.3}, and is negligible under the  condition $\sqrt{T}h^{p+d/2}\to 0$. For the term $\mathbf{A}_{\mathbf{i,3}}$, we can further write

\begin{eqnarray*}
\mathbf{A}_{\mathbf{i,3}} &=&\left( \sum_{t=1}^T\widetilde{\mathbf{x}}_{\mathbf{i},t} \widetilde{\mathbf{x}}_{\mathbf{i},t} ^\top \right)^{-1}   \sum_{t=1}^T\widetilde{\mathbf{x}}_{\mathbf{i},t}  \widetilde{\mathbf{x}}_{\mathbf{i},t} ^\top \eta_t \cdot (\widetilde{\pmb\lambda}_{\mathbf{i}} - \widehat{\pmb \theta}_{\mathbf{i}}) = o_P(\widetilde{\pmb\lambda}_{\mathbf{i}} - \widehat{\pmb \theta}_{\mathbf{i}}),
\end{eqnarray*}
where the second equality follows from the fact that $\{\eta_t \}$ are i.i.d. draws from $N(0,1)$ and are independent of the sample. 

Therefore, we only need to pay attention to $\mathbf{A}_{\mathbf{i},1}$ below. It suffices to consider $\frac{1}{\sqrt{T}}\sum_{t=1}^T\mathbf{\pmb\xi}_t^*$,
where $\mathbf{\pmb\xi}_t^* =\frac{1}{\sqrt{h^d}}\widetilde{\mathbf{x}}_{\mathbf{i},t} \varepsilon_t \eta_t$. As $\{\eta_t \}$ are i.i.d. draws from $N(0,1)$, it is easy to know that 

\begin{eqnarray*}
\text{Var}^*\left(\frac{1}{\sqrt{T}}\sum_{t=1}^T\mathbf{\pmb\xi}_t^*\right) \simeq \text{Var}\left(\frac{1}{\sqrt{T}}\sum_{t=1}^T\frac{1}{\sqrt{h^d}}\widetilde{\mathbf{x}}_{\mathbf{i},t}\varepsilon_t\right)
\end{eqnarray*}
in view of the proof of Lemma \ref{LemA.3}.

Below, we consider $E^*[\|\pmb{\xi}_1^*\|^2 \cdot I(\|\pmb{\xi}_1^*\| \ge \epsilon \sqrt{T})]$. Write

\begin{eqnarray*}
E^*[\|\pmb{\xi}_1^*\|^2 \cdot I(\|\pmb{\xi}_1^*\| \ge \epsilon \sqrt{T})]&\le &\{E^*\|\pmb{\xi}_1^*\|^{2\cdot \frac{2+\nu}{2}}\}^{\frac{2}{2+\nu}} \left\{ E^*[ I(\|\pmb{\xi}_1^*\| \ge \epsilon \sqrt{T})] \right\}^{\frac{\nu}{2+\nu}}\nonumber \\
&\le  &\{E^*\|\pmb{\xi}_1^*\|^{2+\nu}\}^{\frac{2}{2+\nu}} \left\{ \frac{E^*\|\pmb{\xi}_1^*\|^{2+\nu}}{\epsilon^{2+\nu} T^{\frac{2+\nu}{2}}}\right\}^{\frac{\nu}{2+\nu}} =  \frac{1}{\epsilon^{\nu} T^{\frac{\nu}{2}}}\left\{ E^*\|\pmb{\xi}_1^*\|^{2+\nu}\right\}^{\frac{1}{2+\nu} \cdot (2+\nu)} \nonumber \\
&\le &  O(1) \frac{1 }{\epsilon^{\nu} T^{\frac{\nu}{2}}}  \| \widetilde{\mathbf{m}}(\mathbf{x}_1\, |\, \mathbf{x}_{\mathbf{i},0})\varepsilon_1   \|^{2+\nu}\cdot E|\eta_1|^{2+\nu}\cdot \frac{1}{h^{\frac{d}{2} \cdot (2+\nu) }} 
\nonumber\\
& = &  O(1) \frac{1 }{\epsilon^{\nu} (Th^d)^{\frac{\nu}{2}}}  \cdot  \frac{1}{h^d}  \| \widetilde{\mathbf{m}}(\mathbf{x}_1\, |\, \mathbf{x}_{\mathbf{i},0}) \varepsilon_1 \|^{2+\nu} =  O_P(1) \frac{1}{ (Th^d)^{\frac{\nu}{2}}} ,
\end{eqnarray*}
where the first inequality follows from H\"older inequality, the second inequality follows from  Chebyshev's inequality,  the third inequality follows from the definition of $E^*$ and Minkowski inequality, and the last step follows from $\frac{1}{h^d}E  \| \widetilde{\mathbf{m}}(\mathbf{x}_1\, |\, \mathbf{x}_{\mathbf{i},0}) \varepsilon_1 \|^{2+\nu}=O(1)$ by the proof of Lemma \ref{LemA.3}. Consequently,

\begin{eqnarray*}
\frac{1}{T}\sum_{t=1}^{T}E^*[\|\pmb{\xi}_t^*\|^2 \cdot I(\|\pmb{\xi}_t^*\| \ge \epsilon \sqrt{T})] =O_P(1) \frac{1 }{ (Th^d)^{\frac{\nu}{2}}}=o_P(1),
\end{eqnarray*}
where the last step follows from $Th^d\to 0$. Therefore, the Lindberg condition is justified. Then the result follows. \hspace*{\fill}{$\blacksquare$}

\bigskip

\noindent \textbf{Proof of Corollary \ref{CoroA1}:}

The proof is a simpler version of that presented for Lemma \ref{LemA.3} and Theorem \ref{Thm3.1}, therefore it is omitted. \hspace*{\fill}{$\blacksquare$}

\subsection{Proofs for the Binary Model}\label{App.45}

\noindent \textbf{Proof of Lemma \ref{Lem34}:}

(1). First, note that provided $0<x, x_0<1$, we have the following two expressions by the following Taylor expansions:

\begin{eqnarray}
\log x &=&\log x_0 + (x -x_0)\frac{1}{x_0} - (x-x_0)^2\frac{1}{2(x^*)^2},\label{EqA.18} \\
\log (1-x) &=&\log (1-x_0) - (x -x_0)\frac{1}{1-x_0} -(x -x_0)^2\frac{1}{2(1-x^\dagger)^2} , \label{EqA.19}
\end{eqnarray}
where both $x^*$ and $x^\dagger$ lie between $x$ and $x_0$.

We are now ready to start our investigation. By \eqref{EqA.18} and \eqref{EqA.19}, write
\begin{eqnarray}\label{EqA.20}
&&\frac{1}{T}\log L(g(\cdot)) - \frac{1}{T}\log L(\widetilde{s}(\cdot\, | \, \pmb{\Theta}))\nonumber \\
&=& -\frac{1}{T} \sum_{t=1}^T(1-z_t) \left\{ \log  [1-\Phi_\eta(\widetilde{s}(\mathbf{x}_t\, | \, \pmb{\Theta}))]-\log  [1-\Phi_\eta(g(\mathbf{x}_t))]\right\}\nonumber \\
&&-\frac{1}{T} \sum_{t=1}^T z_t \left\{ \log \Phi_\eta(\widetilde{s}(\mathbf{x}_t\, | \, \pmb{\Theta})) -\log \Phi_\eta(g(\mathbf{x}_t)) \right\} \nonumber \\
&=& \frac{1}{T} \sum_{t=1}^T[\Phi_\eta(\widetilde{s}(\mathbf{x}_t\, | \, \pmb{\Theta})) - \Phi_\eta(g(\mathbf{x}_t)) ] \cdot  \frac{1-z_t}{1- \Phi_\eta(g(\mathbf{x}_t))}\nonumber \\
&&+\frac{1}{T} \sum_{t=1}^T[\Phi_\eta(\widetilde{s}(\mathbf{x}_t\, | \, \pmb{\Theta}))  - \Phi_\eta(g(\mathbf{x}_t)) ]^2 \cdot  \frac{1-z_t}{2(1-\Phi_t^\dagger)^2} \nonumber \\
&&- \frac{1}{T} \sum_{t=1}^T[\Phi_\eta(\widetilde{s}(\mathbf{x}_t\, | \, \pmb{\Theta})) - \Phi_\eta(g(\mathbf{x}_t)) ] \cdot  \frac{z_t}{\Phi_\eta(g(\mathbf{x}_t))}\nonumber \\
&&+\frac{1}{T} \sum_{t=1}^T[\Phi_\eta(\widetilde{s}(\mathbf{x}_t\, | \, \pmb{\Theta}))  - \Phi_\eta(g(\mathbf{x}_t)) ]^2 \cdot  \frac{z_t}{2(\Phi_t^*)^2}  \nonumber \\
&=& \frac{1}{T} \sum_{t=1}^T[\Phi_\eta(\widetilde{s}(\mathbf{x}_t\, | \, \pmb{\Theta}))  - \Phi_\eta(g(\mathbf{x}_t)) ] \cdot  \left[\frac{1-z_t}{1- \Phi_\eta(g(\mathbf{x}_t))}- \frac{z_t}{\Phi_\eta(g(\mathbf{x}_t))}\right]\nonumber \\
&&+\frac{1}{T} \sum_{t=1}^T[\Phi_\eta(\widetilde{s}(\mathbf{x}_t\, | \, \pmb{\Theta}))  - \Phi_\eta(g(\mathbf{x}_t)) ]^2 \cdot  \left[ \frac{1-z_t}{2(1-\Phi_t^\dagger)^2} + \frac{z_t}{2(\Phi_t^*)^2} \right]\nonumber \\
&\coloneqq &\mathbb{L}_{T,1} +\mathbb{L}_{T,2},
\end{eqnarray}
where both $\Phi_t^*$ and $\Phi_t^\dagger$ lie between $\Phi_\eta(\widetilde{s}(\mathbf{x}_t\, | \, \pmb{\Theta}))$ and $\Phi_\eta(g(\mathbf{x}_t))$, and the definitions of $\mathbb{L}_{T,1}$ and $\mathbb{L}_{T,2}$ are obvious. 

\bigskip

We then consider $\mathbb{L}_{T,1}$ and $\mathbb{L}_{T,2}$ respectively, and start with $\mathbb{L}_{T,1}$. For notational simplicity, we let

\begin{eqnarray*}
&&e_t =\frac{1-y_t}{1- \Phi_\eta(g(\mathbf{x}_t))}- \frac{y_t}{\Phi_\eta(g(\mathbf{x}_t))} ,\nonumber \\
&&\Delta \Phi_\eta(g(\mathbf{x}_t)) =\Phi_\eta(\widetilde{s}(\mathbf{x}_t\, | \, \pmb{\Theta}))  - \Phi_\eta(g(\mathbf{x}_t)).
\end{eqnarray*}
Simple algebra shows that  

\begin{eqnarray}\label{EqA.21}
&&E[e_t\, |\, \mathbf{x}_t]=0,\nonumber \\  
&& E[e_t^2\, |\, \mathbf{x}_t] = \frac{1}{\Phi_\eta(g(\mathbf{x}_t))[1-\Phi_\eta(g(\mathbf{x}_t))]},\nonumber \\
&&|\Delta \Phi_\eta(g(\mathbf{x}_t)) |\le 2.
\end{eqnarray}

For any given $\Delta \Phi_\eta(g(\cdot))$, we then consider

\begin{eqnarray}\label{EqA.22}
&&E\left\| \frac{1}{T} \sum_{t=1}^T \Delta \Phi_\eta(g(\mathbf{x}_t))  e_t\right\|^2\nonumber \\
&= & \frac{1}{T^2} \sum_{t=1}^T  E[[\Delta \Phi_\eta(g(\mathbf{x}_t))  ]^2\cdot e_t^2] \nonumber \\
&&+\frac{1}{T^2} \sum_{t=1}^T (1-t/T) E[[\Delta \Phi_\eta(g(\mathbf{x}_1)) ][\Delta \Phi_\eta(g(\mathbf{x}_{t+1})) ]\cdot e_1e_{t+1}] \nonumber \\
&\le & \frac{4}{T^2} \sum_{t=1}^T  E\left[ \frac{1}{\Phi_\eta(g(\mathbf{x}_t))[1-\Phi_\eta(g(\mathbf{x}_t))] } \right]\nonumber \\
&&+\frac{1}{T^2} \sum_{t=1}^T (1-t/T)\alpha(t)^{\nu/(2+\nu)} \{E[e_1^{2+\nu}]\}^{2/(2+\nu)} = O\left(\frac{1}{T} \right),
\end{eqnarray}
where the inequality follows from Assumption \ref{Ass.3}.1, and Davydov's inequality and \eqref{EqA.21}. By Lemmas A1 and A2 of \cite{WP2003}, we immediately obtain that

\begin{eqnarray}\label{EqA.23}
\sup_{\widetilde{s}\in \mathcal{S}}|\mathbb{L}_{T,1}| =o_P(1).
\end{eqnarray}

\medskip

We next investigate $\mathbb{L}_{T,2}$. Write

\begin{eqnarray*}
\mathbb{L}_{T,2}&=&\frac{1}{T} \sum_{t=1}^T[\Delta \Phi_\eta(g(\mathbf{x}_t))  ]^2 \cdot  \left[ \frac{1-z_t}{2(1-\Phi_t^\dagger)^2} + \frac{z_t}{2(\Phi_t^*)^2} \right]\nonumber \\
&\ge &\frac{1}{T} \sum_{t=1}^T[\Delta \Phi_\eta(g(\mathbf{x}_t))  ]^2 \cdot  \left\{ \frac{1-z_t}{4[1+(\Phi_t^\dagger)^2]} + \frac{z_t}{2(\Phi_t^*)^2} \right\}  \nonumber \\
&\ge &\frac{1}{T} \sum_{t=1}^T[\Delta \Phi_\eta(g(\mathbf{x}_t)) ]^2 \cdot  \left\{ \frac{1-z_t}{4\cdot 2} + \frac{z_t}{2} \right\} \nonumber \\
&\ge &\frac{1}{8}\cdot \frac{1}{T}\sum_{t=1}^T[\Delta \Phi_\eta(g(\mathbf{x}_t))  ]^2  ,
\end{eqnarray*}
where the first inequality follows from $\frac{1}{(a+b)^2} \ge \frac{1}{2a^2 + 2b^2}$ because of $(a+b)^2\le 2a^2 + 2b^2$, the second inequality follows from the fact that $\Phi_t^*$ and $\Phi_t^\dagger$ lie between $\Phi_\eta(\widetilde{s}(\mathbf{x}_t\, | \, \pmb{\Theta}))$ and $\Phi_\eta(g(\mathbf{x}_t))$, and the third inequality follows from that $ \frac{1-y_t}{4\cdot 2} + \frac{y_t}{2} \ge \frac{1}{8}$ because of $z_t$ taking the value of 1 or 0 only.

By the fact that $0\ge\frac{1}{T}\log L(g(\cdot)) - \frac{1}{T}\log L(\widetilde{s}(\cdot\, | \, \widehat{\pmb{\Theta}}))$, and \eqref{EqA.20} and \eqref{EqA.23}, we now conclude that
\begin{eqnarray*}
o_P(1)=\frac{1}{T}\sum_{t=1}^T[\Phi_\eta(\widetilde{s}(\mathbf{x}_t\, | \, \widehat{\pmb{\Theta}}))  - \Phi_\eta(g(\mathbf{x}_t)) ]^2 \asymp \frac{1}{T}\sum_{t=1}^T(g(\mathbf{x}_t)-\widetilde{s}(\mathbf{x}_t\, | \, \widehat{\pmb{\Theta}}) )^2 ,
\end{eqnarray*}
which completes the proof of this lemma.  \hspace*{\fill}{$\blacksquare$}

\bigskip

By Lemma \ref{Lem34}, it is obvious that

\begin{eqnarray*}
&&\frac{1}{T}\sum_{t=1}^T(\widetilde{s}(\mathbf{x}_t\, | \, \widetilde{\pmb{\Lambda}}) -\widetilde{s}(\mathbf{x}_t\, | \, \widehat{\pmb{\Theta}}) )^2 \nonumber \\
&=&\frac{1}{T}\sum_{t=1}^T(\widetilde{s}(\mathbf{x}_t\, | \, \widetilde{\pmb{\Lambda}}) - g(\mathbf{x}_t)+g(\mathbf{x}_t)-\widetilde{s}(\mathbf{x}_t\, | \, \widehat{\pmb{\Theta}}) )^2\nonumber \\
&\le & \frac{2}{T}\sum_{t=1}^T(\widetilde{s}(\mathbf{x}_t\, | \, \widetilde{\pmb{\Lambda}}) - g(\mathbf{x}_t)  )^2+\frac{2}{T}\sum_{t=1}^T( g(\mathbf{x}_t)-\widetilde{s}(\mathbf{x}_t\, | \, \widehat{\pmb{\Theta}}) )^2\nonumber \\
&=& O_P\left(h^{2p}  \right) +o_P(1)=o_P(1),
\end{eqnarray*}
where the third equality follows from the first result of Lemma \ref{Lem34} and Lemma \ref{LEM2.3}. 

\bigskip

\noindent \textbf{Proof of Lemma \ref{LemA.4}:}

(1). Note that by Lemma \ref{LemA.4}, we can write

\begin{eqnarray*}
o_P(1) &=& \frac{1}{T}\sum_{t=1}^T(g(\mathbf{x}_t)-\widetilde{s}(\mathbf{x}_t\, | \, \widehat{\pmb{\Theta}}) )^2= \frac{1}{T}\sum_{t=1}^T\left(\sum_{\mathbf{x}_t\in C_{\mathbf{x}_{0\mathbf{i}}, h}} [g(\mathbf{x}_t)-s(\mathbf{x}_t \, | \, \mathbf{x}_{\mathbf{i},0}, \widehat{\pmb{\theta}}_{\mathbf{i}} )] \right)^2 \nonumber \\
&=&\sum_{\mathbf{i}\in [M]^d} \frac{1}{T}\sum_{t=1}^TI_{\mathbf{i},h}(\mathbf{x}_t) [g(\mathbf{x}_t) - s(\mathbf{x}_t \, | \, \mathbf{x}_{\mathbf{i},0}, \widehat{\pmb{\theta}}_{\mathbf{i}} )]^2\ge 0,
\end{eqnarray*}
where the third equality follows from the fact that $\mathbf{x}_t$ can not simultaneous belong to $C_{\mathbf{x}_{0\mathbf{i}},h}$ and $C_{\mathbf{x}_{0\mathbf{j}},h}$ for $\mathbf{i}\ne \mathbf{j}$. Thus, we must have

\begin{eqnarray*}
 \frac{1}{T}\sum_{t=1}^TI_{\mathbf{i},h}(\mathbf{x}_t) [g(\mathbf{x}_t) - s(\mathbf{x}_t \, | \, \mathbf{x}_{\mathbf{i},0}, \widehat{\pmb{\theta}}_{\mathbf{i}} )]^2 =o_P(1),
\end{eqnarray*}
which completes the proof of the first result.

\medskip

(2). By \eqref{EqA.24}, we denote

\begin{eqnarray*}
\frac{\partial^2 \log L(\widetilde{s}(\cdot\, | \, \pmb{\Theta})) }{\partial \pmb{\theta}_{\mathbf{i}} \partial \pmb{\theta}_{\mathbf{i}}^\top } \coloneqq-\mathbf{L}_1(s(\cdot \, | \, \mathbf{x}_{\mathbf{i},0},\pmb{\theta}_{\mathbf{i}} )) + \mathbf{L}_2( s(\cdot \, | \, \mathbf{x}_{\mathbf{i},0},\pmb{\theta}_{\mathbf{i}} ) ) -\mathbf{L}_3( s(\cdot \, | \, \mathbf{x}_{\mathbf{i},0},\pmb{\theta}_{\mathbf{i}} )),
\end{eqnarray*}
where the definitions of $ \mathbf{L}_j(\cdot)$ for $j=1,2,3$ should be obvious.

First, we consider $\mathbf{L}_2(s(\cdot \, | \, \mathbf{x}_{\mathbf{i},0},\pmb{\theta}_{\mathbf{i}} ))$ and $\mathbf{L}_3(s(\cdot \, | \, \mathbf{x}_{\mathbf{i},0},\pmb{\theta}_{\mathbf{i}} ))$. For $\mathbf{L}_2(s(\cdot \, | \, \mathbf{x}_{\mathbf{i},0},\pmb{\theta}_{\mathbf{i}} ))$, we write

\begin{eqnarray}\label{EqA.25}
&&\frac{1}{T}\widetilde{\mathbf{L}}_2(s(\cdot \, | \, \mathbf{x}_{\mathbf{i},0},\pmb{\theta}_{\mathbf{i}} )) =\frac{1}{T}\mathbf{H}\mathbf{D}\mathbf{L}_2(s(\cdot \, | \, \mathbf{x}_{\mathbf{i},0},\pmb{\theta}_{\mathbf{i}} )) \mathbf{D}^\top \mathbf{H}\nonumber \\
&=&\frac{1}{T}\sum_{t=1}^T[z_t -\Phi_\eta(s(\mathbf{x}_t \, | \, \mathbf{x}_{\mathbf{i},0},\pmb{\theta}_{\mathbf{i}} ) )] \mathbf{H}\mathbf{D} \widetilde{\mathbf{X}}_{\mathbf{i}, t} (s(\mathbf{x}_t \, | \, \mathbf{x}_{\mathbf{i},0},\pmb{\theta}_{\mathbf{i}} )) \mathbf{D}^\top \mathbf{H}\nonumber \\
&= & \frac{1}{T}\sum_{t=1}^T[\Phi_\eta(g(\mathbf{x}_t)) -\Phi_\eta(s(\mathbf{x}_t \, | \, \mathbf{x}_{\mathbf{i},0},\pmb{\theta}_{\mathbf{i}} ))] \mathbf{H}\mathbf{D} \widetilde{\mathbf{X}}_{\mathbf{i}, t} (s(\mathbf{x}_t \, | \, \mathbf{x}_{\mathbf{i},0},\pmb{\theta}_{\mathbf{i}} )) \mathbf{D}^\top \mathbf{H} \nonumber \\
&&+  \frac{1}{T}\sum_{t=1}^T[z_t -\Phi_\eta(g(\mathbf{x}_t))] \mathbf{H}\mathbf{D} \widetilde{\mathbf{X}}_{\mathbf{i}, t} (s(\mathbf{x}_t\, | \, \mathbf{x}_{\mathbf{i},0},\pmb{\theta}_{\mathbf{i}} )) \mathbf{D}^\top \mathbf{H},
\end{eqnarray}
where the definition of $ \widetilde{\mathbf{X}}_{\mathbf{i}, t} (\cdot) $ is obvious.

For the first term on the right hand side of \eqref{EqA.25}, we write

\begin{eqnarray*}
&&\left\|  \frac{1}{T}\sum_{t=1}^T[\Phi_\eta(g(\mathbf{x}_t)) -\Phi_\eta(s(\mathbf{x}_t \, | \, \mathbf{x}_{\mathbf{i},0}, \widehat{\pmb{\theta}}_{\mathbf{i}} ) )] \mathbf{H}\mathbf{D} \widetilde{\mathbf{X}}_{\mathbf{i}, t} (s(\mathbf{x}_t \, | \, \mathbf{x}_{\mathbf{i},0}, \widehat{\pmb{\theta}}_{\mathbf{i}} )  ) \mathbf{D}^\top \mathbf{H}\right\| \nonumber \\
&\le &\left\{  \frac{1}{T}\sum_{t=1}^TI_{\mathbf{i},h}(\mathbf{x}_t) [\Phi_\eta(g(\mathbf{x}_t)) -\Phi_\eta(s(\mathbf{x}_t \, | \, \mathbf{x}_{\mathbf{i},0}, \widehat{\pmb{\theta}}_{\mathbf{i}} ) )]^2 \right\}^{1/2}\nonumber \\
&&\cdot\left\{  \frac{1}{T}\sum_{t=1}^T\| \mathbf{H}\mathbf{D} \widetilde{\mathbf{X}}_{\mathbf{i}, t} (s(\mathbf{x}_t \, | \, \mathbf{x}_{\mathbf{i},0}, \widehat{\pmb{\theta}}_{\mathbf{i}} ) ) \mathbf{D}^\top \mathbf{H} \|^2\right\}^{1/2} \nonumber \\
&\le &O(1)\left\{  \frac{1}{T}\sum_{t=1}^TI_{\mathbf{i},h}(\mathbf{x}_t) [ g(\mathbf{x}_t) - s(\mathbf{x}_t \, | \, \mathbf{x}_{\mathbf{i},0}, \widehat{\pmb{\theta}}_{\mathbf{i}} )]^2 \right\}^{1/2} \nonumber \\
&&\cdot\left\{  \frac{1}{T}\sum_{t=1}^T\| \mathbf{H}\mathbf{D} \widetilde{\mathbf{X}}_{\mathbf{i}, t} (s(\mathbf{x}_t \, | \, \mathbf{x}_{\mathbf{i},0}, \widehat{\pmb{\theta}}_{\mathbf{i}} ) ) \mathbf{D}^\top \mathbf{H} \|^2\right\}^{1/2} \nonumber \\
&=&o_P(1),
\end{eqnarray*}
where the first inequality follows from Cauchy-Schwarz inequality, the second inequality follows from Mean-Value Theorem and the fact that $\phi_\eta(\cdot)$ is uniformly bounded, and the last step follows from the first result of this lemma and \eqref{EqA.10}.

For the second term on the right hand side of \eqref{EqA.25}, by some tedious algebra and the first result of this lemma, it is not hard to see that

\begin{eqnarray*}
&&\frac{1}{T}\sum_{t=1}^T[z_t -\Phi_\eta(g(\mathbf{x}_t))] \mathbf{H}\mathbf{D} \widetilde{\mathbf{X}}_{\mathbf{i}, t} (s(\mathbf{x}_t\, | \, \mathbf{x}_{\mathbf{i},0}, \widehat{\pmb{\theta}}_{\mathbf{i}} )) \mathbf{D}^\top \mathbf{H}\nonumber \\
&=& \frac{1}{T}\sum_{t=1}^T[z_t -\Phi_\eta(g(\mathbf{x}_t))]  \mathbf{H}\mathbf{D} \widetilde{\mathbf{X}}_{\mathbf{i}, t} (g(\mathbf{x}_t))\mathbf{D}^\top \mathbf{H}\cdot (1+o_P(1)).
\end{eqnarray*}
Further, using Assumption \ref{Ass.3} and Billingsley's inequality following a  procedure similar (but simplified)  as in \eqref{EqA.22}, and in connection with \eqref{EqA.10}, we can show that  

\begin{eqnarray*}
&& \left\|  \frac{1}{T}\sum_{t=1}^T[z_t -\Phi_\eta(g(\mathbf{x}_t))] \mathbf{H}\mathbf{D} \widetilde{\mathbf{X}}_{ \mathbf{i}, t} (g(\mathbf{x}_t))\mathbf{D}^\top \mathbf{H}\right\|  =o_P(1).
\end{eqnarray*}
Based on the above development, we are readily to conclude that

\begin{eqnarray*}
\frac{1}{T}\| \widetilde{\mathbf{L}}_2(s(\mathbf{x}_t\, | \, \mathbf{x}_{\mathbf{i},0}, \widehat{\pmb{\theta}}_{\mathbf{i}} ))  \|=o_P(1).
\end{eqnarray*}

Similar to the analysis of $ \widetilde{\mathbf{L}}_2(s(\mathbf{x}_t\, | \, \mathbf{x}_{\mathbf{i},0}, \widehat{\pmb{\theta}}_{\mathbf{i}} ))$, we can also obtain that

\begin{eqnarray*}
\frac{1}{T}\|  \mathbf{H}\mathbf{D}\mathbf{L}_3(s(\mathbf{x}_t\, | \, \mathbf{x}_{\mathbf{i},0}, \widehat{\pmb{\theta}}_{\mathbf{i}} ))\mathbf{D}^\top \mathbf{H} \|=o_P(1).
\end{eqnarray*}

Below, we focus on $\mathbf{L}_1(s(\mathbf{x}_t\, | \, \mathbf{x}_{\mathbf{i},0}, \widehat{\pmb{\theta}}_{\mathbf{i}} ))$, and write

\begin{eqnarray*}
&&\frac{1}{T} \mathbf{H}\mathbf{D}\mathbf{L}_1(s(\mathbf{x}_t\, | \, \mathbf{x}_{\mathbf{i},0}, \widehat{\pmb{\theta}}_{\mathbf{i}} ) ) \mathbf{D}^\top \mathbf{H}\nonumber \\
&=& \frac{1}{T}\mathbf{H}\mathbf{D}\mathbf{L}_1(g(\mathbf{x}_t)) \mathbf{D}^\top \mathbf{H}+\frac{1}{T}\mathbf{H}\mathbf{D}[ \mathbf{L}_1(s(\mathbf{x}_t\, | \, \mathbf{x}_{\mathbf{i},0}, \widehat{\pmb{\theta}}_{\mathbf{i}} ) ) -\mathbf{L}_1(g(\mathbf{x}_t))] \mathbf{D}^\top \mathbf{H}\nonumber \\
&=&\frac{1}{T}\mathbf{H}\mathbf{D}\mathbf{L}_1(g(\mathbf{x}_t))\mathbf{D}^\top \mathbf{H} +o_P(1) \nonumber \\
&=& E\left[ \frac{[\phi_\eta(g(\mathbf{x}_1))]^2 }{[1- \Phi_\eta(g(\mathbf{x}_1))] \Phi_\eta(g(\mathbf{x}_1))} I_{\mathbf{i},h}(\mathbf{x}_1) \mathbf{H}\mathbf{m}(\mathbf{x}_1\, |\, \mathbf{x}_{\mathbf{i},0})\mathbf{m}(\mathbf{x}_1\, |\, \mathbf{x}_{\mathbf{i},0})^\top \mathbf{H} \right] +o_P(1)\nonumber \\
&=&   \frac{f_{\mathbf{x}}(\mathbf{x}_{\mathbf{i},0}) \phi_\eta(g(\mathbf{x}_{\mathbf{i},0}))^2 }{[1- \Phi_\eta(g(\mathbf{x}_{\mathbf{i},0}))] \Phi_\eta(g(\mathbf{x}_{\mathbf{i},0}))} \int_{[-1,1]^d}  \mathbf{m}(\mathbf{x}\, |\, \mathbf{0})  \mathbf{m}(\mathbf{x}\, |\, \mathbf{0})^\top  \mathrm{d} \mathbf{x}  +o_P(1),
\end{eqnarray*}
where the second equality follows from similar steps as those for $\widetilde{\mathbf{L}}_2(s(\mathbf{x}_t\, | \, \mathbf{x}_{\mathbf{i},0}, \widehat{\pmb{\theta}}_{\mathbf{i}} )) $, the third equality follows from a proof similar to those for \eqref{EqA.13}, and the fourth equality follows from a development similar to \eqref{EqA.12}.

Thus, we can now conclude that for each $\mathbf{i}$

\begin{eqnarray*}
\left\|\frac{1}{T}\mathbf{H}\mathbf{D}\frac{\partial \log L(\widetilde{s}(\cdot\, | \, \widehat{\pmb{\Theta}})) }{\partial \pmb{\theta}_{\mathbf{i}} \partial \pmb{\theta}_{\mathbf{i}}^\top }\mathbf{D}^\top \mathbf{H} -\widetilde{\pmb{\Sigma}}_{\mathbf{i}}\right\| =o_P(1),
\end{eqnarray*}
where $\widetilde{\pmb{\Sigma}}_{\mathbf{i}}$ is defined in the body of this lemma.

The proof of the second result is now completed.

\medskip

(3). In view of the fact that $\pmb{\theta}_{\mathbf{i}}^*$ lies between $\widehat{\pmb{\theta}}_{\mathbf{i}}$ and $ \widetilde{\pmb{\lambda}}_{\mathbf{i}}$, the result follows immediately by going through the same procedure as the second result of this lemma. 

\medskip

(4). Write

\begin{eqnarray*}
&&\frac{1}{\sqrt{Th^d}} \mathbf{H} \mathbf{D}\frac{\partial \log L(g(\cdot) ) }{\partial \pmb{\theta}_{\mathbf{i}}  }  \nonumber \\
&=&\frac{1}{\sqrt{Th^d}}  \sum_{t=1}^T  \frac{[ z_t - \Phi_\eta(g(\mathbf{x}_t))]\cdot \phi_\eta(g(\mathbf{x}_t))}{\Phi_\eta(g(\mathbf{x}_t)) [1-\Phi_\eta(g(\mathbf{x}_t) )]}   \mathbf{H} \mathbf{D}\widetilde{\mathbf{x}}_{\mathbf{i}, t} \nonumber \\
&=&\frac{1}{\sqrt{Th^d}}  \sum_{t=1}^T   \frac{[ z_t - \Phi_\eta(g(\mathbf{x}_t))]\cdot \phi_\eta(g(\mathbf{x}_t))}{\Phi_\eta(g(\mathbf{x}_t)) [1-\Phi_\eta(g(\mathbf{x}_t) )]}   \widetilde{\mathbf{m} }(\mathbf{x}_t\, |\, \mathbf{x}_{\mathbf{i},0}) \cdot (1+o_P(1))\nonumber \\
&=&\frac{1}{\sqrt{Th^d}}  \sum_{t=1}^T  u_t \widetilde{\mathbf{m} }(\mathbf{x}_t\, |\, \mathbf{x}_{\mathbf{i},0}) \cdot (1+o_P(1))
\end{eqnarray*}
where $\widetilde{\mathbf{m} }(\mathbf{x}_t\, |\, \mathbf{x}_{\mathbf{i},0})$ is defined in \eqref{EqA.11},  the second equality follows from \eqref{EqA.10}, and in the third equality we let  

\begin{eqnarray*}
\frac{[ z_t - \Phi_\eta(g(\mathbf{x}_t))]\cdot \phi_\eta(g(\mathbf{x}_t))}{\Phi_\eta(g(\mathbf{x}_t)) [1-\Phi_\eta(g(\mathbf{x}_t) )]} \coloneqq u_t 
\end{eqnarray*}
for notational simplicity. Moreover, as $g$ is defined on $[-a, a]$, it is easy to know that $ 0<\mathtt{c}\le \Phi_\eta(g(\mathbf{x}_t))\le \mathtt{C}<1$. Thus, we can further write

\begin{eqnarray*}
|u_t|\le  \phi_\eta(g(\mathbf{x}_t))\left( \frac{ 1}{\Phi_\eta(g(\mathbf{x}_t))  }  \vee \frac{1}{ 1-\Phi_\eta(g(\mathbf{x}_t) )}  \right) =O(1).
\end{eqnarray*}
Also, simple algebra shows that

\begin{eqnarray*}
E[u_t^2\, | \, \mathbf{x}_t] &=&\frac{ \phi_\eta(g(\mathbf{x}_t))^2 }{\Phi_\eta(g(\mathbf{x}_t))^2 [1-\Phi_\eta(g(\mathbf{x}_t) )]^2}  E\left[z_t^2 -2z_t  \Phi_\eta(g(\mathbf{x}_t))+\Phi_\eta(g(\mathbf{x}_t))^2  \, | \, \mathbf{x}_t\right] \nonumber \\
&=&\frac{ \phi_\eta(g(\mathbf{x}_t))^2 }{\Phi_\eta(g(\mathbf{x}_t))  [1-\Phi_\eta(g(\mathbf{x}_t) )] }  .
\end{eqnarray*}

\medskip

We now move on and write

\begin{eqnarray*}
&&E\left[\left(\frac{1}{\sqrt{Th^d}}  \sum_{t=1}^T   u_t   \widetilde{\mathbf{m} }(\mathbf{x}_t\, |\, \mathbf{x}_{\mathbf{i},0}) \right)\left(\frac{1}{\sqrt{Th^d}}  \sum_{t=1}^T   u_t   \widetilde{\mathbf{m} }(\mathbf{x}_t\, |\, \mathbf{x}_{\mathbf{i},0}) \right)^\top \right] \nonumber \\
&=&\frac{1}{Th^d} \sum_{t=1}^T   E\left[  \widetilde{\mathbf{m} }(\mathbf{x}_t\, |\, \mathbf{x}_{\mathbf{i},0})  \widetilde{\mathbf{m} }(\mathbf{x}_t\, |\, \mathbf{x}_{\mathbf{i},0})^\top u_t^2 \right] \nonumber \\
&&+ \frac{1}{Th^d}\sum_{t\ne s}^{T}E [\widetilde{\mathbf{m}}(\mathbf{x}_t\, |\, \mathbf{x}_{\mathbf{i},0})  \widetilde{\mathbf{m}}(\mathbf{x}_{s}\, |\, \mathbf{x}_{\mathbf{i},0})  ^\top u_tu_{s} ]  .
\end{eqnarray*}
Similar to \eqref{jitith21}, we have

\begin{eqnarray*}
\left\| \frac{1}{Th^d}\sum_{t\ne s}^{T}E [\widetilde{\mathbf{m}}(\mathbf{x}_t\, |\, \mathbf{x}_{\mathbf{i},0})  \widetilde{\mathbf{m}}(\mathbf{x}_{s}\, |\, \mathbf{x}_{\mathbf{i},0})  ^\top u_tu_{s} ] \right\|=o(1) .
\end{eqnarray*} 
Thus , we can conclude that

\begin{eqnarray*} 
&&E\left[ \left(\frac{1}{\sqrt{T}}\sum_{t=1}^T\widetilde{\mathbf{m}}(\mathbf{x}_t\, |\, \mathbf{x}_{\mathbf{i},0}) u_t\right) \left(\frac{1}{\sqrt{T}}\sum_{t=1}^T\widetilde{\mathbf{m}}(\mathbf{x}_t\, |\, \mathbf{x}_{\mathbf{i},0}) u_t\right)^\top \right]\nonumber \\
&=&\frac{1}{T}\sum_{t=1}^T  E \left[\frac{ \phi_\eta(g(\mathbf{x}_t))^2 }{\Phi_\eta(g(\mathbf{x}_t))  [1-\Phi_\eta(g(\mathbf{x}_t) )] } \widetilde{\mathbf{m}}(\mathbf{x}_t\, |\, \mathbf{x}_{\mathbf{i},0})  \widetilde{\mathbf{m}}(\mathbf{x}_t\, |\, \mathbf{x}_{\mathbf{i},0})  ^\top \right] +o(1)\nonumber \\
&\to &  \frac{  f_{\mathbf{x}}(\mathbf{x}_{\mathbf{i},0})  \phi_\eta(g(\mathbf{x}_{\mathbf{i},0}))^2 }{\Phi_\eta(g(\mathbf{x}_{\mathbf{i},0}))  [1-\Phi_\eta(g(\mathbf{x}_{\mathbf{i},0}) )] } \int_{[-1,1]^d}  \mathbf{m}(\mathbf{x}\, |\, \mathbf{0})  \mathbf{m}(\mathbf{x}\, |\, \mathbf{0})^\top  \mathrm{d} \mathbf{x},
\end{eqnarray*}
where the last step follows from a procedure similar to \eqref{EqA.12}.

\medskip

Below, we further use small-block and large-block to prove the normality. To employ the small-block and large-block arguments, we partition the set $\{1,\ldots, T \}$ into $2k_T+1$ subsets with large blocks of size $l_T$ and small blocks of size $s_T$ and the last remaining set of size $T-k_T(l_T+s_T)$, where $l_T$ and $s_T$ are selected such that

\begin{eqnarray*}
s_T\to \infty,\quad  \frac{s_T}{l_T}\to 0,\quad  \frac{l_T^3 }{ Th^d }\to 0,\quad\text{and}\quad k_T\equiv \left\lfloor \frac{T}{l_T+s_T} \right\rfloor,
\end{eqnarray*}
where $\nu$ is defined in Assumption \ref{Ass.3}.1

For $j=1,\ldots, k_T$, define

\begin{eqnarray*}
&&\pmb{\xi}_{j,1} = \sum_{t=(j-1)(l_T+s_T)+1}^{jl_T+(j-1)s_T} \frac{1}{\sqrt{h^d}}\widetilde{\mathbf{m}}(\mathbf{x}_t\, |\, \mathbf{x}_{\mathbf{i},0}) u_t,\quad \pmb{\xi}_{j,2} = \sum_{t=jl_T+ (j-1)s_T+1}^{j(l_T+s_T)} \frac{1}{\sqrt{h^d}}\widetilde{\mathbf{m}}(\mathbf{x}_t\, |\, \mathbf{x}_{\mathbf{i},0})  u_t ,\nonumber \\
&&\pmb{\xi}_{0} =\sum_{t=k_T(l_T+s_T)+1}^T \frac{1}{\sqrt{h^d}}\widetilde{\mathbf{m}}(\mathbf{x}_t\, |\, \mathbf{x}_{\mathbf{i},0}) u_t.
\end{eqnarray*}
Note that $\alpha(T) = o(1/T)$ and $k_Ts_T/T\to 0$. By direct calculation, we immediately obtain that

\begin{eqnarray*}
\frac{1}{T}E\left\|\sum_{j=1}^{k_T} \pmb{\xi}_{j,2}\right\|^2\to 0\quad \text{and}\quad \frac{1}{T}E\left\| \pmb{\xi}_{0}\right\|^2\to 0.
\end{eqnarray*}
Therefore,

\begin{eqnarray*}
\frac{1}{\sqrt{Th^d}}\sum_{t=1}^T\widetilde{\mathbf{m}}(\mathbf{x}_t\, |\, \mathbf{x}_{\mathbf{i},0})   u_t = \frac{1}{\sqrt{T}}\sum_{j=1}^{k_T}\pmb{\xi}_{j,1} +o_P(1).
\end{eqnarray*}
By Proposition 2.6 of \cite{FanYao}, we have as $T\to 0$

\begin{eqnarray*}
&&\left| E\left[\exp\left( \frac{iw}{\sqrt{T}}\sum_{j=1}^{k_T}\pmb{\xi}_{j,1} \right) \right] -\prod_{j=1}^{k_T}E\left[\exp\left( \frac{iw \pmb{\xi}_{j,1} }{\sqrt{T}} \right) \right]\right| \nonumber \\
&\le & 16(k_T-1) \alpha(s_T)\to 0,
\end{eqnarray*}
where $i$ is the imaginary unit. Thus, the Feller condition is fulfilled as follows:

\begin{eqnarray*}
\frac{1}{T}\sum_{j=1}^{k_T}E[\pmb{\xi}_{j,1} \pmb{\xi}_{j,1} ^\top]\to  \frac{  f_{\mathbf{x}}(\mathbf{x}_{\mathbf{i},0})  \phi_\eta(g(\mathbf{x}_{\mathbf{i},0}))^2 }{\Phi_\eta(g(\mathbf{x}_{\mathbf{i},0}))  [1-\Phi_\eta(g(\mathbf{x}_{\mathbf{i},0}) )] } \int_{[-1,1]^d}  \mathbf{m}(\mathbf{x}\, |\, \mathbf{0})  \mathbf{m}(\mathbf{x}\, |\, \mathbf{0})^\top  \mathrm{d} \mathbf{x}.
\end{eqnarray*}

\medskip

Also, we note that

\begin{eqnarray*}
E[\|\pmb{\xi}_{j,1} \|^2 \cdot I(\|\pmb{\xi}_{j,1} \| \ge \epsilon \sqrt{T})] &\le &\{E\|\pmb{\xi}_{j,1} \|^4\}^{\frac{1}{2}} \left\{ E[ I(\|\pmb{\xi}_{j,1} \| \ge \epsilon \sqrt{T})] \right\}^{\frac{1}{2}}\nonumber \\
&\le  &\{E\|\pmb{\xi}_{j,1} \|^4\}^{\frac{1}{2}}  \left\{ \frac{E\|\pmb{\xi}_{j,1} \|^{4}}{\epsilon^{4} T^2}\right\}^{\frac{1}{2}}\nonumber \\
&= &  \frac{1}{\epsilon^2 T }\left\{ E\|\pmb{\xi}_{j,1} \|^4\right\}^{\frac{1}{4} \cdot 4}=  O(1) \frac{l_T^4 }{\epsilon^2 Th^d } ,
\end{eqnarray*}
where the first inequality follows from H\"older inequality, the second inequality follows from  Chebyshev's inequality, and the last step follows from Minkowski inequality. Consequently,

\begin{eqnarray*}
\frac{1}{T}\sum_{j=1}^{k_T}E[\|\pmb{\xi}_{j,1} \|^2 \cdot I(\|\pmb{\xi}_{j,1} \| \ge \epsilon \sqrt{T})] =O\left( \frac{k_Tl_T^4 }{ T \cdot Th^d }\right) = O\left( \frac{l_T^3 }{ Th^d }\right)=o(1),
\end{eqnarray*}
which is the Lindberg condition.  Using a Cram{\'e}r-Wold device, the CLT follows immediately by the standard argument. \hspace*{\fill}{$\blacksquare$}

\bigskip

\noindent \textbf{Proof of Theorem \ref{Thm3.3}:}

By the first order condition, we have

\begin{eqnarray*}
0 =\frac{\partial \log L(\widetilde{s}(\cdot\, | \, \widehat{\pmb{\Theta}})) }{\partial \pmb{\theta}_{\mathbf{i}}  }=\frac{\partial \log L(s(\cdot\, | \,  \mathbf{x}_{\mathbf{i},0}, \widehat{\pmb{\theta}}_{\mathbf{i}})) }{\partial \pmb{\theta}_{\mathbf{i}}  },
\end{eqnarray*}
where the second equality follows from \eqref{EqA.240}.

Using Taylor expansion, we have

\begin{eqnarray*} 
0 &=&\frac{\partial \log L( s(\cdot\, | \,\mathbf{x}_{\mathbf{i},0}, \widetilde{\pmb{\lambda}}_{\mathbf{i}}) ) }{\partial \pmb{\theta}_{\mathbf{i}}  }  +\frac{\partial^2 \log L(s(\cdot\, | \,  \mathbf{x}_{\mathbf{i},0}, \pmb{\theta}_{\mathbf{i}}^*)) }{\partial \pmb{\theta}_{\mathbf{i}} \partial \pmb{\theta}_{\mathbf{i}}^\top } ( \widehat{\pmb{\theta}}_{\mathbf{i}} - \widetilde{\pmb{\lambda}}_{\mathbf{i}})
\end{eqnarray*}
where $\pmb{\theta}_{\mathbf{i}}^*$ lies between $\widehat{\pmb{\theta}}_{\mathbf{i}}$ and $ \widetilde{\pmb{\lambda}}_{\mathbf{i}}$, and the second equality follows from Lemma \ref{LEM2.3} and the continuity of $\phi_\eta$ and $\Phi_\eta$.

Thus, by Lemma \ref{LemA.4}, the result follows immediately.  \hspace*{\fill}{$\blacksquare$}

\bigskip

\noindent \textbf{Proof of Theorem \ref{Thm3.4}:}

The proof follows from  \eqref{EqA.240} and \eqref{EqA.24} and a procedure very similar to that given in Theorem \ref{Thm3.2}.  \hspace*{\fill}{$\blacksquare$}
\newpage

\begin{figure}[htp!] 
\hspace*{-1cm}\includegraphics[scale=0.19]{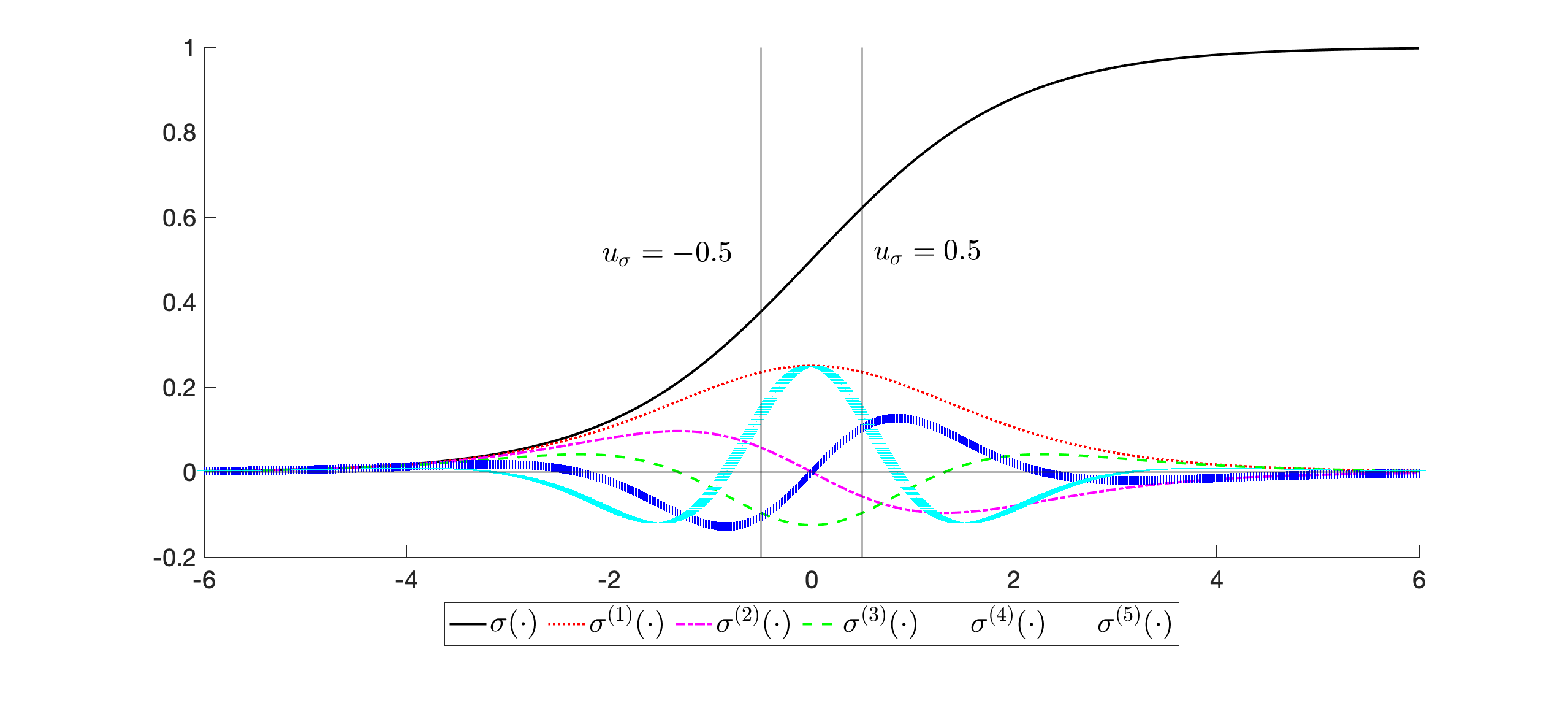}
\vspace*{-1.0cm}\caption{Plots of Sigmoidal Squasher with Its Derivatives}\label{FigSig}
\end{figure}

\begin{figure}[htp!] 
\hspace*{-2cm}\vspace*{-1cm}\includegraphics[scale=0.3]{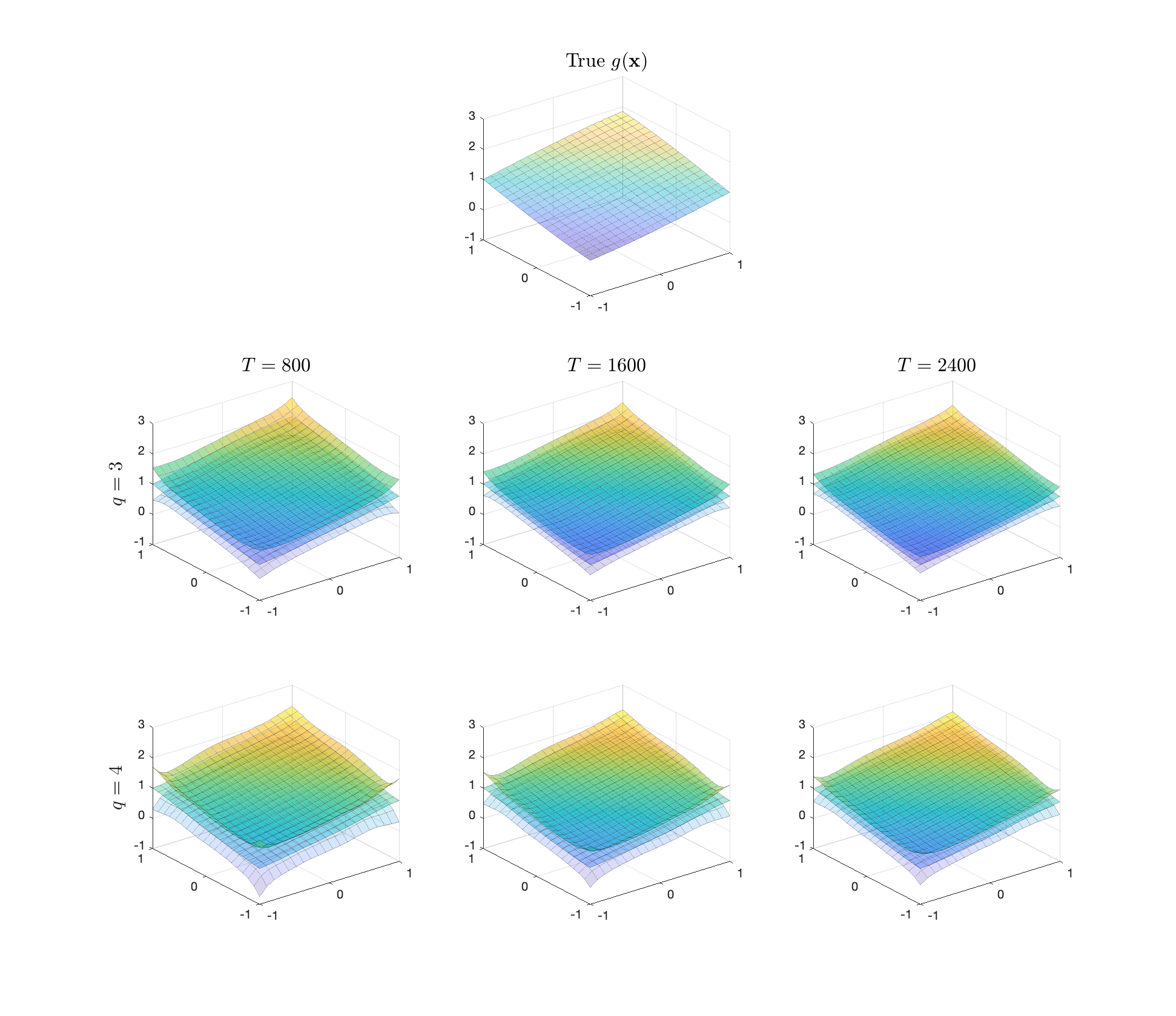}
\vspace*{-1cm}\caption{Simulation Results of Fully Nonparametric Model ($u_{\sigma}=-0.5, d=2$)}\label{FigSim1a}
\end{figure}

\begin{figure}[htp!] 
\hspace*{-2cm}\vspace*{-1cm}\includegraphics[scale=0.3]{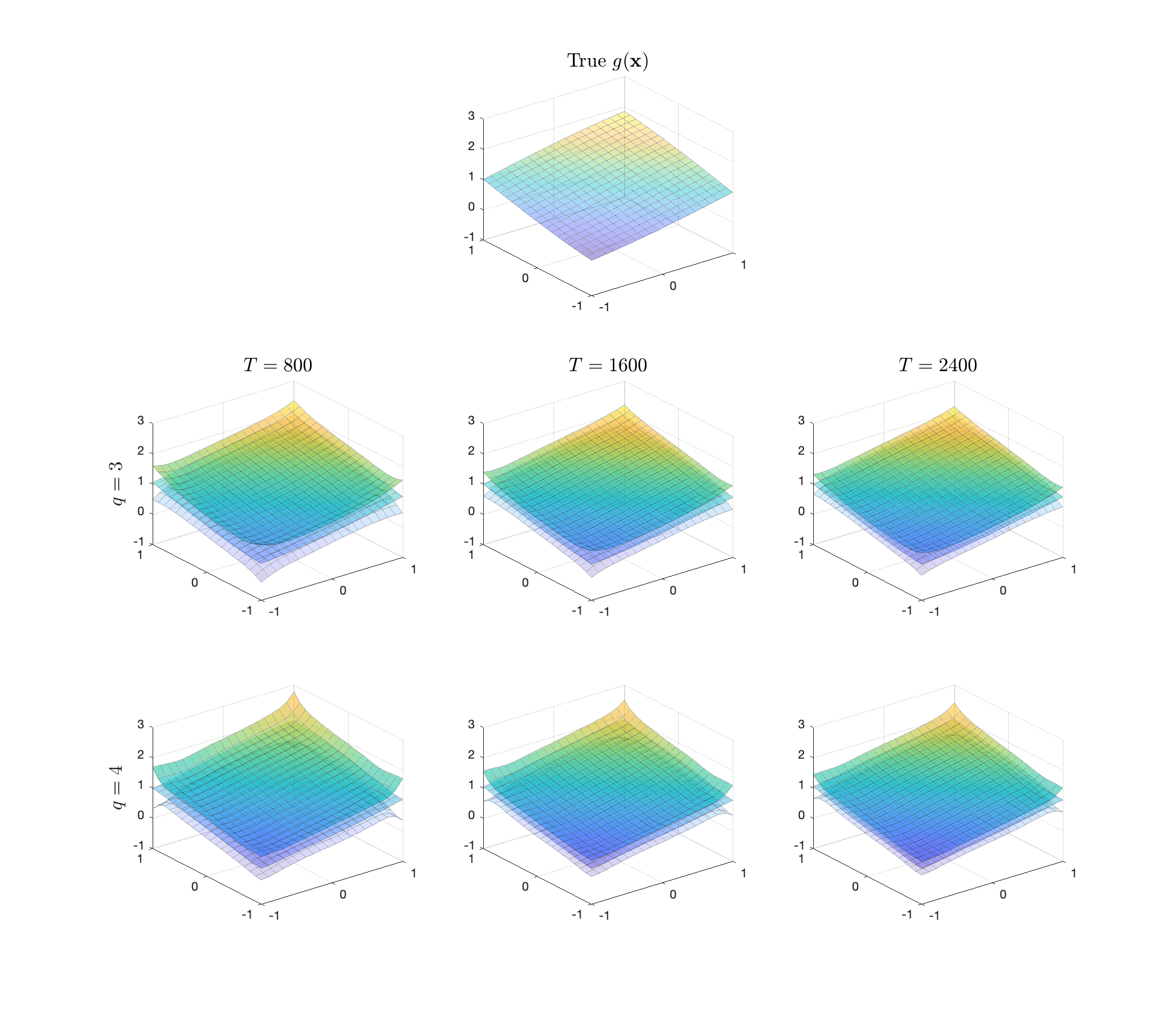}
\vspace*{-1cm}\caption{Simulation Results of Fully Nonparametric Model ($u_{\sigma}=0.5, d=2$)}\label{FigSim2a}
\end{figure}

\begin{table}[htp!]\renewcommand{\arraystretch}{0.8}
\centering
\caption{RMSE and CR (for the Fully Nonparametric Model)}\label{Tab1a}
\begin{tabular}{llrrrrrr}
\hline\hline
 &  &  & \multicolumn{2}{c}{RMSE} & \multicolumn{1}{l}{} & \multicolumn{2}{c}{CR} \\
 &  & $T\setminus d$ & 2 & 8 &  & 2 & 8 \\
$u_\sigma =-0.5$ & $q=3$ & 800 & 0.1327 & 0.8468 &  & 0.9073  & 0.9012 \\
 &  & 1600 & 0.0925  & 0.5929 &  & 0.9097  & 0.9020 \\
 &  & 2400 & 0.0778  & 0.4201 &  & 0.9260  & 0.9351 \\
 & $q=4$ & 800 & 0.1649  & 3.6023 &  & 0.9185  & 0.8141 \\
 &  & 1600 & 0.1184  & 1.7751 &  & 0.9234  & 0.8927 \\
 &  & 2400 & 0.0935  & 1.3810 &  & 0.9248  & 0.8929 \\
 &  &  &  &  &  &   &  \\
$u_\sigma =0.5$ & $q=3$ & 800 & 0.1353 & 0.8154 &  & 0.9023 & 0.8956 \\
 &  & 1600 & 0.0947 & 0.5985 &  & 0.9098 & 0.9095 \\
 &  & 2400 & 0.0795 & 0.4570 &  & 0.9169 & 0.9251 \\
 & $q=4$ & 800 & 0.1614 & 3.7447 &  & 0.9159 & 0.8195 \\
 &  & 1600 & 0.1158 & 1.6810 &  & 0.9224 & 0.9039 \\
 &  & 2400 & 0.0915 & 1.1334 &  & 0.9289 & 0.9129 \\
 \hline\hline
\end{tabular}
\end{table}

\begin{figure}[htp!] 
\hspace*{-2cm}\vspace*{-1cm}\includegraphics[scale=0.27]{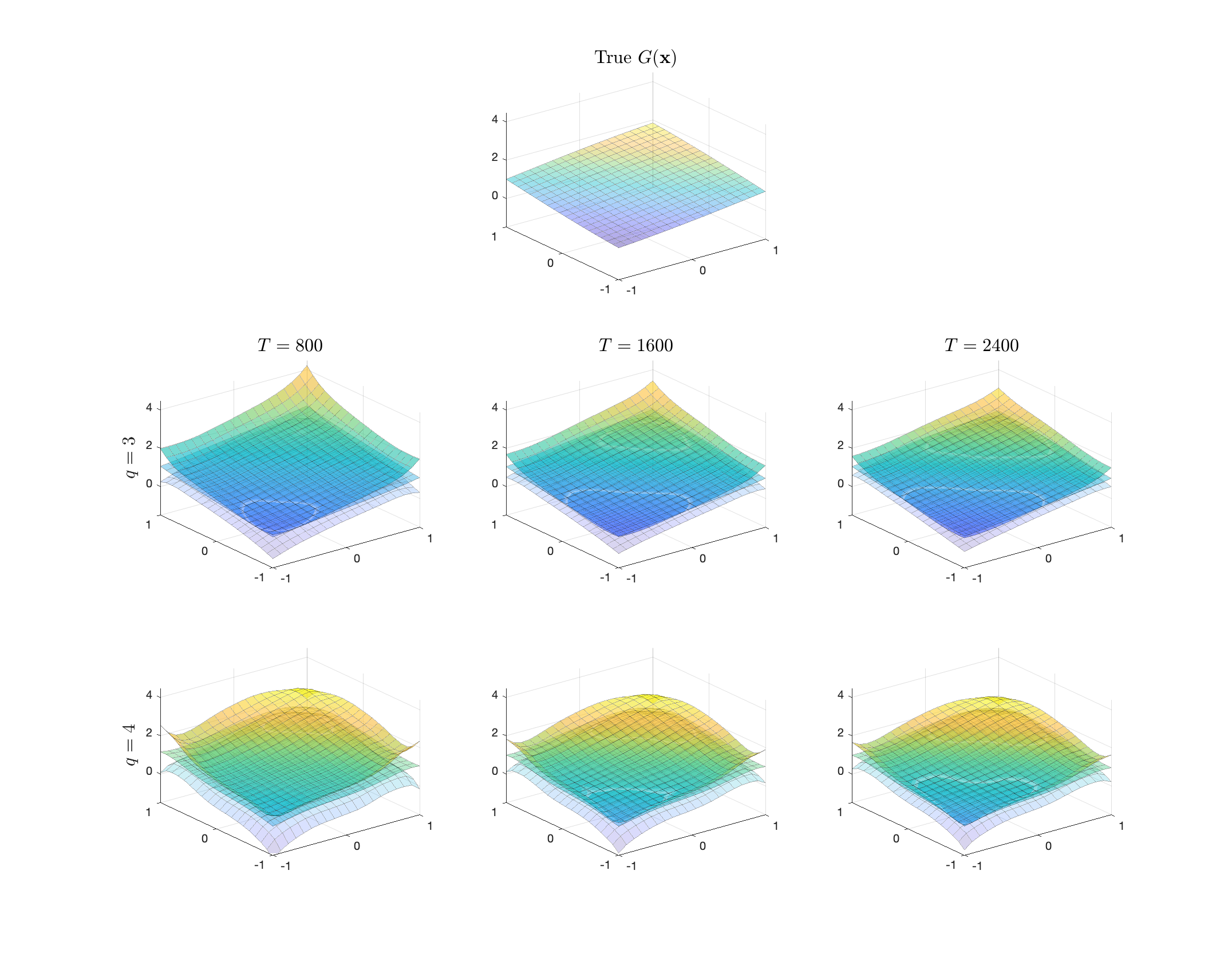}
\vspace*{-1cm}\caption{Simulation Results (for Binary Structure) with $u_{\sigma}=-0.5, d=2$}\label{FigSim3}
\end{figure}

\begin{figure}[htp!] 
\hspace*{-2cm}\vspace*{-1cm}\includegraphics[scale=0.27]{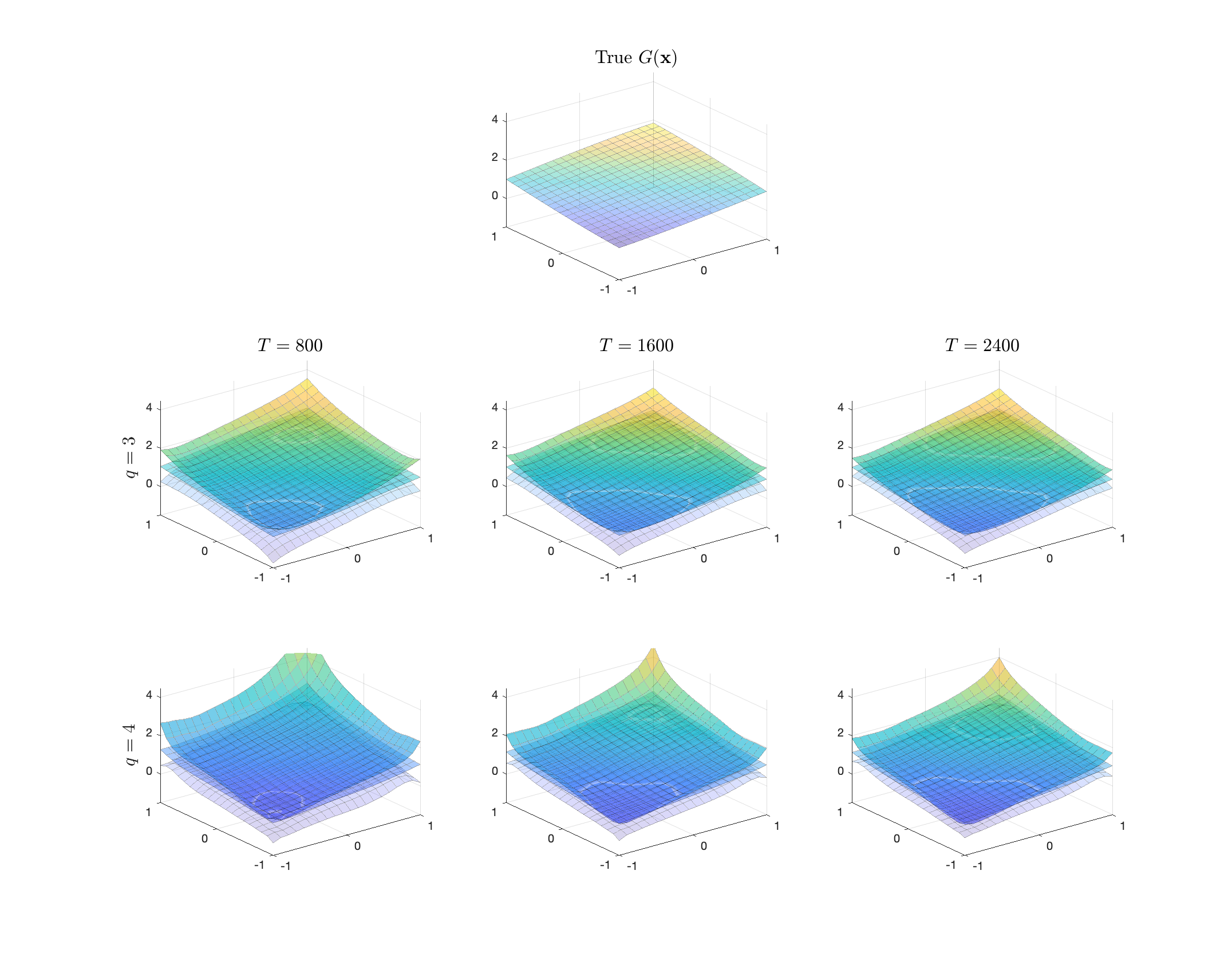}
\vspace*{-1cm}\caption{Simulation Results (for Binary Structure) with with $u_{\sigma}=0.5, d=2$}\label{FigSim4}
\end{figure}

\begin{table}[htp!]\renewcommand{\arraystretch}{0.8}
\begin{center}
\caption{RMSE and CR (for the Binary Structure)}\label{Tab2}
\begin{tabular}{llrrrrrr}
\hline\hline
 &  & \multicolumn{1}{l}{} & \multicolumn{2}{c}{RMSE} & \multicolumn{1}{l}{} & \multicolumn{2}{c}{CR} \\
 &  & $T\setminus d$ & 2 & 8 &  & 2 & 8 \\
$u_\sigma =-0.5$ & $q=3$ & 800 & 0.2380 & 2.3085 &  & 0.9308 & 0.8512 \\
 &  & 1600 & 0.1528  & 1.1258 &  & 0.9433 & 0.8852 \\
 &  & 2400 & 0.1231  & 0.8208 &  & 0.9398 & 0.9140 \\
 & $q=4$ & 800 & 0.2800 & 3.2359 &  & 0.9191 & 0.9957 \\
 &  & 1600 & 0.1901 & 2.6372 &  & 0.9274 & 0.9799 \\
 &  & 2400 & 0.1575 & 1.8722 &  & 0.9282 & 0.9773 \\
 &  &  &  &  &  &  &  \\
$u_\sigma =0.5$ & $q=3$ & 800 & 0.2200 & 2.5309 &  & 0.9346 & 0.8177 \\
 &  & 1600 & 0.1452 & 1.3872 &  & 0.9353 & 0.8411 \\
 &  & 2400 & 0.1190 & 0.8427 &  & 0.9397 & 0.9135 \\
 & $q=4$ & 800 & 0.3553 & 3.4827 &  & 0.9308 & 0.9952 \\
 &  & 1600 & 0.2194 & 2.4127 &  & 0.9402 & 0.9890 \\
 &  & 2400 & 0.1631 & 2.2000 &  & 0.9417 & 0.9704 \\
 \hline\hline
\end{tabular}
\end{center}
\end{table}

}

}
\end{document}